\documentclass[12pt]{report}   
\usepackage{graphicx}  
\usepackage[letterpaper, left=1in, right=1in, top=1in, bottom=1in]{geometry}
\usepackage{setspace}  
\usepackage{times}  
\usepackage[explicit]{titlesec}  
\usepackage[titles]{tocloft}  
\usepackage[backend=bibtex, sorting=none, bibstyle=ieee]{biblatex}  
\usepackage{appendix}  
\usepackage{rotating}  
\usepackage[normalem]{ulem}  
\usepackage{textcomp} 
\usepackage{indentfirst} 
\usepackage{booktabs,array,arydshln} 
\usepackage{amsmath} 
\usepackage[T1]{fontenc} 
\usepackage[utf8]{inputenc} 
\usepackage[hidelinks]{hyperref}

\usepackage{algorithm}
\usepackage{algorithmic}
\usepackage{multirow}
\usepackage{multicol}
\usepackage{bbm}
\usepackage{array}
\usepackage{comment}
\usepackage{hyperref}
\usepackage{bibentry}

\usepackage{caption}
\DeclareCaptionLabelFormat{adja-page}{\hrulefill\\#1 #2 \emph{(previous page)}}

\usepackage{amsmath,amsfonts,bm}
\usepackage{latexsym} 
\usepackage{booktabs}
\usepackage{amsmath}
\usepackage{verbatim}
\usepackage{amssymb}
\usepackage{amsthm}
\usepackage{algorithm}
\usepackage{algorithmic}
\usepackage{graphicx}
\usepackage{bbm}

\usepackage{booktabs}       
\usepackage{amsfonts}       
\usepackage{nicefrac}       
\usepackage{microtype}      

\def\eqref#1{equation~\ref{#1}}

\def\1{\bm{1}}

\DeclareMathAlphabet{\mathsfit}{\encodingdefault}{\sfdefault}{m}{sl}
\SetMathAlphabet{\mathsfit}{bold}{\encodingdefault}{\sfdefault}{bx}{n}

\makeatletter
\newcommand{\thickhline}{%
    \noalign {\ifnum 0=`}\fi \hrule height 1pt
    \futurelet \reserved@a \@xhline
}

\DeclareMathOperator{\EX}{\mathbb{E}}
\newcommand\norm[1]{\left\lVert#1\right\rVert}

\bibliography{references}

\AtEveryBibitem{\clearfield{issn}}
\AtEveryBibitem{\clearlist{issn}}

\AtEveryBibitem{\clearfield{language}}
\AtEveryBibitem{\clearlist{language}}

\AtEveryBibitem{\clearfield{doi}}
\AtEveryBibitem{\clearlist{doi}}

\AtEveryBibitem{\clearfield{url}}
\AtEveryBibitem{\clearlist{url}}

\AtEveryBibitem{%
  \ifentrytype{online}
    {}
    {\clearfield{urlyear}\clearfield{urlmonth}\clearfield{urlday}}}

\begin{document}
\doublespacing  


\begin{titlepage}
\begin{center}

\begin{singlespacing}
\vspace*{6\baselineskip}
\textbf{\large Topological Representational Similarity Analysis in Brains and Beyond} \\
\vspace{3\baselineskip}
{Baihan Lin}\\
\vspace{18\baselineskip}
Submitted in partial fulfillment of the\\
requirements for the degree of\\
Doctor of Philosophy\\
under the Executive Committee\\
of the Graduate School of Arts and Sciences\\
\vspace{3\baselineskip}
COLUMBIA UNIVERSITY\\
\vspace{3\baselineskip}
2023
\vfill

\end{singlespacing}

\end{center}
\end{titlepage}


\begin{titlepage}
\begin{singlespacing}
\begin{center}

\vspace*{35\baselineskip}

\textcopyright  \,  
2023 \\
\vspace{\baselineskip}	
Baihan Lin\\
\vspace{\baselineskip}	
All Rights Reserved
\end{center}
\vfill

\end{singlespacing}
\end{titlepage}

\pagenumbering{gobble}

\begin{titlepage}
\begin{center}

\vspace*{5\baselineskip}
\textbf{\large Abstract}

Topological Representational Similarity Analysis in Brains and Beyond

Baihan Lin
\end{center}

Understanding how the brain represents and processes information is crucial for advancing neuroscience and artificial intelligence. Representational similarity analysis (RSA) has been instrumental in characterizing neural representations by comparing multivariate response patterns elicited by sensory stimuli. However, traditional RSA relies solely on geometric properties, overlooking crucial topological information. This thesis introduces topological RSA (tRSA), a novel framework that combines geometric and topological properties of neural representations.

tRSA applies nonlinear monotonic transforms to representational dissimilarities, emphasizing local topology while retaining intermediate-scale geometry. The resulting geo-topological matrices enable model comparisons that are robust to noise and individual idiosyncrasies. This thesis introduces several key methodological advances:

\begin{enumerate}
    \item \textit{Topological RSA (tRSA)}: Identifies computational signatures as accurately as RSA while compressing unnecessary variation with capabilities to test topological hypotheses;
\item \textit{Adaptive Geo-Topological Dependence Measure (AGTDM)}: Provides a robust statistical test for detecting complex multivariate relationships;
\item \textit{Procrustes-aligned Multidimensional Scaling (pMDS)}: Aligns time-resolved representational geometries to illuminate processing stages in neural computation;
\item \textit{Temporal Topological Data Analysis (tTDA)}: Applies spatio-temporal filtration techniques to reveal developmental trajectories in biological systems;
\item \textit{Single-cell Topological Simplicial Analysis (scTSA)}: Characterizes higher-order cell population complexity across different stages of development.
\end{enumerate}

Through analyses of neural recordings, biological data, and simulations of neural network models, this thesis demonstrates the power and versatility of these new methods. By advancing RSA with topological techniques, this work provides a powerful new lens for understanding brains, computational models, and complex biological systems. These methods not only offer robust approaches for adjudicating among competing models but also reveal novel theoretical insights into the nature of neural computation.

This thesis lays the foundation for future investigations at the intersection of topology, neuroscience, and time series data analysis, promising to deepen our understanding of how information is represented and processed in biological and artificial neural networks. The methods developed here have potential applications in fields ranging from cognitive neuroscience to clinical diagnosis and AI development, paving the way for more nuanced understanding of brain function and dysfunction.

\vspace*{\fill}
\end{titlepage}

\pagenumbering{roman}
\setcounter{page}{1} 
\renewcommand{\cftchapdotsep}{\cftdotsep}  
\renewcommand{\cftchapfont}{\normalfont}  
\renewcommand{\cftchappagefont}{}  
\renewcommand{\cftchappresnum}{Chapter }
\renewcommand{\cftchapaftersnum}{:}
\renewcommand{\cftchapnumwidth}{5em}
\renewcommand{\cftchapafterpnum}{\vskip\baselineskip} 
\renewcommand{\cftsecafterpnum}{\vskip\baselineskip}  
\renewcommand{\cftsubsecafterpnum}{\vskip\baselineskip} 
\renewcommand{\cftsubsubsecafterpnum}{\vskip\baselineskip} 

\titleformat{\chapter}[display]
{\normalfont\bfseries\filcenter}{\chaptertitlename\ \thechapter}{0pt}{\large{#1}}

\renewcommand\contentsname{Table of Contents}

\begin{singlespace}
\tableofcontents
\setlength{\cftparskip}{\baselineskip}
\listoffigures
\listoftables
\end{singlespace}

\clearpage

\phantomsection
\addcontentsline{toc}{chapter}{Acknowledgments}

\clearpage
\begin{center}

\vspace*{5\baselineskip}
\textbf{\large Acknowledgements}
\end{center}


As I reach the culmination of this transformative journey, my heart overflows with gratitude for those who have illuminated my path and shaped my growth.

First and foremost, I owe an immeasurable debt of gratitude to my family. To my mom, dad, and grandparents: your love and support have spanned oceans, forging an unbreakable bond that has been my anchor throughout this journey. Your unwavering belief in me has been a constant source of strength and inspiration.

To my darling Esther, you've been my muse and steadfast partner in this adventure from Seattle to NYC. Your wit, insights, and unwavering support have been my daily inspiration. This achievement is as much yours as it is mine.

My furry companions, Neptune, Jupiter, and Cosmo, have been more than pets; they've been my loyal allies through the ebb and flow of research life. Neptune, you'll always hold a special place in my heart. And to Lava, my trusty car, our shared adventures and your battle scars are a testament to the roads we've traveled together.

In the realm of academia, I am profoundly grateful to Niko Kriegeskorte, my PhD advisor at Columbia University. Niko, you've been the sagacious captain of my research ship, indulging my wildest scientific fantasies with a blend of enthusiasm and discernment that taught me the true essence of scientific inquiry. Your mentorship has been an intellectual compass, guiding me through the nebulous waters of computational neuroscience and charting my course in academia.

My heartfelt thanks go to my thesis committee members - Itsik Pe'er, Tian Zheng, Ning Qian, Mark Churchland, and Andrea Califano. Your collective wisdom and inspiring guidance have been pivotal in shaping both my research and career trajectory. Your generous time, warm support, and invaluable insights are deeply appreciated.

I extend my gratitude to Raul Rabadan, Peter Sims, Nick Arpaia, Ron Liem, Beth Kauderer, Jeffrey Sears, Sherry Bermeo, and Donna Farber for their instrumental role in my professional development at Columbia. Your diverse perspectives have enriched my journey immensely. A special thank you to Zaia Sivo, our PhD program coordinator, whose unwavering support has been a beacon throughout this journey.

Columbia University, particularly the neuroscience and systems biology departments, and the vibrant machine learning communities, have provided fertile ground for my intellectual growth. To New York City: your pulsating energy has been a constant source of inspiration and resilience.

I'm deeply indebted to my long-term collaborators at IBM Research and Mila - Quebec AI Institute, especially Guillermo Cecchi, Irina Rish, Djallel Bouneffouf, Jenna Reinen, Kush Varshney, Stefan Zecevic, and Lydia Yu. Our collaborations have left indelible marks on my research trajectory.

To my industry mentors and collaborators: Mar Gonzalez-Franco at Microsoft, and Sarah Laszlo, Garrett Honke, Lam Nguyen, Anu Thubagere, and Sam Khamis at Google X, your guidance and support have been invaluable in bridging academia and industry.

I'm grateful for the insights gained from my clinical collaborators and colleagues: Cheryl Corcoran, Yulia Landa, Rachel Jespersen, Daniel Kimmel, Chris Sidey-Gibbons, John Weinstein, and David Jaffray, who brought the intersection of medicine, neuroscience, and psychiatry to life.

To Richard The and Daniel Sauter of Parsons School of Design, your advice on faculty applications was crucial. I'm also thankful for the foundational guidance from my early mentors at the University of Washington, Seattle and BGI Genomics, Jaime Olavarria, David Baker, Hong Qian, Shwetak Patel, Rachel Chapman, Chris Vogl, Henry Huanming Yang, Zhaozheng Guo, and Yingrui Li, who first ignited my passion for neuroscience, genomics, and AI.

I've been fortunate to mentor bright minds like Sultana Yeasmin and Aylin Gunal, who've taught me as much as I've guided them. Your potential reassures me that the future of science is in capable hands.

My gratitude extends to Pierre-Yves Oudeyer, Felipe Leno da Silva, Karen Feigh, Malte Schilling, and Cory Brunson for their support in my career growth and immigration journey. To Neil Parikh, Paul Buttenwieser, Maritsa Patrinos, Lucas Dileo, and Robert Goulburn: thank you for broadening my horizons across various fields and uncharted territories of life.

To my friends scattered across the globe, in New York, Seattle, Texas, New England, Europe, China, and beyond: your companionship has been a source of joy and comfort throughout this journey.

This PhD odyssey has been a collective effort, and this thesis is as much a product of your support, guidance, and inspiration as it is of my work. From the depths of my heart, thank you all for being part of this transformative chapter of my life.

\clearpage


\phantomsection
\addcontentsline{toc}{chapter}{Dedication}

\begin{center}

\vspace*{5\baselineskip}
\textbf{\large Dedication}
\end{center}


To my mom Xinpei and my family, whose love and wisdom reach me across oceans,


And in loving memory of Neptune, whose silent presence was my solace through countless hours of contemplation and creation.




\clearpage
\pagenumbering{arabic}
\setcounter{page}{1} 

\phantomsection
\addcontentsline{toc}{chapter}{Preface}

\begin{center}
\vspace*{5\baselineskip}
\textbf{\large Preface}
\end{center}


Representational similarity analysis (RSA) offers a powerful framework for understanding neural computations by comparing representational geometries between brains, models, and tasks. However, traditional RSA often relies solely on pairwise distances between response patterns, ignoring topological relationships critical for a complete characterization. This thesis introduces topological RSA (tRSA), integrating geometric and topological properties to more deeply probe neural representations.

Through five chapters, the readers will explore techniques that adapt RSA to leverage topological data analysis methods. The chapters showcase applications to neural recordings, simulations of neural network models, time-series data, and beyond, demonstrating enhanced model comparison, independence testing, and visualizations of representational dynamics.

After an introductory review of RSA, core chapters detail the development of tRSA and its evaluations on neural data. Analyses suggest tRSA accurately recovers models while ignoring unnecessary geometric variations, on par with the performance of geometric RSA, but provides additional interpretable insights. This approach not only enhances our ability to compare brain regions and neural network layers but also offers a unique tool for testing topological hypotheses about neural representations.

Additional chapters showcase temporal techniques to reveal processing sequences and developmental trajectories. By aligning time-resolved representational geometries, we unveil the dynamic nature of neural computations, offering new insights into how the brain processes information over time. Our temporal topological data analysis techniques further extend this approach to developmental biology, revealing hidden patterns in cellular differentiation and characterizing higher-order cell population complexity.

An important extension of our work is the development of an Adaptive Geo-Topological Dependence Measure (AGTDM). This novel statistical tool, inspired by the principles of tRSA, provides a robust method for detecting complex dependencies in multivariate data. By adapting to the structure of relationships, AGTDM offers enhanced sensitivity across a wide range of scenarios, demonstrating the broad applicability of our topological approach beyond neuroscience.

This thesis provides both practical tools and theoretical context to analyze complex networks. The overarching theme is incorporating topological invariants to better understand system dynamics. Diverse applications highlight the techniques' versatility for probing geometry, topology, and time across various domains of science.

For readers seeking advanced methods to compare multidimensional datasets, this thesis offers an expansive toolkit. Through numerous examples, it demonstrates enhanced characterization of systems from neuroscience to biology and beyond. Both novices and experts will find topological extensions that advance representational analysis.

Broadly, this work redresses the limitations of geometry-solely techniques by integrating topological considerations. The presented techniques empower deeper interrogation of neural and complex network representations. With these methods in hand, readers can explore how topological principles underlie computations in both artificial and biological networks.

As we journey through the chapters, we will see how tRSA and its extensions not only advance our understanding of neural computations but also open new avenues for data analysis across scientific disciplines. From uncovering the hidden structure of neural representations to revealing developmental trajectories in biological systems, this thesis demonstrates the power of topology in unlocking new insights from complex data.

By the end of this journey, readers will have gained a new perspective on how to analyze and interpret high-dimensional data, equipped with tools that bridge geometry and topology. This work lays the foundation for future investigations at the intersection of topology, neuroscience, and data analysis, promising to deepen our understanding of how information is represented and processed in biological and artificial systems alike.

\titleformat{\chapter}[display]
{\normalfont\bfseries\filcenter}{}{0pt}{\large\chaptertitlename\ \large\thechapter : \large\bfseries\filcenter{#1}}  
\titlespacing*{\chapter}
  {0pt}{0pt}{30pt}	
  
\titleformat{\section}{\normalfont\bfseries}{\thesection}{1em}{#1}

\titleformat{\subsection}{\normalfont}{\thesubsection}{0em}{\hspace{1em}#1}



\chapter{Introduction to Representational Similarity Analysis (RSA)}

Neural representations are the fundamental building blocks of cognition, yet understanding how the brain encodes and processes information remains a central challenge in neuroscience. This thesis introduces novel methodologies to probe and analyze these representations, building upon the foundational framework of Representational Similarity Analysis (RSA).

As we embark on this journey, we'll first explore the principles and applications of classical RSA. This powerful technique has revolutionized our ability to compare neural activity patterns across different measurement modalities, species, and computational models. However, as we'll see, RSA also has limitations that motivate the new approaches developed in this thesis.

In this chapter, we'll delve into the core concepts of RSA, its mathematical foundations, and its practical applications. This background will set the stage for the innovative methods introduced in subsequent chapters, which extend RSA to incorporate topological information, temporal dynamics, and adaptive optimization principles. By the end of this chapter, readers will understand both the power of classical RSA and the open questions that drive the novel contributions of this thesis.



\section{From Representational Pattern to Representational Geometry}

A representational pattern refers to the pattern of activity elicited in a brain region in response to a stimulus or experimental condition. This activity pattern captures how the stimulus is represented in certain region of interest (ROI). For example, in fMRI studies, the activity pattern could consist of the blood-oxygen-level dependent (BOLD) response measured across multiple voxels in a ROI for a given stimulus. In electrophysiology studies, it could consist of the spiking activity or local field potentials recorded from multiple electrodes or channels in an ROI for a given stimulus. In magnetoencephalography (MEG) or electroencephalogram (EEG) studies, it could consist of the voltage fluctuations measured across multiple sensors on the scalp for a given stimulus.

The activity pattern is interpreted as reflecting the information represented about the stimulus or condition in that particular brain region \cite{kriegeskorte2011pattern}. The central idea is that stimuli are not simply encoded in the overall activation level of a region, but in the relative activation levels across the neuronal population. This is known as a distributed code or population code \cite{haxby2001distributed}.

RSA falls under the broader category of multivariate pattern analysis (MVPA) methods \cite{haxby2012multivariate}, which aim to analyze these distributed activity patterns to understand neural representations and information. Other common MVPA techniques include pattern classification (decoding stimulus category from activity patterns) \cite{haxby2014decoding}, pattern component modeling (decomposing patterns into component features) \cite{diedrichsen2011comparing,diedrichsen2018pattern}, encoding models (predicting patterns from stimulus features) \cite{van2017primer}, and pattern information analysis (estimating information content of patterns) \cite{kriegeskorte2011pattern}. RSA is a special kind of MVPA method to investigate the \textit{content} and \textit{format} of the representations.

A key advantage of MVPA over classical univariate activation analyses is that MVPA targets fine-grained distributed information contained in activity patterns, rather than just overall regional activation. However, RSA provides a particularly powerful approach for characterizing what is termed the \textit{representational geometry} \cite{kriegeskorte2013representational} and testing computational models in terms of the congruence of their representational geometries with those from the neural data.

A brain region's representation can be envisioned as a high-dimensional space, where each dimension aligns with one neuron's activity. The collective activation pattern across all neurons determines a location in this representational space. Any perceived object or mental content maps to a distinct point in the space corresponding to that brain region. The entirety of possible perceptual and cognitive experiences traces out a large set of points filling the representation. It is the spatial configuration and relationships between these points - the \textit{geometry} - that reveals the representational structure. How points cluster, the dimensions along which they spread, and the distances between them characterize how the brain carves up possible experiences and distinguishes mental contents. Examining the representational geometry provides insight into the function and computations unfolding in a brain region. The principles and constraints shaping the geometry offer clues to how the brain solves its core computational challenges.

This definition of representational geometry gives an additional benefit, which is a comparable summary statistic agnostic of the data modalities. As introduced earlier, the activity patterns collected from different data modalities can vary in terms of their dimensionalities. fMRI studies code activity profiles in terms of 3D voxels, while single-cell recording or EEG studies use recording channels. Relating these activity patterns and comparing them to computational models with different numbers of parameters faces a spatial correspondency challenge. Finding one-to-one mappings between model units and recorded neurons or voxels is often infeasible, as brain sensors reflect many neurons, not individuated units. Even when possible, fitting a full linear transform requires impractically large parameter spaces. Absent a clear mapping, directly comparing representations can be ill-motivated. However, the geometry of these representational patterns, characterized as a second-order isomorphism between stimuli and their representations, can skirt this correspondence issue as long as the stimulus space is shared. The representational geometry provides a modality-agnostic summary statistic that enables comparing representations across different measurements and models.

RSA aims to characterize and compare the geometries of the representational patterns elicited by different stimuli/conditions \cite{kriegeskorte2008representational,kriegeskorte2013representational}. The similarity of the activity patterns reflects how similarly the stimuli are represented in the region. Regions that emphasize different distinctions between stimuli will have different representational similarity structures.

As an early summary, several key considerations of representational patterns include:
\begin{itemize}
\item A representational pattern refers to the multivariate activity across a neuronal population, not just overall activation level.
\item Patterns capture how stimuli are encoded in a distributed fashion across  a set of channels (voxels, electrodes, etc).
\item RSA analyzes the geometry and similarity structure of patterns to understand representations.
\item The similarity between patterns for two stimuli indicates how similarly they are represented.
\item The similarities among patterns reveal the representational geometry of a region.
\item Representational patterns can be compared between regions of interest, subjects, and models.
\end{itemize}


There are some caveats to keep in mind when interpreting representational patterns:
The measured activity may not directly reflect neural firing patterns due to nonlinear measurement physics;
Noise and limited resolution may distort fine-grained neural codes;
Downstream regions may not have access to the full measured pattern;
Overall regional activation is lost when focusing on pattern similarities, but it is sometimes a trade off worthwhile \cite{kriegeskorte2019peeling}. 
As we will see in the coming sections, RSA offers several key advantages over traditional univariate analyses:
\begin{enumerate}
    \item It captures fine-grained information distributed across many channels or voxels.
    \item It allows direct comparison between different measurement modalities (e.g., fMRI, electrophysiology, behavioral data).
    \item It facilitates testing of computational models against brain data.
    \item It can reveal representational distinctions that may be invisible to activation-based analyses.
\end{enumerate}

We will discuss some of these considerations in more details in the coming sections. However, RSA provides an exploratory tool to discover the representational geometry. This can guide developing focused computational models to explain the measurement patterns and predict effects of perturbations. Multivariate patterns provide a rich source of information to constrain and test computational theories.


\section{Understanding and Preprocessing Neural Representation Data}

RSA provides a framework for analyzing the representational patterns measured with various neuroimaging modalities. The goal is to understand what information these activity patterns carry about the presented stimuli. However, directly analyzing and comparing the raw multivariate patterns can be challenging. RSA provides a powerful framework for summarizing and analyzing representational patterns in a principled way.

The key idea is to characterize each representational pattern not individually, but in terms of the dissimilarities between patterns. This abstracts the pattern information into a format that enables comparisons between representations in different modalities, regions, and models. Some examples of how RSA has been applied to analyze representations across modalities:
    comparing fMRI and cell recording response patterns in monkey IT cortex \cite{kriegeskorte2008matching};
    relating MEG sensor patterns to EEG and fMRI patterns \cite{cichy2014resolving};
    comparing computational models to brain representations \cite{khaligh2014deep,kriegeskorte2008representational}.

\begin{figure}[tb]
\centering
\includegraphics[width=.8\linewidth]{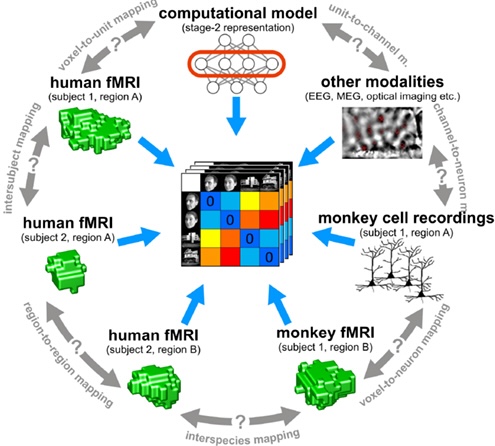}
\caption{\textbf{Application scenarios enabled by representational comparison.} RSA relates representations across species, modalities, and models. (Extracted from \cite{kriegeskorte2008representational}).}
\label{fig:rsa_modes}
\end{figure}

More generally, we see that, by abstracting the information into their dissimilarity matrices, RSA enables comparing representational geometries between different datasets and models (Fig. \ref{fig:rsa_modes}). This allows testing computational theories and relating brain measurements across individuals, species, and measurement modalities, depending on the research questions in mind:

\begin{itemize}
\item \textit{Integrating models into brain data analysis}: RSA allows directly testing computational models against brain activity patterns. Statistical tests can assess model fits to brain RDMs and compare multiple models.

\item \textit{Relating regions}: RSA can quantitatively compare representations between different regions within a brain using representational connectivity analyses.

\item \textit{Relating subjects}: RSA provides subject-invariant summaries that enable comparing representations across individuals.

\item \textit{Relating species}: RSA can relate representations in analogous regions across species, like monkey and human visual cortex.

\item \textit{Relating modalities}: Abstracting representations as RDMs enables comparing fMRI, EEG, MEG, and invasive recordings.

\item \textit{Linking brain and behavior}: RSA can relate brain-activity measurements to behavioral dissimilarities like perceptual confusions.

\item \textit{Rich experimental designs}: Condition-rich designs with many stimuli can efficiently test many questions within one experiment.

\end{itemize}

\subsection{Preprocessing considerations}

As illustrated above, the major advantage of RSA is providing a common representational space to relate brain activity, models, behavior, and computations. This allows testing theories and translating insights across measurement modalities, individuals, and species. RSA offers a powerful and flexible framework for systems neuroscience. In practice, 
before applying RSA, some of the common preprocessing steps applied to the raw neural activity measurements include:
{trial averaging} (averaging together repeats of the same condition to estimate the pattern for each condition),
{temporal windowing} (extracting patterns from specific time windows of interest, say, the stimulus onset),
{spatial normalization} (for fMRI, normalizing anatomy across subjects to facilitate comparisons),
{multivariate noise normalization} (using spatial whitening to account for noise correlations between measurement channels),
and
{dimensionality reduction} (optionally, reducing high-dimensional patterns using PCA or related techniques).

These steps help isolate the stable condition-related patterns and account for noise structure. The resulting cleaned patterns can then be compared to construct the representational dissimilarity matrix (RDM), which is the topic of our next section.

\section{Computing Representational Dissimilarity Matrices (RDM)}

In RSA, the representational pattern elicited by each stimulus is compared to the pattern of every other stimulus by computing some measure of dissimilarity between the two patterns.

\begin{figure}[tb]
\centering
\includegraphics[width=\linewidth]{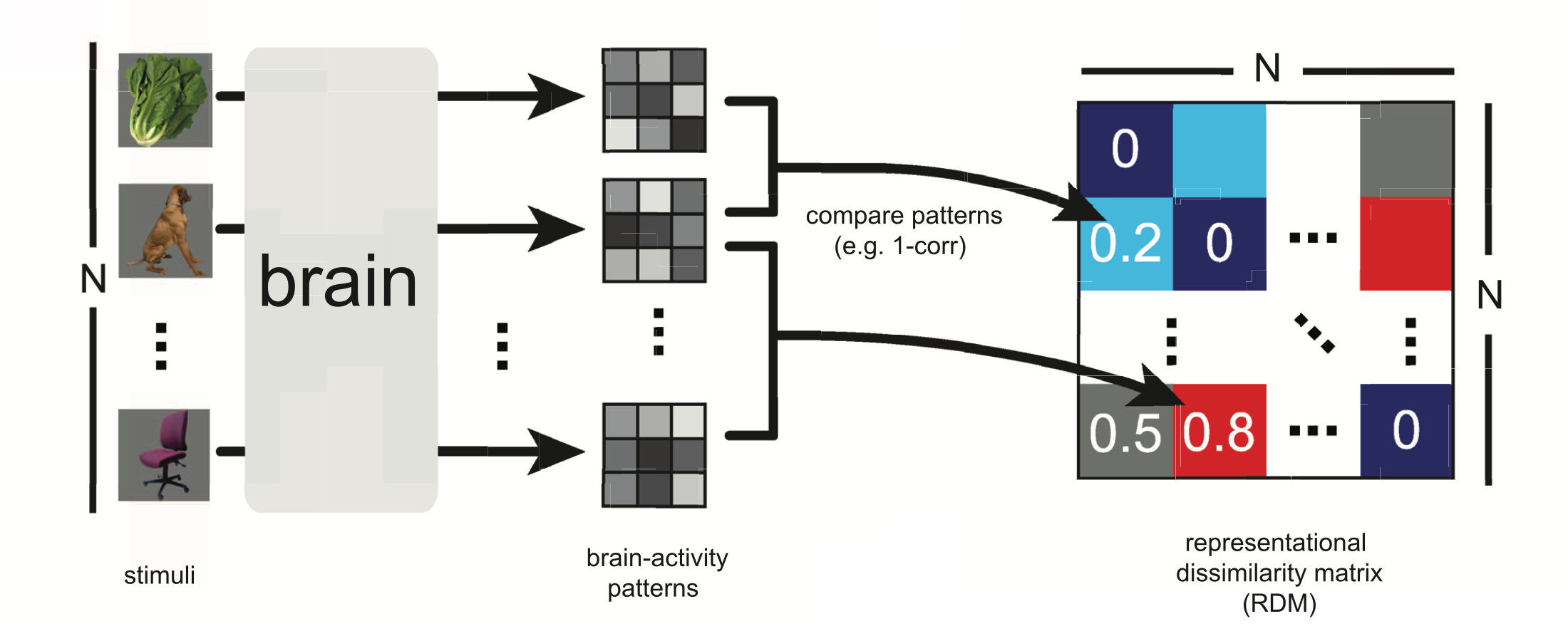}
\caption{\textbf{Computing the representational dissimilarity matrix (RDM).} Dissimilarities are computed between all pairs of activity patterns and assembled into a representational dissimilarity matrix (RDM). The RDM is symmetric about a diagonal of zeros. (Extracted from \cite{nili2014toolbox}).}
\label{fig:rdm_compute}
\end{figure}

During an RSA experiment, each subject's brain activity is measured while they experience a set of conditions (e.g. stimuli). For each brain region, an activity pattern is estimated for each condition. For example, given multivoxel fMRI patterns for a set of stimuli, the dissimilarity between the patterns for stimulus A and B could be measured as some sort of distance. Iterating through each pair of stimuli, these pairwise dissimilarities of the activity patterns are assembled into a representational dissimilarity matrix (RDM), which fully characterizes the representational geometry for that brain region (Fig. \ref{fig:rdm_compute}).

At its core, RSA relies on the concept of a RDM. Let $X$ be an $n × m$ matrix of neural response patterns, where $n$ is the number of stimuli or conditions and $m$ is the number of measurement channels (e.g., voxels, electrodes). The RDM $R$ is an $n × n$ symmetric matrix where each element $R_ij$ represents the dissimilarity between the response patterns for stimuli $i$ and $j$:

\begin{equation}
    R_{ij} = d(X_i, X_j)
\end{equation}

Here, $d()$ is a distance function, commonly chosen to be correlation distance (1 - Pearson correlation) or Euclidean distance. The choice of distance metric can affect the results and should be considered carefully based on the specific research question and data properties, which we will discuss next.


\subsection{Dissimilarity or distance measures}

Some commonly used dissimilarity measures include correlation distance (1 minus the Pearson correlation between activity patterns), Euclidean distance \cite{edelman1998toward} (the square root of the sum of squared differences between the two pattern), Mahalanobis distance (the Euclidean distance measured after linearly recoding the space so as to whiten the noise \cite{mahalanobis2018generalized,kriegeskorte2006information}) and pattern classifier distance (1 minus cross-validated classifier accuracy for discriminating two patterns). To especially note that, the correlation distance normalizes for the mean and variance of the activity patterns \cite{haxby2001distributed,kiani2007object,aguirre2007continuous}, and thus, is invariant to scaling differences but affected by shifts in the origin. The correlation distance is equivalent to the cosine distance after subtracting the mean value from each voxel pattern. The Euclidean distance on the other hand, is invariant to baseline shifts, which is sometimes beneficial if the baseline was not reliably estimated or meaningfully defined \cite{walther2016reliability}.


\subsection{Positively biased and unbiased distance measures}

It is also noted that these dissimilarity or \textit{distance estimates are positively biased}, due to the measurement noise leading to non-zero deviations despite a null hypothesis assuming true distances of zero. To correct for this effect, one can perform cross-validation on the dataset to obtain the dissimilarity. One example of such is the multivariate separation measure, the linear-discriminant t (LD-t) value \cite{nili2014toolbox}. LD-t is effectively a crossvalidated, normalized variation on the Mahalanobis distance using Fisher linear discriminant, and thus, also called the crossnobis distance. 

Because this cross-validated distance measure is unbiased, their distribution would be centered on 0 given a true distance of 0, and thus, can introduce negative distances. As such, the crossnobis distance is no longer a distance metric in mathematical sense. However, this positive bias of distance estimates should only matter if we want to interpret relative scales of the distances or test against zero distances. In most cases (``classical'' RSA), we are only interested in the order or the rank of the distances.

The choice of dissimilarity measure depends on the nature of the representations and activity patterns. The RDM abstracts the key information about the representation into a format that enables comparison between different datasets. As we briefly covered in the last section, the RDMs can be considered as a common language to study representations by comparing them between species, subjects, and brain imaging modalities.

Several factors influence the choice of dissimilarity measure for RSA:

\begin{itemize}
\item \textit{Noise normalization}: measures like correlation distance or Mahalanobis distance account for noise variability. For instance, Mahalanobis distance, as the multivariate noise normalized version of the Euclidean distance, downsize channels with high noise covariance. 
\item \textit{Mean response normalization}: correlation distance discounts the differences of mean activation levels.
\item \textit{Nonlinearities}: rank-based dissimilarities like 1 minus the Spearman's rho may be preferred if nonlinear measurement effects are expected.
\item \textit{Interpretability}: Euclidean distances relate directly to multivariate pattern separability.
\item \textit{Efficiency}: measures like dot product or cosine distance are simple and fast to compute.
\end{itemize}

When selecting the dissimilarity measure, the RDM characterizes the representational geometry and serves as a signature of a region's representational geometry that abstracts key information from the high-dimensional activity patterns. Since the RDMs are usually symmetrical, we only pick the  upper (or equivalently the lower) triangle of the matrices, excluding the diagonal cells. This enables testing computational models and relating representations across diverse measurements.

\section{Visualizing and Comparing Representation Geometries}

A key application of RSA is to test computational models of brain information processing. RSA can test different kinds of brain-computational models, from conceptual models with experimental priors to complex ones with deep neural networks, as long as they can have an input correspondence to the same set of stimuli or experimental conditions. The computational models, like our brain activity data, can be measured in terms of their response patterns corresponding to the same set of inputs, and hence, have their RDMs computed. Finally, to test the computational models, we compare the RDM of a brain region to the RDM predicted by the model.

Different models can be compared in terms of how well they explain the representation in multiple routes: (1)
visualize model and brain RDMs using multidimensional scaling (MDS);
(2) quantify similarity between RDMs using rank correlation;
(3) perform statistical inference to test if models are significantly related;
(4) compare models using ranking or inference tests of differential relatedness.

\begin{figure}[tb]
\centering
\includegraphics[width=\linewidth]{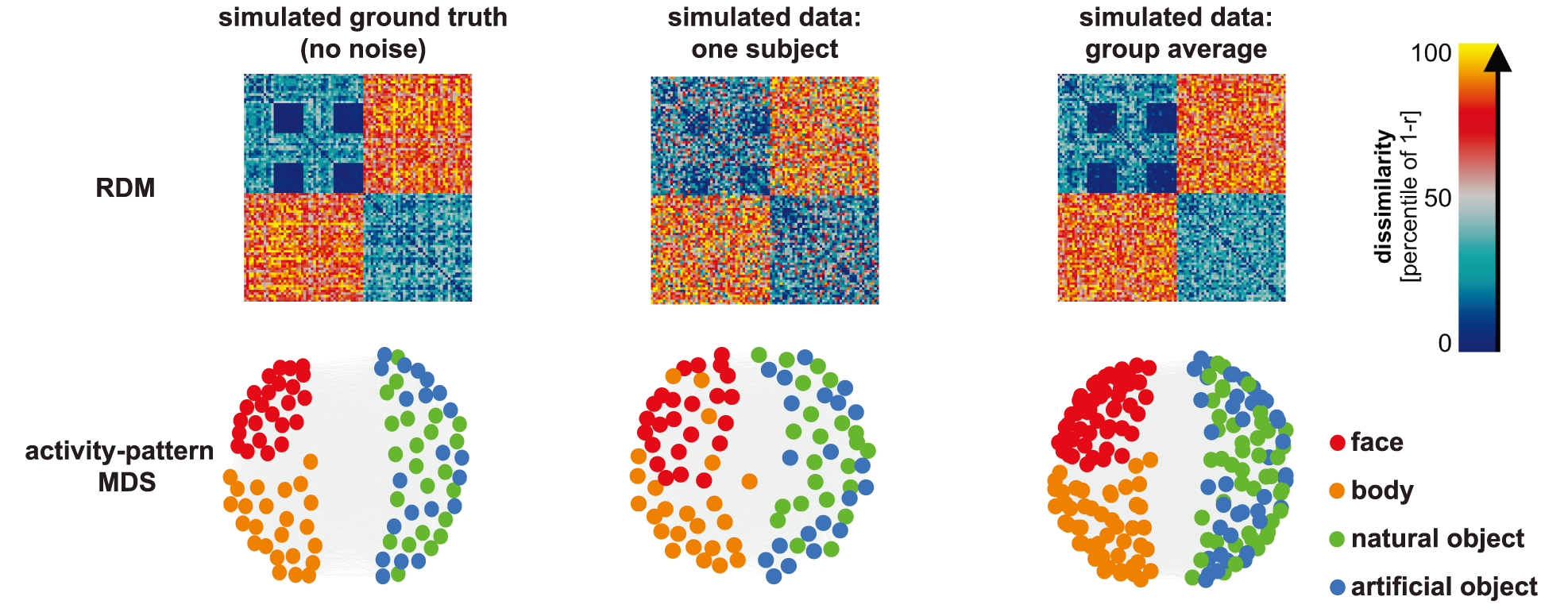}
\caption{\textbf{Visualizing RDMs with multi-dimensional scaling (MDS).} The three eample RDMs can be visualized in 2d arrangements via unsupervised learning methods such as MDS, with each dot as a representation pattern colored by their categorical information. (Extracted from \cite{nili2014toolbox}).}
\label{fig:mds}
\end{figure}

\subsection{RDM visualizations}

While RDMs provide a comprehensive summary of representational structure, they can be challenging to interpret directly. Several visualization techniques can help researchers gain intuition about the underlying representational geometry:

\begin{itemize}
    \item \textit{MDS plots}: multidimensional scaling (MDS) can arrange stimuli in 2D or 3D space to approximately reflect their dissimilarities.
    \item \textit{Hierarchical clustering}: Dendrograms can reveal hierarchical structure in the representations.
\item \textit{t-SNE \cite{van2008visualizing} and UMAP \cite{mcinnes2018umap}}: These nonlinear dimensionality reduction techniques can sometimes capture more complex structures than MDS.
\item \textit{Direct visualization}: For smaller stimulus sets, the RDM itself can be visualized as a color-coded matrix or heatmap.
\end{itemize}

Each of these techniques offers different insights, and researchers often use multiple visualizations to gain a comprehensive understanding of their data. The direct visualization and MDS are the most common visualization methods for RDM, which we will discuss in more detail next.

Multidimensional scaling (MDS) arranges the stimuli in 2D based on their modeled representational distances. This reveals the the major continuums that structure the representational geometry (Fig. \ref{fig:mds}) and an intuitive visualization of the major dimensions of variation emphasized and deemphasized in the representation. Additional techniques like hierarchical clustering and color labeling of the stimuli-relevant information can also help reveal categorical divisions that may be present in the representational geometry. 
This enables visually inspecting the similarities among the representational geometries of different brain regions and models. However, these visualizations do not indicate statistical significance. Proper statistical inference is still required to draw definitive conclusions about relationships between RDMs. MDS provides an exploratory tool for observing patterns and generating hypotheses, which can then be tested using inferential statistics.

\subsection{RDM comparators}

To compare RDMs, we need a metric that quantifies the agreement between two matrices. As briefly discussed in the distance metrics used for dissimilarity among activity patterns, we also need to tread carefully to decide if a linear relationship between the dissimilarities in different RDMs is required to assume. As pointed out in prior literatures \cite{kriegeskorte2008representational,nili2014toolbox}, unless we are confident a model captures the neuronal geometry and its nonlinear reflection in the measurements, a linear assumption seems questionable.

We therefore prefer metrics like rank correlation that only assume the model predicts the rank order of dissimilarities. Recommended rank correlations include Spearman's rho and Kendall's tau-a. Kendall's tau-a is more robust when models predict tied ranks, as it avoids favoring simplified models with fewer unique dissimilarity values. However, Kendall's tau-a is slower to compute than Spearman's rho for large RDMs.

\begin{figure}[H]
\centering
\includegraphics[width=\linewidth]{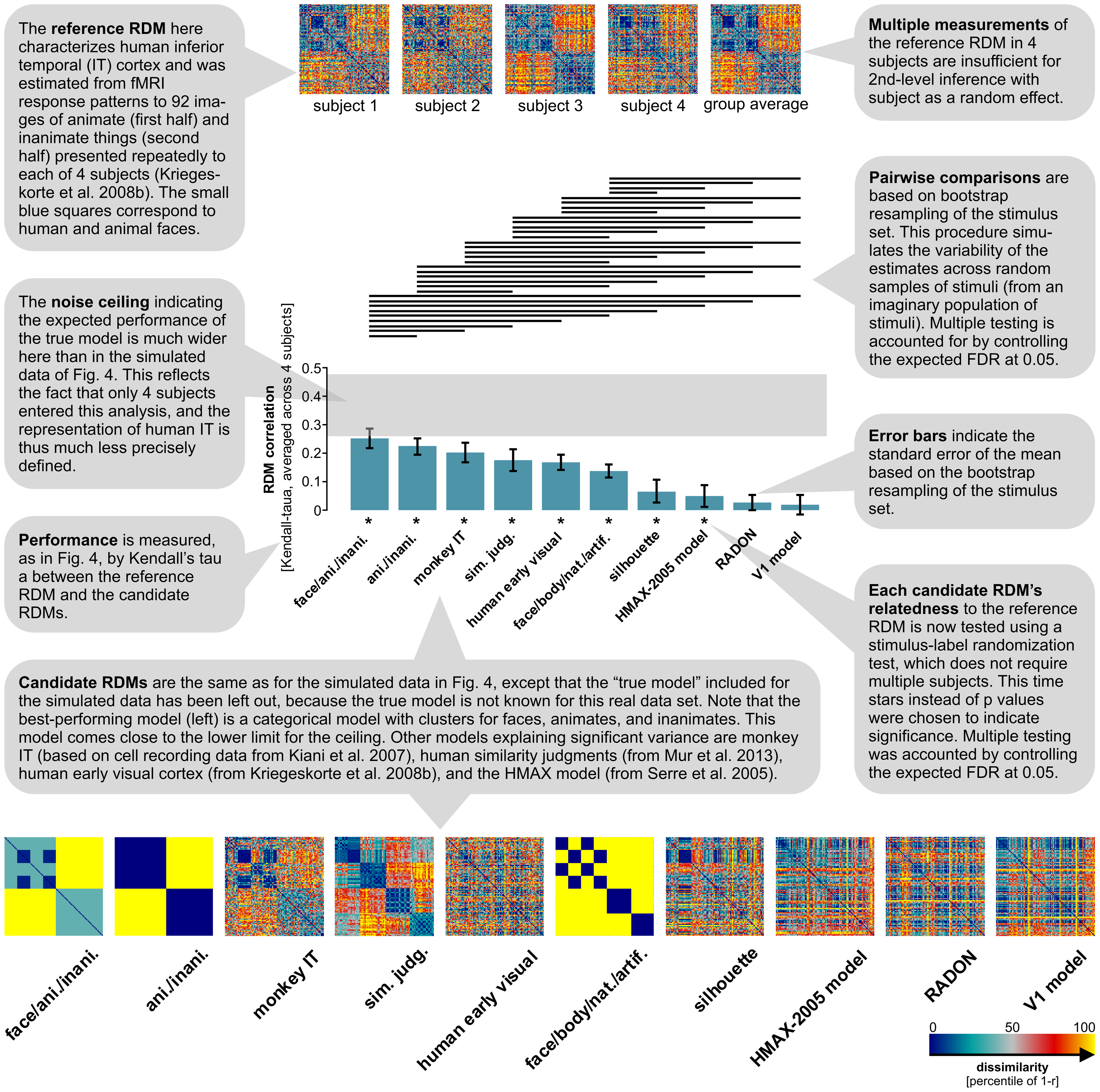}
\caption{\textbf{Inferential comparison of multiple model representations.} Demonstrated here is an example model comparison results enabled by the Matlab RSAToolbox \cite{nili2014toolbox}. The RDMs of 10 candidate models are compared with the reference RDM (computed from the brain-activity data) and ranked in terms of their performance to predict the data RDM. (Extracted from \cite{nili2014toolbox}).}
\label{fig:inference}
\end{figure}

In the statistical inference step, we relate one reference RDM (e.g. the data RDM of a brain region) to multiple candidate RDMs (e.g. candidate models to explain the cognitive functions in question). Other than computing the agreement or relateness between the reference RDM and a candidate RDM with the aforementioned rank-based measure, to draw conclusions with statistical significance as in the example in Fig. \ref{fig:inference}, there are several considerations 
\cite{schutt2023statistical,nili2014toolbox}:

\subsection{Statistical inference and bootstrapping}

To draw valid conclusions from RSA, rigorous statistical testing is crucial. Common approaches include:

\begin{itemize}
    \item \textit{Randomization tests}: Shuffling condition labels to create a null distribution.
    \item \textit{Bootstrap resampling}: Resampling with replacement to estimate confidence intervals.
    \item \textit{Crossvalidation}: Splitting data to test generalization of representational structure.
\end{itemize}

The choice of statistical approach depends on the specific research question and experimental design. Researchers should be cautious of multiple comparisons issues when testing many brain regions or model comparisons.

For instance, to estimate the variability of model RDM fits, it is recommended to use a bootstrap resampling approach. This involves repeatedly sampling a random subset of conditions with replacement from the full condition set. The RDM fits are recomputed for each bootstrap sample.

The variability of the model fits across bootstrap iterations reflects the dependency on the particular sample of conditions tested. This provides a data-driven way to estimate error bars and generalize the results to the broader condition population.

Bootstrap resampling provides an attractive approach for estimating variability in RSA model fits while requiring few assumptions. By resampleing conditions, it reveals the robustness of results to changes in condition sampling. An alternative way would be to use the Bayesian or variational methods to infer model probabilities and compare evidence for different models \cite{cai2016bayesian,cai2019representational,friston2019variational}.

Proper statistical procedures are important for drawing valid inferences from RSA model comparisons. Methods like noise ceilings and model flexibility help avoid overinterpreting results and guard against overfitting.

\subsection{Noise ceiling}

The \textit{noise ceiling} provides an estimate of the expected RDM correlation for the unknown true model, given the noise in the data. In another word, noise ceilings inform whether a non-perfect model correlations are related to the model or are they related to the noise in the data. It indicates the maximum possible performance.
If the best model does not reach the noise ceiling, a better model may exist. If the best model reaches the ceiling but the ceiling is low, experimental improvements to increase sensitivity may be needed.

The noise ceiling is estimated using the variability between subject RDMs. Operationally, the group mean RDM can be used as an estimate of the best performance any model could achieve on the data. The upper bound estimates the expected correlation of the true model RDM fitted to the subject RDMs -- i.e., each subject is compared to the full group mean RDM (positively biased since that subject is included in the mean). The lower bound uses a leave-one-subject-out approach to avoid overfitting -- i.e., each subject is compared to the group mean excluding that subject (negatively biased since less data is used).

The noise ceiling accounts for noise limitations and helps avoid overinterpreting model failures. If a model reaches ceiling, it likely explains the representational geometry as well as possible given the noise. The noise ceiling provides an important benchmark for RSA model evaluations.

\subsection{Model flexibility}

When comparing RSA models, it is important to account for model flexibility such as the number of free parameters. More flexible models with more parameters tend to fit data better, even if they do not reflect the true underlying processes.

One approach is to use quantitative model flexibility measures like the Akaike Information Criterion (AIC) or Bayesian Information Criterion (BIC) to penalize models with more free parameters. The model with the best fit after penalizing for flexibility can be selected.
Another approach is to fit computational models by optimizing parameters to match an RDM, then validate on independent test data. For example, a neural network model could be trained to match one RDM, then tested on its fit to another RDM from separate data. This avoids circularity and ensures the model truly captures the representation, not just fits the noise.

Accounting for model flexibility is important when comparing RSA models to ensure inferences reflect the underlying neural representations and computations, not just model overfitting. Quantitative flexibility measures or cross-validation on independent data help address this issue.

\subsection{Additional options and considerations of RSA}

Beyond the core RSA workflow, there are several extensions and variants that enable addressing additional questions:

\begin{itemize}
\item \textit{Representational connectivity}: inter-region RDM correlations can reveal relationships between brain areas, indicating which areas have similar representational geometries.

\item \textit{Temporal RSA}: Applying RSA in sliding time windows can reveal the temporal evolution of representations by analyzing how RDMs change over time.

\item \textit{Searchlight RSA}: RSA can be applied in sliding searchlight spheres to map local RDM model fits throughout the brain volume. This can localize regions that best match the representational geometry of a particular model.
\end{itemize}

These analyses demonstrate how the core RSA approach can be extended and adapted to address a diverse array of questions about neural representations and computations. RSA provides a flexible framework that can be tailored to different experimental questions and datasets.

Lastly, here are some additional considerations for properly interpreting the results of RSA model comparisons:
(1) A model matching the brain RDM may reflect the underlying neural code, intervening readout mechanisms, or both. Further analyses are needed to dissociate these factors.
(2) Model comparisons reveal relative rather than absolute model performance.
Low model correlation could indicate an inadequate model or noise/measurement limitations.
(3) RSA reveals linear decodability of model features but not their single-unit neural implementation.
These features are not necessarily unique -- different feature sets can span the same representational geometry.
(4) Related models may make similar RDM predictions. Comparisons should discount shared variance.
Thoughtful interpretation is important for leveraging RSA to gain computational insights into the nature of neural representations and computations.

\section{From Classical RSA to Topological and Dynamic Representations}

As we will see in the following chapters, RSA converts complex neural activity patterns into an interpretable representational similarity space that reveals key properties of neural representations and computations. This enables testing computational theories against rich multivariate measurements of brain activity. However, this framework also has its limitations, such as only emphasizing on the geometric or metric information of the dissimilarity. 
While RSA has proven to be a powerful tool for analyzing neural representations, several open problems and limitations have emerged:

\begin{enumerate}
    \item \textit{Geometry vs. Topology}: Classical RSA focuses primarily on the geometric relationships between neural activity patterns. However, the topological structure of these representations, such as the presence of clusters, holes, or more complex manifolds, may contain crucial information about neural computations that is not captured by geometry alone.
    \item \textit{Sensitivity to Noise}: The distance measures used in traditional RSA can be sensitive to noise in neural recordings, potentially obscuring the true representational structure.
    \item \textit{Static Representations}: Most RSA applications analyze neural activity patterns at a single time point or averaged over a time window, missing the dynamic nature of neural computations.
\end{enumerate}

This thesis addresses these limitations through several novel contributions:

\begin{enumerate}
    \item \textit{Topological RSA (tRSA)}: We introduce methods to incorporate topological information into RSA, providing a more comprehensive characterization of neural representations that is robust to geometric distortions.
    \item \textit{Temporal Dynamics}: We develop techniques to analyze how representational geometries and topologies evolve over time, revealing the temporal structure of neural computations.
    \item \textit{Adaptive Methods}: We introduce adaptive techniques that can automatically determine the most informative aspects of representational geometry or topology for a given dataset.
\end{enumerate}


These advances not only enhance our ability to analyze neural data but also provide new insights into the fundamental principles of neural computation. As we progress through the following chapters, we'll explore each of these contributions in detail, demonstrating how they address the limitations of classical RSA and open new avenues for understanding the brain's representational spaces.
By bridging classical RSA with topological data analysis, and time series analytical approaches, this thesis aims to provide a more comprehensive framework for decoding the neural code and understanding how the brain represents and processes information.



\clearpage

\chapter{Topological Representational Similarity Analysis (tRSA)}


As we journey deeper into the realm of neural representations, we now venture beyond the geometric landscape of classical RSA into the rich topological terrain of neural computations. This chapter introduces Topological Representational Similarity Analysis (tRSA), a novel framework that marries the strengths of RSA with the insights of topological data analysis.

The brain's representational spaces are not merely defined by distances between neural activity patterns, but also by their intrinsic shape and structure. tRSA allows us to capture these crucial topological features, providing a more robust and comprehensive characterization of neural representations.

In this chapter, we'll explore the motivations behind tRSA, its mathematical foundations, and its practical implementation. We'll see how tRSA can reveal representational structures that are invisible to classical RSA, and how it can provide more reliable comparisons between brain regions, individuals, and computational models. This chapter introduces the theoretical foundations and methodology of tRSA. For a detailed evaluation and application of tRSA to neural data and simulations, please see Chapter 3.

As we unfold the concepts of tRSA, keep in mind that this approach is not just an incremental improvement, but a fundamental shift in how we think about and analyze neural representations. It opens new avenues for understanding the brain's computational architecture and provides a bridge between the geometric intuitions of classical RSA and the rich world of algebraic topology.

This chapter leads to the following publications: 

\cite{lin2023topology} \fullcite{lin2023topology}.

\cite{van2023rsa} \fullcite{van2023rsa}.

\section{From Representational Geometry to Representational Topology} 

\begin{figure}[tb]
\centering 
    \includegraphics[width=\linewidth]{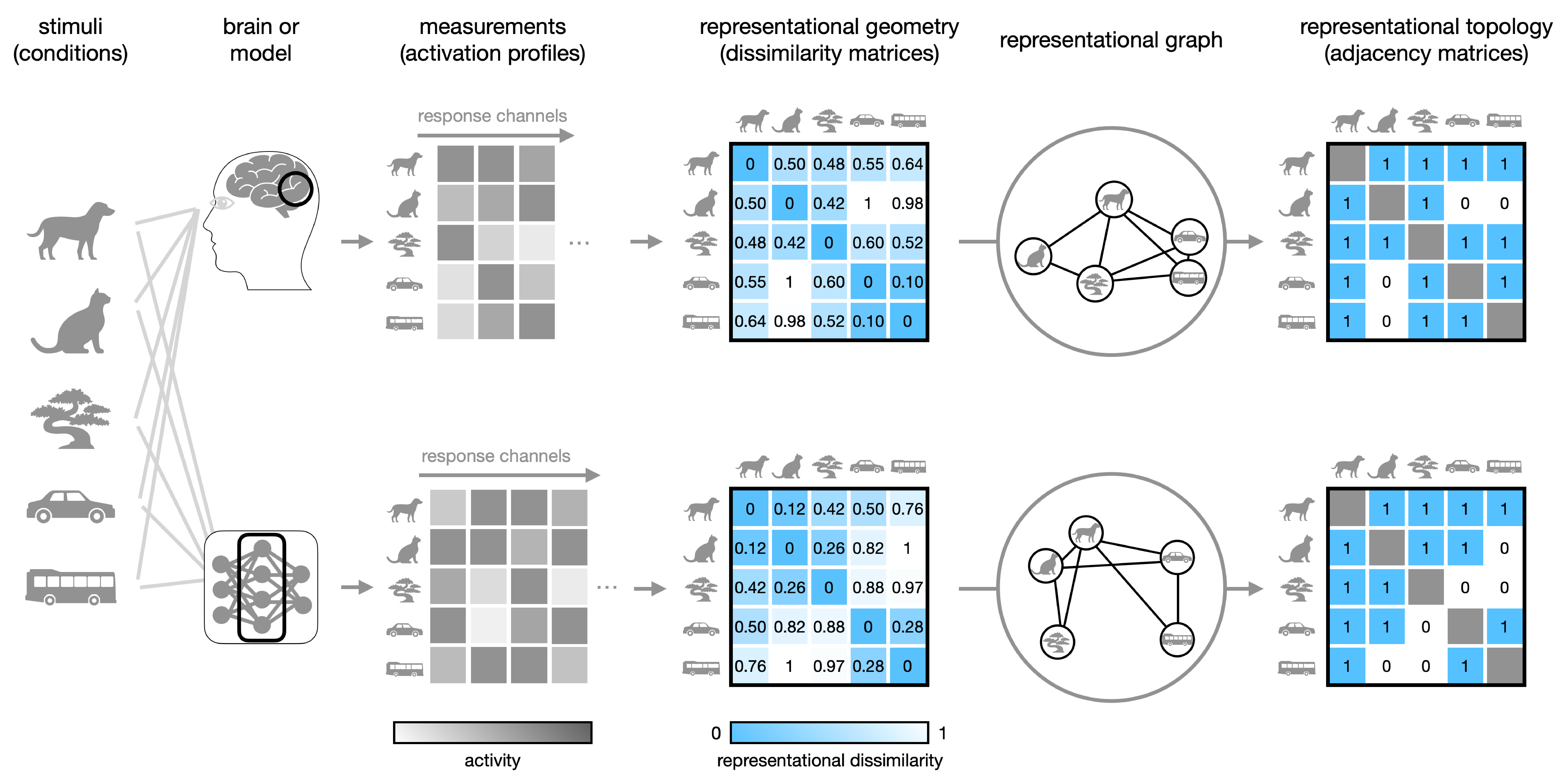}
\caption{\textbf{Comparing representations between brains and models.} To understand the degree to which a computational model can account for the cognitive process of a certain brain region, the same set of stimuli is presented to both the model and the biological system. The response patterns across measured response channels (e.g. neurons or voxels) are then characterized by a summary statistic, the representational dissimilarity matrix (RDM, center), which defines the metric configuration of the stimuli in the neural population response space. However, the metric configuration can be sensitive to measurement noise and idiosyncrasies of individual brains that do not reflect computational function. An alternative summary statistic that captures the topology would be the adjacency matrix (right), which defines the unweighted graph of neighborhood relationships in the population response space. This summary statistic promises to be more robust to noise and idiosyncrasies, but may discard too much information. Considering the geometry (RDM) and topology (adjacency matrix) as extremes of a continuum suggests that it may be possible to get the best of both (Fig. \ref{trsa_fig2}).
}\label{trsa_fig1}
\end{figure}

Geometrical and topological analyses can be applied profitably to the structure and connectivity of brains, on the one hand, and to neural population code representations on the other. Let us first consider the realm of brain structure and connectivity. Geometrical characterizations have been used to study the physical structure of brains (e.g. the geometry of the cortical surface \cite{dale1999cortical,desikan2006automated,fischl2002whole,kriegeskorte2001efficient}) and anatomical connectivity \cite{bastos2016tutorial,honey2009predicting,van2010exploring}. Topological and graph-based characterizations have been extensively used in network neuroscience \cite{bassett2017network,medaglia2017brain,gu2015controllability,srivastava2020models,bullmore2012economy,farahani2019application,sporns2022graph} to investigate the anatomical connections and functional correlations between brain regions and how structural and functional network topology is related to cognition. This paper is not about either the physical structure or the connectivity of the brain, but about the topology and geometry of neural representations.

In the realm of the neural representations, geometrical characterizations have been used to study the relationships between neural population activity patterns: the representational geometry \cite{kriegeskorte2008matching,kriegeskorte2008representational,kriegeskorte2013representational,kriegeskorte2021neural,sorscher2022neural,nieh2021geometry,freeman2018neural,chung2021neural} (Fig. \ref{trsa_fig1}, left, middle). The representational geometry provides a useful intermediate level of description capturing the information represented in a neuronal population code, while abstracting from the roles of individual measured responses (reflecting neurons or voxels) \cite{kriegeskorte2019peeling}. Considering the representational dissimilarities, rather than the representational patterns, enables direct comparisons of population-code representations between different individuals and species, as well as between brains and computational models. The analysis of representational geometries, known as representational similarity analysis (RSA, \cite{kriegeskorte2008representational}), has been successfully applied to understand diverse functions \cite{kriegeskorte2013representational}, including perception in the visual \cite{kriegeskorte2008matching}, auditory \cite{sievers2021visual} and other modalities \cite{fournel2016multidimensional} and higher cognitive functions such as abstraction \cite{nieh2021geometry,chung2021neural}, decision-making \cite{van2019computational}, working memory, social cognition \cite{thornton2018theories, thornton2023brain, tamir2016neural, tamir2018modeling}, and planning \cite{ehrlich2022geometry}. Representational geometries can be visualized by arranging stimuli in two or three dimensions, such that their distances approximately reflect the corresponding distances in the high-dimensional neural response space. Representational geometries, captured by the matrix of pairwise distances (the representational dissimilarity matrix, RDM), can also be used as a basis for model comparison \cite{kriegeskorte2008representational,nili2014toolbox,schutt2023statistical}, an approach that has enabled researchers to adjudicate among competing models of brain representations \cite{khaligh2014deep,kietzmann2019recurrence,cichy2017multivariate,konkle2022can}.

When we investigate the representational geometry by considering distances among neural activity patterns, we abstract from the roles of individual neurons. The representational topology provides a further step of abstraction. We may care less about the precise distances among the points in the high-dimensional response space that define the geometry than about the way the points hang together in what is sometimes called the neural manifold \cite{low2018probing,kriegeskorte2021neural,chung2021neural}. We may hypothesize, for example, that the overall geometry of the representation in a given cortical area or layer of a neural network model may vary across individual people or instances of a neural network model trained from different random seeds \cite{mehrer2020individual}. If the corresponding cortical areas in two people or the corresponding layers in two model instances served the same computational purpose, however, we may expect that stimuli that are neighbors in one individual's (or model instance's) representation remain neighbors in the other individual's (or instance's) representation. A graph of representational neighborhood relationships can be obtained by thresholding the distance matrix (Fig. \ref{trsa_fig1}, right). The thresholding operation is well-motivated when we care only about whether two points are in the same neighborhood or not. If they are in the same neighborhood, we consider them related and do not care whether they are close or very close. If they are not in the same neighborhood, we consider them unrelated and do not care whether they are very far apart or merely far enough not to count as neighbors. The neighborhood graph characterizes the representational topology.

An important question is to what extent the further step of abstraction involved in going from the distance matrix (geometry) to the neighbor graph (topology) is desirable or undesirable. It could be desirable for providing a more robust reduced signature of a region's computational function. However, it could be undesirable if it removes geometrical information important for discerning regions that implement distinct computational functions. Here we address this question empirically, using human functional MRI data and simulations based on neural network models.

Topological data analysis techniques \cite{wasserman2018topological} such as the persistent homology \cite{edelsbrunner2008persistent} and the Mapper algorithm \cite{singh2007topological}  are popular in many fields of biology \cite{rizvi2017single,ttda,geniesse2019generating,bibm,ellis2019feasibility,pike2020topological} and have also been used to directly analyze the representational space of the population activity \cite{chaudhuri2019intrinsic}. For instance, a study has discovered that the structure of spontaneous and evoked activity patterns in V1 can be mapped onto a manifold that has the topology of a sphere, whose two dimensions may reflect orientation and spatial frequency \cite{singh2008topological}, with the population response selective to the extremes of spatial frequency mapped towards the two poles of the sphere. Similarly, a recent study applied topological analysis techniques to study the population activity of grid cells, which are thought to be involved in spatial navigation and orientation \cite{gardner2022toroidal}. This study found that the population activity of grid cells has a toroidal topology, meaning that the manifold wraps around like a donut, whose two surface dimensions correspond to the 2d space navigated, implementing a cyclic representation. These examples illustrate the power of topological data analysis techniques to reveal the structure of the neighborhood graph of neural population representations.

These inspiring studies notwithstanding, topological characterizations are more widely used in network neuroscience and only beginning to impact investigations of the relationships among neural population activity patterns. Here we build on the early topological analyses of neural population activity patterns \cite{singh2008topological,ellis2019feasibility,gardner2022toroidal,low2018probing} and introduce a new family of summary statistics that can characterize the geometry as well as the topology of neural activity patterns. These geo-topological summary statistics enable researchers to calibrate the geometric and topological sensitivity of the analysis, so as to define a good signature of the computational role of each brain region. The new representational signatures can then be used not only for visualization of the representational geometry and topology, but also as a basis for formal inferential model comparison in the framework of RSA \cite{schutt2023statistical}, where our geo-topological summary statistics can replace the RDM, which characterizes the geometry.

Consider the example of visual perception (Fig. \ref{trsa_fig1}). We begin by measuring the brain-activity patterns elicited by each of a set of stimuli in a brain region or computational model. By estimating the distances among the stimulus representations (with full metric information), we can gain insights into distinctions the brain region or model layer emphasizes. Metric distance estimates promise detailed geometrical information, but are sensitive to noise and individual idiosyncrasies. Thresholding the distances provides a graph with binary edges, which captures how the neural manifold hangs together and also promises to be more resilient to nuisance variation. The methods we introduce here share a focus on neighborhood relationships with popular visualization techniques like Isomap \cite{tenenbaum2000global}, locally linear embedding \cite{roweis2000nonlinear}, $t$-SNE \cite{vanDerMaaten2008tSNE}, and UMAP \cite{mcinnes2018umap}. However, while these techniques aim to visualize a single representation, our aims are to characterize multiple representations in models and brains, quantify their similarity, and statistically compare models in terms of their ability to account for the topology and geometry of brain representations. To combine the benefits of detailed metric information and a binary edge description, we seek to define a representational graph that captures aspects of both the representational geometry and topology using weighted edges.

To integrate geometric (distance-based) and topological (graph-based) characteristics, we define the edge weights of the graph by a nonlinear monotonic transformation of the distances that (1) emphasizes the distinction between small and large distances, (2) compresses very small distances, thus disregarding the distinctions among them, (3) represents a continuum of intermediate distances, and (4) compresses very large distances. An example of the effectiveness of this transformation is an adaptive generalization of distance correlation based on proposed transformed distances, which provides a dependence measure with robust sensitivity to geometric and topological characteristics \cite{lin2018adaptive,www2022}. 

There are two motivations for defining the edge weights as a nonlinear monotonic transform of the distances, one theoretical and one data-analytical. From a theoretical perspective, differences among very large representational distances may not provide the most useful signature of the computational function of a brain region. It is the local geometry that determines which stimuli the representation renders indiscriminable, which it discriminates, but places together in a cluster, and which it places in different neighborhoods. The global geometry of the clusters (whether two stimuli are far or very far from each other in the representational space) may be less relevant to computation for two reasons. First, once two stimuli are perfectly discriminable, moving them even further apart does not improve discriminability. Second, in a high-dimensional space, a set of randomly placed clusters will tend to afford linear separability of arbitrary dichotomies among the clusters \cite{rigotti2013importance,kushnir2018neural,cover1965geometrical} independent of the exact global geometry. 

Like in a storage room, related things may need to be placed together in a representational space and unrelated things in different locations. The requirement of co-localization strongly constrains the local geometry because there is only one direction toward a given location. The requirement that two things be far from each other, by contrast, only weakly constrains the global geometry, because there are many directions away from a given location, especially in a high-dimensional space. This argument suggests the hypothesis that variations among large distances are idiosyncratic to an individual brain or model instance \cite{mehrer2020individual}. If this hypothesis were true, then compressing this variation may help us focus on more functionally relevant features of the representation that are less variable across individuals and model instances.
\\

\begin{figure}[H]
\centering
    \includegraphics[width=\linewidth]{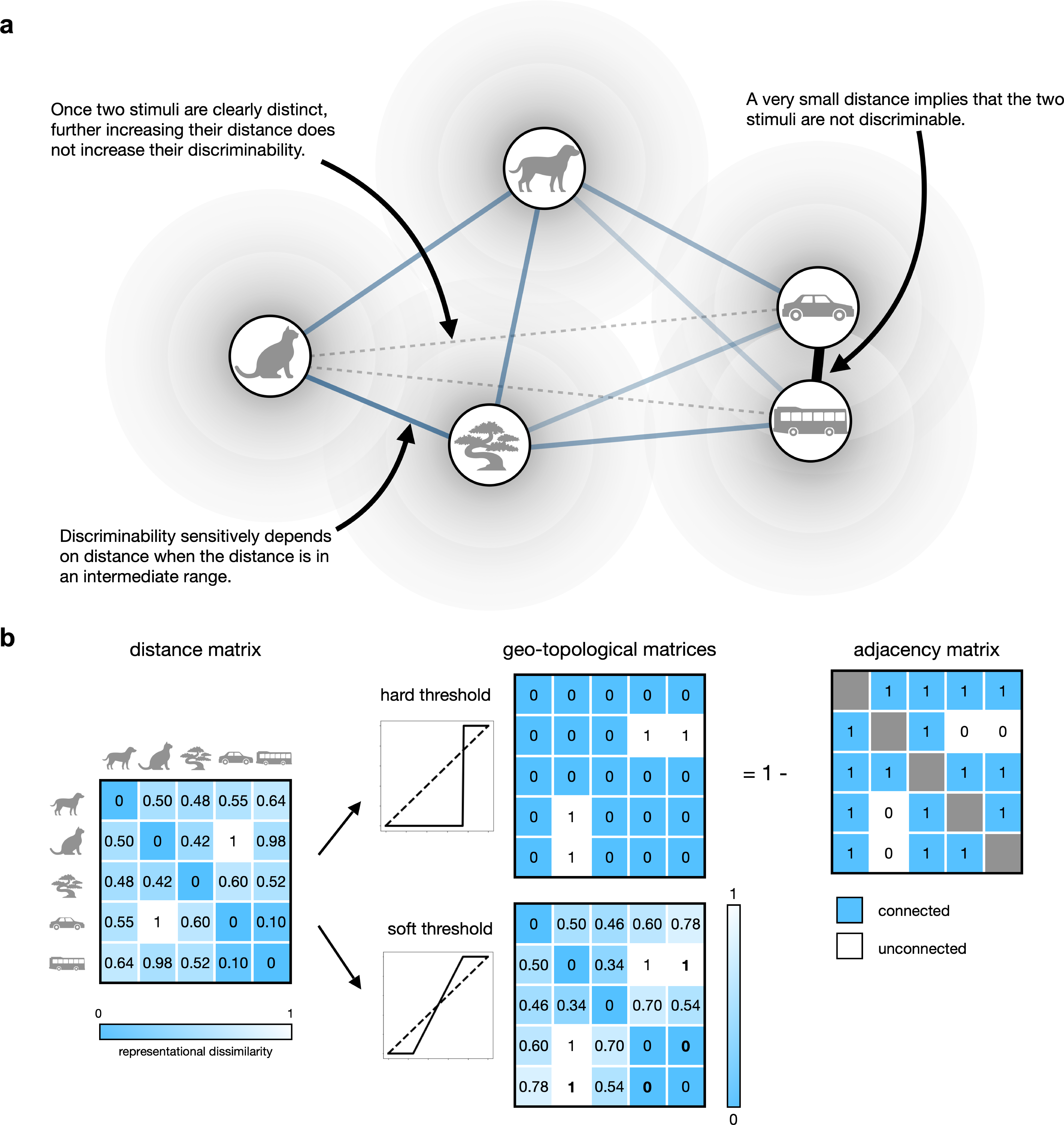}
  \captionsetup{labelformat=empty}
  \caption{}
\end{figure}
\clearpage
\begin{figure}
  \captionsetup{labelformat=adja-page}
  \ContinuedFloat
  \caption{\textbf{Intuition of the geo-topological transform of distances.} \textbf{(a)} Consider the five visual stimuli from Fig. \ref{trsa_fig1}, whose representation in a visual cortical area can be characterized by its representational geometry. If the response patterns are all affected by isotropic noise (or the noise has been whitened by a transform), then the Euclidean distances monotonically reflect the discriminabilities. Variation among large distances, however, is not associated with great differences in discriminability, because all pairs of well-separated stimuli are nearly perfectly discriminable (dashed lines). Similarly, variation among very small distances is not associated with great differences in discriminability, because all pairs of neighboring stimuli are indiscriminable (thick black edge). This suggests that variation among small distances and variation among large distances can be suppressed in favor of emphasizing the transition from small to large distances.  \textbf{(b)} To emphasize the transition from small to large distances while suppressing variation among small distances and variation among large distances, we can threshold the distances, such that small distances are pushed to zero and large distances are pushed to the maximum. We can either use a hard threshold (upper row) or a soft threshold (lower row). A hard threshold yields a binary matrix whose complement is the adjacency matrix of a graph that connects neighboring stimuli. A soft threshold creates a continuous transition, reflecting the graded increase in discriminability as the distance grows, and defines a weighted graph, where the weights reflect distances, but the pairs of stimuli that are furthest from each other are not directly connected (dashed lines in (a)).
}\label{trsa_fig2}
\end{figure}

From a data-analytical perspective, conversely, very small distances may be unreliable given the various noise sources that affect the measurements. Compressing small distances, thus, promises to reduce the influence of measurement noise on visualizations and inferential results. 
Compressing small and large distances is achieved by thresholding of the distances. However, thresholding may be too aggressive in that it completely removes all continuous information reflecting the geometry. As illustrated in Fig. \ref{trsa_fig2}, on one end, we have the distance matrix with the full metric information, and on the other, we simply have an adjacency matrix telling us whether two stimuli are neighbors in the representational space or not. To get the best of both worlds, it seems attractive to focus our sensitivity on a particular intermediate range of distances, so as to maintain reliable geometric information, while reducing the influence of noise (by compressing variation among small distances) and the influence of individual idiosyncrasies (by compressing variation among large distances). We show that this can be accomplished by a monotonic transform of the representational distances and that the resulting geo-topological representational summary statistics robustly reveal the functional distinctions among human brain regions and DNN layers. We introduce a family of geo-topological summary statistics that generalizes the RDM and provides a basis for topological RSA (tRSA), a generalization of RSA that balances sensitivity to the topology and geometry of neural representations.

\section{Topological RSA (tRSA) Framework}

\subsection{Nonlinear monotonic transforms of representational dissimilarities provide a family of geo-topological descriptors}

Topological RSA builds on the literature on topological methods (e.g., persistent homology \cite{zomorodian2005topology} and TDA mapping \cite{carlsson2009topology}). In order to suppress noise, we would like to find a lower threshold $l$ below which we consider stimuli as co-localized (i.e. the distance is $0$). In order to abstract from idiosyncrasies of individual brains and highlight the representational properties that are key to their computational function, we would like to find an optimal upper distance threshold $u$ above which we consider stimuli maximally distinct (i.e. we do not consider differences between larger distances meaningful). Two stimuli whose distance is larger than $u$ are disconnected in the graph capturing the topology.

\begin{figure}[H]
\centering
    \includegraphics[width=\linewidth]{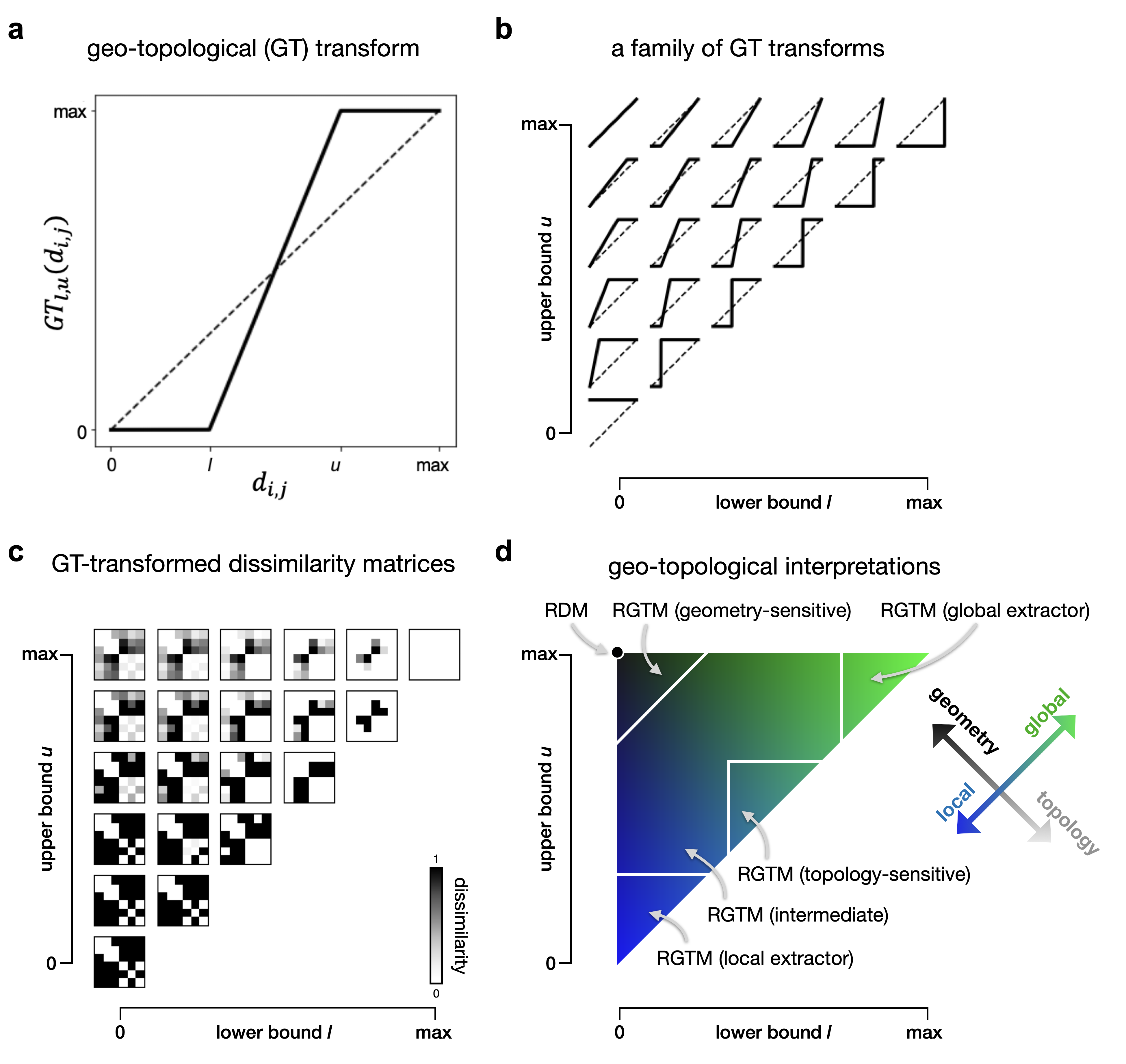}
\caption{\textbf{A family of geo-topological transforms of the RDM.} \textbf{(a)} The geo-topological (GT) transform is formulated as a linear piecewise function, such that any distances smaller than a lower bound $l$ will be mapped to zero, and any distances bigger than an upper bound $u$ will be mapped to 1. Between $l$ and $u$, the transition is linear. \textbf{(b)} By varying the thresholds $l$ and $u$, we select among a family of GT transforms. \textbf{(c)} By applying different GT transforms to the RDM, we obtain so-called representational geo-topological matrices (RGTMs). \textbf{(d)} To interpret the way the GT transforms reflect geometric and topological properties of the representation, we group the family members in different zones of the plane spanned by $l$ and $u$. The closer a GT transform is to the upper left corner ($l=0, u=max$), the more similar the RGTM is to the RDM. As we approach the diagonal line ($l=u$), the GT transform approaches a hard threshold, emphasizing the topology rather than the geometry. As we move diagonally from the bottom left to the upper right, the RGTMs go from emphasizing the local neighbor relationships among the stimuli to emphasizing the global structure of the representation.
}\label{trsa_fig3}
\end{figure}

Between the two thresholds we place a continuous transition so as to retain geometrical sensitivity in the range where it is meaningful (Fig. \ref{trsa_fig3}a). For simplicity, we propose a piecewise linear function as the monotonic distance transform. Note, however, that alternative monotonic transforms, such as the logistic function or a cosine transition, could be applied here. Given an original distance $d_{i,j}$ between the two neural signatures of two objects in the representational space, the piecewise linear geo-topological (GT) transform is defined as:

\begin{equation}
GT_{l,u}(d_{i,j}) = 
\begin{cases}
    0,& \text{if } d_{i,j} \leq l\\
    \frac{d_{i,j}-l}{u-l}, & \text{if } l < d_{i,j} < u\\
    1, & \text{if } u \leq d_{i,j} 
\end{cases}
\label{gt}
\end{equation}

By varying the lower bound $l$ and upper bound $u$, we obtain a family of GT transforms (Fig. \ref{trsa_fig3}b). Each member of this family transforms the original dissimilarity matrix (RDM) into a \textit{representational geo-topological matrix (RGTM)}, which provides a multivariate summary statistic with particular degrees of topological and geometrical sensitivity (Fig. \ref{trsa_fig3}c). The RGTM replaces the RDM in topological RSA. Note that the RDM is itself a member of the family, where $l=0$ and $u$ is set to the maximum (upper left corner in Fig. \ref{trsa_fig3}b, c, d). The RGTM, thus, generalizes the RDM.

When applying topological RSA to neural data to understand a brain representation, we can benefit from considering not only the RDM, but also other members of the RGTM family to gain an understanding of the geometrical and topological features of a brain representation. 
For the particular purpose of model selection, we aim to choose thresholds $l$ and $u$, so as to best recover the true (data-generating) model representation if it is among our models, or the best approximation among the models we are testing. 

\subsection{Geo-topological descriptor family captures both geometric and multi-scale topological information}

For interpretation of the family of GT transforms, we can consider the two diagonal dimensions of the triangular set spanned by $l$ and $u$ (Fig. \ref{trsa_fig3}d): From the upper left corner (RDM) toward the lower right, we go from geometrical to topological sensitivity, approaching the thresholding transforms ($l=u$) along the diagonal. From the lower left to the upper right, we go from local to global distance sensitivity.

For instance, the original RDM is in the top left corner ($l=0, u=1$), and around it, we have a region of RGTMs that is \textit{geometry-sensitive} where $l$ is small and $u$ is big. On the bottom left corner, where both $l$ and $u$ are small, we have \textit{local extractors}, which are RGTMs that are sensitive to whether or not two items are very close neighbors. On the upper right corner, where both $l$ and $u$ are big, we have \textit{global extractors}, which are RGTMs that are sensitive to whether or not two items are on opposite ends of the ensemble. If we go along the diagonal line from lower left to upper right, we have a belt zone of RGTMs which are close to binary (i.e. $l$ and $u$ are close to each other).

Formally, we require $l<u$, so the diagonal, where the transform is a simple thresholding function, is excluded from the family. Allowing $l=u$, so as to formally include simple thresholding functions, is possible but would complicate Eq. \ref{gt}. In either variant, the GT transform approaches a binary thresholding function for points approaching the diagonal. The thresholding functions along the diagonal relate our approach here to the mathematical filtration process used to reveal persistent homology in topological data analysis.

By exploring the choices of $l$ and $u$, we can identify the GT transforms that best enable us to match functionally corresponding cortical areas between different individuals. Similarly, we can generate data (with simulated measurement noise) from a layer in a deep neural network model, and determine which choices for $l$ and $u$ best enable us to identify the data-generating layer when using a range of layers from other instances of the DNN architecture (trained from different random seeds) as the models in analyses. 

One possibility is that the ideal setting is $l=0$, $u=max$, i.e., the original RDM, which characterizes the geometry. Other settings of $l$ and $u$ remove information about the geometry. Whether removing information by choosing a larger $l$ or a smaller $u$ helps or hurts depends on the relative extent to which it reduces signal and noise (nuisance variation) in the context of a particular data-analytical objective. If a topology-sensitive summary statistic reduced the variation caused by measurement noise and individual idiosyncrasies (i.e. nuisance variation) more than variation reflecting computational roles of different representations, then inferential comparisons of deep neural network models would benefit from topological RSA.

\subsection{Geodesic distances in representational sets provide an alternative geo-topological descriptor}

In addition to RGTMs, we propose the use of geodesic distances in the representation. Let us first consider the theoretical notion of a geodesic and then the practical analyses it motivates. Theoretically, the ``representation'' of a population of possible stimuli can be defined as the set of response patterns the stimuli elicit. If the stimulus population is a continuous set, we may hypothesize that so is the set of corresponding neural response patterns in our brain region of interest. This set of neural response patterns is often referred to as the neural manifold. (Note, however, that the set of response patterns would need to be locally homeomorphic to a Euclidean space to conform to the definition of manifold. Whether this is the case for a particular neural population is an empirical question.) A geodesic distance between two stimulus representations is the length of a shortest path traversing the representational set from one to the other. Unless the straight line between the two points is a subset of the representation, we will traverse a longer, curved shortest path through the representational set. The length of the geodesic path then will be larger than the Euclidean distance.

In practice, we will have data for a finite sample from the population of stimuli. To estimate the geodesic distance on the manifold, we can measure geodesics in the discrete graph characterizing our representation. In a graph, a geodesic distance is the length of the shortest path between two nodes. We define a representational graph for each member of the RGTM family. For each node $i$, edges exist only to other nodes $j$ with dissimilarity $d_{i,j}<u$. The edge weights are defined by the transform $GT(l,u)$ and are interpreted as distances between nodes. Note that edge weights, thus, can be zero. Zero-distance edges can be motivated as a correction of small positive distance estimates resulting from off-manifold noise displacements of patterns whose locations on the manifold are not significantly distinct. In order to maintain direct comparability of the edge-weight matrices of different brain and model representations, we do not collapse nodes with zero distance, but maintain one node for each stimulus.

The geodesic distance between two nodes is defined as the length of the shortest path that leads from one node to the other, where the length of a path is the sum of the internode distances (i.e. the edge weights) along it. The geodesic distance is infinite if there is no path connecting the two nodes through the edges. The shortest paths for all pairs of nodes (each corresponding to a stimulus) can be found using Dijkstra's algorithm \cite{dijkstra2022note}. The result is a stimulus-by-stimulus matrix, which we refer to as the \textit{representational geodesic-distance matrix (RGDM)}. The RGDM can be used in place of the RGTM or the RDM for our visualization and model-comparative inference procedures. As an alternative to an RGTM-based graph, we could use a binary graph (edge weights $\in {0, 1}$) to compute the RGDM. For example, we could use the binary graph in which each node is connected to its $k$ nearest neighbors or the graph containing connections for the $k$ smallest representational dissimilarities in the RDM. Our analyses here, however, use RGDMs computed from graphs with continuous internode distances, based on members of the RGTM family as described above.

\subsection{Leave-one-out evaluation on human fMRI data and DNN models can quantitatively evaluate the region identification power of the summary statistics}

We would like to quantitatively evaluate the power of topological RSA in the context of model selection, where each model predicts a representational geometry.
We therefore consider cases, where the ground-truth model is known. This enables us to objectively evaluate the impact of different choices of $l$ and $u$ on the accuracy of model selection. If conventional RSA were optimal, then setting $l=0$ and $u=max$ would be the best settings. If other settings afforded equal or better accuracy for model selection, then topological signatures would deserve consideration in future studies applying RSA to adjudicate even among models that predict not just representational topologies, but full representational geometries.

\textbf{Evaluating topological RSA's brain-region-identification accuracy (fMRI).} The fMRI evaluation was performed on pre-existing data from a human fMRI experiment \cite{walther2015beyond,walther2016sudden}, in which 24 subjects were presented  with 62 colored images depicting faces, objects, and places. We use 8 regions of interests (ROIs) here: the primary visual cortex (V1), the secondary visual cortex (V2), the extrastriate visual cortex (V3), the lateral occipital complex (LOC), the occipital face area (OFA), the fusiform face area (FFA), the parahippocampal place area (PPA), and the anterior temporal lobe (aIT).  

We investigate the brain-region-identification accuracy (RIA), where each brain region is considered a model. The region labels provide the ground truth: For data from each brain region in a held-out subject, we would like to identify which region the data came from on the basis of the data for all the regions from the other subjects. We therefore perform leave-one-subject-out (LOSO) RIA evaluation. First, we randomly sample 10 sets of $l$'s and $u$'s in each of the five interpretable RGTM zones: topology-sensitive (TS), geometry-sensitive (GS), local extractor (LE), global extractor (GE), and intermediate (I). The 10 samples of $l$ and $u$ for each region define 10 different GT transforms and provide an estimate of the RIA, averaged across regions and subjects, that reflects the performance at region identification of different zones of the family of RGTMs.

For each region in a held-out subject, we assign the region label of the average RGTM from the other subjects that is closest (in terms of Euclidean distance) to the RGTM being identified. Each subject is held out once in a full crossvalidation cycle and the RIA is the average identification accuracy.

In order to inferentially compare the different zones of the RGTM family, we perform frequentist comparisons. We would like to consider a difference in performance as significant if we expect it to generalize to experiments performed with different samples of subjects and stimuli drawn at random from the same populations of subjects and stimuli. We therefore perform a 2-factor bootstrap procedure, resampling both subjects and stimuli simultaneously \cite{nili2014toolbox, schutt2023statistical}. The standard deviation of the RIA estimates across 1,000 bootstrap samples of subjects and stimuli serves as our estimate of the standard error of the RIA estimate. Two-sided t-tests are then applied to assess the significance of differences between RIA estimates for different choices of $l$ and $u$. The degrees of freedom in this approach correspond to the smaller factor, which is the number of subjects (24) in this case.

\textbf{Evaluating topological RSA's layer identification accuracy (DNNs).} The DNN evaluation was performed on a convolutional neural network architecture \cite{lecun1995convolutional} called the All Convolutional Neural Net \cite{springenberg2014striving}. We investigate the layer identification accuracy (LIA), where each layer is considered a candidate brain-computational model. We perform leave-one-instance-out (LOIO) LIA evaluation. We trained 10 model instances, starting from 10 different random seeds, of the All-CNN-C network architecture \cite{springenberg2014striving}, a 9-layer fully convolutional network
that exhibits state-of-the-art performance on a well-known small object-classification benchmark
task (CIFAR-10 \cite{krizhevsky2009learning}).

We used the same numbers of feature maps (96, 96, 96, 192, 192, 192, 192, 192, 10) and kernel dimensions (3, 3, 3, 3, 3, 3, 3, 1, 1) as in the original paper. The training of All-CNN-C network instances involved 350 epochs using the ADAM optimizer with a momentum value of 0.9 and a batch size of 128. A preliminary learning rate of 0.01 was employed, along with an L2 regularization coefficient of $10^{-5}$ and gradient norm-clipping value of 500. Following \cite{mehrer2020individual}, we trained the DNNs on the complete CIFAR-10 image dataset (both training and test sets), which comprises 10 distinct object categories, each represented by 5000 training and 1000 test images, implemented with TensorFlow (version 1.3.0) and Python 3.5.4.

Like different individual subjects, these instances differ in their detailed connectivity, but perform the recognition task at similar levels of accuracy \cite{mehrer2020individual}. In addition to evaluating the LIA across instances for different choices of $l$ and $u$, we study the effect on the LIA of injecting Gaussian noise of a variety of variances $\sigma^2$ into the dissimilarity estimates.

The DNN-simulation-based evaluations follow the same procedures as the human-fMRI-based evaluations: The neural networks were presented with the same set of 62 object images to define the representational geometries. We perform LOIO LIA evaluation. Each layer in a held-out instance is identified as the layer whose average RGTM across the other instances is closest (in terms of Euclidean distance) to the RGTM being identified. Each instance is held out once in a full crossvalidation cycle and the LIA is the average identification accuracy. The inference, likewise, employs the same 2-factor (instance and stimulus) bootstrap method.

To briefly summarize the evaluation approach, we assess the performance of tRSA compared to classical RSA by introducing two key metrics:

\begin{itemize}
    \item \textit{Region Identification Accuracy (RIA)}: This measure quantifies how well a method can identify corresponding brain regions across different subjects. A high RIA indicates that the method captures consistent representational features across individuals.
    \item \textit{Layer Identification Accuracy (LIA)}: Similarly, LIA measures how accurately a method can identify corresponding layers in different instances of a neural network model. A high LIA suggests that the method is capturing fundamental computational properties of each layer, rather than idiosyncratic features.
\end{itemize}

\subsection{Model-comparative statistical inference for tRSA}

Topological RSA (tRSA) can use the well-developed inferential techniques of RSA for comparison between representational models \cite{kriegeskorte2008representational, diedrichsen2020comparing, schutt2023statistical}. The nonparametric inference methods of RSA3 \cite{schutt2023statistical} can simply use the geo-topological statistics (RGTMs and RGDMs) in place of RDMs. As in conventional RSA, the representational similarity of brain regions and model layers can be quantified using various comparators, such as cosine similarity or correlation, but applied to geo-topological statistics rather than RDMs.

Here we used Euclidean distance as the comparator for the geo-topological summary statistics. The $l$ and $u$ are defined as quantiles (expressed as percentiles or ranks) relative to the set of dissimilarities in a given RDM. Defining $l$ and $u$ as quantiles enables matching choices for model and brain representations whose dissimilarities may have different magnitudes and may lack a common unit that would render them commensurable. In addition to defining $l$ and $u$ as quantiles, we use the ranks within each RDM to define the dissimilarities entering the GT transform. This has the benefit that the resulting RGTMs have identical distributions of values. In this scenario, the squared Euclidean distance as a comparator of representational summary statistics is proportional to the Pearson correlation distance and to the cosine distance. These comparators thus would have yielded identical model-selection results, rendering the analyses relevant to the most common choices of RDM comparator in RSA.

Another popular choice for the RDM comparator is a rank correlation coefficient, such as Kendall's $\tau_a$ \cite{nili2014toolbox} or the more computationally efficient Spearman-type coefficient $\rho_a$ \cite{schutt2023statistical}. In the present study, the RDMs are rank-transformed before computing the RGTMs. Our comparators therefore benefit from the robustness afforded by the rank transform, obviating the need for another rank-transform at the level of the RGTMs, as would happen if we chose a rank correlation coefficient as the comparator. Using a rank correlation coefficient is closely related (but not mathematically equivalent) to the present analyses.

\subsection{Topological RSA offers greater robustness to noise and intersubject variability}

Different modalities of brain-activity measurement are affected by different kinds and levels of noise. For example, functional magnetic resonance imaging (fMRI) is a widely used method for measuring the patterns of hemodynamic responses associated with neural activity, which is sensitive to many sources of noise such as physiological noise (e.g. heart rate, respiration), motion artifacts, and low signal-to-noise ratio \cite{greve2013survey}. Electroencephalography (EEG) is another widely used method for measuring neural activity, which is sensitive to different types of noise such as electrical noise from the environment, and muscle artifacts \cite{pijn1991chaos,reddy2013artifact}. Similarly, magnetoencephalography (MEG) is a method for measuring neural activity using magnetic fields, which is sensitive to noise from external magnetic fields and from the subject's head movement \cite{gonzalez2014signal,chholak2021event}. Invasive neural recordings suffer less from external noise sources and reflect the internal neural variability of repeated responses to the same experimental conditions. The neural variability may reflect stochasticity and/or neural computational mechanisms we do not understand yet \cite{churchland2010stimulus, churchland2012two, goris2014partitioning, renart2014variability, orban2016neural}. 

Different modalities of brain-activity measurement also vary in their temporal and spatial resolution and in their susceptibility to intersubject variation. For instance, functional MRI (fMRI) using blood-oxygen-level-dependent (BOLD) contrast \cite{ogawa1990brain,ogawa1992intrinsic, bandettini1992time, logothetis2004nature, logothetis2008we} has low temporal resolution, capturing brain activity at a scale of seconds. However, compared to other noninvasive whole-brain human neuroimaging techniques, BOLD fMRI has relatively high spatial resolution, enabling the localization of brain activity to specific regions. with voxels ranging from $500 \mu$ to $5 mm$, each measurement channel reflects tens or hundreds of thousands of neurons. BOLD fMRI patterns can reflect both anatomical and physiological individual differences, introducing intersubject variation that needs to be accounted for in analyses. Electroencephalography (EEG), in contrast to fMRI, offers high temporal resolution, recording neural activity with millisecond precision. While EEG can detect rapid changes in brain dynamics, its spatial resolution is coarse, which makes the precise localization of neural sources challenging. Additionally, EEG data can be sensitive to variations in scalp thickness, skull conductivity, and electrode positioning, further contributing to intersubject differences that contribute nuisance variation to data analysis.

All these factors can affect the ability of a summary statistical descriptor to reveal the signatures of specific computations in neural activity. By choosing a suitable $l$ and $u$ to define the nonlinear monotonic transform, the geo-topological descriptors can be adapted to the structure of the noise and intersubject variability of the particular measurement modality used. This flexibility promises to reveal the invariant structure of brain representations when visualizing representational spaces and could lend topological RSA improved accuracy in the context of inferential model comparisons.

\section{RSAToolbox Python Package}

Although RSA is conceptually straightforward, the array-based computations required to apply RSA to complex datasets necessitate robust, custom software solutions. Historically, an early MATLAB-based solution, the Matlab RSAToolbox \cite{nili2014toolbox}, has been widely adopted within the neuroimaging community. This toolbox laid the groundwork for implementing RSA but was tailored to the MATLAB environment, which may not be accessible to all researchers.

Parallel developments in multivariate pattern analysis (MVPA) have led to the creation of a suite of tools designed for extracting multivariate patterns, including pyMVPA \cite{hanke2009pymvpa}, CosmoMVPA \cite{oosterhof2016cosmomvpa}, and MVPA-light \cite{treder2020mvpa}, as well as approaches for Pattern Component Modeling (PCM) \cite{diedrichsen2018pattern}. While these tools provide components that are essential for RSA, they are not exclusively designed for it, often requiring a patchwork of additional coding to conduct a complete RSA workflow.

The RSAToolbox for Python emerges as a cohesive and comprehensive package that addresses this fragmentation. As illustrated in Fig. \ref{fig:pipeline}, the RSAToolbox encapsulates a full pipeline of RSA-related functionalities—spanning from data simulation (\texttt{pyrsa.sim}) to model evaluation (\texttt{pyrsa.inference}) -- all within a single, coherent software environment. This integration reduces the need for ad-hoc scripting and bridges various aspects of RSA, from preprocessing neural representation data to evaluating representational models.

\begin{figure}[tb]
    \centering
    \includegraphics[width=\columnwidth]{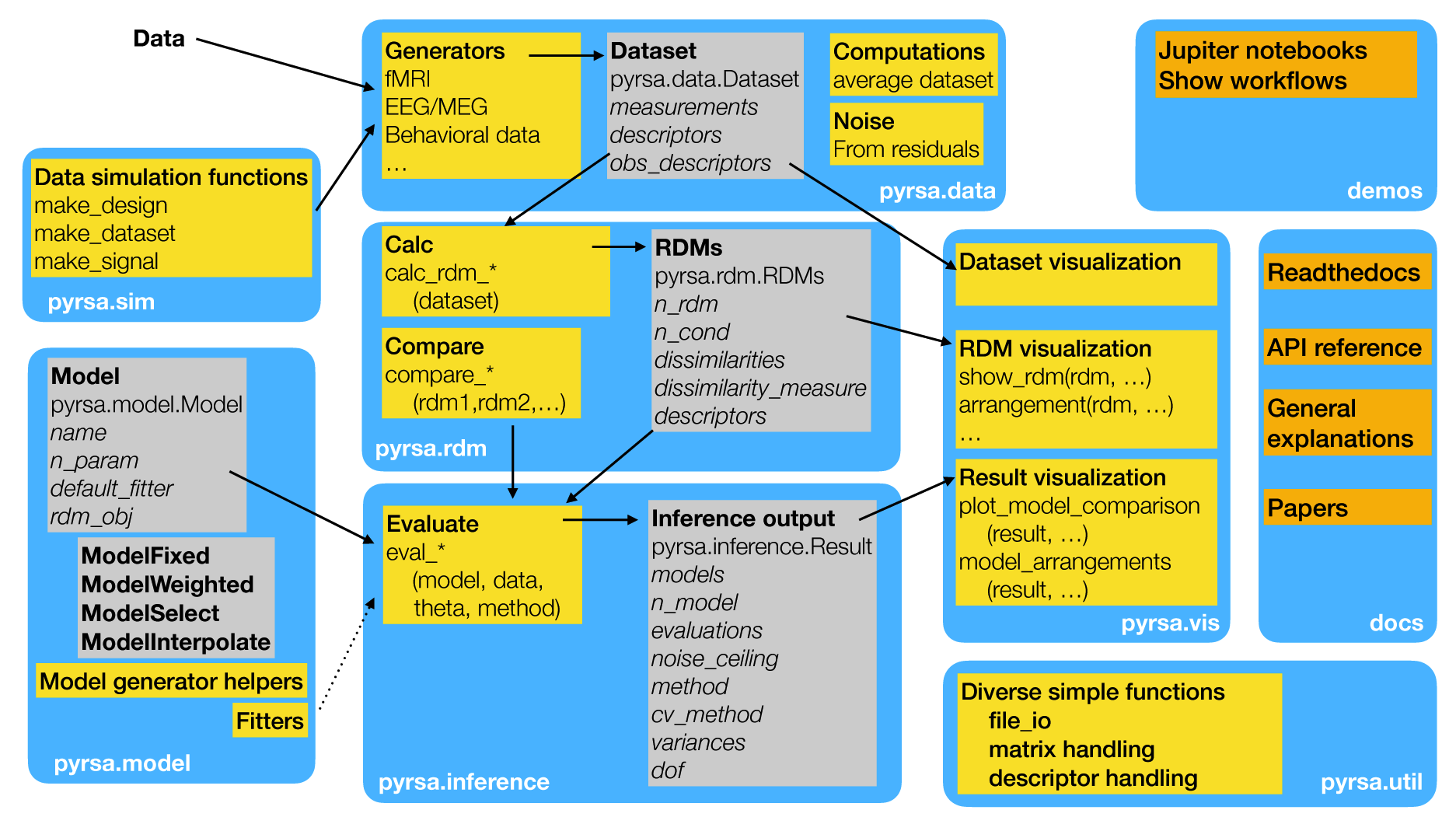}
    \caption{\textbf{Modules and pipelines in RSAToolbox Python package.} Shown is the comprehensive workflow facilitated by the package, from data ingestion to result visualization.}
    \label{fig:pipeline}
\end{figure}

The RSAToolbox enables seamless computation of RDMs, model fitting, comparison, and visualization, offering tools that are not inherently provided by other MVPA packages. Furthermore, the package includes functionalities for noise estimation and statistical testing, critical steps for robust RSA. The Python library also supports a variety of visualizations for RDMs and model comparisons, providing a user-friendly interface to inspect and interpret the results (module \texttt{pyrsa.vis}).

In essence, the RSAToolbox represents a democratization of RSA techniques, lowering the barrier to entry for new researchers and enhancing the reproducibility of results. Its utility is not only in consolidating existing RSA methods but also in extending the methodology to accommodate novel representational models, such as those arising from topological data analysis, thereby fostering innovation within the field.

In the next chapter, we'll put tRSA into practice, applying it to both neural recordings and simulations of neural network models. This empirical evaluation will demonstrate the power of tRSA in identifying unique computational signatures of different brain regions and neural network layers.

\chapter{Topology and Geometry of Neural Representations}


Having established the theoretical foundations of topological Representational Similarity Analysis (tRSA) in the previous chapter, we now embark on an empirical journey to evaluate its effectiveness and explore its implications. This chapter serves as the experimental counterpart to our theoretical framework, putting tRSA to the test in real-world scenarios.

As we navigate through this chapter, we'll examine how tRSA performs when applied to both neural recordings and simulations of neural network models. Our goal is twofold: first, to demonstrate the practical utility of tRSA in identifying unique computational signatures of different brain regions and neural network layers; and second, to compare its performance against classical RSA methods.

This exploration will not only validate the tRSA approach but also provide new insights into the topological and geometric properties of neural representations. By the end of this chapter, we'll have a clearer understanding of how tRSA can reveal aspects of neural computation that remain hidden to traditional analysis methods.

This chapter builds directly upon the theoretical foundations of tRSA introduced in Chapter 2. Here, we put those concepts into practice, applying tRSA to both neural recordings and simulations of neural network models. If you need a refresher on the tRSA methodology, please refer back to Section 2.2. 

This chapter leads to the following publication: 

\cite{lin2023topology} \fullcite{lin2023topology}.

\section{Geo-Topological Descriptors Can Discern Topological and Not Just Geometric Distinctions}

As a proof of concept, consider the geometric and topological similarities among the four hypothetical representations in Fig. \ref{trsa_fig4}. We call the four representations the ``flat 8'', the ``bent 8'', the ``untangled flat 8'' and the ``untangled bent 8''. Imagine an experiment in which the idealized continuous representational set is sampled using of 40 stimuli (balls in Fig. \ref{trsa_fig4}a). The flat 8 and the bent 8 share the self-intersection, which creates two holes, rendering the shapes topologically similar, although the bending of the latter greatly changes the geometry. Similarly, the untangled flat 8 and the untangled bent 8 are topologically similar (in that neither has a self-intersection and both have one hole) and geometrically dissimilar because of the bending.
By contrast,  the flat 8 and the untangled flat 8 are geometrically similar (both flat and similar in their RDM), and topologically distinct (one versus two holes). Likewise, the bent 8 and untangled bent 8 are geometrically similar (both bent) and topologically distinct (double arrows in Fig. \ref{trsa_fig4}a).

The different summary statistics reflect the topology and geometry of these hypothetical neural representations to different degrees (Fig. \ref{trsa_fig4}b). As expected, the RDM reflects the geometric distinctions but is not very sensitive to the topological distinctions, which are implemented here through minimal metric displacements that determine whether or not there is a self-intersection. The multi-dimensional scaling (MDS) arrangement of the four RDMs (Fig. \ref{trsa_fig4}c) shows the dominance of the flat versus bent distinction when characterizing the representations with RDMs.

The representational geo-topological matrices (RGTMs) were obtained using small values for both $l$ and $u$, yielding almost binary matrices with local topological sensitivity, which reveal whether or not there is a self-intersection. The MDS shows that the RGTM here balances sensitivity to the topology and the geometry of the representation. Note the prominent blue ``eyes'', which reflect the self-intersection where the contour of the 8 crosses itself. The blue ``eyes'' are present for the two-hole representations with self-intersection (left two representations in Fig. \ref{trsa_fig4}b), but absent for the untangled one-hole representations. 

\begin{figure}[H]
\centering
    \includegraphics[width=\linewidth]{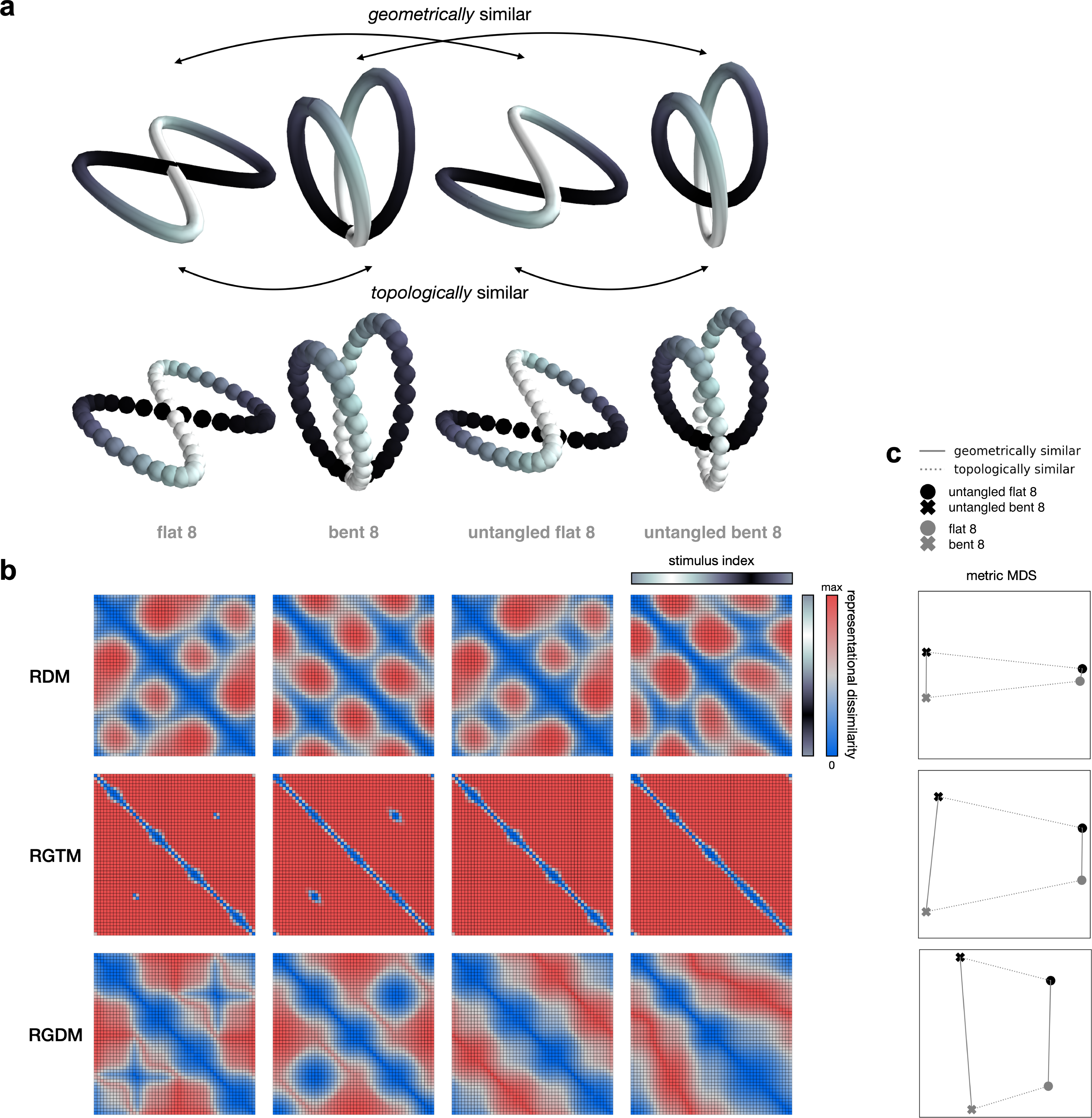}
  \captionsetup{labelformat=empty}
  \caption{}
\end{figure}
\clearpage
\begin{figure}
  \captionsetup{labelformat=adja-page}
  \ContinuedFloat
  \caption{\textbf{The geometric and topological similarities between hypothetical neural representations (proof of concept).} \textbf{(a)} We consider four hypothetical representations of 40 stimuli (balls in lower row of a). The response patterns are sampled from idealized continuous sets of neural response patterns (top row in a). Two of these continuous sets are manifolds (the untangled shapes) and the other two are not (the tangled shapes, where the set self-intersects, forming a neighborhood that is not homeomorphic to a Euclidean space). From left to right, we label the four representations the flat 8, the bent 8, the untangled flat 8, and the untangled bent 8. The flat 8 and the untangled flat 8 are geometrically similar, while being topologically dissimilar (with the former self-intersecting). Their bent versions, as well, are geometrically similar and topologically dissimilar. The flat 8 and the bent 8, on the other hand, are topologically similar (with the self-intersection creating two holes) and geometrically dissimilar (because the twisting substantially changes the metric geometry). Their untangled versions, as well, are topologically similar and geometrically dissimilar. \textbf{(b)} The RDMs (Euclidean distance, top row) do a good job characterizing the geometric relationships but are insensitive to the topological relationships. The RGTMs (middle row) are more sensitive to topology. To achieve local topological sensitivity, we chose  $l=0$ and $u=0.075$, revealing the self-intersection in the two leftmost representations. The RGDMs (bottom row) are exquisitely sensitive to the topological relationships, while de-emphasizing geometrical relationships between the representations. Note the ``eyes'' in the two leftmost RGTMs, representing the self-intersections. Each RGDM captures the lengths of the shortest paths in the graph of the RGTM shown above it. The RGDMs more prominently reflect the topology as the ``shortcut'' paths enabled by the self-intersection affects shortest-path lengths for a broad swath of stimuli. \textbf{(c)} Multi-dimensional scaling (MDS) on the four representations shows the pairwise similarities among the four matrices of each row, confirming the increasing sensitivity to topological differences as we go from RDM to RGTM and on to RGDM.   
}\label{trsa_fig4}
\end{figure}

Finally, the representational geodesic matrices (RGDMs) are even more sensitive to the topological distinctions. The RGDMs were defined on the basis of the graphs of the RGTMs in the row above. The shortest-path lengths reflect the existence of ``shortcut'' routes through the self-intersection. For many pairs of stimuli, the shortcut (when available) provides a shorter path than going the straight route around the 8. The small geometrical distortion causing the self-intersection is therefore reflected across a large portion of the RGDM. The MDS plots show that the RGTMs and RGDMs can effectively characterize both the geometric and topological similarities, whereas the RDMs mostly reflect the geometry of the representation.

\section{Geo-Topological Descriptors Reveal What Aspects of Geometry and Topology Enable Accurate Identification of Brain Regions Across Subjects (fMRI)}

Using human fMRI data, we quantitatively evaluated the performance of the RGTM summary statistics (including the RDM as a special case) at revealing the correspondences among ventral-stream cortical regions across subjects. A summary statistic will succeed in this evaluation, if it is robust to noise and interindividual variability while maintaining sensitivity to differences between different cortical areas.

\begin{figure}[H]
\centering
    \includegraphics[width=\linewidth]{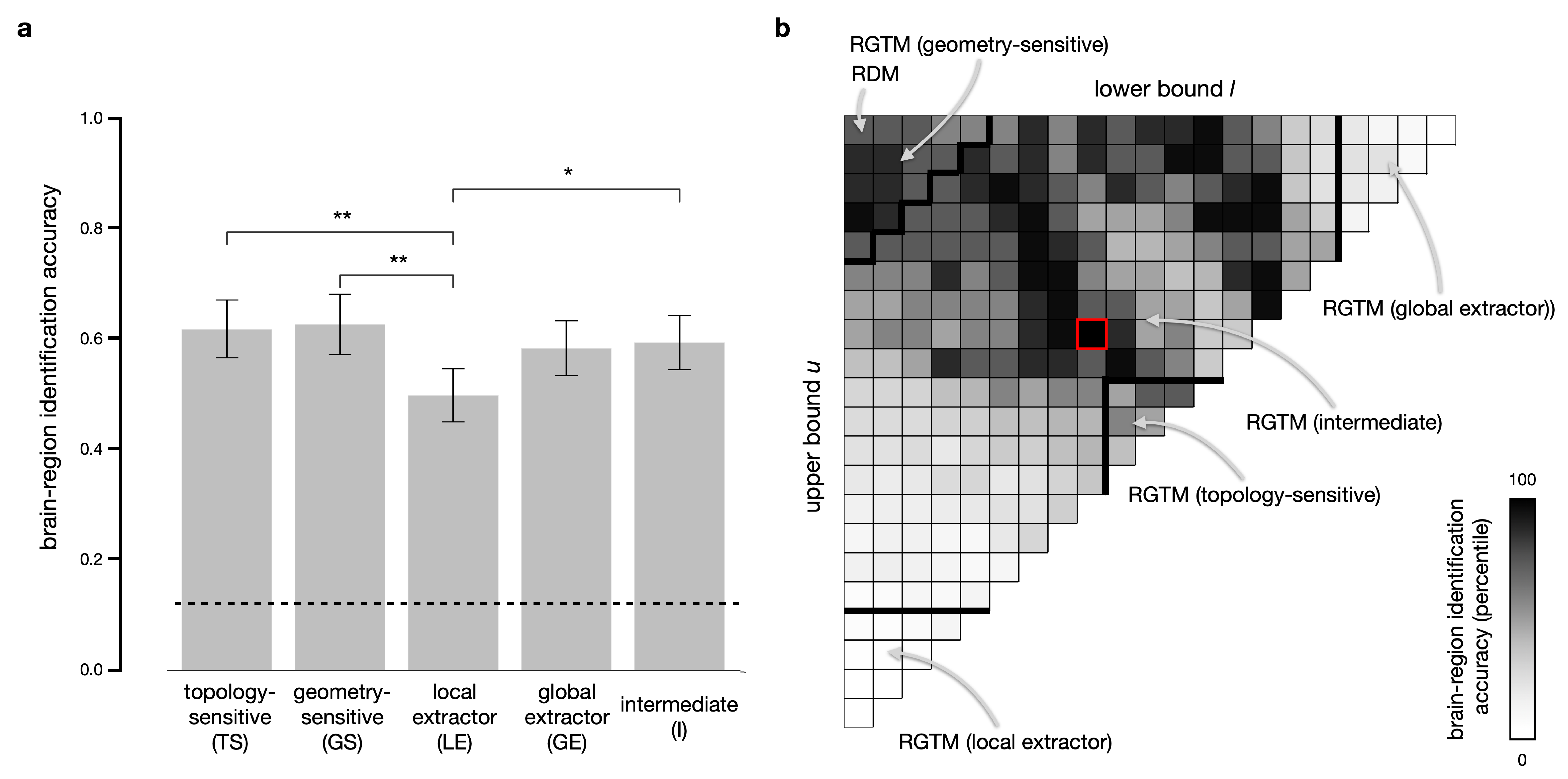}
\caption{\textbf{Brain-region identification accuracy across the family of geo-topological descriptors.} Our analysis of human brain regions used fMRI data from 24 subjects, collected from 8 regions of interest, including the primary and secondary visual cortices, the lateral occipital complex, the occipital face area the fusiform face area, the parahippocampal place area, and the anterior temporal lobe. (\textbf{a}) Region-identification accuracy (RIA) is evaluated using leave-one-subject-out crossvalidation, where a classifier (on which we compute RIA) is trained on all available data except for one subject and then tested on that left-out subject. This process is repeated for each subject, with the final performance measure being the average across all iterations. We used bootstrapping to obtain an unbiased estimate of the standard error as the error bound for region identification accuracy and and used crossvalidation to prevent overfitting.  We randomly sampled 10 sets of lower bounds $l$'s and upper bounds $u$'s in each of the five interpretable RGTM zones as defined in Fig. \ref{trsa_fig3} and applied a paired \textit{t}-test to compare the RIA between different RGTM zones as defined in panel b) and Fig. \ref{trsa_fig3} (**** p < 1e-4, *** p < 1e-3, ** p < 1e-2, * p < 0.05). \textbf{(b}) RIA percentile (gray colorscale) as a function of the combination of upper and lower bounds, $l$ and $u$, defining the RGTM (layout as in Fig. \ref{trsa_fig3}) with the best-performing RGTM marked in red.}\label{trsa_fig5}
\end{figure}

We grouped the members of the RGTM family into interpretable zones in Fig. \ref{trsa_fig3}d. Results for each zone are shown in Fig. \ref{trsa_fig5}a). We plotted the region identification accuracy (RIA) as a heatmap across choices for $l$ and $u$ (Fig. \ref{trsa_fig5}b). The highest RIA was achieved for $l=0.40$ and $u=0.65$ (red square in Fig. \ref{trsa_fig5}b). However, other settings yielded similar RIA. Results suggest that $l$ and $u$ should not be both low or both high. The RDM is an effective summary statistic. However, it contains information not needed for region identification. This result suggests that compressing less informative distance variation is possible without a reduction in model-selection accuracy. In this data set, we did not find that the benefits of noise reduction significantly outweighed the loss of information.

We inferentially compared RIA between different zones within the RGTM family using simultaneous bootstrapping of both subjects and stimuli (2-factor bootstrap, \cite{schutt2023statistical}) for frequentist pairwise comparisons. The 2-factor approach serves to test for differences expected to generalize across random samples of subjects and stimuli drawn from the same populations. The summary statistics in the topology-sensitive (TS) and geometry-sensitive (GS) zones did not perform significantly differently ($p = 0.434$). The topology-sensitive (TS), geometry-sensitive (GS), and intermediate RGTMs all performed better than those in local extractor (LE) zone ($p = 4.776e-03$, $p = 4.972e-03$,  and $p = 4.454e-02$, respectively). We found no other significant differences. Overall, geometry-sensitive and topology-sensitive summary statistics performed similarly, but the latter suffer when the upper bound is in the lower third of the distribution.

These findings suggest that a substantial portion of the smallest and largest dissimilarities can be compressed without reducing the RIA. Using the RGTM to focus on the intermediate range of dissimilarity variation (perhaps the middle third of the distribution of dissimilarities) while compressing the lower and upper third of the dissimilarities yields optimal model selection in this data set.

\section{Geo-Topological Descriptors Reveal What Aspects of Geometry and Topology Enable Accurate Identification of DNN Layers Across Instances}

Deep neural networks (DNNs) have emerged as important models of vision and brain-information processing in recent years \cite{yamins2016using,cadieu2014deep,long2018mid,kriegeskorte2015deep,dwivedi2021unveiling}. DNNs can learn non-linear representational transformations through feedforward and recurrent processing, enabling them to capture the computations underlying task performance. When a neural network architecture is trained repeatedly, starting from different random connection weights, the resulting trained model instances have distinct parameters and distinct detailed unit tuning, despite performing the same task roughly equally well. Just like individual humans differ, thus, so do instances of DNNs trained from different random seeds \cite{mehrer2020individual}. 

Analogously to the brain-region identification analysis in the previous section, we investigated the usefulness of RGTMs in identifying which layer of a neural network model a given RGTM was computed from, given RGTMs for all layers in other instances of the architecture. Like the human fMRI analysis, this analysis informs us about the best choice of representational summary statistic when performing DNN model comparisons using human data. We are able to objectively evaluate the summary statistics, because we know the ground-truth data-generating model. The DNN simulation analyses are complementary to the human fMRI analyses: On the one hand, they afford complete knowledge of the computational mechanisms (an advantage). On the other hand, they are less realistic with respect to the biology and structure of the noise in the data used as a basis for model comparison (a disadvantage).

Our layer-identification analysis reveals to what extent different RGTMs (including geometry-sensitive and topology-sensitive members of the family of GT transforms) abstract from nuisance variation across the same layer in different model instances, while capturing computationally meaningful variation between different layers (which are thought to play distinct computational roles).

\begin{figure}[H]
\centering
    \includegraphics[width=\linewidth]{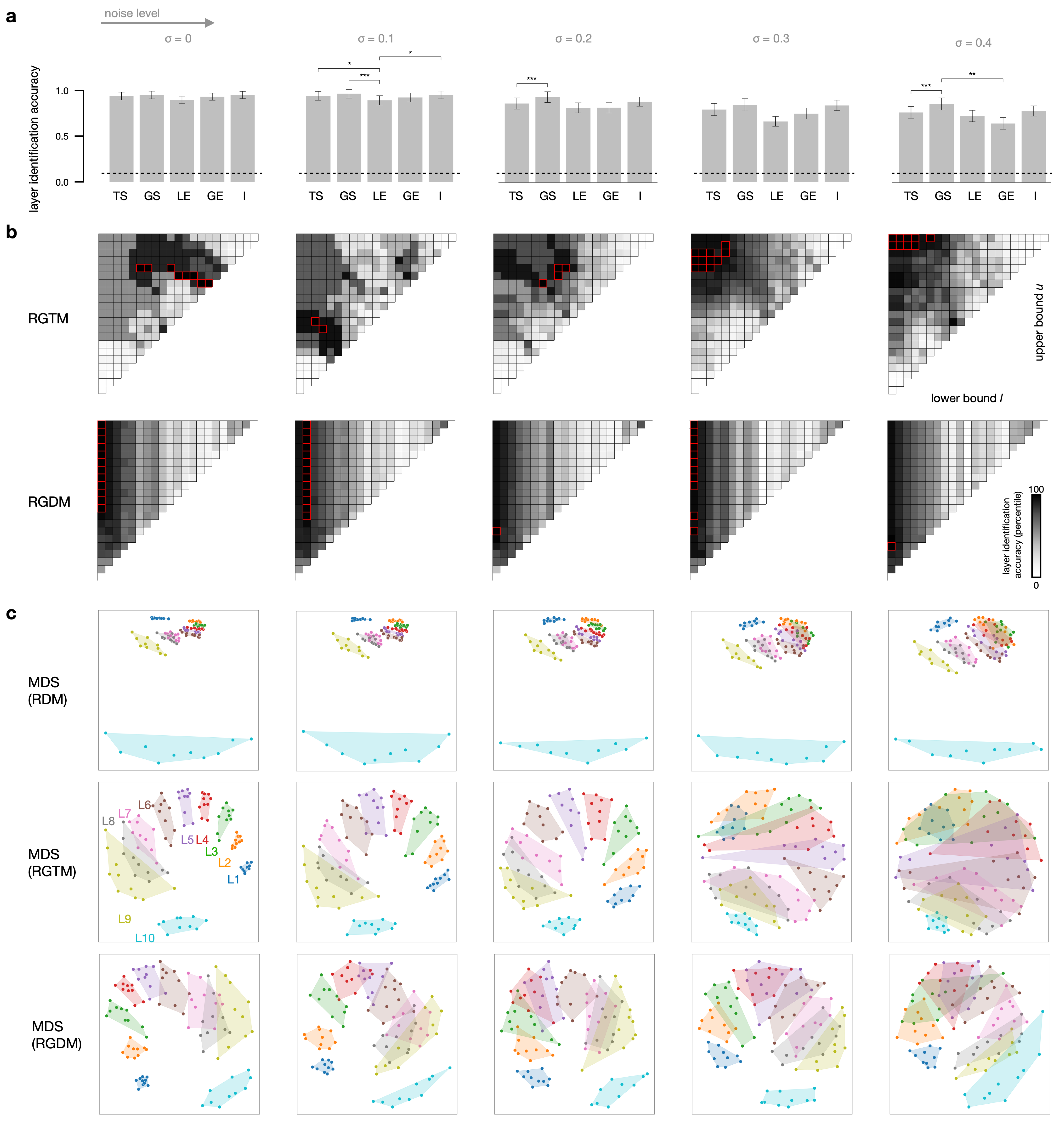}
  \captionsetup{labelformat=empty}
  \caption{}
\end{figure}
\clearpage
\begin{figure}
  \captionsetup{labelformat=adja-page}
  \ContinuedFloat
  \caption{\textbf{Neural-network-layer identification accuracy across the family of geo-topological descriptors.} The performance of different RGTMs at neural-network layer identification with gradually increasing Gaussian noise $\sigma$ on the dissimilarity estimates (columns). Results are for All Convolutional Neural Nets, a simplified model architecture with many convolutional layers (with 10 separate layers of interest in total). Layer identification accuracy (LIA) was estimated in leave-one-instance-out cross-validation, where a classifier is trained on all available model instances except for one  and then tested on that left-out model instance. This process is repeated for each instance, with the final performance measure being the average across all iterations. To enable crossvalidated evaluation similar to the brain-activity data in the previous figure, for each model, we trained 10 instances (analog of individual subjects) from different random seeds. Cross-validation was performed within a bootstrapping process to estimate LIA and its estimation error. We randomly sampled 10 sets of $l$'s and $u$'s in each of the five interpretable RGTM zones as defined in Fig. \ref{trsa_fig3} and applied a paired \textit{t}-test to compare the LIAs afforded by different RGTM zones. \textbf{(a)} Bar graphs showing the LIA for each RGTM zone and noise level with statistical comparisons. LIA is similar for topology-sensitive RGTMs and geometry-sensitive RGTMs. \textbf{(b)} The LIA of RGTMs and corresponding RGDMs across the $(l,u)$ threshold pairs are represented in a percentile heatmaps, with the maximum performance marked in red. \textbf{(c)} The MDS plots of the RDMs (top row) show geometrical similarities among layer representations. Each dot is the RDM for one layer in one model instance. For each layer, a convex hull is drawn to group its representations across all instances. MDS plots reflecting geo-topological similarities are shown for the best-performing RGTMs and RGDMs (red in b), revealing the good discrimination of layers these geo-topological summary statistics provide (middle and lower rows, respectively).
}\label{trsa_fig6}
\end{figure}

\begin{figure}[tb]
\centering
    \includegraphics[width=\linewidth]{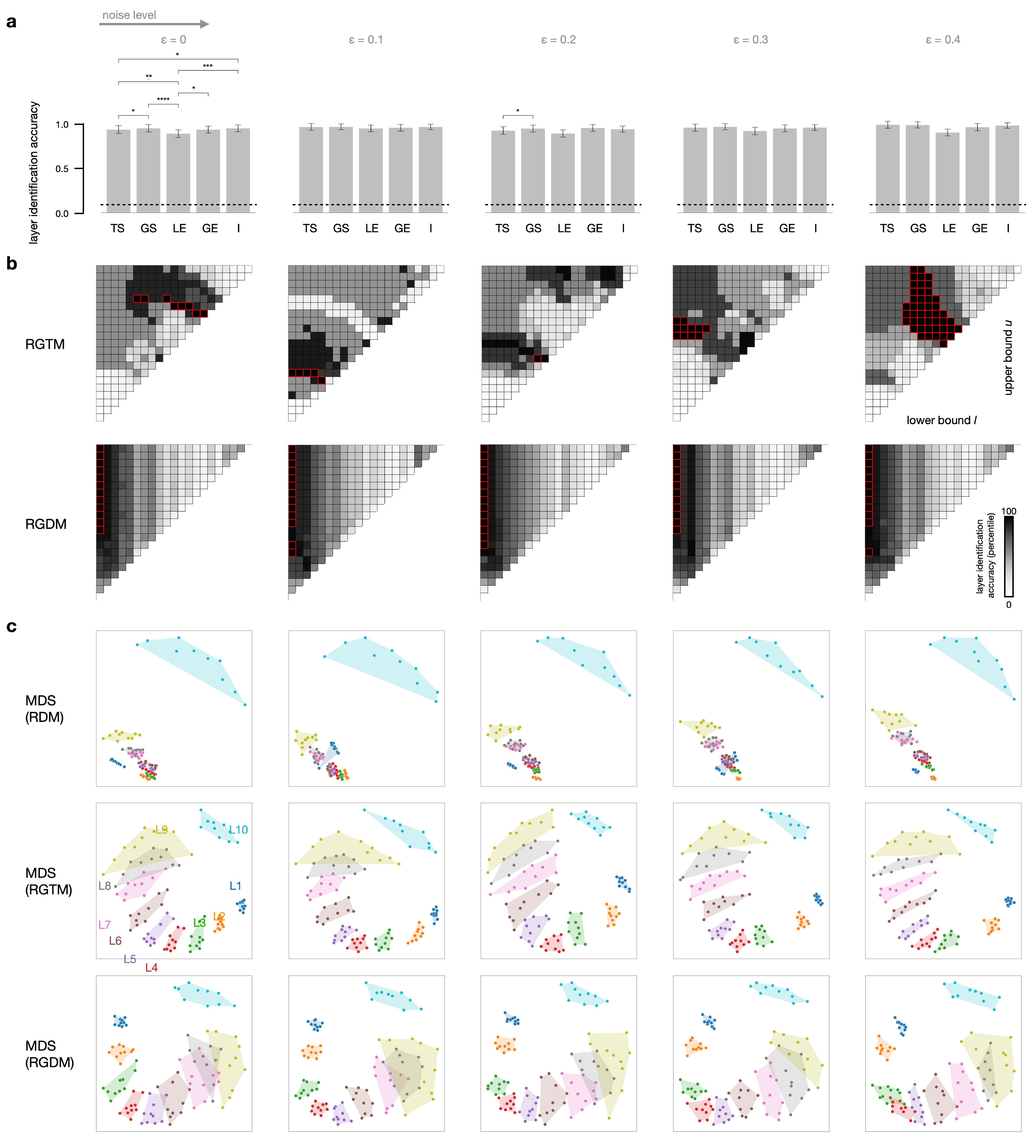}
\caption{\textbf{The performance of different RGTMs and their corresponding RGDMs of deep neural networks with gradually increasing Bernoulli noise $\epsilon$ during training.} Similar to Figure \ref{trsa_fig6}, the performance is evaluated by the layer identification accuracy (LIA) among the neural network layers in a leave-one-out cross-validation process, on the dataset of the same neural net architecture as Figure \ref{trsa_fig6}, but with gradually increasing Bernoulli noise $\epsilon$ instead of Gaussian noise.  
}\label{trsa_fig7}
\end{figure}

As for the human brain regions, we quantitatively evaluated the performance of the summary statistics from the RGTM family (including the RDM). Results are shown in Fig. \ref{trsa_fig6}, grouped by noise level (Fig. \ref{trsa_fig3}d). We present results for two types of noise: Gaussian noise $\sigma$ (added to the dissimilarity estimates, Fig. \ref{trsa_fig6}) and Bernoulli noise $\epsilon$ (applied during training as dropout rates, Fig. \ref{trsa_fig7}).

Statistical comparisons of layer identification accuracy (LIA) for GT-transform zones showed that geometry-sensitive statistics outperformed topology-sensitive statistics at higher noise levels ($\sigma \geq 0.2$). The heatmaps showing LIA as a function of $l$ and $u$ in RGTMs (Fig. \ref{trsa_fig6}b) likewise show that as the noise level increases, the optimal statistics (red frames) go from more topology-sensitive to more geometry-sensitive. The LIA heatmaps based on the RGDMs show a prominent fall-off of accuracy with $l$ (horizontal axis, as in Fig. \ref{trsa_fig3}) and little dependence on $u$: As $l$ grows, more and more pairs of neighboring points collapse to $0$ distance in the RGTM, and thus more and more shortest-path lengths between points connected by paths of $0$-distance edges collapse to $0$ in the RGDM. The weak dependence of LIA on $u$ reflects the fact that more separated points can be connected either through a smaller number of larger steps (when $u$ is large) or a larger number of smaller steps (when $u$ is small), rendering the RGDM less sensitive than the RGTM to $u$. Overall, in RGTMs, there is no significant LIA difference between the geometry-sensitive and the topology-sensitive zone when $\sigma < 0.2$. Selecting a summary statistic that emphasizes only local neighborhood (local extractor) can be significantly worse than selecting the geometry-sensitive or intermediate statistics, consistent with what we observed for the human fMRI data. This suggests that, in a low-noise regime, some of the information in the RDM is redundant and the extremes of the distance distribution can be compressed without compromising our power to distinguish representations of different computational modules (layers here). In the high-noise regime, having the full RDM information can better characterize the subtle distinctions among computational modules.

To further interpret the distributions of representations associated with different layers and instances, we performed multi-dimensional scaling (MDS) on the RDMs, RGTMs, and RGDMs (using the optimal threshold pairs $(l,u)$ marked in red in the corresponding heatmaps in Fig. \ref{trsa_fig6}b). We used the Procrustes alignment \cite{procrustes} to obtain a consistent orientation and reflection of the MDS plots (Fig. \ref{trsa_fig6}c). We observe that the RDMs cluster the representations of several layers together, whereas the RGTMs and RGDMs manage to separate out layers in a smoother progression from the early layers to the later layers. This suggests that the geo-topological and geodesic transforms can help emphasize topological invariants that consistently characterize layer representations across variation among model instances.

\section{Understanding Empirical Results}

Theories about neural codes and computations can predict how the neural manifold hangs together (i.e. the topology) without predicting a specific representational geometry. Such theories are consistent with an infinite number of distinct geometries, and so do not predict a specific RDM. Topological RSA enables researchers to evaluate and statistically compare such theories as expressed by a representational graph and associated RGTM. Our results do not support the conclusion that topological descriptors should replace geometrical descriptors in RSA in general. Studies that aim to adjudicate among brain-computational models that predict representational geometries can continue to use geometrical descriptors. Topology may not be needed as a tool to suppress nuisance variation in RSA analyses of representational geometries. However, the topology of neural representations is interesting in its own right.

For theories and brain-computational models that do predict geometries, evaluation on the basis of their topological predictions provides an alternative and complementary perspective. If our models' RDM predictions do not reach the noise ceiling, the model that best predicts the geometry may not be the model that best predicts the topology. We can simply apply a geo-topological transform to the brain RDM before performing model-comparative RSA inference with a rank-based comparator. The choice of parameters $l$ and $u$ (which can be defined as percentiles within the dissimilarities of each RDM) enables us to calibrate the relative sensitivity to the topological and geometrical properties of the representation. This approach enables comparisons of both topology-predicting and geometry-predicting models within the same tRSA framework. The fact that region identification and layer identification do not suffer when we drastically reduce the information in the RDM suggests that the bulk of the information distinguishing the representations in brain regions and NN layers is captured by topological descriptors.

\subsection{Characterizing topology and geometry provides a comprehensive view of neural representations} 

Considering topology-sensitive representational summary statistics alongside geometrical ones can provide a more comprehensive and nuanced understanding of the data. We introduced geo-topological descriptors as a new class of summary statistics for RSA. These descriptors emphasize the topology by compressing variation among the smallest representational distances and among the largest representational distances. Variation among small representational distances is sensitive to noise, and variation among very large distances may reflect individual idiosyncrasies more than computationally meaningful information. While tRSA did not outperform conventional RSA at region or layer identification, it matched the model-selection performance of RSA while substantially reducing the information, thus revealing what range of variation among representational distances captures the discriminable computational signatures of different representations. Topological RSA provides a robust way to identify  neural representations in spite of noise and interindividual variation. We emphasize the synergistic potential of combining the topological and geometrical perspectives. 

\subsection{Testing topological representational hypotheses requires tRSA} 

From a theoretical perspective, the choice of using topological or geometrical summary statistics in RSA depends on the hypotheses about the neural representations that are to be tested. If the hypothesis is that a neural representation conforms to a particular topology, then topological RSA provides a straightforward method to test the hypothesis. Note that a hypothesis about the representational topology cannot be straightforwardly tested with conventional RSA because the hypothesis corresponds to a complex set of RDMs any of which conforms to the hypothesized topology. This is a major motivation for tRSA independent of the question of whether it can help reduce nuisance variation in adjudicating among models that predict geometries.

Do theories about neural mechanisms imply specific predictions about the representational topology? Investigating the relationship between brain-computational theories and the topology of neural population representations is an important direction for future computational work. 
Here we focused on visual representations in both the model-based simulations and the fMRI dataset, where the representational space has many dimensions and we do not have simple topological hypotheses about the structure of the representations. In the absence of simple topological hypotheses, a data-driven approach can reveal to what extent the representational topologies are consistent between individuals and distinct across stages of processing or across stages of learning or development. The relationships among brain-representational topologies can be visualized using MDS of the RGTMs or RGDMs as we show for DNN layers in Fig. \ref{trsa_fig6}c. Such analyses have been performed previously for RDMs (e.g. \cite{kriegeskorte2008representational, visconti2017neural} and require independent data for each region to prevent correlated noise fluctuations from confounding the RDM estimates as shown in \cite{henriksson2015visual}. Future studies could use tRSA to test strong predictions about the topology of lower-dimensional stimulus sets sampling, e.g., orientations and spatial frequencies of gratings, or directions and velocities of visual motion, which are known to be represented at different levels of the primate visual hierarchy. Another example where tRSA could support a strongly theory-driven approach seeking to adjudicate among alternative topological hypotheses is the head and travelling direction system in mice and fruit flies \cite{chaudhuri2019intrinsic,langdon2023unifying,kim2019generation,lyu2022building}.

\subsection{Testing geometrical representational hypotheses may benefit from tRSA}

If the models to be tested predict specific representational geometries, conventional RSA can be used to adjudicate among them. Topological descriptors are functions of geometrical ones and reduce the information, so may appear undesirable if geometrical hypotheses are to be tested. Consider the example of a set of visual gratings covering the entire cycle of orientations with equal spacing. For a wide variety of hypothetical representations, we expect the topology to be that of a circle, whereas the geometry may not be that of a circle, for example, if the neural code is anisotropic, more precisely representing cardinal orientations (which are more frequent in natural images) \cite{kriegeskorte2021neural,appelle1972perception,girshick2011cardinal}. The topology, in that case, fails to capture an important feature of the neural code. In other scenarios, however, a topological descriptor may capture the essential features of the neural code. If our topological descriptor reduces nuisance variation in the data more than the representational signatures that distinguish different models, it can support improved model-adjudication accuracy. For the visual representations investigated here in models (deep neural networks) and humans (fMRI), we did not find evidence for such an advantage. Future studies will reveal if a focus on the topology reduces the noise more than the signal in in other circumstances. 
 
However, even if our models make predictions about representational geometries (as do DNNs) and tRSA does not improve model-adjudication power, it can still provide complementary information by revealing to what extent the topology implicit to the geometrical predictions enables model adjudication and which model best predicts the representational topology. If the best model fully explains the geometry, it will also correctly predict the topology. However, the relative performance of models that do not fully explain the geometry can change when we evaluate models by their topological predictions. Our analyses here showed that tRSA, using reduced topological summaries, can match the performance of geometrical RSA at model adjudication. This suggests that topological descriptors that compress the variation among small and among large representational distances capture the information essential to the computational signature of different representations.



\subsection{Practical considerations for implementing tRSA}

Topological RSA model comparison can rely on the RSA3 statistical inference framework \cite{schutt2023statistical}, which has been implemented in Python in the open-source \href{https://github.com/rsagroup/rsatoolbox}{RSA Toolbox}). The geo-topological descriptors introduced here and the RSA3 inference methods enable analyses based on a wide range of brain-activity data, including invasive neural recordings, fMRI, EEG, and MEG. A topological representational hypothesis can be expressed directly in a representational graph and associated RGTM. Alternatively, topological hypotheses can be derived from representational models that predict geometries, i.e., any descriptors of the experimental conditions. If the conditions correspond to visual stimuli, for example, representational models can be derived from properties of the images or of the objects depicted and their categories \cite{kriegeskorte2008matching}. In particular, topological hypotheses can be derived from brain-computational models, such as neural network models that implement candidate hypotheses about the computations performed by the brain and predict a geometry for each stage of representation \cite{kriegeskorte2015deep, yamins2016using, kriegeskorte2018cognitive}.

If the theories to be compared predict representational topologies (RGTMs), then we can avoid having to choose $l$ and $u$. We can compare the RGTMs predicted by the theories to the brain RDM directly using the $\rho_a$ rank correlation coefficient \cite{schutt2023statistical}. If the theories to be compared predict geometries (RDMs), but we intend to compare models in terms of their predictions of the topology of the brain representation, then we can convert the brain RDM to a brain RGTM by selecting $l$ and $u$ to choose the desired calibration of the summary statistic to the geometry and the topology of the brain representation. The brain RGTM can then be compared to the model RDMs using the $\rho_a$ rank correlation coefficient and model-comparative inference can proceed on this basis. Finally, if the goal is to assess whether the geometrical predictions of a model (the model RDM) significantly exceed its topological predictions (model RGTM, based on a choice of $l$ and $u$), we can use the $\rho_a$ rank correlations with the brain RDM to statistically compare the model's topological and geometrical predictions.

The values for $l$ and $u$ can be set by a priori considerations, e.g., on the basis of the level of graph connectivity predicted by the theories to be evaluated, or on the basis of results of earlier studies that use a different data set (e.g., this study). If different settings of $l$ and $u$ are to be explored in the analyses, it is important to avoid biases to the analyses that can result from selecting these two parameters \cite{kriegeskorte2009circular, hosseini2020tried}. One approach is to choose a small number of settings for $l$ and $u$ and add each resulting RGTM as a separate model in the model-comparative inference, where standard adjustments for multiple testing can then be used (controlling the family-wise error rate or the false-discovery rate as implemented in the \href{https://github.com/rsagroup/rsatoolbox}{RSA Toolbox}). Alternatively, an independent analysis, for layer- or brain-region identification accuracy, could inform the choice of $l$ and $u$. Importantly, whatever exploration or search procedures were employed in selecting $l$ and $u$ must be fully reported.

It is important to note that both topological and geometrical descriptors depend on the experimental conditions (e.g. stimuli) for which activity patterns are included in the analysis. However, topological descriptors can be even more sensitive to the set of conditions included. For example, the geodesic distance between the representations of two conditions depends on the other conditions included (which determine the shortest path), whereas the Mahalanobis distance does not.

\subsection{Future directions: Topological RSA for model adjudication}

While this study has demonstrated the effectiveness of tRSA in ground-truth-known region and layer identification tasks, the application of tRSA to the problem of statistically comparing competing brain-computational models on the basis of brain-activity data is still outstanding. This application represents an important next step and crucial frontier for tRSA research. Future studies should focus on:

\begin{itemize}
    \item Developing formal frameworks for using tRSA in model comparison, including appropriate statistical tests and model selection criteria.
\item Comparing the sensitivity of tRSA and classical RSA in detecting subtle differences between competing models.
\item Exploring how tRSA can be integrated with other model evaluation techniques to provide a more comprehensive assessment of brain-computational models.

\end{itemize}

This line of research is essential for fully realizing the potential of tRSA as a tool for advancing our understanding of neural computation.

In conclusion, tRSA is essential for testing topological representational hypotheses. Even when testing models that predict representational geometries, tRSA can reveal to what extent the models capture, in particular, the representational topology observed in a neural population. The combination of topological and geometrical descriptors offers a promising comprehensive approach for analyzing neural representations.

\chapter{From Geo-Topological Summary Statistics to Adaptive Dependence Measure}


As we continue our exploration of neural representations, we now turn our attention to a powerful extension of the concepts introduced in the previous chapters. This chapter bridges the gap between the geo-topological summary statistics of tRSA and the fundamental statistical concept of detecting the statistical dependence between random variables.

The journey we embark on in this chapter takes us from the specialized domain of neuroscience into the broader realm of statistical analysis. We'll see how the insights gained from tRSA can be generalized to create a novel dependence measure that is both adaptive and sensitive to complex relationships in multivariate data.

This chapter represents a significant leap in the applicability of our work, demonstrating how techniques developed for analyzing neural representations can have far-reaching implications in data science and statistics. As we progress, we'll explore the mathematical foundations of this new measure
and its empirical performance across a range of scenarios.

By the end of this chapter, readers will understand not only how this dependence measure works, but also its connections to the geo-topological concepts introduced earlier in the thesis. This connection underscores the broader impact of our research beyond neuroscience, opening new avenues for detecting and characterizing statistical dependencies in diverse fields.

This chapter leads to the following publication: 

\cite{lin2018adgtic} \fullcite{lin2018adgtic}.


\section{Search for the Optimal Statistical Dependence Measure for Independence Testing}
\label{sec:intro}

Detecting statistical dependence between random variables is a fundamental problem of statistics. The simplest scenario is detecting linear or monotonic univariate relationships, where Pearson's r, Spearman's $\rho$, or Kendall's $\tau$ can serve as test statistics. Often researchers need to detect nonlinear relationships between multivariate variables. In recent years, many nonlinear statistical dependence indicators have been developed: distance-based methods such as distance or Brownian correlation (dCor) \cite{szekely2007measuring,szekely2009brownian}, mutual information (MI)-based methods with different estimators \cite{kraskov2004estimating, pal2010estimation,steuer2002mutual}, kernel-based methods such as the Hilbert-Schmidt Independence Criterion (HSIC) \cite{gretton2005measuring, gretton2008kernel} and Finite Set Independence Criterion (FSIC) \cite{jitkrittum2016adaptive}, and other dependence measures including Maximal Information Coefficient (MIC) \cite{reshef2011detecting,reshef2013equitability}, Multiscale Graph Correlation (MGC) \cite{vogelstein2019discovering} and HHG’s test (HHG) \cite{heller2013consistent}. 

There's no free lunch: any indicator will outperform any other indicator given data whose dependence structure it is better suited to detect. However, it is desirable to develop indicators that adapt to the grain of the dependency structure and to the amount of data available to maintain robust power across relationships found in real applications. Except for FSIC, the established methods are not adaptive. Some of them are sensitive to the setting of hyperparameters, or have low statistical power for detecting important nonlinear or high-dimensional relationships \cite{simon2014comment}. 

\begin{figure}[tb]
\centering
\includegraphics[width=0.24\linewidth]{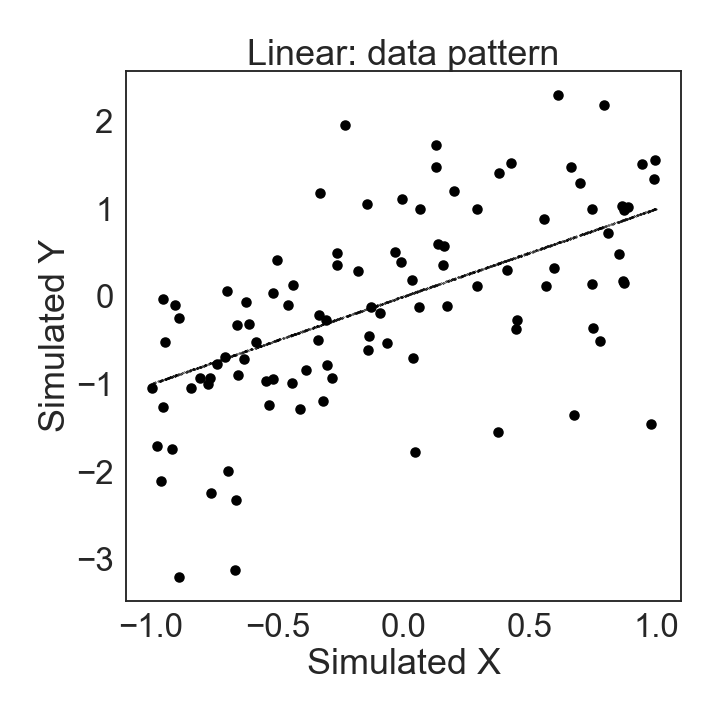}
\includegraphics[width=0.24\linewidth]{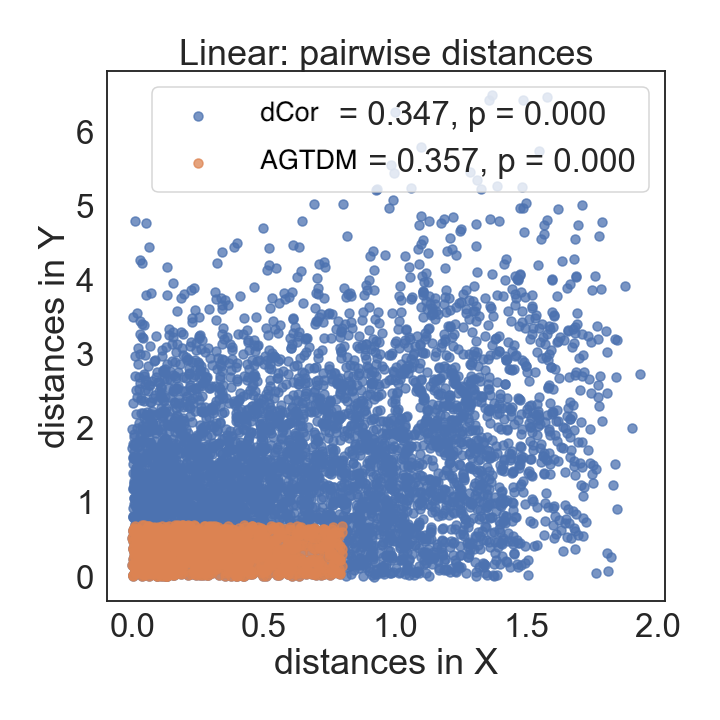}
\includegraphics[width=0.24\linewidth]{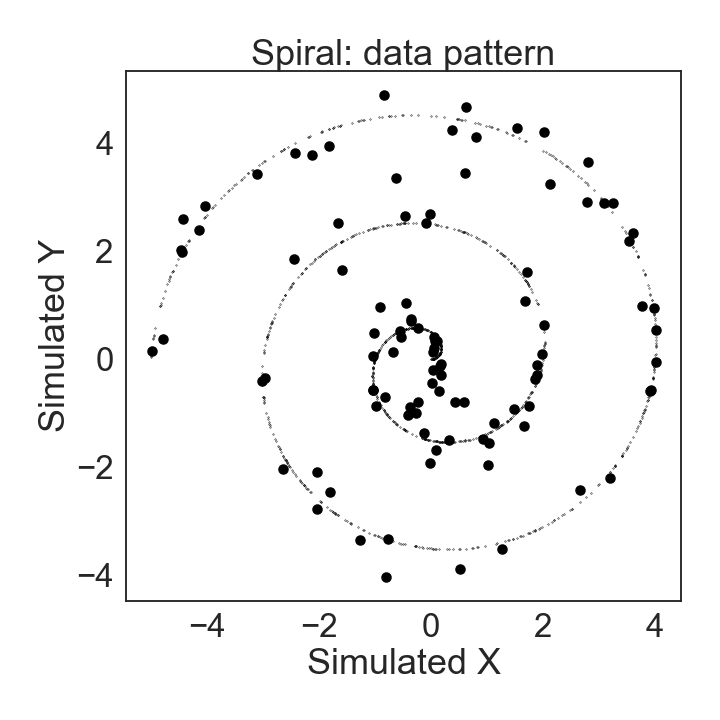}
\includegraphics[width=0.24\linewidth]{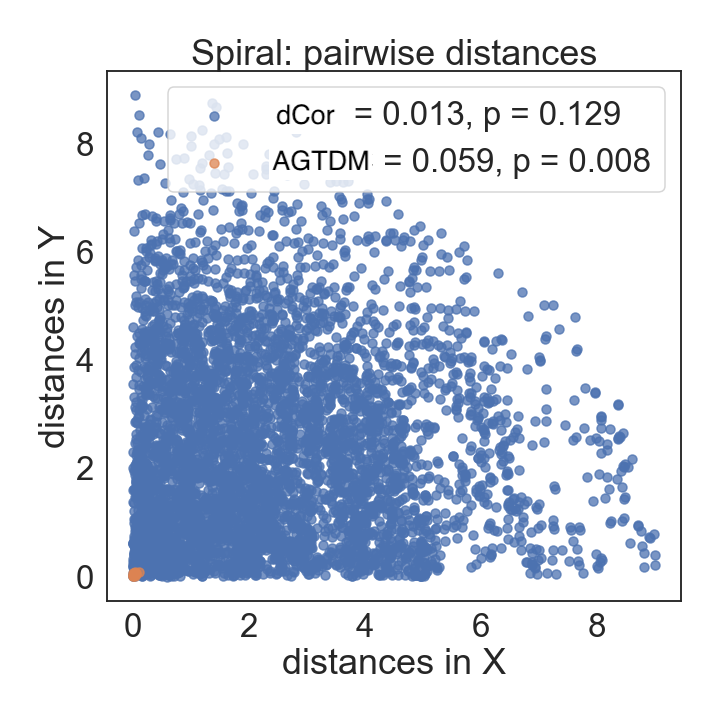}
\caption{\textbf{Motivation of a geo-topological dependence measure.} In this example, we present three multivariate patterns. The distance correlation, dCor uses all the original distances (blue dots) to compute its test statistic and p-value, whereas AGTDM transforms the smaller distances to zero, the larger distances to the maximum, and the rest (orange dots) to be distinctive from one another in the original scale. By reshaping the original distances and emphasizing certain distances, AGTDM discovers the dependency in the spiral pattern that dCor misses.}\label{fig:example}
\end{figure}

Here we propose a family of adaptive distance-based dependence measure inspired by two ideas: (1) Representational geometries can be compared by correlating distance matrices \cite{kriegeskorte2013representational}. (2) We can relax the constraint of linear correlation of the distances by nonlinearly transforming distance matrices, such that they capture primarily neighbor relationships. Such a transformed (e.g. thresholded) distance matrix captures the topology, rather than the geometry. Detecting matching topologies between two spaces $\mathcal{X}$ and $\mathcal{Y}$ will indicate statistical dependency. As illustrated in Fig. \ref{fig:example}, given a specific multivariate associate pattern, the proposed Adaptive Geo-Topological Dependence Measure (AGTDM) transforms the pairwise distances that are too small or too big, only keeping a subset of the original distances as the matching topology. In the presented case, smaller distances are the most distinctive topological edges in the spiral pattern, which is not the case in the linear pattern. 

The geo-topological indicators are based on the distance correlation, computed after a parameterized monotonic transformation of the distance matrices for spaces $\mathcal{X}$ and $\mathcal{Y}$. We use an adaptive search framework to automatically select the parameters of the monotonic transform so as to maximize the distance correlation. 
The adaptive threshold search renders the dependence test robustly powerful across a wide spectrum of scenarios and across different noise amplitudes and sample sizes, while guaranteeing (via permutation test) the specificity at a false positive rate of 5\%. 


\subsection{Similar Insights into the Geometric and Topological Properties of Data}

The motivation of this line of work is similar to that of the topological RSA framework introduced before. As in many representational data, the multivariate relationship we wish to investigate here is high-dimensional, noisy, nonlinear, and most likely non-stationary. Neuroscience research, for instance, generates high-dimensional functional magnetic resonance imaging (fMRI) with systematic noises from head movement, the heart beats, or breathing, in different textures across subjects. The specialized application of independence testing in cognitive neuroscience is the Representational similarity analysis (RSA), which aims to find dependence patterns within brain-activity measurement, behavioral measurement, and computational models \cite{kriegeskorte2008representational}. Like traditional statistics such as dCor, RSA computed pairwise distances (or dissimilarities) between neural activities among different stimuli (e.g. images shown to a subject). These dissimilary matrices are usually considered as the representational \textit{geometry}. 

\textit{Topology}, on the other hand, has different definitions in different contexts. In the context of computational topology, the analysis is often accomplished by topological data analysis (TDA), which is a successful method to discover patterns and meanings in the shape of the data \cite{epstein2011topological}. For example, the persistence homology diagrams can help reveal the most fundamental shape of the data; the Mapper algorithm is able to transform any data (such as point-cloud data) or function output (such as a similarity measure) into a graph which provides a compressed summary of the data distributions and association patterns \cite{singh2007topological}. In the context of our investigation, topology is defined as the consistent dissimilarities that carry multivariate dependence, an abstraction of the representational space independent of systematic noises from data collection procedures, while geometry is defined as the dissimilarity distances. We define the \textit{Geo-Topology}, as the transformed geometry ``denoised'' to capture only the dependence-relevant dissimilarity, to be conceptually considered a topology.  

There are two motivations for this in applications like neural data, one theoretical and one data-analytical. From a theoretical perspective, the computational function of a brain region might depend more on the local than on the global representational geometry, i.e. on differences among small representational distances rather than differences among large representational distances. The local geometry determines, which stimuli the representation renders indiscriminable, which it discriminates, but places together in a cluster, and which it places in different neighborhoods. The global geometry of the clusters (whether two stimuli are far or very far from each other in the representational space) may be less relevant to computation: In a high-dimensional space a set of randomly placed clusters will tend to afford linear separation of arbitrary dichotomies among clusters \cite{kushnir2018} independent of the exact global geometry. Like a storage room, a representational space may need to collocalize related things, while the global location of these categories may be arbitrary.

\begin{figure}[tb]
\centering
\includegraphics[width=1.0\linewidth]{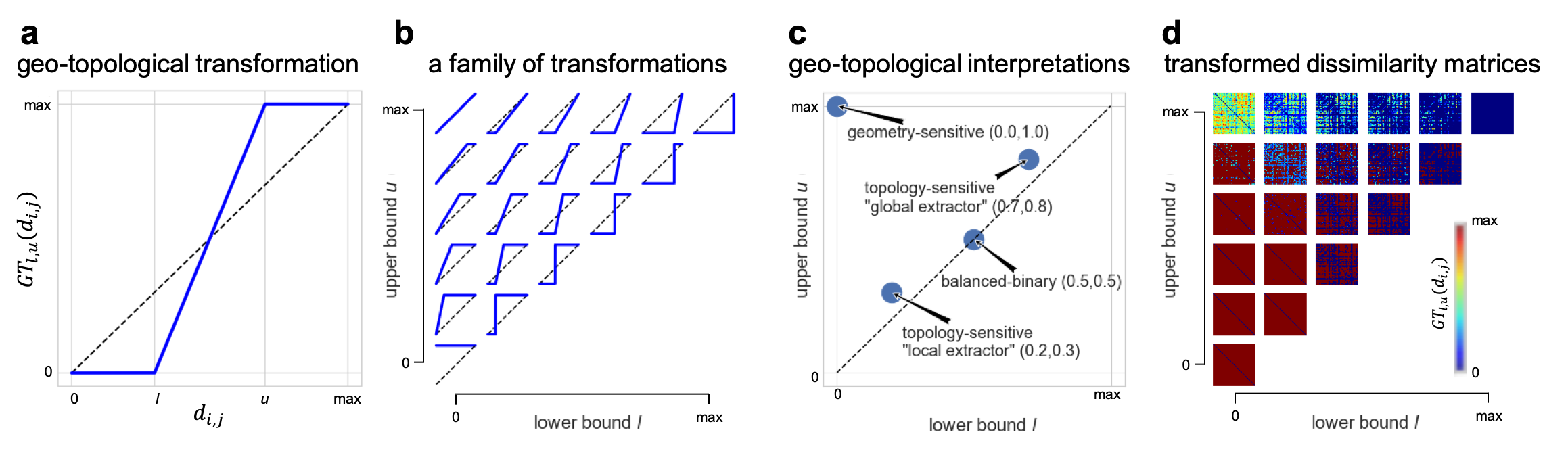}
\par\caption{\textbf{Geo-Topological (GT) Transforms in AGTDM}. (\textbf{a}) Similar to the GT Transforms in tRSA, to discard geometrical information that is either not meaningful or unreliable, we subject each distance $d_{i,j}$ for stimuli $i$ and $j$ to a \textit{monotonic transform} $GT_{l,u} (d_{i,j})$. (\textbf{b}) We refer to this family of transforms as geo-topological, because it combines aspects of geometry and topology. (\textbf{c}) Depending on the choice of a lower and upper bound, the transform can threshold ($l =u$) at an arbitrary level, adapting to the granularity of a data set. It can also preserve all ($l =0$, $u=\max$) or some ($l<u$) geometrical information. (\textbf{d}) monotonically transformed distance matrices under a set of threshold pairs.}\label{fig:RGTA}
\end{figure}

From a data-analytical perspective, conversely, small distances may be unreliable given the various noise sources that may affect the measurements. From both theoretical and data-analytical perspectives, it seems possible that focusing our sensitivity on a particular range of distances turns out to be advantageous because is reduces the influence of noise and/or arbitrary variability (e.g., of the global geometry) that does not reflect dependency function. In order to suppress these noise, we would like to find a lower distance threshold $l$ below which we consider certain data points as co-localized (i.e., the points have collapsed into the same node in the graph). Between the two thresholds we place a continuous linear transition to retain some geometrical sensitivity in the range where it matters, as in Fig. \ref{fig:RGTA}. This formulation encapsulates the special case of full geometry: one possibility is that the ideal setting is $l =0$, $u=max$, i.e., the original distances.

The novel contribution is the idea to extract the geometry and topology from pairwise distances for independence testing. In the following sections, we introduce AGTDM, an attractive adaptive dependence measure, and the empirical merit of its robust power to detect different kinds of statistical dependency.

\section{Adaptive Geo-Topological Dependence Measure (AGTDM)}
\subsection{Primers and definitions of AGTDM}

\textbf{Problem description.} Let $\mathbbm{P}_x$ and $\mathbbm{P}_y$ be the marginal distributions on space $\mathcal{X}$ and $\mathcal{Y}$ and $\mathbbm{P}_{xy}$ be a Borel probability measure defined on their domain $\mathcal{X} \times \mathcal{Y}$. Given the independent and identically distributed (i.i.d.) sample $Z:=(X, Y)=\{(x_1,y_1),\cdots,(x_m,y_m)\}$ of size $m$ drawn independently and identically distributed according to $\mathbbm{P}_{xy}$, with each row corresponding to an observation of both variables, the statistical test $\mathcal{T}(Z): (\mathcal{X} \times \mathcal{Y} \mapsto \{0,1\})$ is used to distinguish between the null hypothesis $\mathcal{H}_0: \mathbbm{P}_{xy}=\mathbbm{P}_x\mathbbm{P}_y$ and the alternative hypothesis $\mathcal{H}_1: \mathbbm{P}_{xy}\not=\mathbbm{P}_x\mathbbm{P}_y$. 

\textbf{Distance correlation (dCor).}
Distance covariance was introduced by \cite{szekely2007measuring} to test dependence between random variables $X$ and $Y$ with finite first moments  in space $\mathcal{X}$ and $\mathcal{Y}$, computed in terms of a weighted $L_2$ norm between the characteristic functions of the joint distribution of $X$ and $Y$ and the product of their marginals, computed in terms of certain expectations of pairwise Euclidean distances: 

\begin{align}
\label{eq:dCov}
\begin{split}
\mathcal{V}^2(X,Y) =& \EX[\norm{X-X'}\norm{Y-Y'}] \\
&+ \EX[\norm{X-X'}]\EX[\norm{Y-Y'}] \\
&- 2\EX[\norm{X-X'}\norm{Y-Y''}]
\end{split}
\end{align}

where $\EX$ denotes expected values, $(X,Y)$ and $(X',Y')$ are drawn i.i.d from $\mathbbm{P}_{xy}$, primed random variables $(X',Y')$ and $(X'',Y'')$ are i.i.d. copies of the variables $X$ and $Y$. Distance correlation (dCor) is obtained by dividing $\mathcal{V}^2(X,Y)$ by the product of their distance standard deviations:

\begin{equation}
\label{eq:dCor}
dCor(X,Y) = \frac{\mathcal{V}^2(X,Y)}{\sqrt{\mathcal{V}^2(X,X)\mathcal{V}^2(Y,Y)}}
\end{equation}

\cite{lyons2013distance} showed that if metrics $\rho \mathcal{X}$ and $\rho \mathcal{Y}$ satisfy strong negative type, the distance correlation in a metric space characterizes independence: $\mathcal{V}^2_{\rho \mathcal{X},\rho\mathcal{Y}}(X,Y)=0 \Leftrightarrow X$ and $Y$ are independent.


\textbf{Geo-Topological transform (GT).} Suppose $d_{i,j}$ is the distance between two sample observation $(x_i,x_j) \overset{i.i.d}\sim X$. Let $GT(d_{i,j})$ be the general form of a nonlinear monotonic transformation parameterized by two positive real numbers $l$ and $u$ satisfying $l<u$. Let $d_{\max} = \max_{i,j\in (1,\cdots,m)} d_{i,j}$ be the largest pairwise distance in space $X$. Here we define $GT_{l,u}(d_{i,j})$ to be the simplified version of the general GT transform as a continuous nonlinear \textit{bounded} \textit{functional} $f(d;l,u,d_{\max})$ onto $L^2[0,1]$:

\begin{equation}
f(d;l,u) =
  \begin{cases}
0 & \text{if $0 \leq d < l$} \\
d_{\max}\cdot\frac{d-l}{u-l} & \text{if $l \leq d < u$} \\
d_{\max} & \text{if $u \leq d$}
  \end{cases}
\end{equation}

as the empirical choice for our test statistics. 
Fig. \ref{fig:RGTA} offered an illustration of the effect of a set of parameter pairs $(l,u)$ on the $GT_{l,u}(d_{i,j})$ function, the distance matrices as well as data-driven interpretations of the lower bound $l$ and the upper bound $u$ for this stepwise function.

\textbf{Adaptive Geo-Topological Dependence Measure (AGTDM).} 

Before we define our statistics, we always assume the following regularity conditions: (1) $(X, Y)$ have finite second moments, (2) neither random variable is a constant and (3) $(X, Y)$ are continuous random variables, which are also required by dCor to establish convergence and consistency. Since we are using the population definition of the distance correlation, the nonconstant condition ensures a more stable behavior and avoids the trivial case. Given a Geo-Topological Transform $f(\cdot) := GT(\cdot;l,u)$, the population expression for the GT-transformed distance covariance can be defined as:

\begin{align}
\label{eq:dCovf}
\begin{split}
\mathcal{V}^{2*}(X,Y;f) = &\EX[f(\norm{X-X'})f(\norm{Y-Y'})] \\
&+\EX[f(\norm{X-X'})]\EX[f(\norm{Y-Y'})]  \\
&- 2\EX[f(\norm{X-X'})f(\norm{Y-Y''})]
\end{split}
\end{align}

where the same set of parameters $l$ and $u$ applies to the monotonic transforms on all the distances. The GT-transformed distance correlation is then:

\begin{equation}
\label{eq:dCovf}
dCor^*(X,Y;f) = \frac{\mathcal{V}^{2*}(X,Y;f)}{\sqrt{\mathcal{V}^{2*}(X,X;f)\mathcal{V}^{2*}(Y,Y;f)}}
\end{equation}

We can naturally define AGTDM to be the maximum GT-transformed distance correlation within the parameter domain $\mathcal{S} := \{l\in[0,1), u\in(0,1], l<u\}$:

\begin{equation}
\label{eq:AGTDM}
AGTDM(X,Y) = \max_{(l,u)\in\mathcal{S}} dCor^*(X,Y;GT(\cdot;l,u))
\end{equation}

\textbf{Test description.} 
The statistical test of independence can be performed by locating the test statistic in its distribution under $\mathcal{H}_0$ using a permutation procedure \cite{gretton2008kernel}. As the preliminary, given the i.i.d. sample $\mathcal{Z}=(X,Y)=\{(x_1,y_1),\cdots,(x_m,y_m)\}$ defined earlier, the statistical test $\mathcal{T}(Z):=(\mathcal{X}\times\mathcal{Y})^m\rightarrow \{0,1\}$ distinguishes between the null hypothesis $\mathcal{H}_0: \mathbbm{P}_{xy}=\mathbbm{P}_x\mathbbm{P}_y$ and the alternative hypothesis $\mathcal{H}_1: \mathbbm{P}_{xy}\not=\mathbbm{P}_x\mathbbm{P}_y$. This is achieved by comparing the test statistic, in our case $AGTDM(Z)$, with a particular threshold: if the threshold is exceeded, then the test rejects the null hypothesis. The permutation test involves the following steps. Based on a finite sample, incorrect answers can yield two kinds of errors: the Type I error is the probability of rejecting $\mathcal{H}_0$ when $x$ and $y$ are in fact independent, and the Type II error is the probability of accepting $\mathbbm{P}_{xy}\not=\mathbbm{P}_x\mathbbm{P}_y$ when in fact the underlying variables are dependent. To obtain an estimate of the Type I and Type II error, we need to create a null distribution of $Z$, for instance, by shuffling the labels of $X$ or $Y$ such that their one-to-one correspondence are now disconnected, and thus, independent. By computing the statistics multiple times on the null distribution the original dataset, we obtain the null distribution for our test statistics $AGTDM(\text{null}(Z))$, where we can align our test statistics computed for the real data $AGTDM(Z)$ to get Type I error and Type II error. For instance, if our computed $AGTDM(Z)$ is larger than 97\% of the $AGTDM(\text{null}(Z))$, then the Type I error. In practice, we specify a cutoff $\alpha$ for the false positive rates to be the upper bound on the Type I error. We can further define the empirical statistical power as the fraction of true datasets yielding a statistic value greater than 95\% of the values yielded by the corresponding null datasets, with a theoretical guarantee that the false positive rate is below $\alpha =$ 5\%.

\begin{figure}[tb]
\centering
    \includegraphics[width=\linewidth]{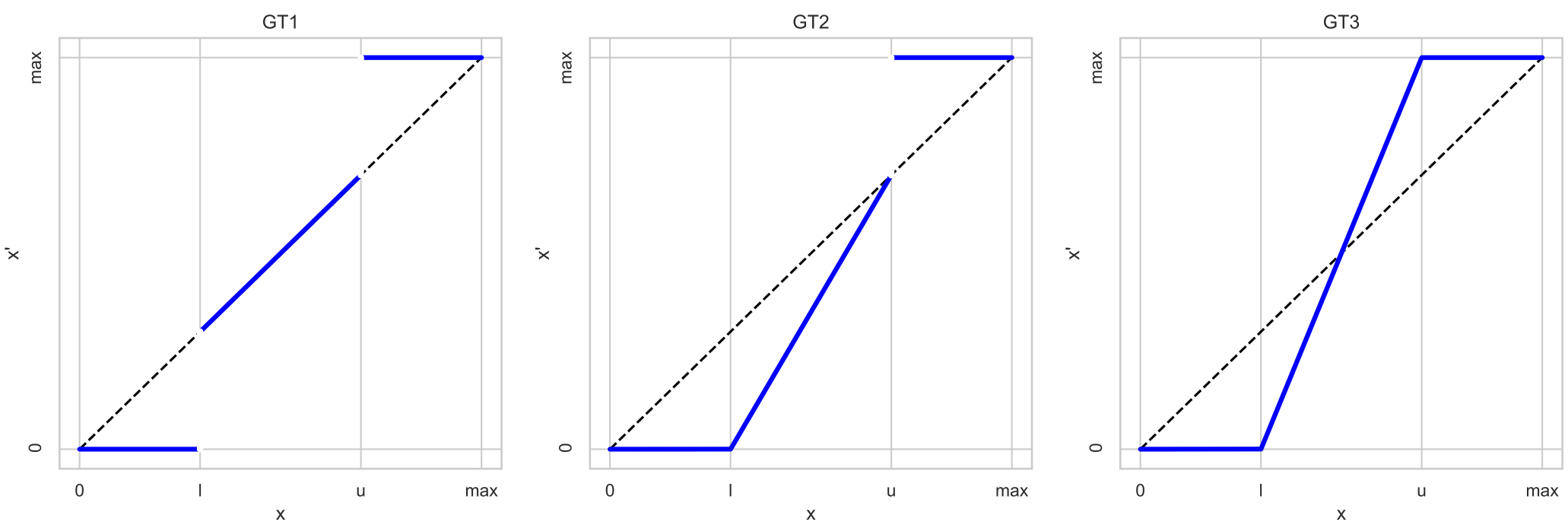}
\par\caption{\textbf{Different variants of GT transforms.} Demonstrated are several options of GT transforms in AGTDM.}\label{fig:GT}
\end{figure}

\textbf{Algorithmic variants.}
There are several algorithmic decisions one can make to compute AGTDM. Here we briefly describe the four types of geo-topological transforms and three types of subroutines used in the empirical evaluation. Instead of manually setting the parameters for the monotonic transformation, we adaptively selects threshold parameters, using the maximum of the statistics over the parameter domain to compute the test statistic. As primed earlier, the nonlinear monotonic transforms can be specified in different formats. One working hypothesis is that only local distances contribute to the mutual information, such that larger distances aren't as relevant to the global topology. In this setting, only small distances are counted as neighbors to form an edge within the topological graph. \textit{AGTDM - t0} sets the lower bound $l$ to zero, and only search for the optimal upper bound $u$ (here the ``optimal'' means yielding the maximum statistics). Fig. \ref{fig:GT} illustrated three other types of monotonic transform that we considered, denoted \textit{AGTDM - t1}, \textit{AGTDM - t2}, \textit{AGTDM - t3}. In this setting (also the default one), only pairs with intermediate distance are counted as neighbors. The logic behind this setup is that the dependence between very proximal data points can be more likely attribute to noise, therefore by ``discrediting'' these edges, we have a more stable topology. Other than the transforms, we also compared three variants of our test statistics, denoted \textit{AGTDM - s1}, \textit{AGTDM - s2}, \textit{AGTDM - s3} (``s'' stands for subroutine):  

\begin{equation}
\label{eq:AGTDMv}
\begin{split}
\mathcal{V}^{2*}_{s1} (X,Y) &= \max_{(l,u)\in\mathcal{S}} dCor^*(X,Y;GT(\cdot;l,u)) \\
\mathcal{V}^{2*}_{s2} (X,Y) &= \max_{(l,u)\in\mathcal{S}} \frac{dCor^*(X,Y;GT(\cdot;l,u))}{dCor^*(\text{null}(X),\text{null}(Y);GT(\cdot;l,u))} \\
\mathcal{V}^{2*}_{s3} (X,Y) &= \frac{\max_{(l,u)\in\mathcal{S}} dCor^*(X,Y;GT(\cdot;l,u))}{\sqrt{\text{Var}(\forall_{(l,u)\in\mathcal{S}} dCor^*(X,Y;GT(\cdot;l,u)))}}\\
\end{split}
\end{equation}

where \textit{AGTDM - s2} selects the maximum of the ratio of the transformed dCor for the dataset over the transformed dCor for the null dataset, and \textit{AGTDM - s3} includes an additional noise normalization procedure (dividing the maxmimum test statistics by the standard deviation of all the test statistics in the parameter domain). Last but not least, we also considered the case where the upper and lower thresholds were not cut-offs of a specific distance value, but the cut-off of a specific fraction of the ranked data. In another word, we set the $l$ and $u$ as percentiles instead of scales. We denote this variant \textit{pAGTDM} (where ``p'' stands for ``percentile'').

\textbf{Other weighting schemes.} 
It is worth exploring whether replacing the piece-wise linear transform with an alternative (e.g. differentiable) monotonic transform brings further improvements. However, we might expect such functions to behave similarly and prefer the simplicity of piece-wise linear functions and the fact that they make subsetting distances straightforward.




\textbf{Computational complexity.}
Here, we consider the most computationally demanding 
of the family, \textit{AGTDM - s3}, which consist of a combinatorial threshold space and a noise normalization. In the typical setup (very large sample size $m$ and small number of thresholds $k$), the computational complexity is dominated by the threshold searching with two variables. Hence, we achieve a cost in terms of the sample size of $\mathcal{O}(m^2(k(k-1)/2)^2) \approx \mathcal{O}(m^2k^4)$. In the special case of the distance covariance with univariate real-valued variables, \cite{huo2016fast} achieve an $\mathcal{O}(m\log m)$ cost for dCor computation, thus potentially reducing complexity for AGTDM to $\mathcal{O}(m\log m(k(k-1)/2)^2) \approx \mathcal{O}(mk^4\log m)$.

\begin{figure}[H]
\centering
    \includegraphics[width=0.24\linewidth]{Figures_agtic/Linear_data}
    \includegraphics[width=0.24\linewidth]{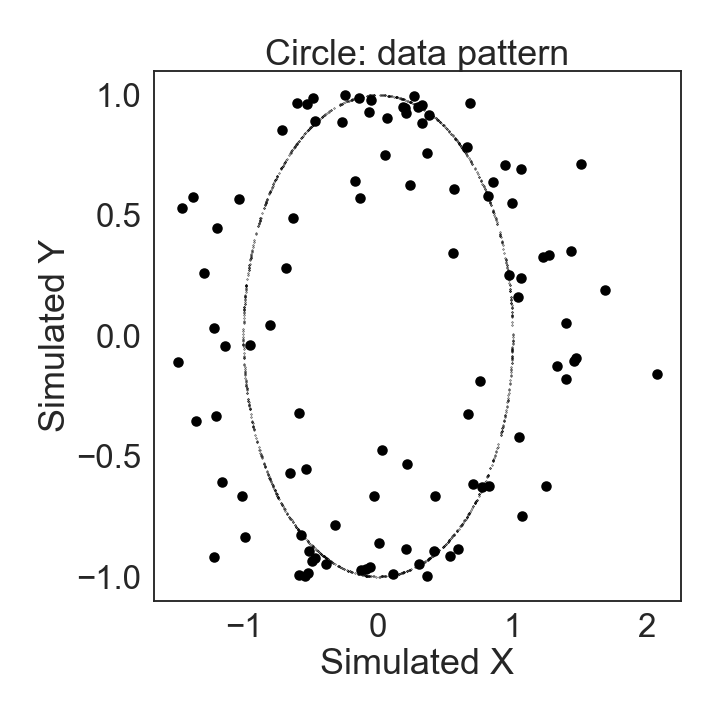}
    \includegraphics[width=0.24\linewidth]{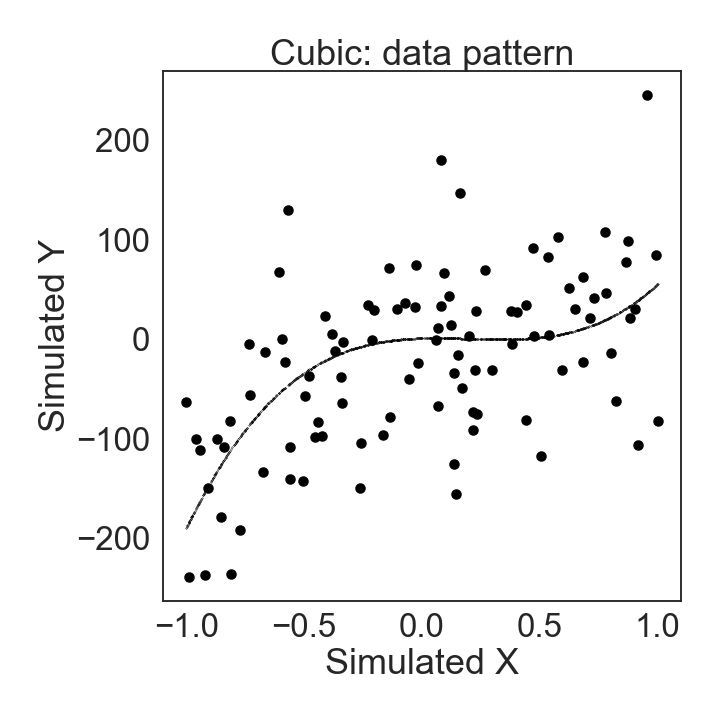}
    \includegraphics[width=0.24\linewidth]{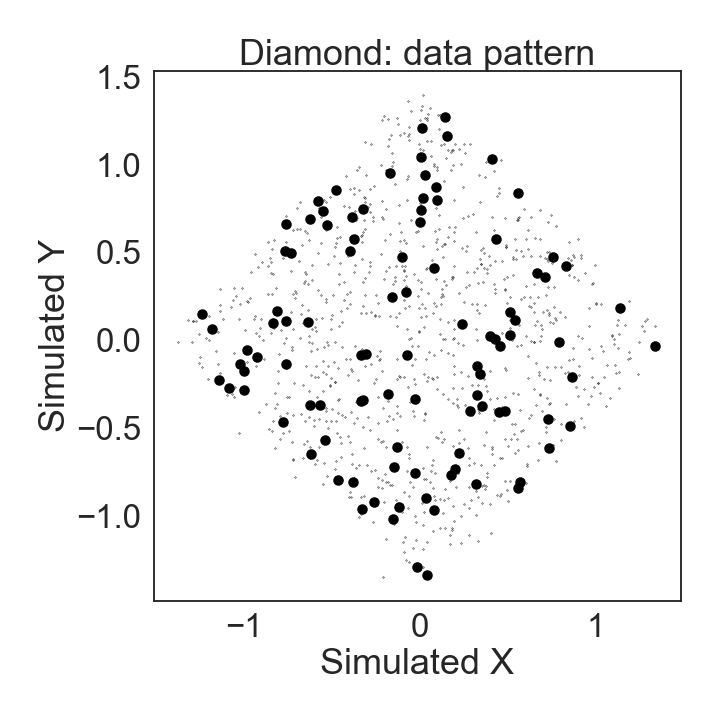}
    \includegraphics[width=0.24\linewidth]{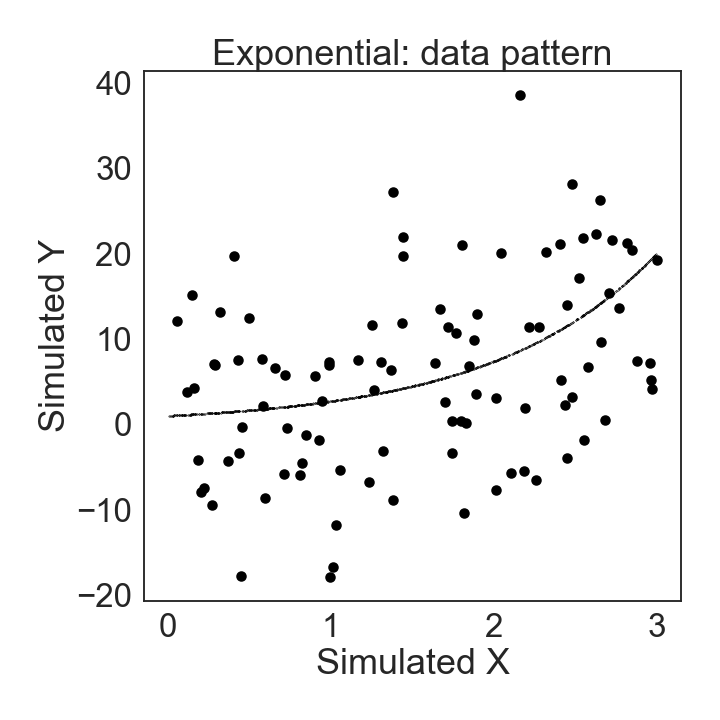}
    \includegraphics[width=0.24\linewidth]{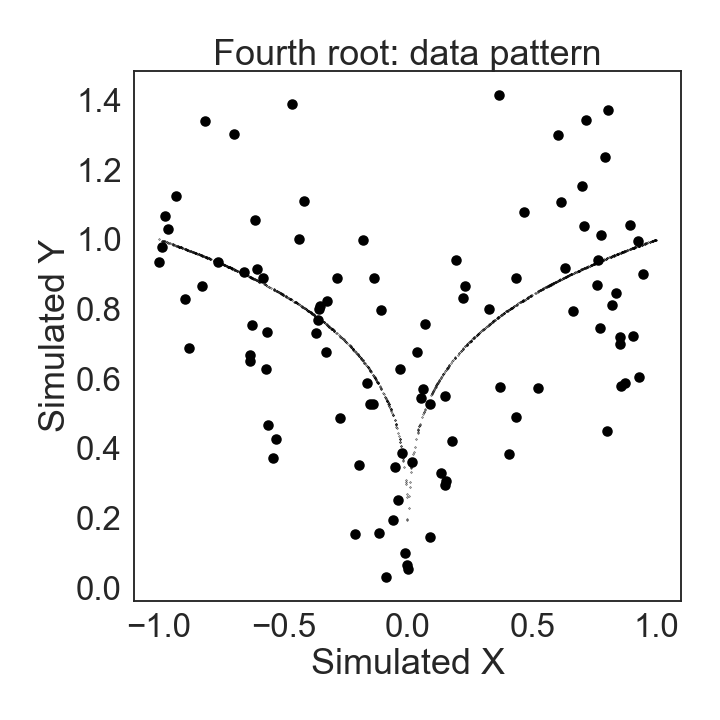}
    \includegraphics[width=0.24\linewidth]{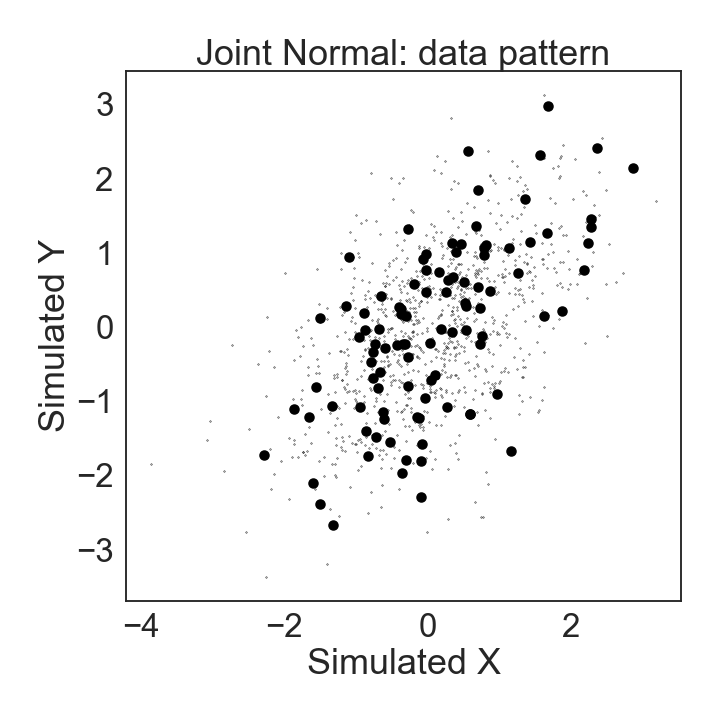}
    \includegraphics[width=0.24\linewidth]{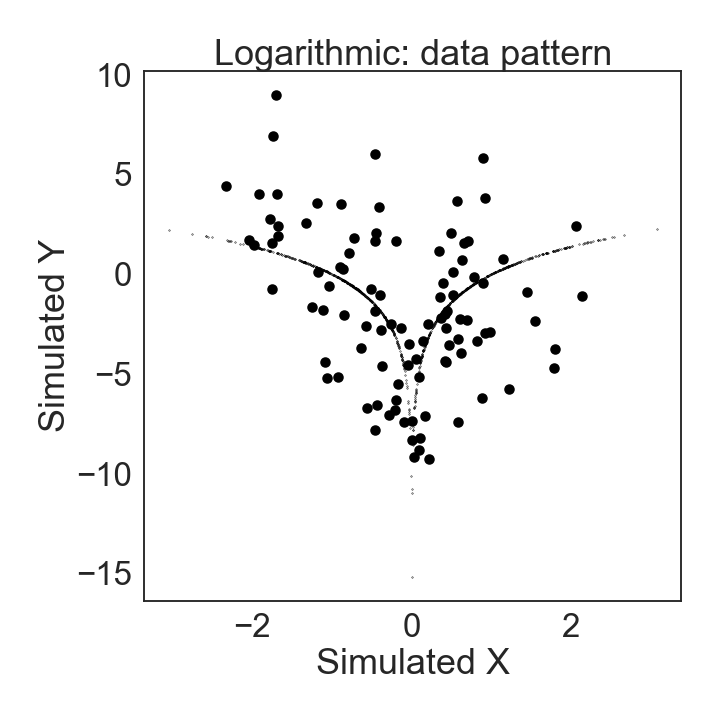}
    \includegraphics[width=0.24\linewidth]{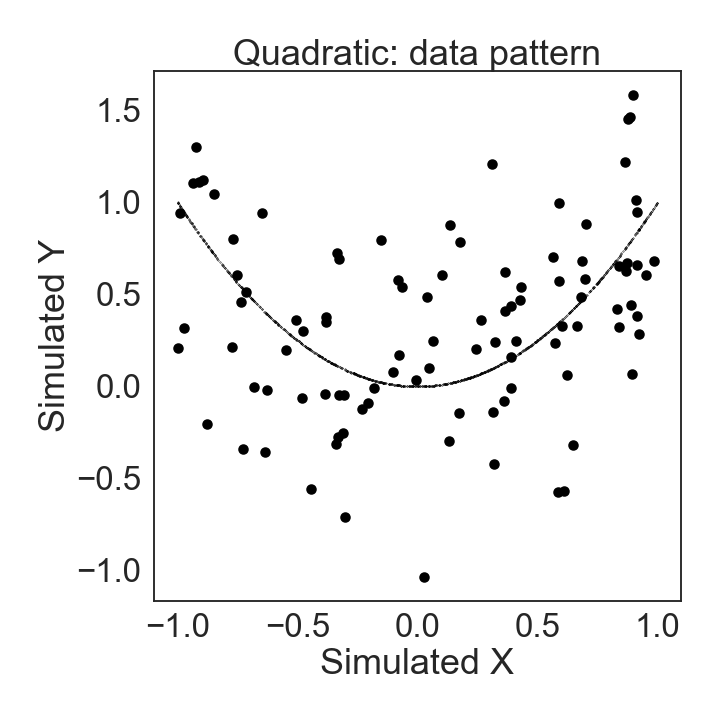}
    \includegraphics[width=0.24\linewidth]{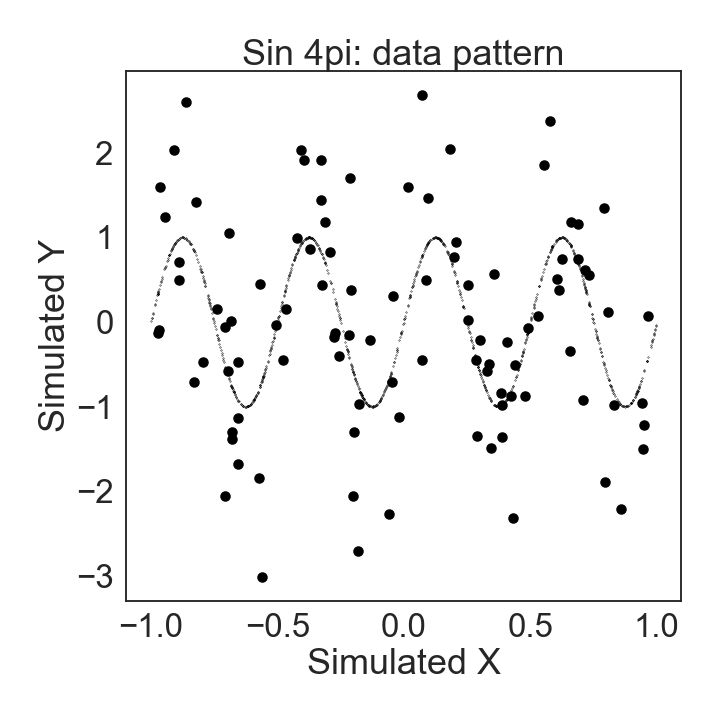}
    \includegraphics[width=0.24\linewidth]{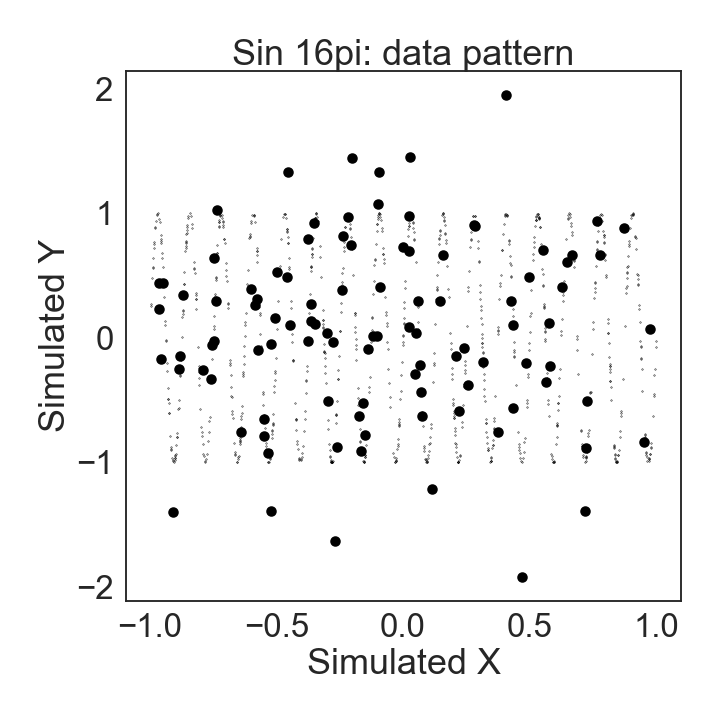}
    \includegraphics[width=0.24\linewidth]{Figures_agtic/Spiral_data}
    \includegraphics[width=0.24\linewidth]{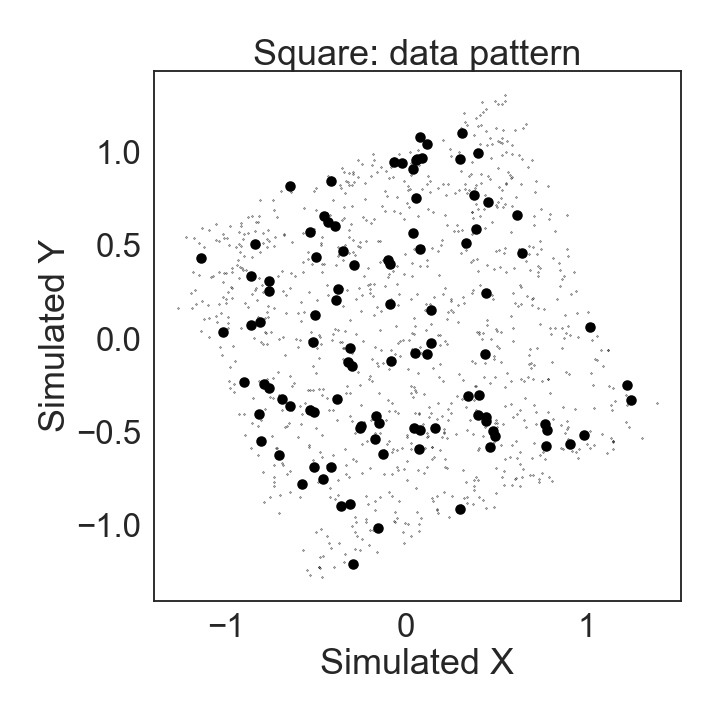}
    \includegraphics[width=0.24\linewidth]{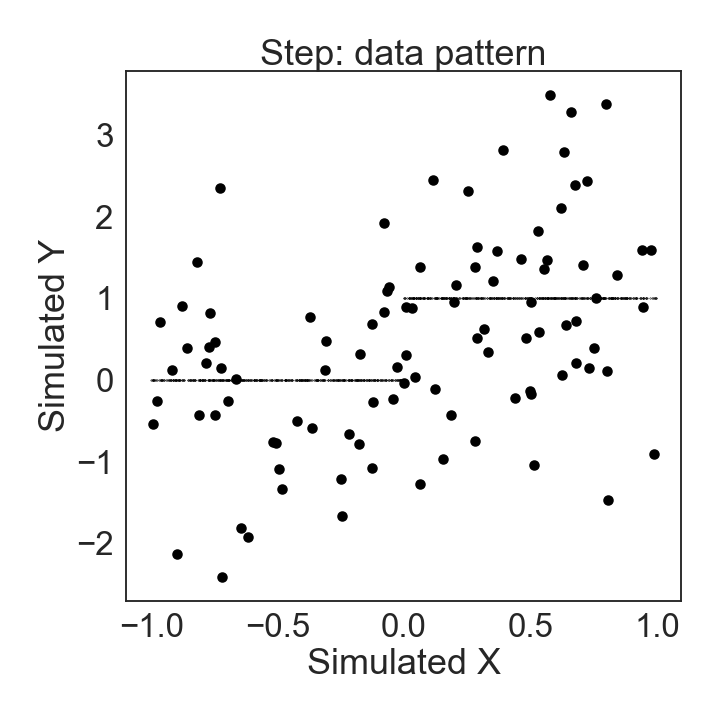}
    \includegraphics[width=0.24\linewidth]{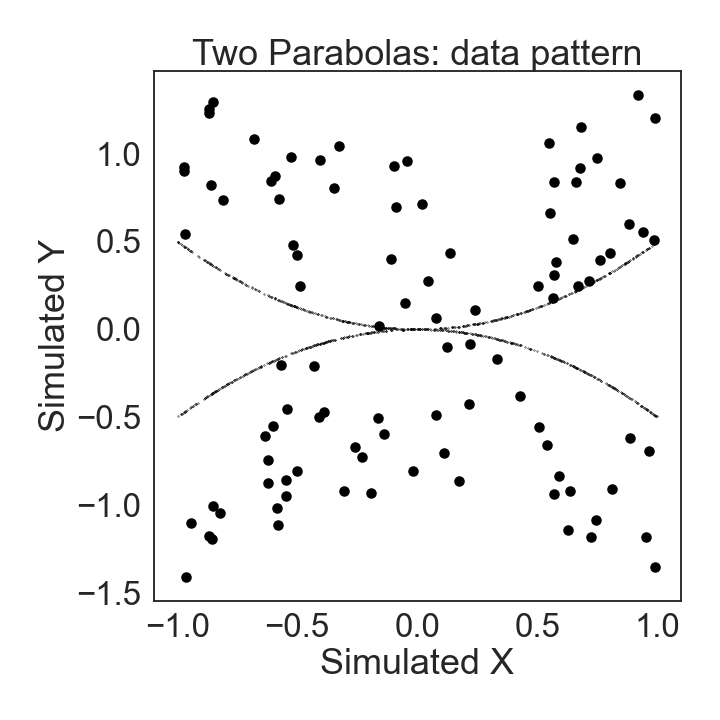}
    \includegraphics[width=0.24\linewidth]{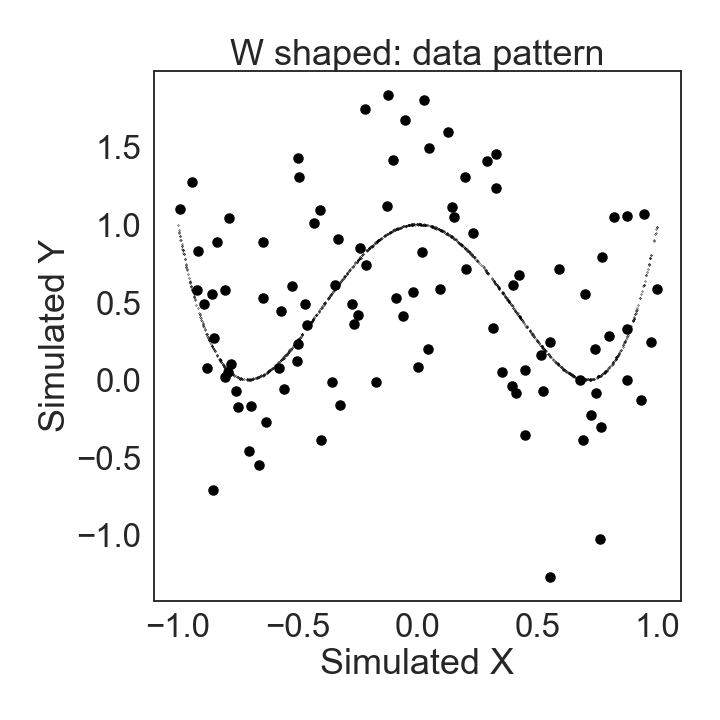}
\par\caption{\textbf{Multivariate association patterns.} Demonstrated are the 16 association patterns considered in this work.}\label{fig:data}
\end{figure}

\textbf{Cheaper options via sampling with convergence.}
Despite the fact that the number of thresholds $k$ is generally small (around 5 to 10), $\mathcal{O}(k^4)$ is still considerably larger than the vanilla dCor. \cite{szekely2007measuring,szekely2014partial} showed that sample dCor can be easily computed to converge to the population dCor via properly centering the Euclidean matrices. Similarly as \cite{szekely2014partial,shen2019distance}, sample AGTDM can also be computed via Euclidean distance matrices after the monotonic transform and the sample version converges to the respective AGTDM up to a difference of $\mathcal{O}(\frac{1}{n})$ where n is the number of sample of threshold sets.

\textbf{Hypothesis. }
We hypothesize that the monotonic transforms of distance matrices associated with the two variables offers an additional attention mechanism to the traditional distance correlation, such that data points with small distances are treated as identical (one collapsed node within a topological graph) and data points with large distances are considered disconnected (no matter how distant they are from each other). Rather than simply thresholding the distance matrix, which replaces a geometrical summary with a topological summary, we explore transforms that can suppress variations among small distances (which tend to be dominated by noise) and among large distances (which may not reflect mutual information between the two variables), while preserving geometrical information (which may boost sensitivity to simple relationships). We refer to these transforms as geo-topological transforms, because they combine aspects of geometry and topology. Depending on the choice of the lower and upper bounds, these transforms can threshold (lower bound $l$ = upper bound $u$) at arbitrary levels, adapting to the granularity of the dependency present in a data set. They can also, optionally, preserve geometrical information (lower bound $l$ < upper bound $u$). 

\section{Empirical Evaluation of AGTDM in Real-World Data}
\label{sec:results}

\begin{table}[tb]
	\centering
	\caption{\textbf{Empirical evaluation of AGTDM.} Statistical power of different statistical tests (rows) for detecting relationships (columns)}
	\resizebox{1\linewidth}{!}{
		\begin{tabular}{ l | l | l | l | l | l | l | l }
EXP & TEST	&  linear & parabolic & sinusoidal & circular & checkerboard & average \\ 
			 \thickhline
\multirow{8}{*}{noise levels} &	AGTDM	&	0.674	$\pm$	0.351	&	\textbf{0.602}	$\pm$	\textbf{0.373}	&	\textbf{0.712}	$\pm$	\textbf{0.384}	&	0.688	$\pm$	0.414	&	\textbf{0.618}	$\pm$	\textbf{0.410}	&	\textbf{0.628}	$\pm$	\textbf{0.396}	\\
& MI	&	0.502	$\pm$	0.404	&	0.534	$\pm$	0.354	&	0.586	$\pm$	0.381	&	0.702	$\pm$	0.391	&	0.602	$\pm$	0.423	&	0.565	$\pm$	0.386	\\
& dCor	&	0.676	$\pm$	0.353	&	0.550	$\pm$	0.425	&	0.452	$\pm$	0.424	&	0.446	$\pm$	0.448	&	0.210	$\pm$	0.140	&	0.467	$\pm$	0.392	\\
& Hoeffding's D	&	0.650	$\pm$	0.356	&	0.460	$\pm$	0.414	&	0.460	$\pm$	0.431	&	0.498	$\pm$	0.475	&	0.200	$\pm$	0.112	&	0.454	$\pm$	0.393	\\
& HSIC	&	0.504	$\pm$	0.416	&	0.556	$\pm$	0.363	&	0.324	$\pm$	0.369	&	0.670	$\pm$	0.439	&	0.196	$\pm$	0.082	&	0.450	$\pm$	0.383	\\
& MIC	&	0.344	$\pm$	0.335	&	0.378	$\pm$	0.299	&	0.586	$\pm$	0.288	&	0.438	$\pm$	0.366	&	0.310	$\pm$	0.193	&	0.411	$\pm$	0.305	\\
& rdmCor	&	0.426	$\pm$	0.406	&	0.534	$\pm$	0.420	&	0.028	$\pm$	0.023	&	\textbf{0.728}	$\pm$	\textbf{0.408}	&	0.036	$\pm$	0.042	&	0.350	$\pm$	0.415	\\
& R$^2$	&	\textbf{0.710}	$\pm$	\textbf{0.325}	&	0.136	$\pm$	0.104	&	0.054	$\pm$	0.053	&	0.028	$\pm$	0.036	&	0.072	$\pm$	0.043	&	0.200	$\pm$	0.300	\\ 
            \hline
        \multirow{8}{*}{sample sizes} &	    AGTDM & \textbf{1.000} $\pm$ \textbf{0.000} & \textbf{1.000} $\pm$ \textbf{0.000} & \textbf{1.000} $\pm$ \textbf{0.000} & \textbf{1.000} $\pm$ \textbf{0.000} & 0.950 $\pm$ 0.218 & \textbf{0.990} $\pm$ \textbf{0.022} \\
& MI & 0.995 $\pm$ 0.011 & 0.985 $\pm$ 0.019 & 0.991 $\pm$ 0.015 & 0.995 $\pm$ 0.010 & \textbf{0.983} $\pm$ \textbf{0.018} & 0.989 $\pm$ 0.005 \\
& Hoeffding's D & \textbf{1.000} $\pm$ \textbf{0.000} & \textbf{1.000} $\pm$ \textbf{0.000} & \textbf{1.000} $\pm$ \textbf{0.000} & \textbf{1.000} $\pm$ \textbf{0.000} & 0.550 $\pm$ 0.497 & 0.910 $\pm$ 0.201 \\
& MIC & 0.984 $\pm$ 0.015 & 0.977 $\pm$ 0.022 & 0.956 $\pm$ 0.160 & 0.891 $\pm$ 0.292 & 0.733 $\pm$ 0.422 & 0.908 $\pm$ 0.105 \\
& dCor & \textbf{1.000} $\pm$ \textbf{0.000} & \textbf{1.000} $\pm$ \textbf{0.000} & 0.900 $\pm$ 0.300 & 0.800 $\pm$ 0.400 & 0.600 $\pm$ 0.490 & 0.860 $\pm$ 0.167 \\
& HSIC & \textbf{1.000} $\pm$ \textbf{0.000} & \textbf{1.000} $\pm$ \textbf{0.000} & 0.850 $\pm$ 0.357 & 0.950 $\pm$ 0.218 & 0.500 $\pm$ 0.500 & 0.860 $\pm$ 0.210 \\
& rdmCor & \textbf{1.000} $\pm$ \textbf{0.000} & \textbf{1.000} $\pm$ \textbf{0.000} & 0.300 $\pm$ 0.458 & 0.000 $\pm$ 0.000 & 0.200 $\pm$ 0.400 & 0.500 $\pm$ 0.469 \\
& R$^2$ & \textbf{1.000} $\pm$ \textbf{0.000} & 0.350 $\pm$ 0.477 & 0.000 $\pm$ 0.000 & 0.000 $\pm$ 0.000 & 0.150 $\pm$ 0.357 & 0.300 $\pm$ 0.417 \\                 
		\end{tabular}
	}  
\label{tab:summary}
\end{table}

\begin{figure}[tb]
\centering
    \includegraphics[width=1\linewidth]{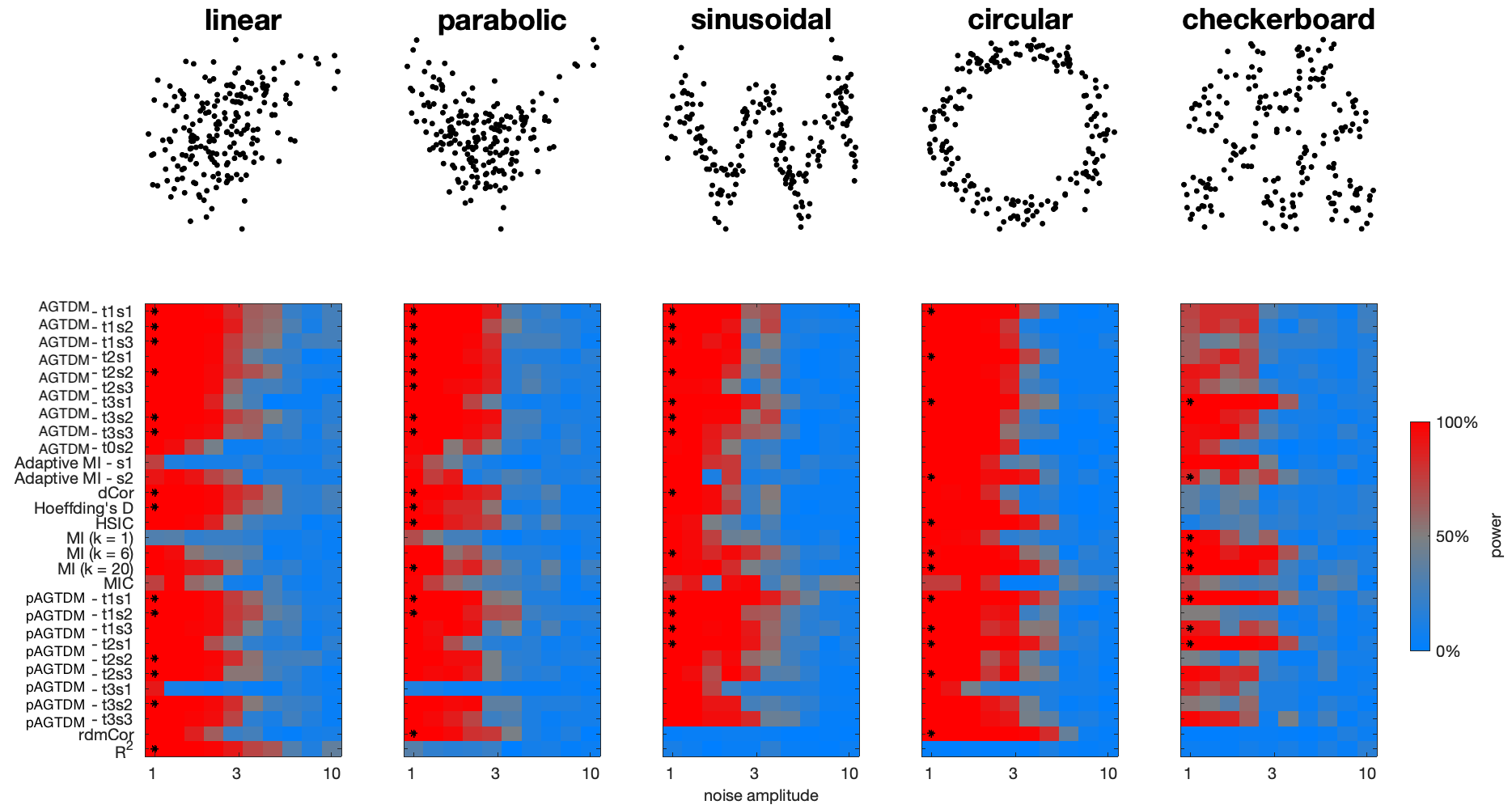}
\par\caption{\textbf{Statistical power of AGTDM in multivariate association patterns.} Power (color-coded) of different tests (rows) for detecting different forms of dependency (panels) over different noise levels  (horizontal axes). For each pattern, an asterisk indicates that the test retains 50\%-power at a noise level within 25\% of the most sensitive test.}\label{fig:power_noise}
\end{figure}

\textbf{Evaluation metric and benchmarks.}
In spirit of \textit{no free lunch} in Statistics, \cite{simon2014comment} stressed the importance of statistical power to evaluate the capacity to detect bivariate association. In our context, the \textit{statistical power} of a dependence measure is the fraction of data sets generated from a dependent joint distribution that yield a significant result (with the false-positives rate controlled at 5\%). \cite{simon2014comment} and \cite{kinney2014equitability} compared several independence measures and showed that dCor \cite{szekely2007measuring, szekely2009brownian} and KNN mutual information estimates (MI) \cite{kraskov2004estimating} have substantially more power than MIC \cite{reshef2011detecting,reshef2013equitability}, but adaptive approaches like ADIC were neither proposed nor tested. To understand the behavior of these adaptive dependence measures, we investigated whether their statistical power can compete with dCor, MIC, MI and representational dissimilarity matrix correlation (rdmCor) \cite{kriegeskorte2008representational}. As a fair comparison, other than the nonparametric or rank-based methods, here we adopted our maximum selection subroutines to one of the popular parametric dependence measure, K-nearest neighbour mutual information estimator, such that the hyperparameter $k$ is adaptively defined, denoted \textit{Adaptive MI - s1} and \textit{Adaptive MI - s2}.


\begin{table}[tb]
	\centering
	\caption{\textbf{Empirical evaluation of AGTDM.} Statistical power of different tests (rows) for detecting relationships (columns) between two univariate variables (averaged over different noise amplitudes, rows ranked by average power)}
	\resizebox{1\linewidth}{!}{
		\begin{tabular}{ l | l | l | l | l | l | l }
			 &  linear & parabolic & sinusoidal & circular & checkerboard & average \\ 
			 \thickhline
pAGTDM - t1s1	&	0.594	$\pm$	0.400	&	0.534	$\pm$	0.431	&	\textbf{0.712}	$\pm$	\textbf{0.384}	&	0.680	$\pm$	0.414	&	\textbf{0.618}	$\pm$	\textbf{0.410}	&	\textbf{0.628}	$\pm$	\textbf{0.396}	\\
MI (k=20)	&	0.502	$\pm$	0.404	&	0.534	$\pm$	0.354	&	0.504	$\pm$	0.409	&	0.702	$\pm$	0.391	&	0.582	$\pm$	0.413	&	0.565	$\pm$	0.386	\\
AGTDM - t1s1	&	0.668	$\pm$	0.349	&	0.580	$\pm$	0.428	&	0.552	$\pm$	0.452	&	0.590	$\pm$	0.464	&	0.420	$\pm$	0.324	&	0.562	$\pm$	0.399	\\
pAGTDM - t2s1	&	0.484	$\pm$	0.399	&	0.366	$\pm$	0.374	&	0.616	$\pm$	0.415	&	0.688	$\pm$	0.414	&	0.616	$\pm$	0.435	&	0.554	$\pm$	0.408	\\
AGTDM - t3s1	&	0.484	$\pm$	0.437	&	0.448	$\pm$	0.433	&	0.594	$\pm$	0.430	&	0.624	$\pm$	0.423	&	0.606	$\pm$	0.441	&	0.551	$\pm$	0.421	\\
pAGTDM - t1s3	&	0.578	$\pm$	0.399	&	0.528	$\pm$	0.401	&	0.624	$\pm$	0.394	&	0.604	$\pm$	0.389	&	0.396	$\pm$	0.358	&	0.546	$\pm$	0.381	\\
AGTDM - t1s2	&	0.674	$\pm$	0.351	&	0.582	$\pm$	0.399	&	0.546	$\pm$	0.422	&	0.502	$\pm$	0.480	&	0.408	$\pm$	0.321	&	0.542	$\pm$	0.392	\\
AGTDM - t3s2	&	0.632	$\pm$	0.371	&	0.580	$\pm$	0.431	&	0.542	$\pm$	0.449	&	0.512	$\pm$	0.468	&	0.432	$\pm$	0.430	&	0.540	$\pm$	0.419	\\
AGTDM - t0s2	&	0.374	$\pm$	0.350	&	0.546	$\pm$	0.393	&	0.578	$\pm$	0.423	&	0.656	$\pm$	0.433	&	0.532	$\pm$	0.440	&	0.537	$\pm$	0.403	\\
AGTDM - t2s2	&	0.636	$\pm$	0.382	&	0.578	$\pm$	0.427	&	0.500	$\pm$	0.432	&	0.540	$\pm$	0.466	&	0.380	$\pm$	0.364	&	0.527	$\pm$	0.408	\\
pAGTDM - t1s2	&	0.636	$\pm$	0.371	&	\textbf{0.602}	$\pm$	\textbf{0.373}	&	0.600	$\pm$	0.396	&	0.560	$\pm$	0.449	&	0.230	$\pm$	0.177	&	0.526	$\pm$	0.381	\\
AGTDM - t3s3	&	0.578	$\pm$	0.387	&	0.560	$\pm$	0.432	&	0.526	$\pm$	0.402	&	0.514	$\pm$	0.451	&	0.448	$\pm$	0.398	&	0.525	$\pm$	0.400	\\
pAGTDM - t2s3	&	0.562	$\pm$	0.404	&	0.494	$\pm$	0.419	&	0.538	$\pm$	0.400	&	0.582	$\pm$	0.422	&	0.440	$\pm$	0.397	&	0.523	$\pm$	0.395	\\
AGTDM - t2s1	&	0.566	$\pm$	0.418	&	0.546	$\pm$	0.439	&	0.500	$\pm$	0.447	&	0.614	$\pm$	0.453	&	0.376	$\pm$	0.333	&	0.520	$\pm$	0.411	\\
MI (k=6)	&	0.380	$\pm$	0.360	&	0.406	$\pm$	0.375	&	0.586	$\pm$	0.381	&	0.620	$\pm$	0.413	&	0.602	$\pm$	0.423	&	0.519	$\pm$	0.389	\\
AGTDM - t1s3	&	0.656	$\pm$	0.337	&	0.582	$\pm$	0.408	&	0.494	$\pm$	0.433	&	0.536	$\pm$	0.461	&	0.318	$\pm$	0.207	&	0.517	$\pm$	0.382	\\
pAGTDM - t3s3	&	0.536	$\pm$	0.424	&	0.442	$\pm$	0.400	&	0.570	$\pm$	0.418	&	0.594	$\pm$	0.417	&	0.442	$\pm$	0.380	&	0.517	$\pm$	0.396	\\
pAGTDM - t2s2	&	0.624	$\pm$	0.375	&	0.504	$\pm$	0.416	&	0.542	$\pm$	0.391	&	0.530	$\pm$	0.424	&	0.214	$\pm$	0.168	&	0.483	$\pm$	0.379	\\
pAGTDM - t3s2	&	0.604	$\pm$	0.392	&	0.514	$\pm$	0.416	&	0.498	$\pm$	0.409	&	0.536	$\pm$	0.450	&	0.250	$\pm$	0.215	&	0.480	$\pm$	0.389	\\
AGTDM - t2s3	&	0.536	$\pm$	0.421	&	0.564	$\pm$	0.415	&	0.410	$\pm$	0.418	&	0.540	$\pm$	0.463	&	0.340	$\pm$	0.278	&	0.478	$\pm$	0.397	\\
dCor	&	0.676	$\pm$	0.353	&	0.550	$\pm$	0.425	&	0.452	$\pm$	0.424	&	0.446	$\pm$	0.448	&	0.210	$\pm$	0.140	&	0.467	$\pm$	0.392	\\
Adaptive MI - s2	&	0.468	$\pm$	0.396	&	0.386	$\pm$	0.355	&	0.354	$\pm$	0.403	&	0.634	$\pm$	0.437	&	0.428	$\pm$	0.324	&	0.454	$\pm$	0.382	\\
Hoeffding's D	&	0.650	$\pm$	0.356	&	0.460	$\pm$	0.414	&	0.460	$\pm$	0.431	&	0.498	$\pm$	0.475	&	0.200	$\pm$	0.112	&	0.454	$\pm$	0.393	\\
HSIC	&	0.504	$\pm$	0.416	&	0.556	$\pm$	0.363	&	0.324	$\pm$	0.369	&	0.670	$\pm$	0.439	&	0.196	$\pm$	0.082	&	0.450	$\pm$	0.383	\\
MIC	&	0.344	$\pm$	0.335	&	0.378	$\pm$	0.299	&	0.586	$\pm$	0.288	&	0.438	$\pm$	0.366	&	0.310	$\pm$	0.193	&	0.411	$\pm$	0.305	\\
AGTDM - t0s1	&	0.368	$\pm$	0.383	&	0.446	$\pm$	0.367	&	0.392	$\pm$	0.414	&	0.492	$\pm$	0.504	&	0.264	$\pm$	0.326	&	0.392	$\pm$	0.394	\\
Adaptive MI - s1	&	0.170	$\pm$	0.206	&	0.262	$\pm$	0.314	&	0.454	$\pm$	0.394	&	0.494	$\pm$	0.424	&	0.482	$\pm$	0.449	&	0.372	$\pm$	0.377	\\
rdmCor	&	0.426	$\pm$	0.406	&	0.534	$\pm$	0.420	&	0.028	$\pm$	0.023	&	\textbf{0.728}	$\pm$	\textbf{0.408}	&	0.036	$\pm$	0.042	&	0.350	$\pm$	0.415	\\
MI (k=1)	&	0.174	$\pm$	0.107	&	0.208	$\pm$	0.228	&	0.402	$\pm$	0.379	&	0.448	$\pm$	0.415	&	0.396	$\pm$	0.390	&	0.326	$\pm$	0.332	\\
pAGTDM - t3s1	&	0.160	$\pm$	0.263	&	0.054	$\pm$	0.040	&	0.344	$\pm$	0.401	&	0.310	$\pm$	0.363	&	0.316	$\pm$	0.330	&	0.237	$\pm$	0.315	\\
R$^2$	&	\textbf{0.710}	$\pm$	\textbf{0.325}	&	0.136	$\pm$	0.104	&	0.054	$\pm$	0.053	&	0.028	$\pm$	0.036	&	0.072	$\pm$	0.043	&	0.200	$\pm$	0.300	\\  
            \hline
		\end{tabular}
	}  
\label{table:1DpowerNoise}
\end{table}

\textbf{Parameter selection.}
For the family of AGTDM, the numbers of possible thresholds for the lower and upper bounds in the geo-topological transforms are set to be 5. The combinatorial search space for the boundary pairs are nchoosek$(5,2) = 10$, since the threshold search itself has a complexity of $O((k(k-1)/2)^2) \approx O(k^4)$. For HSIC, it applies a bootstrap approximation to the test threshold with kernel sizes set to the median distances for $X$ and $Y$ \cite{gretton2008kernel}. For MIC, the user-specified value $B$ was set to be $N^{0.6}$ as advocated by \cite{reshef2011detecting}. For mutual information estimator, three different k ($k=1,6,20$) were used as in \cite{kraskov2004estimating}. 

\textbf{Experimental setting.} 
In the bivariate association experiments, 50 repetitions of 200 samples were generated, in which the input sample was uniformed distributed on the unit interval. Next, we regenerated the input sample randomly in order to generate i.i.d. versions as the null distribution with equal marginals.

\textbf{Multivariate association patterns. }
We considered sixteen common multivariate patterns (Fig. \ref{fig:data}) and evaluate the statistical power of the statistics with different additive noises, sample sizes, dimensions and combinatorial dependencies. Here we report five distinct patterns: linear, parabolic, sinusoidal, circular and checkerboard, and describe the \textit{free lunch}: which is best, where, and how.

\begin{table}[tb]
	\centering
	\caption{\textbf{Empirical evaluation of AGTDM.} Statistical power in 1-dimensional data over different sample sizes (ranked by average power)}
	\resizebox{1\linewidth}{!}{
		\begin{tabular}{ l | l | l | l | l | l | l }
			 &  linear & parabolic & sinusoidal & circular & checkerboard & average \\ \thickhline
AGTDM - t3s1 & \textbf{1.000} $\pm$ \textbf{0.000} & \textbf{1.000} $\pm$ \textbf{0.000} & \textbf{1.000} $\pm$ \textbf{0.000} & \textbf{1.000} $\pm$ \textbf{0.000} & 0.950 $\pm$ 0.218 & \textbf{0.990} $\pm$ \textbf{0.022} \\
MI (k=1) & 0.995 $\pm$ 0.011 & 0.985 $\pm$ 0.019 & 0.991 $\pm$ 0.015 & 0.993 $\pm$ 0.015 & \textbf{0.983} $\pm$ \textbf{0.018} & 0.989 $\pm$ 0.005 \\
AGTDM - t0s2 & \textbf{1.000} $\pm$ \textbf{0.000} & \textbf{1.000} $\pm$ \textbf{0.000} & 0.967 $\pm$ 0.144 & \textbf{1.000} $\pm$ \textbf{0.000} & 0.928 $\pm$ 0.230 & 0.979 $\pm$ 0.032 \\
AGTDM - t0s1 & \textbf{1.000} $\pm$ \textbf{0.000} & \textbf{1.000} $\pm$ \textbf{0.000} & 0.950 $\pm$ 0.218 & \textbf{1.000} $\pm$ \textbf{0.000} & 0.908 $\pm$ 0.279 & 0.972 $\pm$ 0.042 \\
MI (k=6) & 0.992 $\pm$ 0.018 & 0.982 $\pm$ 0.024 & 0.939 $\pm$ 0.211 & 0.995 $\pm$ 0.010 & 0.942 $\pm$ 0.216 & 0.970 $\pm$ 0.027 \\
AGTDM - t3s3 & 0.998 $\pm$ 0.009 & \textbf{1.000} $\pm$ \textbf{0.002} & 0.969 $\pm$ 0.135 & 0.986 $\pm$ 0.041 & 0.889 $\pm$ 0.214 & 0.968 $\pm$ 0.046 \\
AGTDM - t1s1 & \textbf{1.000} $\pm$ \textbf{0.000} & \textbf{1.000} $\pm$ \textbf{0.000} & 0.900 $\pm$ 0.300 & \textbf{1.000} $\pm$ \textbf{0.000} & 0.850 $\pm$ 0.357 & 0.950 $\pm$ 0.071 \\
AGTDM - t3s2 & \textbf{1.000} $\pm$ \textbf{0.000} & \textbf{1.000} $\pm$ \textbf{0.000} & 0.950 $\pm$ 0.218 & 0.932 $\pm$ 0.215 & 0.825 $\pm$ 0.285 & 0.941 $\pm$ 0.072 \\
AGTDM - t2s1 & \textbf{1.000} $\pm$ \textbf{0.000} & \textbf{1.000} $\pm$ \textbf{0.000} & 0.900 $\pm$ 0.300 & \textbf{1.000} $\pm$ \textbf{0.000} & 0.800 $\pm$ 0.400 & 0.940 $\pm$ 0.089 \\
Hoeffding's D & \textbf{1.000} $\pm$ \textbf{0.000} & \textbf{1.000} $\pm$ \textbf{0.000} & \textbf{1.000} $\pm$ \textbf{0.000} & \textbf{1.000} $\pm$ \textbf{0.000} & 0.550 $\pm$ 0.497 & 0.910 $\pm$ 0.201 \\
MIC & 0.984 $\pm$ 0.015 & 0.977 $\pm$ 0.022 & 0.956 $\pm$ 0.160 & 0.891 $\pm$ 0.292 & 0.733 $\pm$ 0.422 & 0.908 $\pm$ 0.105 \\
AGTDM - t2s2 & \textbf{1.000} $\pm$ \textbf{0.000} & \textbf{1.000} $\pm$ \textbf{0.000} & 0.911 $\pm$ 0.269 & 0.905 $\pm$ 0.280 & 0.724 $\pm$ 0.341 & 0.908 $\pm$ 0.113 \\
AGTDM - t2s3 & \textbf{1.000} $\pm$ \textbf{0.000} & 0.997 $\pm$ 0.013 & 0.917 $\pm$ 0.235 & 0.950 $\pm$ 0.162 & 0.639 $\pm$ 0.338 & 0.900 $\pm$ 0.150 \\
AGTDM - t1s2 & \textbf{1.000} $\pm$ \textbf{0.000} & \textbf{1.000} $\pm$ \textbf{0.000} & 0.900 $\pm$ 0.300 & 0.834 $\pm$ 0.336 & 0.766 $\pm$ 0.324 & 0.900 $\pm$ 0.103 \\
AGTDM - t1s3 & \textbf{1.000} $\pm$ \textbf{0.000} & 0.999 $\pm$ 0.003 & 0.901 $\pm$ 0.286 & 0.870 $\pm$ 0.273 & 0.697 $\pm$ 0.301 & 0.893 $\pm$ 0.124 \\
dCor & \textbf{1.000} $\pm$ \textbf{0.000} & \textbf{1.000} $\pm$ \textbf{0.000} & 0.900 $\pm$ 0.300 & 0.800 $\pm$ 0.400 & 0.600 $\pm$ 0.490 & 0.860 $\pm$ 0.167 \\
HSIC & \textbf{1.000} $\pm$ \textbf{0.000} & \textbf{1.000} $\pm$ \textbf{0.000} & 0.850 $\pm$ 0.357 & 0.950 $\pm$ 0.218 & 0.500 $\pm$ 0.500 & 0.860 $\pm$ 0.210 \\
MI (k=20) & 0.945 $\pm$ 0.217 & 0.941 $\pm$ 0.216 & 0.795 $\pm$ 0.396 & 0.895 $\pm$ 0.299 & 0.692 $\pm$ 0.453 & 0.853 $\pm$ 0.108 \\
rdmCor & \textbf{1.000} $\pm$ \textbf{0.000} & \textbf{1.000} $\pm$ \textbf{0.000} & 0.300 $\pm$ 0.458 & 0.000 $\pm$ 0.000 & 0.200 $\pm$ 0.400 & 0.500 $\pm$ 0.469 \\
R$^2$ & \textbf{1.000} $\pm$ \textbf{0.000} & 0.350 $\pm$ 0.477 & 0.000 $\pm$ 0.000 & 0.000 $\pm$ 0.000 & 0.150 $\pm$ 0.357 & 0.300 $\pm$ 0.417 \\                 
            \hline
		\end{tabular}
	}  
\label{table:1DpowerObs}
\end{table}

\begin{figure}[H]
\centering
    \includegraphics[width=\linewidth]{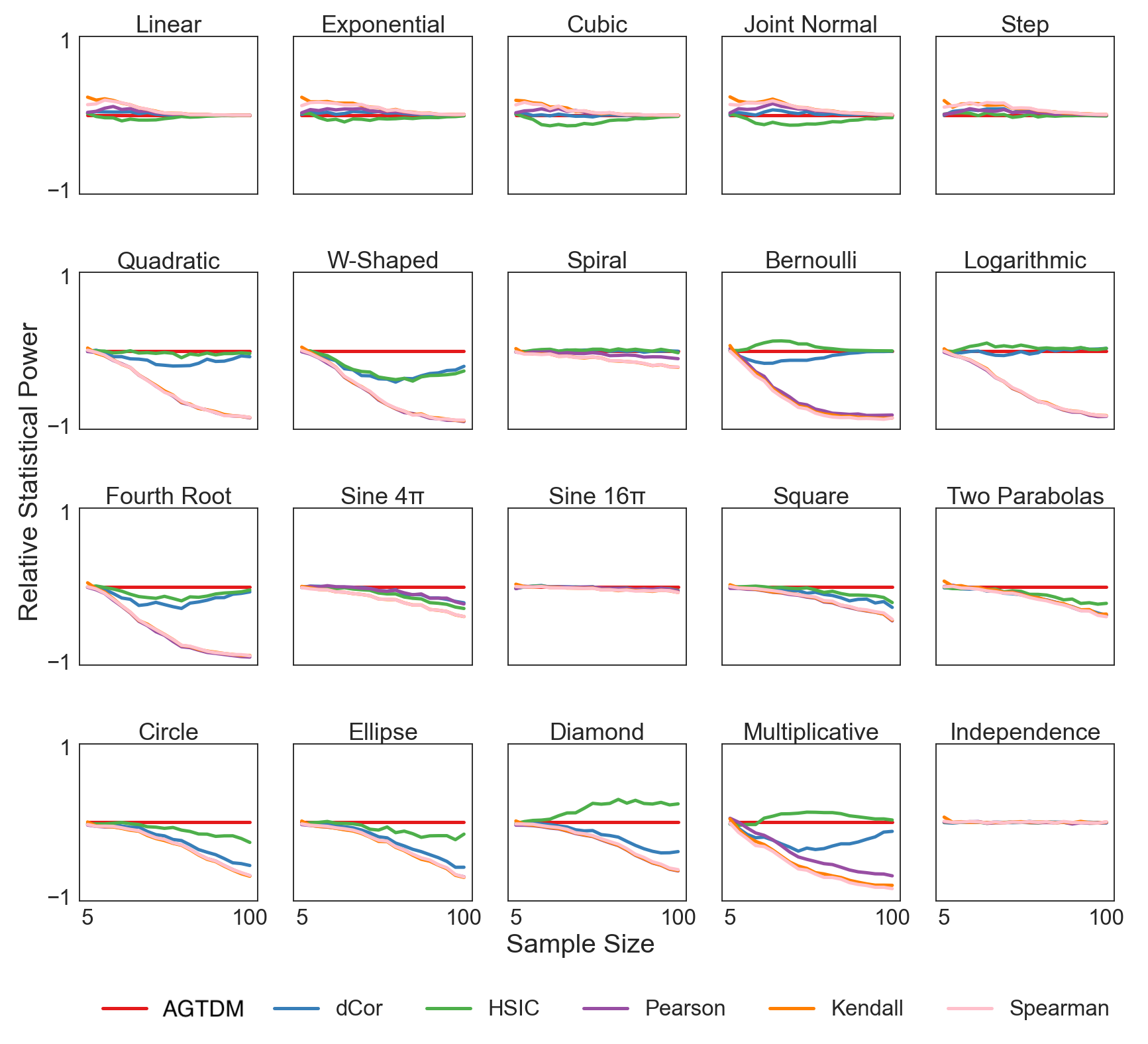}
\par\caption{\textbf{Empirical evaluation of AGTDM.} {Relative statistical power over sample size} in 20 multivariate relationships.}\label{fig:smp}
\end{figure}

\textbf{Resistance to additive noise.} Fig. \ref{fig:power_noise} shows the assessment of statistical power for the competing nonlinear dependence measures as the variance of a series of zero-mean Gaussian noise amplitude which increases logarithmically over a 10-fold range. The heat maps show power values computed for each statistics.
For each pattern, the asterisks indicate that the statistic that have a noise-at-50\%-power that lies within 25\% of this maximum. Among all competing measures, our proposed AGTDM family ranked best in 4 out of 5 relationships (except linear) and best by average (Table \ref{tab:summary}). 

As expected, R$^2$ was observed to have optimal power on the linear relationship, but it is worth noting that all the AGTDM or pAGTDM algorithms adapt to the linear pattern by choosing the most informative threshold pairs to reach a near optimal performance, while R$^2$ shows negligible power on the other relationships which are mirror symmetric as expected. rdmCor as the correlation coefficient on the pairwise distances of the data, shows optimal power in the circular relationship, but poor performance in all others. The behaviors of dCor and Hoeffding's D are very similar across all relationships, and maintained substantial statistical power on all but the checkerboard relationships. On all but the sinusoidal relationship, MIC with $B = N^{0.6}$ as suggested by \cite{reshef2011detecting} was observed to have relatively low statistical power, consistent with the findings of \cite{simon2014comment} and \cite{kinney2014equitability}. The overall performance of the KNN mutual information estimator using k = 1, 6, and 20 differ from case to case: larger k's performed better in complicated relationships like checkerboard  and circular pattern, but they performed poorly comparing the adaptive approaches in linear and parabolic relationships - the two relationships are more representative of many real-world datasets than other relationships. Comparing to our adaptive selection of parameters, the KNN mutual information estimator also has the important parametric disadvantage to demand the user to specify k without any mathematical guidelines, while there is no guarantee larger k's increases the statistical power (as in sinusoidal case). As shown here with three arbitrarily set k's, they can significantly affect the power of one’s mutual information estimates, supporting the discovery of \cite{kinney2014equitability}. The adaptive MI performed slightly better than arbitrarily defined k but the overall performance is not optimal (full results in Table \ref{table:1DpowerNoise}). Table \ref{table:1DpowerStaircase} shows results from the staircase analysis for estimating the noise level at which the power is 80\%. The adaptive geo-topological approach proves quite resilient to noise.

\begin{figure}[H]
\centering
    \includegraphics[width=\linewidth]{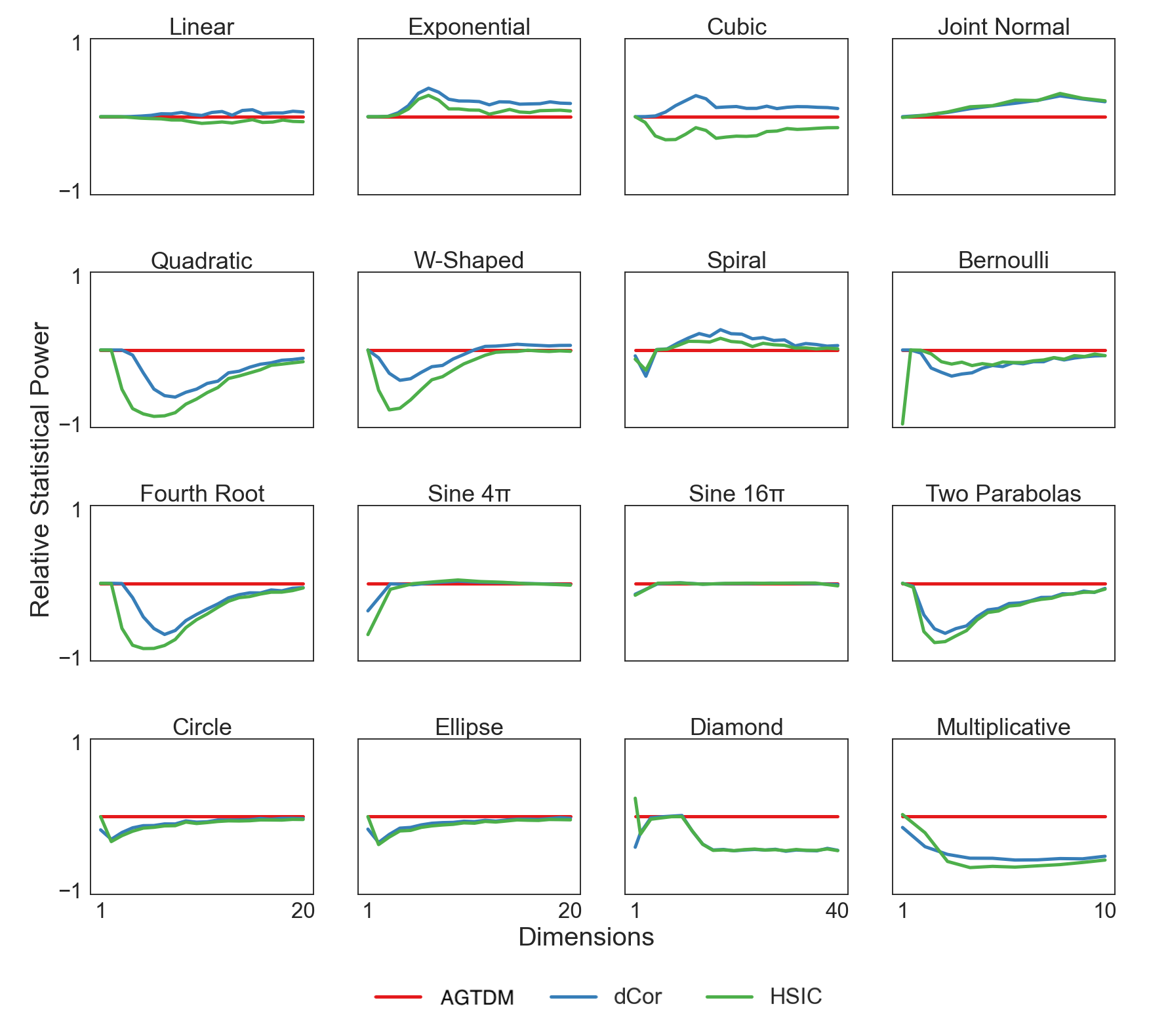}
\par\caption{\textbf{Empirical evaluation of AGTDM.} {Relative statistical power over dimension size} in 16 multivariate relationships.}\label{fig:dim}
\end{figure}

\textbf{Robust in different sample and dimension sizes. } 
100 repetitions of observations with sample size over a 20-fold range from 20 to 400 were generated, in which the input sample was uniformed distributed on the unit interval. 
Table \ref{tab:summary} shows the average statistical power across different sample sizes for different dependence measures in the five relationships. Among all the competing measures, the proposed family of adaptive independence tests demonstrated good robustness in non-functional association patterns (ranked top 1 in all but checkerboard, and top 5 in all relationships). Comparing the three subroutines, \textit{AGTDM - s1} appears more robust than the other two. The three geo-topological transforms each have their advantages for different relationship types (full results in Table \ref{table:1DpowerObs}).  

In the full spectrum of multivariate association patterns, we can also compute the relative statistical power of other methods (taking AGTDM as the baseline). Figs. \ref{fig:smp} and \ref{fig:dim} presents the findings of varying sample sizes and dimension sizes. We observe an overall superiority of AGTDM over other methods in most of the association patterns.

\textbf{Adaptive to combinatorial dependence. } 
50 repetitions of $50 \times 2$ samples were generated, such that each of the two dimensions follows either one of the 5 association patterns (linear, parabolic, sinusoidal, circular, or checkerboard) or random relationship (r), to form a combinatorial two-dimensional dependence. Table \ref{table:2Dpower} shows the statistical power across 20 combinatorial dependence for different statistics. Among all, our methods are top 1 for all but sinusoidal-random (s-r) and checkerboard-random (k-r) relationships, and ranked among top 5 in all relationships. As expected, the statistical power in the pairs of single patterns (l-l, p-p, s-s, c-c, k-k) are higher than the pairs with different patterns, implying certain dependence interference.



\begin{figure}[H]
\centering
    \includegraphics[width=0.24\linewidth]{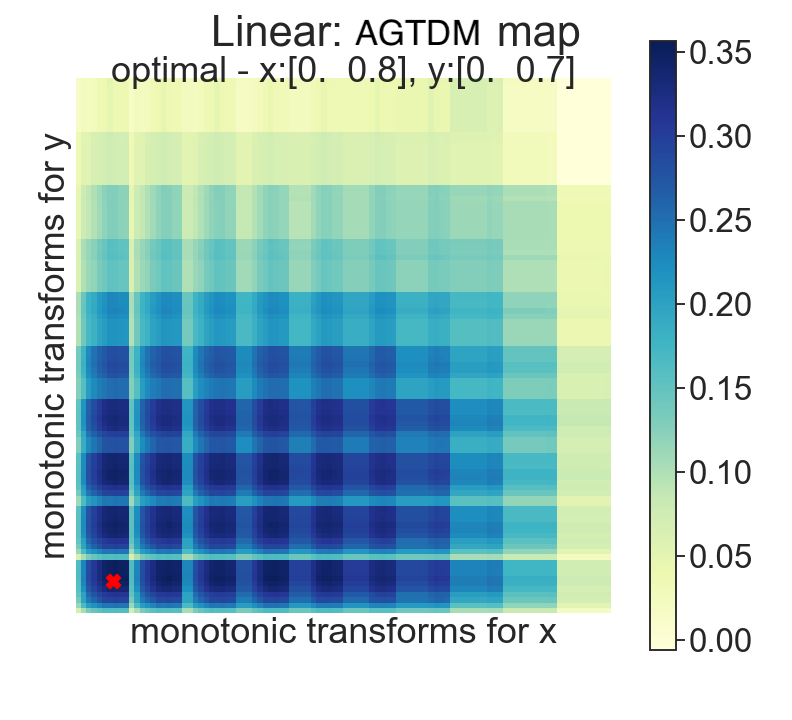}
    \includegraphics[width=0.24\linewidth]{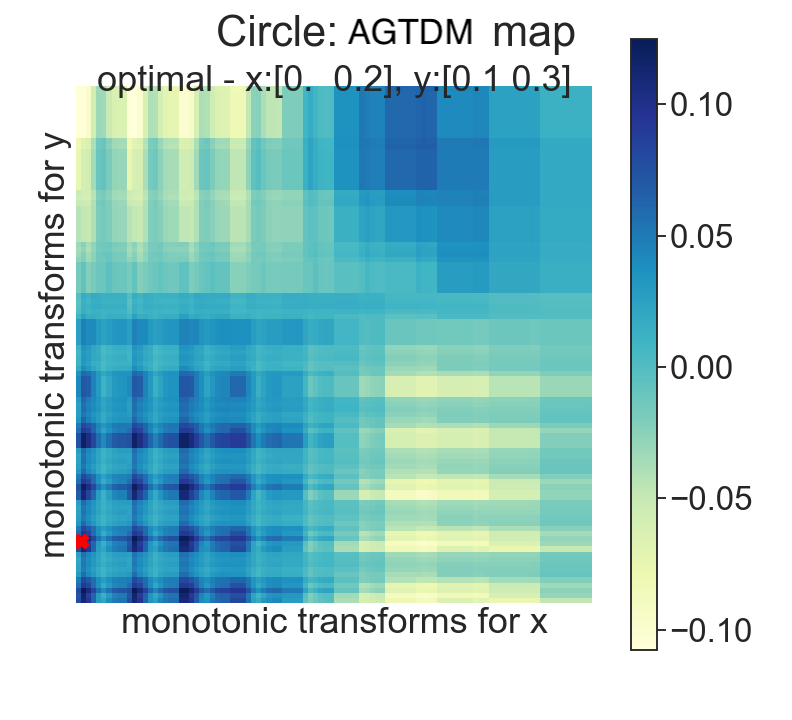}
    \includegraphics[width=0.24\linewidth]{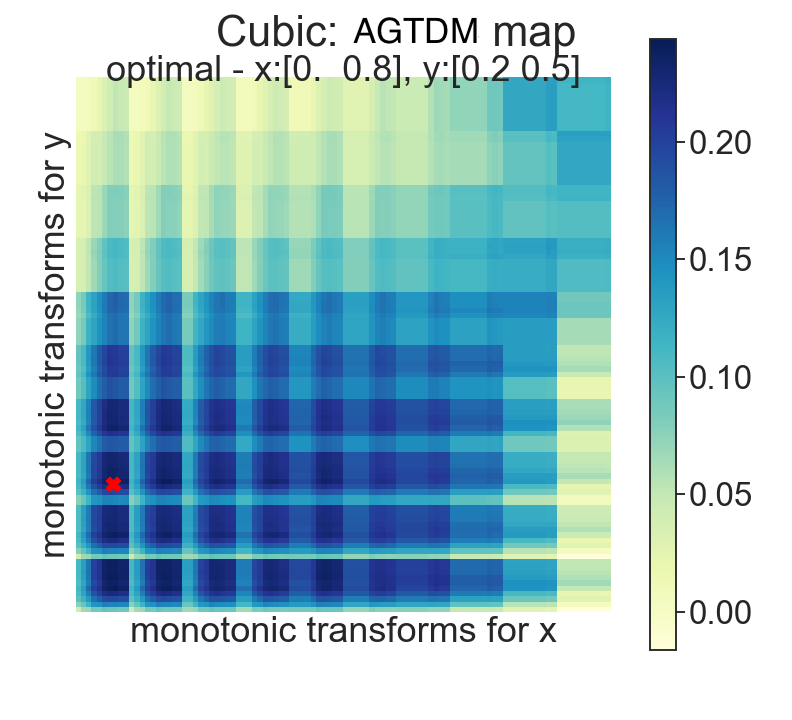}
    \includegraphics[width=0.24\linewidth]{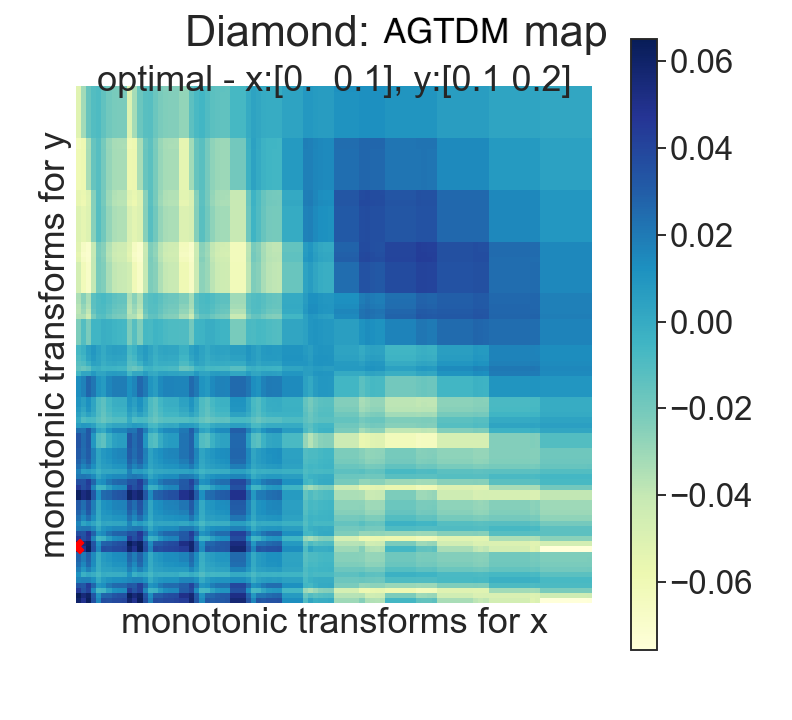}
    \includegraphics[width=0.24\linewidth]{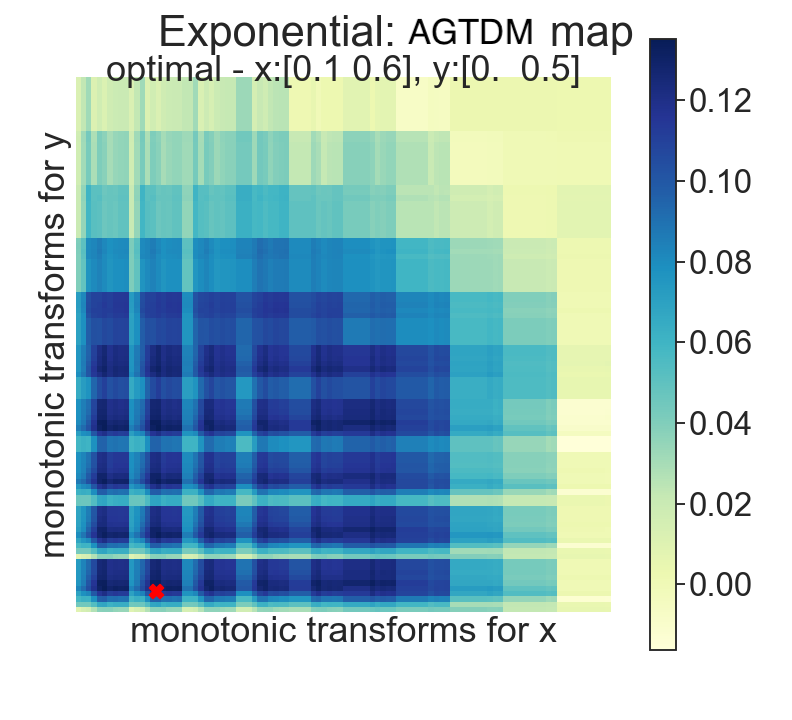}
    \includegraphics[width=0.23\linewidth]{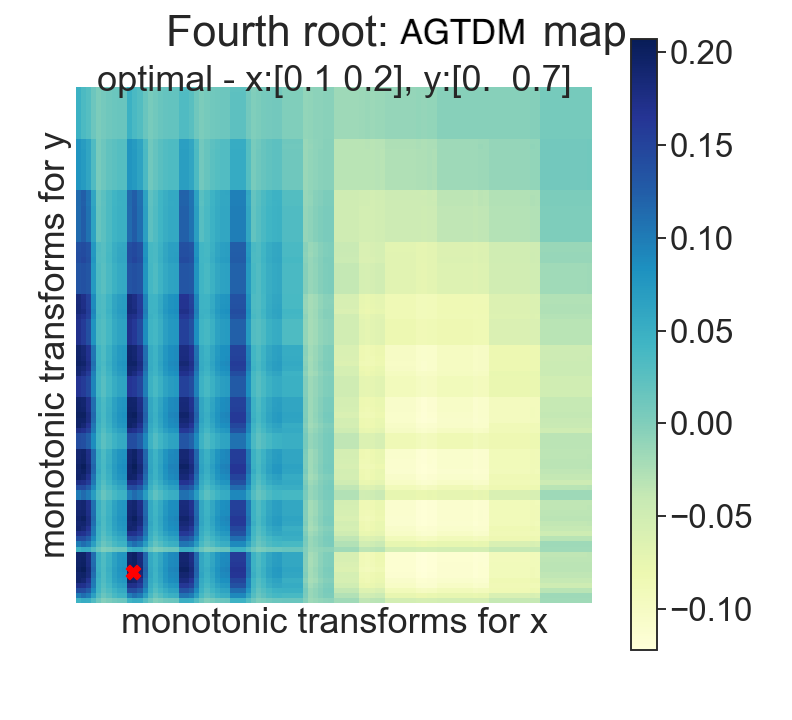}
    \includegraphics[width=0.23\linewidth]{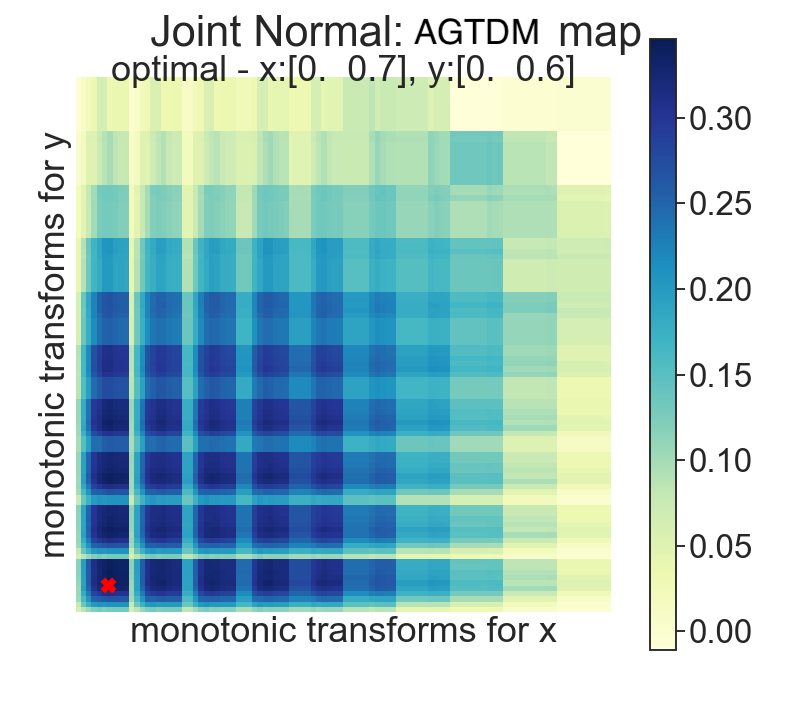}
    \includegraphics[width=0.23\linewidth]{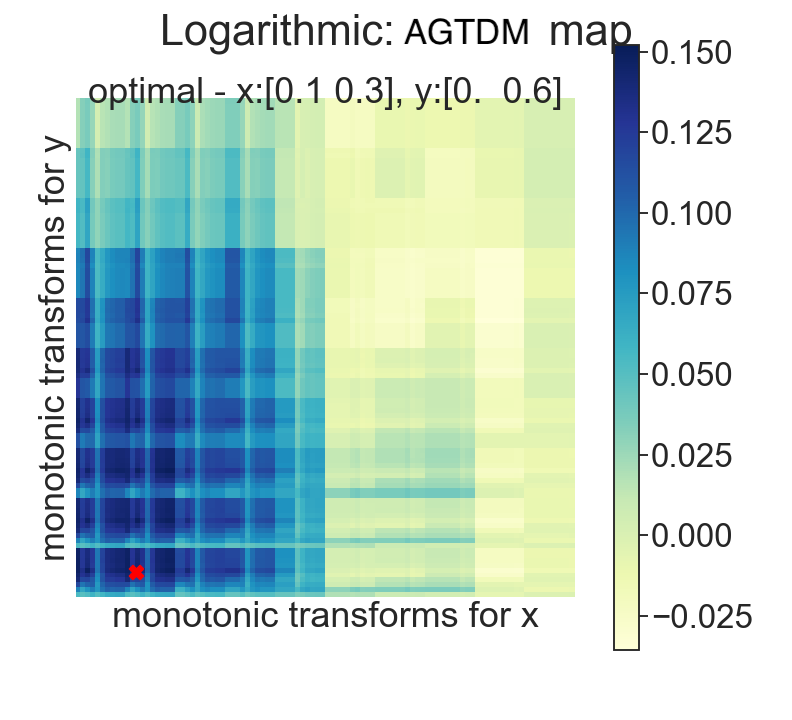}
    \includegraphics[width=0.23\linewidth]{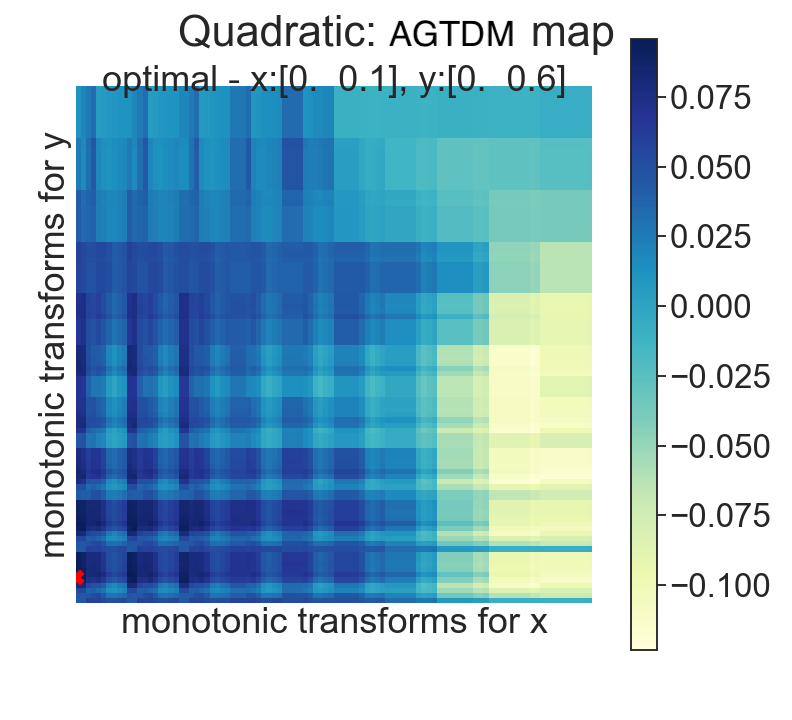}
    \includegraphics[width=0.23\linewidth]{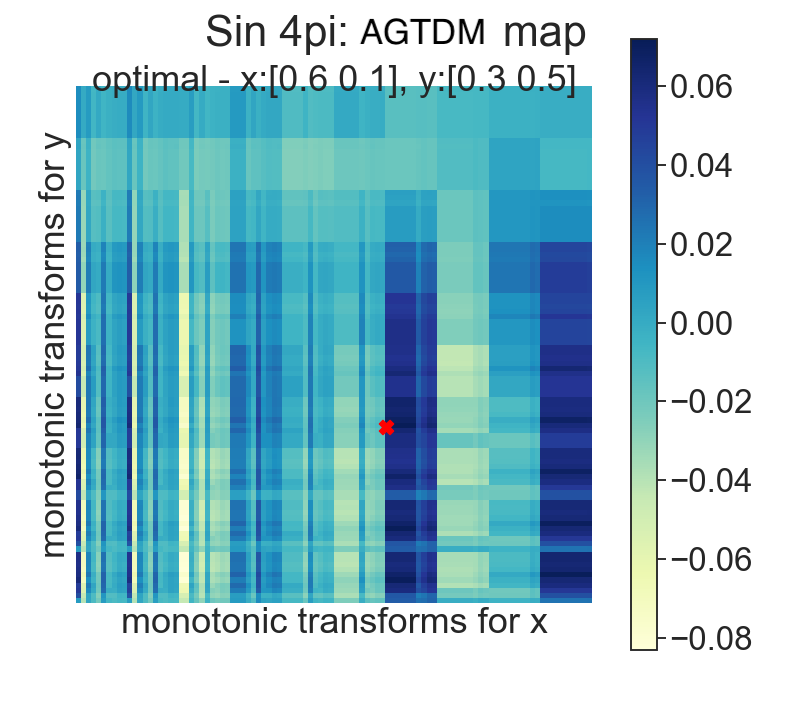}
    \includegraphics[width=0.23\linewidth]{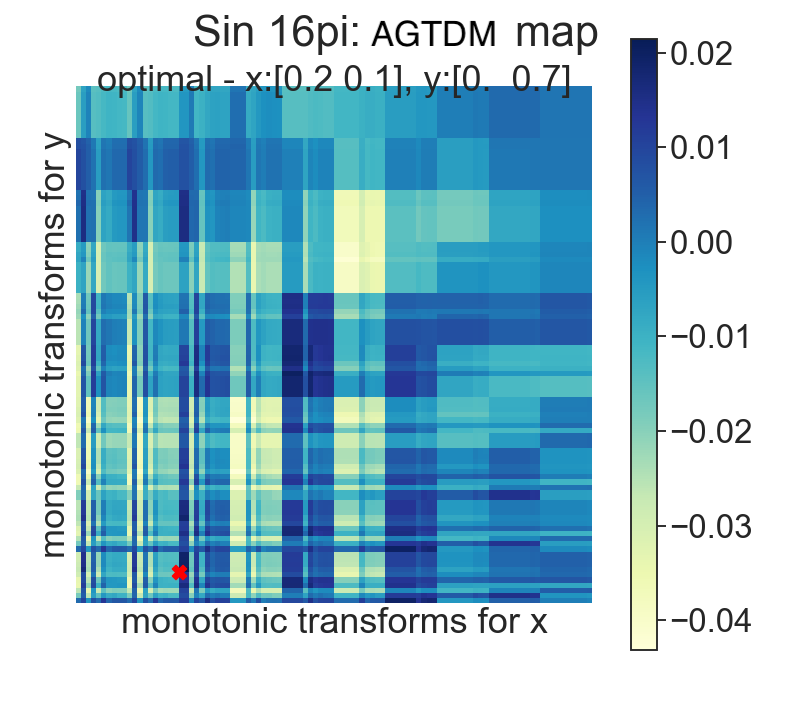}
    \includegraphics[width=0.23\linewidth]{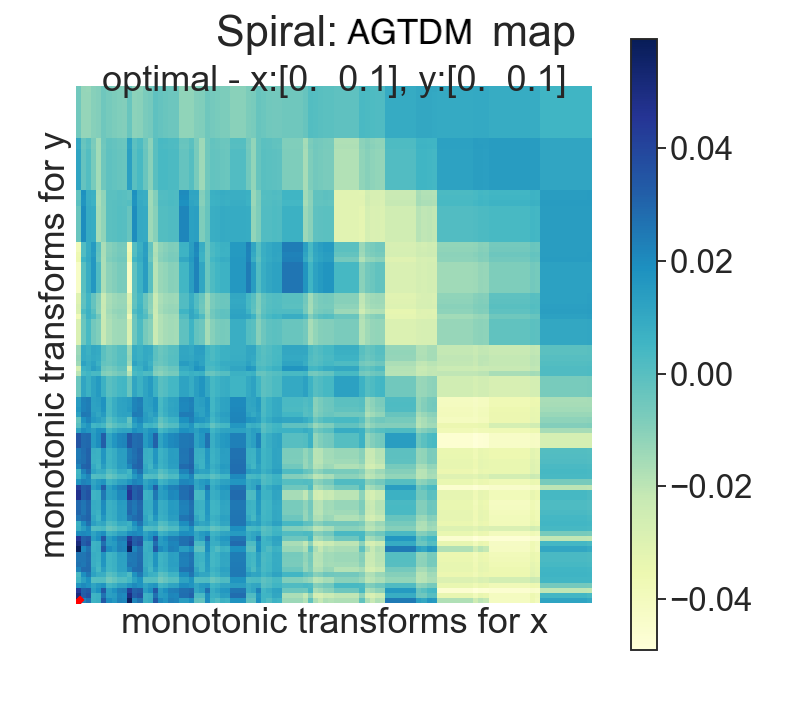}
    \includegraphics[width=0.23\linewidth]{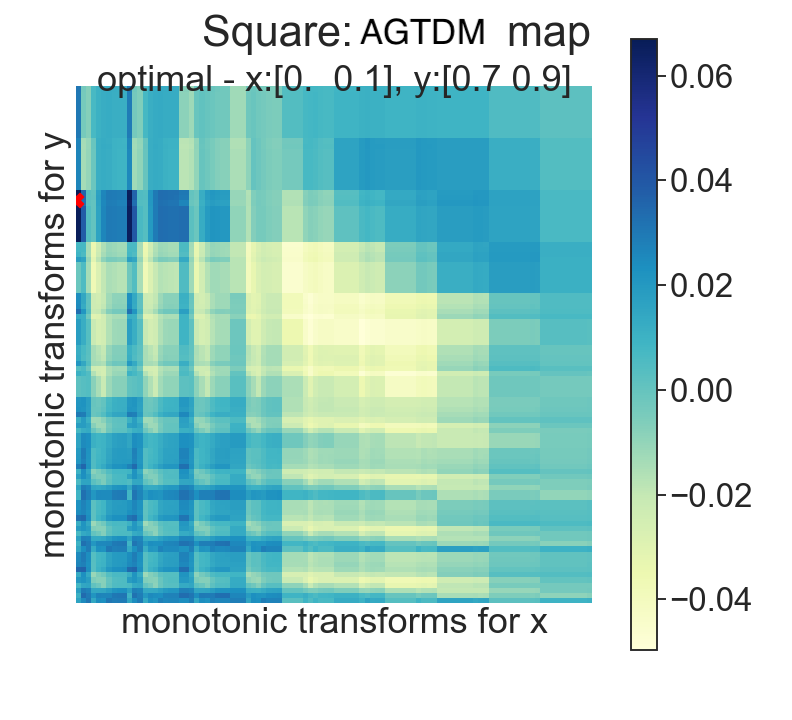}
    \includegraphics[width=0.23\linewidth]{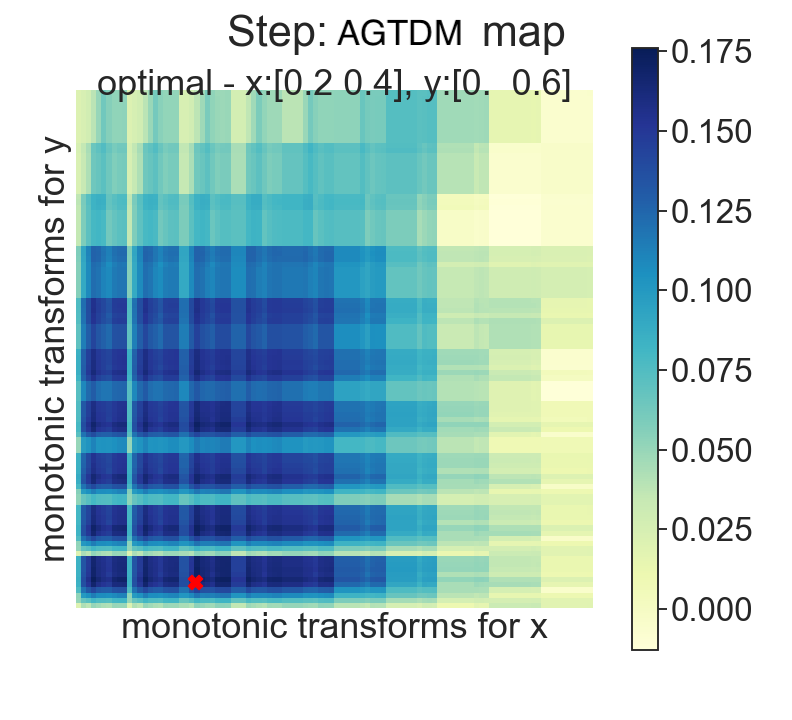}
    \includegraphics[width=0.23\linewidth]{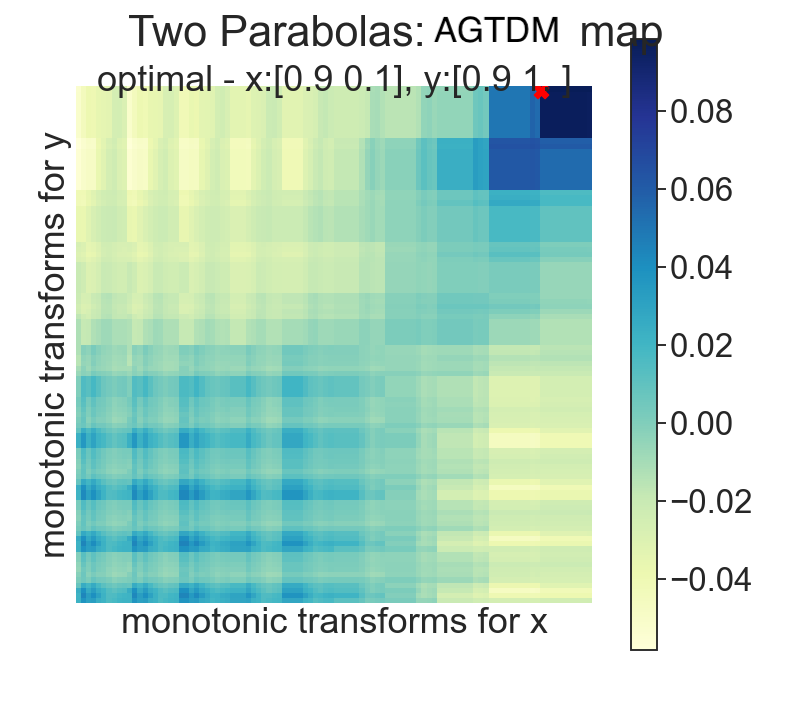}
    \includegraphics[width=0.23\linewidth]{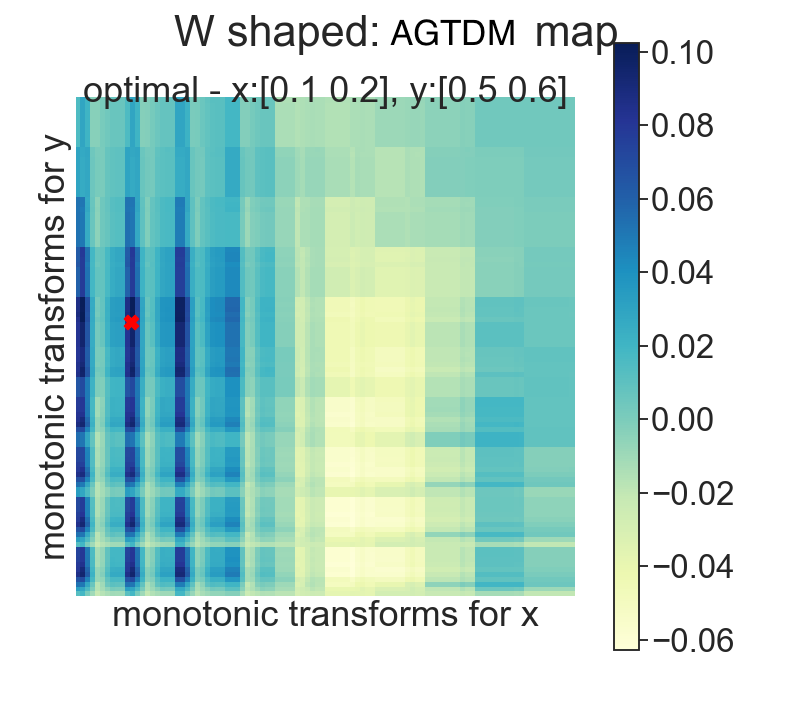}
\par\caption{\textbf{AGTDM maps provide insights to distinctive geo-topologies of each dependency.} The x and y axes of each heatmap denote the sets of threshold pairs $(l,u)$, the color denotes the magnitude of $\mathcal{V}^{2*}(X,Y;GT(\cdot;l,u))$ and the red cross denote the maximum point (i.e. the optimal threshold pairs).}\label{fig:map}
\end{figure}

\textbf{Insightful on granularity of dependence.} 
Fig. \ref{fig:map} reported the AGTDM maps for 16 different 1D relationships where the optimal threshold is marked the red cross. For linear or patch-like patterns, the optimal thresholds usually involved a large distance range ($u-l$ is large), while in skeleton-like patterns (e.g. spiral, circle), this range tends to be a small value, emphasizing finer structures of the data. From the grid-like structures in the maps, we can even decipher the frequency in the sin 4$\pi$ and 16$\pi$. We also noticed that similar dependencies yield similar maps. For instance, the ``Step'' and ``Exponential'' are geometrically similar despite analytically distinct. Thereby, AGTDM can help us understand the relationship in data. 

\begin{table}[H]
	\centering
	\caption{\textbf{Empirical evaluation of AGTDM.} 80\%-power noise level for 1-dimensional data (ranked by avg. power)}
	\resizebox{0.65\linewidth}{!}{
		\begin{tabular}{ l | c | c | c | c | c  }
			 &  linear & parabolic & sinusoidal & circular & checkerboard  \\ \thickhline
pAGTDM - t1s1	&	2.657	&	2.411	&	\textbf{3.975}	&	3.856	&	3.081	\\
MI (k=20)	&	2.302	&	2.289	&	2.191	&	3.987	&	2.839	\\
AGTDM - t1s1	&	2.945	&	2.922	&	2.485	&	3.212	&	1.000	\\
pAGTDM - t2s1	&	1.972	&	1.507	&	3.087	&	4.013	&	\textbf{3.282}	\\
AGTDM - t3s1	&	2.249	&	2.015	&	2.985	&	3.130	&	3.212	\\
pAGTDM - t1s3	&	2.611	&	2.296	&	3.036	&	2.154	&	1.347	\\
AGTDM - t1s2	&	3.032	&	2.573	&	2.305	&	2.720	&	1.000	\\
AGTDM - t3s2	&	2.626	&	\textbf{2.955}	&	2.400	&	2.603	&	2.175	\\
AGTDM - t0s2	&	1.435	&	2.154	&	2.783	&	3.594	&	2.594	\\
AGTDM - t2s2	&	2.713	&	2.895	&	2.182	&	2.907	&	1.417	\\
pAGTDM - t1s2	&	2.713	&	2.434	&	2.524	&	2.945	&	1.000	\\
AGTDM - t3s3	&	2.399	&	2.864	&	2.073	&	2.440	&	1.960	\\
pAGTDM - t2s3	&	2.394	&	2.304	&	2.845	&	2.110	&	2.195	\\
AGTDM - t2s1	&	2.626	&	2.852	&	2.364	&	3.362	&	1.000	\\
MI (k=6)	&	1.423	&	1.463	&	2.856	&	3.081	&	3.239	\\
AGTDM - t1s3	&	2.668	&	2.812	&	2.307	&	2.837	&	1.000	\\
pAGTDM - t3s3	&	2.554	&	1.911	&	2.860	&	2.994	&	1.801	\\
pAGTDM - t2s2	&	2.845	&	2.346	&	2.419	&	1.954	&	1.000	\\
pAGTDM - t3s2	&	2.837	&	2.337	&	2.272	&	2.783	&	1.000	\\
AGTDM - t2s3	&	2.434	&	2.854	&	1.830	&	2.864	&	1.080	\\
dCor	&	3.188	&	2.833	&	1.911	&	2.201	&	1.000	\\
Adaptive MI - s2	&	1.960	&	1.243	&	1.377	&	3.774	&	1.044	\\
Hoeffding's D	&	2.783	&	2.244	&	2.057	&	2.573	&	1.000	\\
HSIC	&	2.265	&	2.280	&	1.406	&	3.987	&	1.000	\\
MIC	&	1.000	&	1.219	&	1.000	&	1.000	&	1.000	\\
AGTDM - t0s1	&	1.531	&	1.448	&	1.770	&	2.822	&	1.204	\\
Adaptive MI - s1	&	1.000	&	1.146	&	1.668	&	2.224	&	2.355	\\
rdmCor	&	1.801	&	2.154	&	1.000	&	\textbf{4.706}	&	1.000	\\
MI (k=1)	&	1.000	&	1.000	&	1.565	&	2.022	&	1.507	\\
pAGTDM - t3s1	&	1.036	&	1.000	&	1.561	&	1.390	&	1.073	\\
R$^2$	&	\textbf{3.233}	&	1.000	&	1.000	&	1.000	&	1.000	\\           
            \hline
		\end{tabular}
	}  
\label{table:1DpowerStaircase}
\end{table}

\begin{table}[tb]
	\centering
	\caption{\textbf{Empirical evaluation of AGTDM.} Statistical power in 2-dimensional data with paired dependency (ranked by average power)}
	\resizebox{1\linewidth}{!}{
		\begin{tabular}{ l | l | l | l | l | l | l | l | l | l | l | l | l | l | l | l | l | l | l | l | l | l }
			 &  l-l & l-p & l-s & l-c & l-k & l-r & p-p & p-s & p-c & p-k & p-r & s-s & s-c & s-k & s-r & c-c & c-k & c-r & k-k & k-r & average \\ \thickhline
            MI (k=6) &  0.88 & 0.98 & 0.98 & 0.88 & \textbf{1.00} & 0.96 & 0.82 & 0.94 & 0.98 & 0.92 & 0.88 & 0.92 & \textbf{1.00} & 0.98 & \textbf{0.98} & 0.94 & 0.88 & 0.94 & 0.98 & \textbf{0.94} & 
\textbf{0.939} $\pm$ \textbf{0.050}       \\
            MI (k=1) & 0.82 & \textbf{1.00} & 0.92 & 0.90 & 0.96 & 0.98 & 0.84 & 0.96 & 0.98 & 0.88 & 0.88 & 0.94 & 0.98 & 0.96 & 0.94 & 0.92 & 0.96 & 0.90 & 0.84 & 0.00 & 
0.878 $\pm$ 0.213     \\
            AGTDM - t3s1 & \textbf{1.00} & \textbf{1.00} & \textbf{1.00} & \textbf{1.00} & \textbf{1.00} & \textbf{1.00} & \textbf{1.00} & \textbf{1.00} & \textbf{1.00} & \textbf{1.00} & \textbf{1.00} & \textbf{1.00} & \textbf{1.00} & 0.00 & 0.00 & \textbf{1.00} & 0.00 & \textbf{1.00} & \textbf{1.00} & 0.00 & 
0.800 $\pm$ 0.410        \\
            Hoeffding's D & \textbf{1.00} & \textbf{1.00} & \textbf{1.00} & \textbf{1.00} & \textbf{1.00} & \textbf{1.00} & \textbf{1.00} & \textbf{1.00} & \textbf{1.00} & \textbf{1.00} & \textbf{1.00} & \textbf{1.00} & \textbf{1.00} & 0.00 & 0.00 & \textbf{1.00} & 0.00 & 0.00 & \textbf{1.00} & 0.00 & 
0.750 $\pm$ 0.444        \\
            MIC & 0.94 & 0.96 & 0.90 & 0.02 & 0.80 & 0.96 & 0.98 & 0.98 & \textbf{1.00} & 0.92 & 0.98 & \textbf{1.00} & 0.80 & 0.00 & 0.00 & 0.90 & 0.64 & 0.12 & 0.08 & 0.00 & 0.649 $\pm$ 0.421     \\
            MI (k=20) & 0.86 & \textbf{1.00} & 0.00 & 0.92 & 0.96 & 0.98 & 0.86 & 0.98 & \textbf{1.00} & 0.88 & 0.86 & \textbf{1.00} & 0.02 & 0.02 & 0.02 & 0.92 & 0.00 & 0.94 & 0.14 & 0.00 & 
0.618 $\pm$ 0.447      \\
            HSIC & \textbf{1.00} & \textbf{1.00} & \textbf{1.00} & 0.00 & 0.00 & \textbf{1.00} & \textbf{1.00} & 0.00 & 0.00 & 0.00 & \textbf{1.00} & \textbf{1.00} & \textbf{1.00} & 0.00 & 0.00 & \textbf{1.00} & 0.00 & \textbf{1.00} & \textbf{1.00} & 0.00 & 
0.550 $\pm$ 0.510          \\
            AGTDM - t0s2 & \textbf{1.00} & \textbf{1.00} & 0.12 & 0.34 & 0.04 & 0.74 & \textbf{1.00} & 0.16 & 0.48 & 0.04 & 0.56 & \textbf{1.00} & \textbf{1.00} & 0.02 & 0.00 & \textbf{1.00} & 0.00 & 0.00 & \textbf{1.00} & 0.00 & 
0.475 $\pm$ 0.444           \\
            dCor &  \textbf{1.00} & \textbf{1.00} & \textbf{1.00} & 0.00 & 0.00 & \textbf{1.00} & \textbf{1.00} & 0.00 & 0.00 & 0.00 & \textbf{1.00} & \textbf{1.00} & 0.00 & 0.00 & 0.00 & \textbf{1.00} & 0.00 & 0.00 & \textbf{1.00} & 0.00 &  
0.450 $\pm$ 0.510      \\
            AGTDM - t1s2 & \textbf{1.00} & \textbf{1.00} & 0.42 & 0.00 & 0.02 & 0.62 & \textbf{1.00} & 0.00 & 0.00 & 0.02 & 0.68 & \textbf{1.00} & 0.02 & 0.00 & 0.74 & \textbf{1.00} & 0.00 & 0.00 & \textbf{1.00} & 0.00 &
 0.426 $\pm$ 0.456         \\
0.423 $\pm$ 0.469            \\
            AGTDM - t2s1 & \textbf{1.00} & \textbf{1.00} & 0.00 & 0.00 & 0.00 & \textbf{1.00} & \textbf{1.00} & 0.00 & 0.00 & 0.00 & \textbf{1.00} & \textbf{1.00} & 0.00 & 0.00 & 0.00 & \textbf{1.00} & 0.00 & 0.00 & \textbf{1.00} & 0.00 & 
0.400 $\pm$ 0.503           \\
            AGTDM - t1s1 & \textbf{1.00} & \textbf{1.00} & 0.48 & 0.00 & 0.00 & 0.00 & \textbf{1.00} & 0.00 & 0.00 & 0.00 & \textbf{1.00} & \textbf{1.00} & 0.00 & 0.00 & 0.00 & \textbf{1.00} & 0.00 & 0.00 & \textbf{1.00} & 0.00 & 
0.374 $\pm$ 0.483       \\
            AGTDM - t3s3 & \textbf{1.00} & \textbf{1.00} & 0.20 & 0.00 & 0.14 & 0.04 & \textbf{1.00} & 0.36 & 0.04 & 0.06 & 0.00 & \textbf{1.00} & 0.08 & 0.00 & 0.06 & \textbf{1.00} & 0.06 & 0.02 & 0.78 & 0.12 & 
0.348 $\pm$ 0.424          \\
            AGTDM - t1s3 & \textbf{1.00} & \textbf{1.00} & 0.12 & 0.02 & 0.16 & 0.12 & \textbf{1.00} & 0.02 & 0.04 & 0.04 & 0.12 & \textbf{1.00} & 0.10 & 0.02 & 0.02 & 0.98 & 0.10 & 0.08 & 0.92 & 0.06 & 
0.346 $\pm$ 0.430         \\
            AGTDM - t2s3 & \textbf{1.00} & \textbf{1.00} & 0.36 & 0.04 & 0.10 & 0.22 & \textbf{1.00} & 0.00 & 0.08 & 0.04 & 0.18 & \textbf{1.00} & 0.02 & 0.00 & 0.02 & 0.98 & 0.08 & 0.00 & 0.68 & 0.12 & 
0.346 $\pm$ 0.415          \\
            AGTDM - t3s2 & \textbf{1.00} & \textbf{1.00} & 0.70 & 0.00 & 0.02 & 0.04 & \textbf{1.00} & 0.00 & 0.00 & 0.00 & 0.06 & \textbf{1.00} & 0.00 & 0.00 & 0.00 & \textbf{1.00} & 0.00 & 0.00 & \textbf{1.00} & 0.00 & 
0.341 $\pm$ 0.468          \\
            AGTDM - t2s2 & \textbf{1.00} & \textbf{1.00} & 0.52 & 0.00 & 0.00 & 0.04 & \textbf{1.00} & 0.00 & 0.00 & 0.00 & 0.04 & \textbf{1.00} & 0.00 & 0.00 & 0.02 & \textbf{1.00} & 0.02 & 0.00 & \textbf{1.00} & 0.00 & 
0.332 $\pm$ 0.463          \\
            AGTDM - t0s1 & \textbf{1.00} & \textbf{1.00} & 0.00 & 0.04 & 0.04 & 0.18 & \textbf{1.00} & 0.00 & 0.10 & 0.00 & 0.00 & \textbf{1.00} & 0.00 & 0.00 & 0.00 & \textbf{1.00} & 0.00 & 0.00 & \textbf{1.00} & 0.00 & 
0.318 $\pm$ 0.460           \\
            rdmCor & \textbf{1.00} & \textbf{1.00} & 0.00 & 0.00 & 0.00 & \textbf{1.00} & \textbf{1.00} & 0.00 & 0.00 & 0.00 & \textbf{1.00} & 0.00 & 0.00 & 0.00 & 0.00 & 0.00 & 0.00 & 0.00 & 0.00 & 0.00 & 
0.250 $\pm$ 0.444           \\
            R$^2$ & \textbf{1.00} & \textbf{1.00} & 0.00 & 0.00 & 0.00 & 0.00 & \textbf{1.00} & 0.00 & 0.00 & 0.00 & 0.00 & 0.00 & 0.00 & 0.00 & 0.00 & 0.00 & 0.00 & 0.00 & \textbf{1.00} & 0.00 &  
0.200 $\pm$ 0.410      \\
            \hline
		\end{tabular}
	}  
\label{table:2Dpower}
\end{table}

\begin{figure}[H]
\centering
    \includegraphics[width=0.48\linewidth]{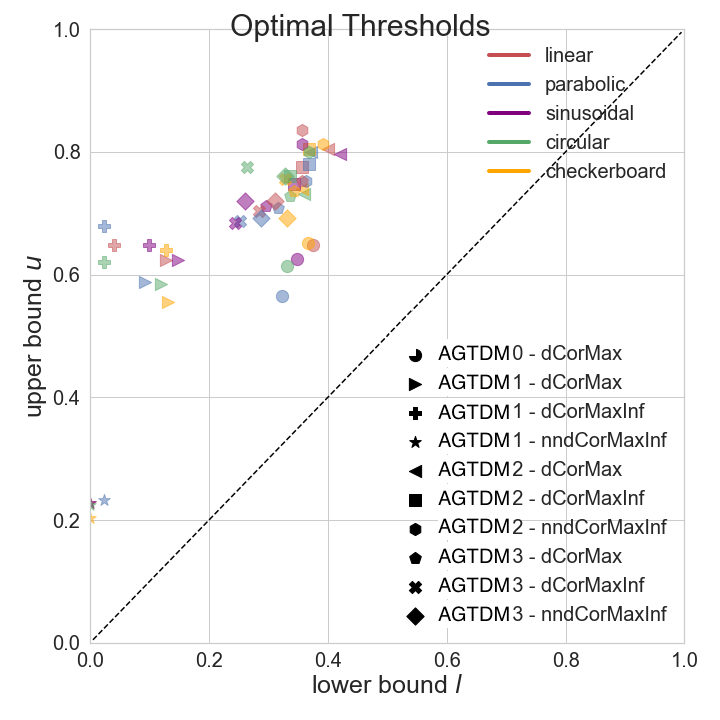}
        \includegraphics[width=0.48\linewidth]{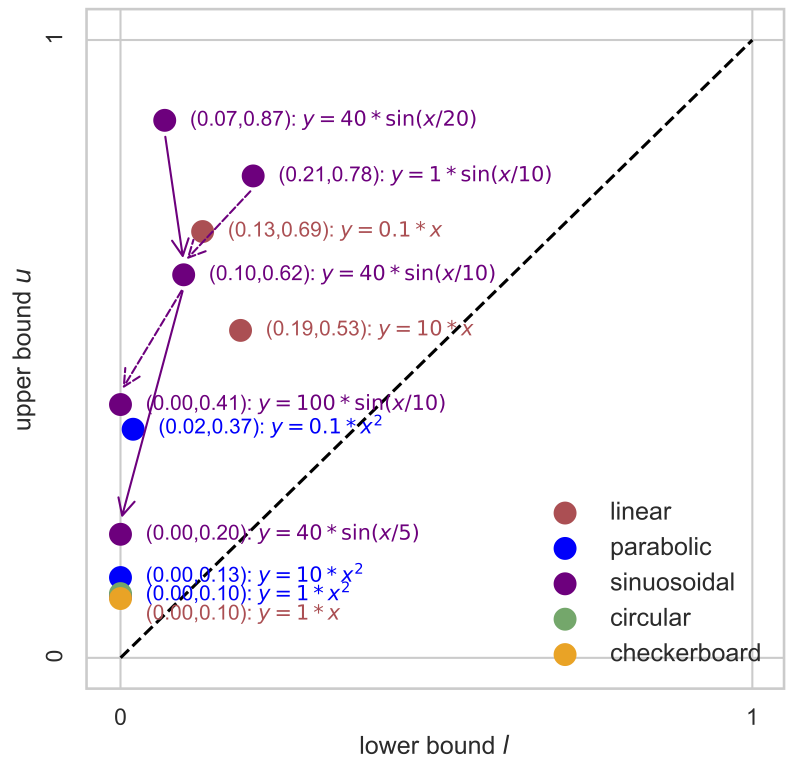}
\par\caption{\textbf{Optimal thresholds of the GT transforms. for different relationships.} Demonstrated are the adaptively chosen lower (horizontal axis) and upper (vertical axis) bounds of the geo-topological transform for tests (shapes) and statistical patterns (colors).}\label{fig:opts}
\end{figure}

Optimal thresholds are also recorded during the bivariate association experiments with increasing noise amplitude (see Table \ref{table:optThresh} and Fig. \ref{fig:opts} for the optimal thresholds). Fig. \ref{fig:opts} maps these optimal thresholds with x as lower bound and y as the upper bounds. While the drifts of optimal thresholds are not obvious, the differences of these thresholds in different scenarios can offer us useful insights about the data quality, structures and properties. For example, checkerboard patterns (yellow) tend to have smaller upper bounds, implying that the distributional periodicity can be better captured by silencing the out-of-phase signals.

\begin{figure}[H]
\centering
    \includegraphics[width=0.9\linewidth]{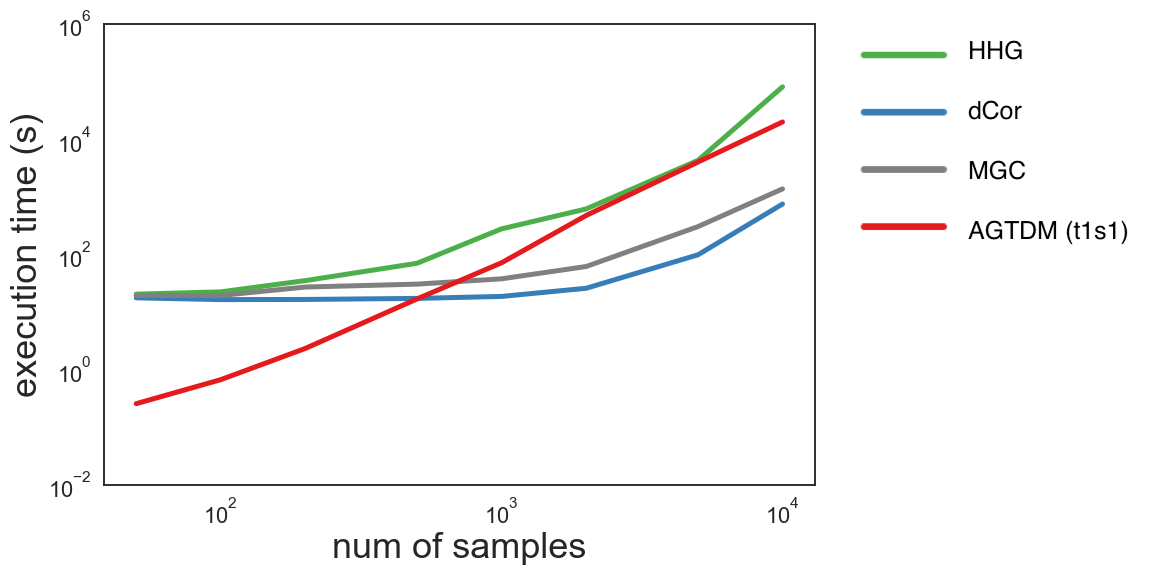}
\par\caption{\textbf{Run time analysis of AGTDM and existing methods.} \textbf{(e)} execution wall time of AGTDM, dCor, MGC and HHG over different numbers sample sizes of 1d linear dependency (with AGTDM in its most time consuming variant).}\label{fig:time} 
\end{figure}

\textbf{Run time analysis.} Fig. \ref{fig:time} records running time for different methods. AGTDM (in its most expensive form) has complexity $O(k^4)$, higher than dCor, but empirically we observed a lower than baseline run time when N is small (partially due to our parallel implementation). When N is large ($>10^4$), AGTDM can be expensive, for which we provided an alternative cheaper solution which replaces exhaustive searching of $k^4$ threshold sets with iterative sampling.

\begin{figure}[tb]
\centering
    \hfill\includegraphics[width=0.42\linewidth]{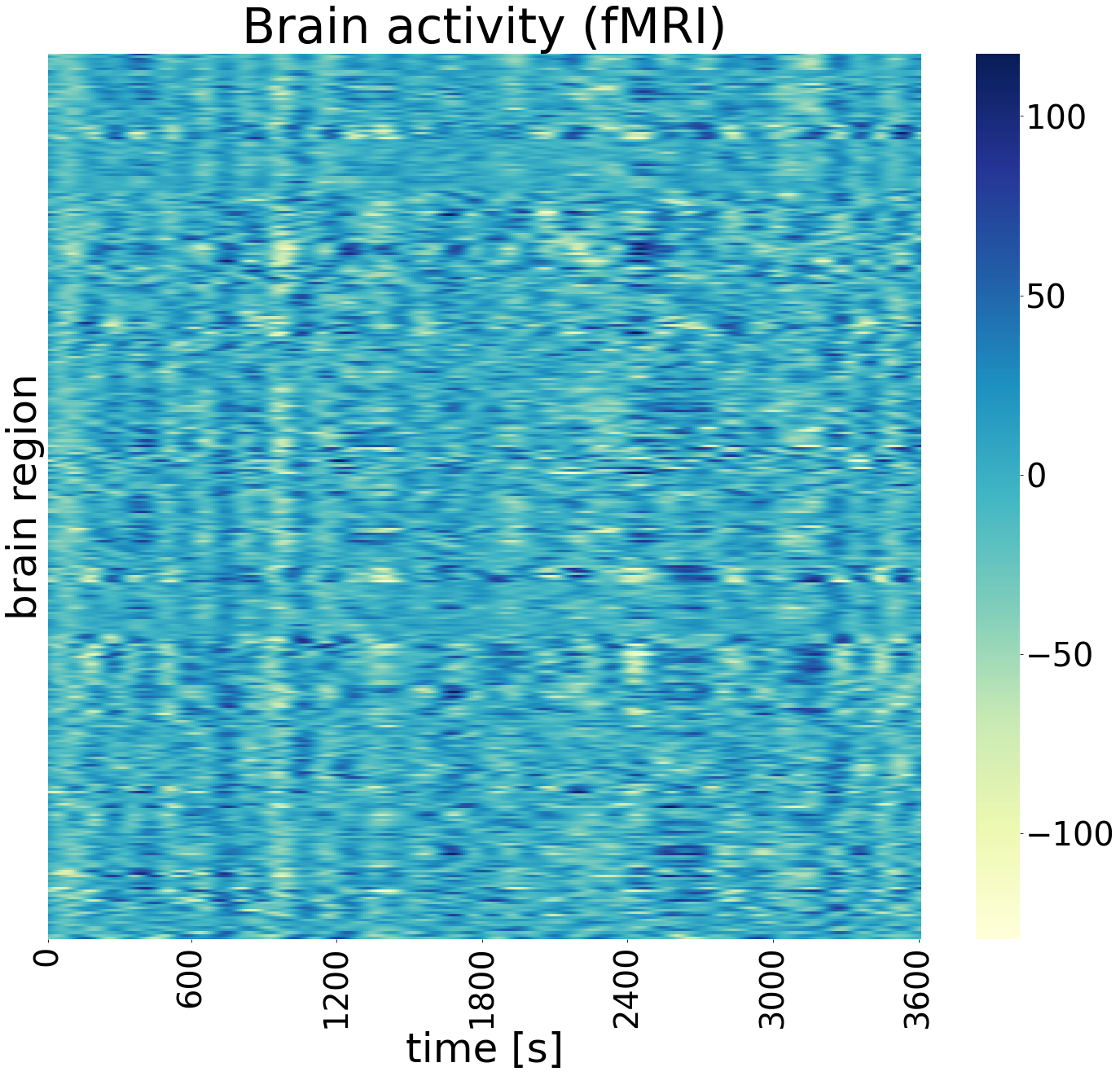}
    \includegraphics[width=0.48\linewidth]{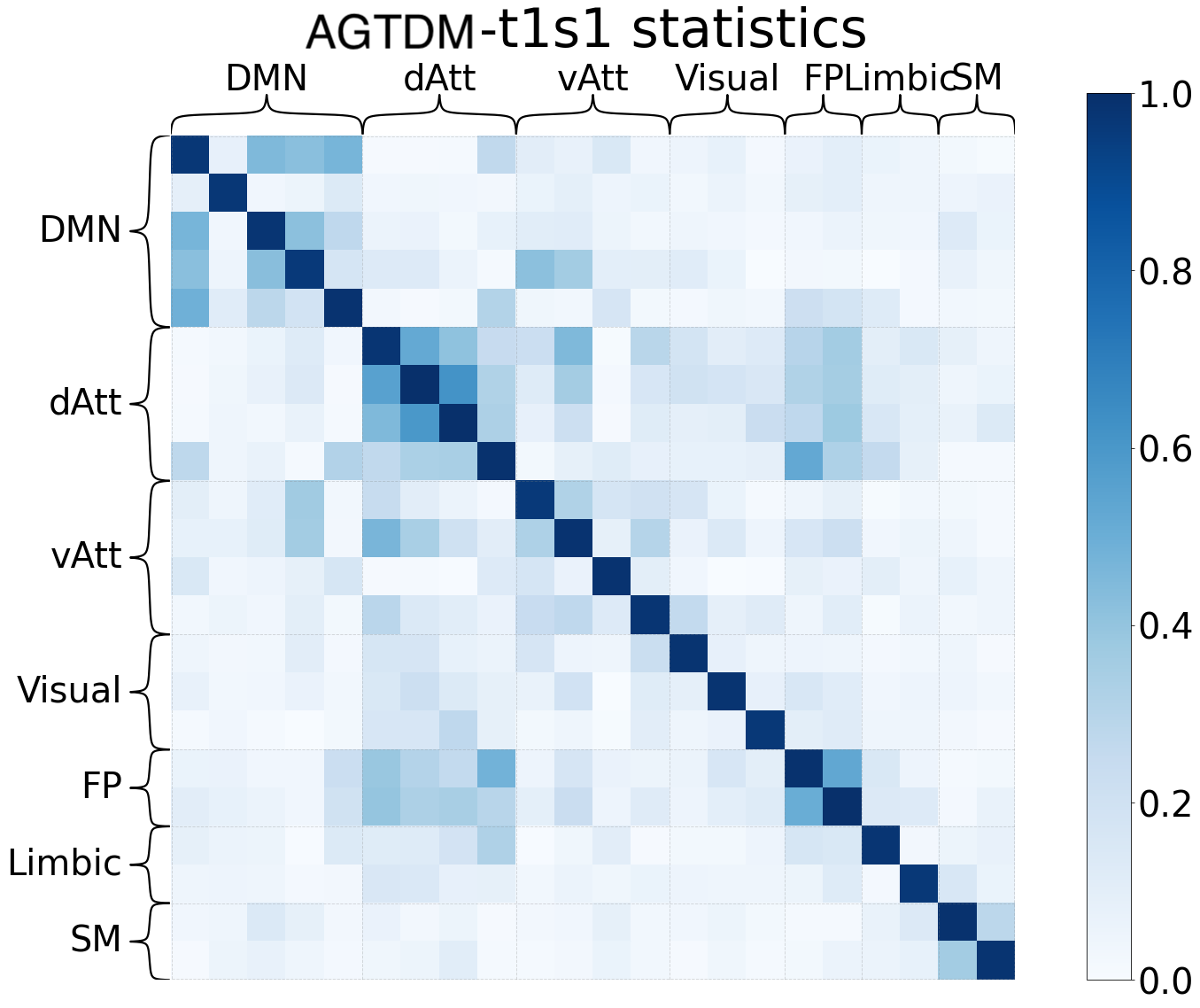}\hfill
    \includegraphics[width=0.48\linewidth]{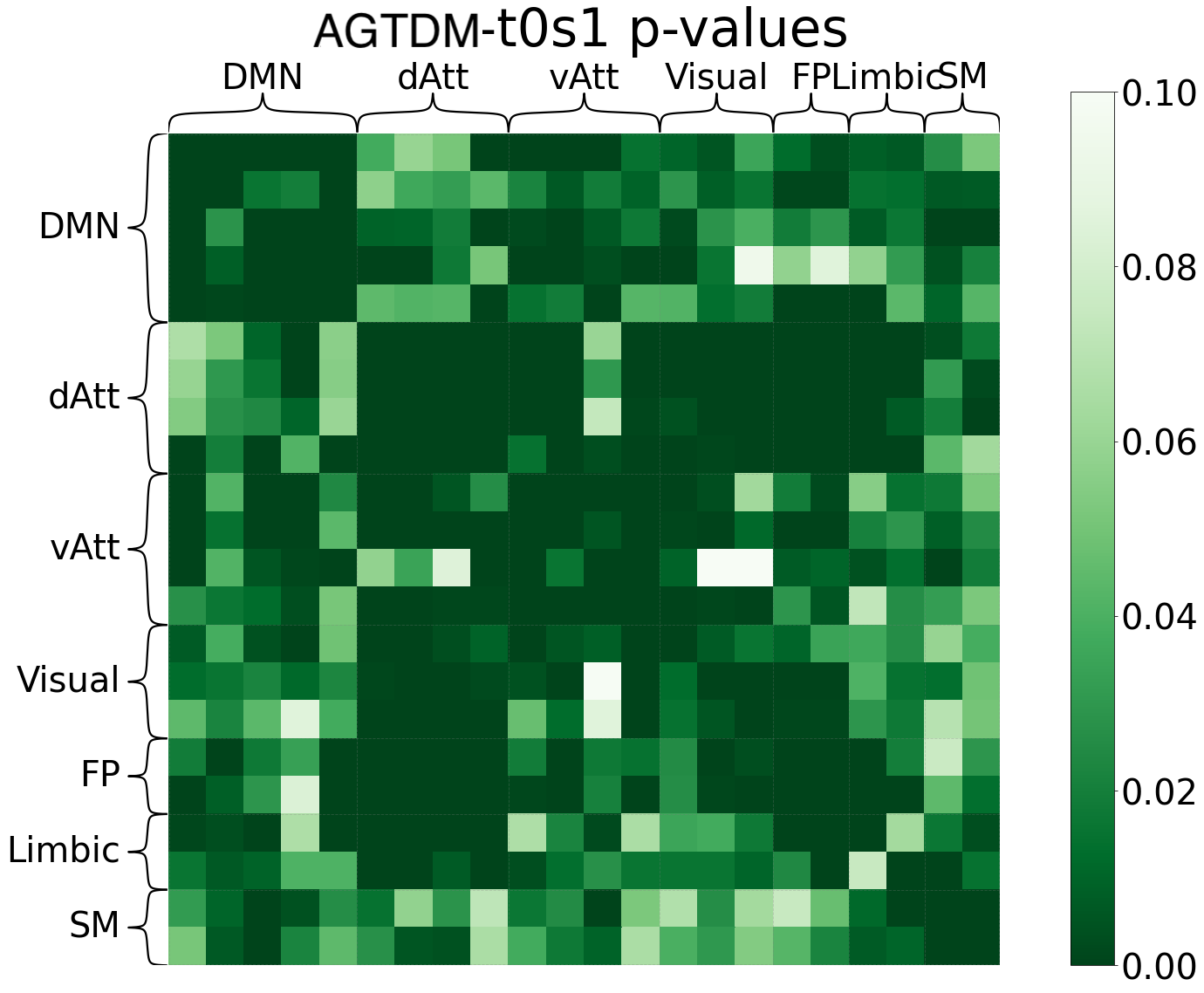}\hfill
    \includegraphics[width=0.48\linewidth]{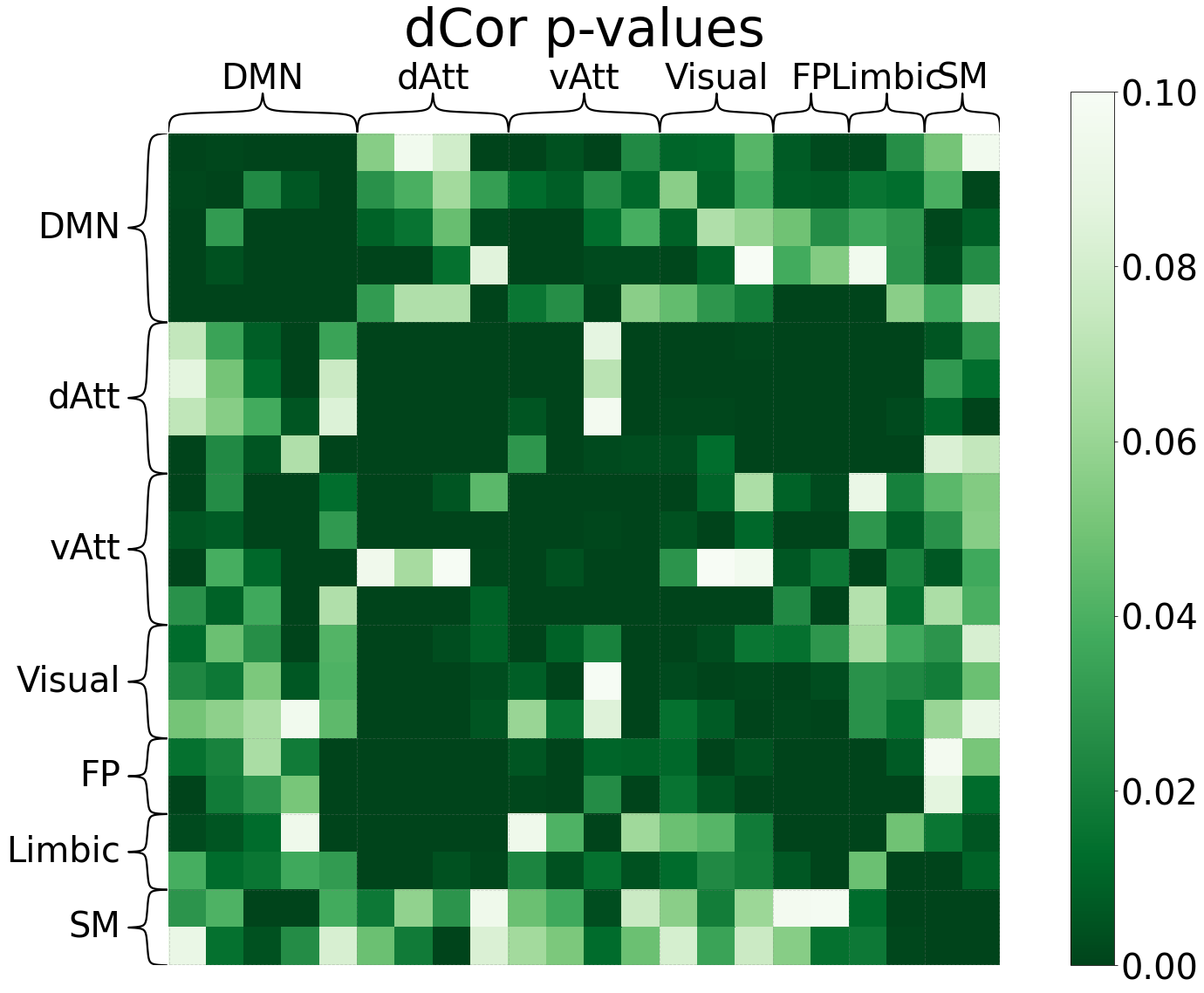}\hfill
\par\caption{\textbf{Empirical results of AGTDM on neuroscience data.} \textbf{(b)} example data format from the human Connectome Program (HCP) dataset \textbf{(b)} AGTDM computed across different brain regions of interest (ROIs); \textbf{(c)} the p-values accompanying the computed AGTDM; \textbf{(d)} the p-values accompanying the computed dCor.}\label{fig:real} 
\end{figure}

\textbf{Proof of concept in a real-world example.} We now apply AGTDM to the analysis of connectivity in the human brain. We analyzed functional MRI brain-activity data from a human subject (ID: 100307) of the Human Connectome Project (HCP). The cortex was parcellated into 180 regions per hemisphere using HCP multi-modal parcellation atlas (\cite{glasser2016multi} and Fig. \ref{fig:real}a), among which 22 parcels were selected as regions-of-interest (ROIs) for connectivity analysis. We computed the AGTDM for each pair of ROIs and report the statistics and their corresponding p-values (Fig. \ref{fig:real}b,c). We see blockwise mutual information among sets of brain regions. The p-value matrix of AGTDM illustrates the improved power compared to dCor in this practical application (Fig. \ref{fig:real}c,d,e). The power advantage is more formally empirically demonstrated in synthetic data using simulations, where the ground-truth is known.

\section{Conclusions and Ongoing Directions for AGTDM}

Distance matrices capture the representational geometry and can be subjected to monotonic nonlinear transforms to capture the representational topology at different granularities. We introduced a novel family of independence tests that adapt the parameters of these geo-topological transforms so as to maximize sensitivity of the distance covariance to statistical dependency between two multivariate variables. The proposed test statistics 
perform well empirically, providing robust sensitivity across a wide range of univariate and multivariate relationships and across different noise levels and amounts of data. 
The present results suggest that it might be useful for a wide range of practical applications such as analyzing biological and societal data where we can (1) detect whether there is any dependency in the data and (2) understand the relationships in structured data.

While initial empirical results suggest that the Adaptive Geo-Topological Dependence Measure (AGTDM) performs well in detecting various types of statistical dependencies in independence testing, it's important to note that a formal proof of its status as an independence criterion is currently lacking. This situation presents both challenges and opportunities for future research.

To summarize the current understanding:

\begin{itemize}
    \item The AGTDM shows promise in empirical tests across various dependency structures.
\item Its adaptive nature allows it to capture both geometric and topological aspects of dependencies.
\item However, without a formal proof, we cannot definitively claim it as an independence criterion.
\end{itemize}

Other than the empirical testing on realistic data and investigating the interpretability of this adaptive method, ongoing research directions on the theoretical side include:

\begin{itemize}
    \item Rigorous mathematical analysis to determine if AGTDM can be proven as an independence criterion or if modifications are needed to achieve this status.
    \item Exploration of the theoretical properties of AGTDM, including its behavior under different types of dependencies and data distributions.
    \item Investigation into potential modifications or extensions that might lead to a provable independence criterion while maintaining the adaptive geo-topological properties.
\end{itemize}

We encourage the research community to engage with these open questions. Resolving the theoretical status of AGTDM or developing related measures with provable properties would significantly advance the field of dependence testing.

\begin{table}[tb]
	\centering
	\caption{\textbf{Adaptively selected optimal thresholds of GT transforms.} AGTDM across noise amplitudes over different relationships}
	\resizebox{1\linewidth}{!}{
		\begin{tabular}{ l | l | l | l | l | l }
			 &  linear & parabolic & sinusoidal & circular & checkerboard \\ \thickhline
            AGTDM - t0s1 & 0.466 & 0.403 & 0.460 & 0.632 & 0.532 \\
            AGTDM - t0s2 & l: 0.375, u: 0.649 & l: 0.323, u: 0.566 & l: 0.349, u: 0.626 & l: 0.332, u: 0.614 & l: 0.366, u: 0.652   \\
            AGTDM - t1s1 & l: 0.128, u: 0.624 & l: 0.092, u: 0.588 & l: 0.148, u: 0.624 & l: 0.120, u: 0.584 & l: 0.132, u: 0.556      \\
            AGTDM - t1s2 & l: 0.400, u: 0.804 & l: 0.372, u: 0.800 & l: 0.420, u: 0.796 & l: 0.360, u: 0.732 & l: 0.356, u: 0.740    \\
            AGTDM - t1s3 & l: 0.356, u: 0.752 & l: 0.316, u: 0.708 & l: 0.296, u: 0.712 & l: 0.336, u: 0.728 & l: 0.344, u: 0.736     \\
            AGTDM - t2s1 & l: 0.040, u: 0.648 & l: 0.024, u: 0.680 & l: 0.100, u: 0.648 & l: 0.024, u: 0.620 & l: 0.128, u: 0.640      \\
            AGTDM - t2s2 & l: 0.356, u: 0.776 & l: 0.368, u: 0.780 & l: 0.344, u: 0.748 & l: 0.336, u: 0.760 & l: 0.368, u: 0.804     \\
            AGTDM - t2s3 & l: 0.284, u: 0.704 & l: 0.252, u: 0.688 & l: 0.244, u: 0.684 & l: 0.264, u: 0.776 & l: 0.328, u: 0.756      \\
            AGTDM - t3s1 & l: 0.000, u: 0.228 & l: 0.024, u: 0.232 & l: 0.000, u: 0.228 & l: 0.000, u: 0.224 & l: 0.000, u: 0.204      \\
            AGTDM - t3s2 & l: 0.356, u: 0.836 & l: 0.364, u: 0.752 & l: 0.356, u: 0.812 & l: 0.368, u: 0.800 & l: 0.392, u: 0.812     \\
            AGTDM - t3s3 & l: 0.312, u: 0.720 & l: 0.288, u: 0.692 & l: 0.260, u: 0.720 & l: 0.328, u: 0.760 & l: 0.332, u: 0.682     \\
            pAGTDM - t1s1 & l: 0.000, u: 0.380 & l: 0.000, u: 0.424 & l: 0.020, u: 0.404 & l: 0.024, u: 0.348 & l: 0.012, u: 0.332       \\
            pAGTDM - t1s2 & l: 0.388, u: 0.756 & l: 0.372, u: 0.788 & l: 0.448, u: 0.816 & l: 0.352, u: 0.760 & l: 0.364, u: 0.748      \\
            pAGTDM - t1s3 & l: 0.216, u: 0.668 & l: 0.280, u: 0.704 & l: 0.368, u: 0.728 & l: 0.232, u: 0.652 & l: 0.324, u: 0.728      \\
            pAGTDM - t2s1 & l: 0.032, u: 0.632 & l: 0.000, u: 0.600 & l: 0.016, u: 0.576 & l: 0.004, u: 0.600 & l: 0.012, u: 0.620  \\
            pAGTDM - t2s2 & l: 0.424, u: 0.820 & l: 0.408, u: 0.807 & l: 0.360, u: 0.796 & l: 0.332, u: 0.788 & l: 0.364, u: 0.840 \\
            pAGTDM - t2s3 & l: 0.280, u: 0.756 & l: 0.252, u: 0.704 & l: 0.304, u: 0.684 & l: 0.296, u: 0.724 & l: 0.276, u: 0.696 \\
            pAGTDM - t3s1 & l: 0.000, u: 0.200 & l: 0.000, u: 0.200 & l: 0.000, u: 0.200 & l: 0.000, u: 0.200 & l: 0.000, u: 0.200 \\
            pAGTDM - t3s2 & l: 0.464, u: 0.864 & l: 0.516, u: 0.888 & l: 0.428, u: 0.872 & l: 0.436, u: 0.816 & l: 0.412, u: 0.816 \\
            pAGTDM - t3s3 & l: 0.344, u: 0.764 & l: 0.356, u: 0.740 & l: 0.352, u: 0.716 & l: 0.256, u: 0.660 & l: 0.280, u: 0.672 \\
            \hline
		\end{tabular}
	}  
\label{table:optThresh}
\end{table}

\chapter{Time as the Third Dimension: Topological Data Analysis in Time Series}


As we delve deeper into the complexities of neural representations, we now confront one of the most fundamental aspects of brain function: its dynamic nature. This chapter marks a significant shift in our analytical approach, as we incorporate time as a crucial third dimension in our topological analysis of neural data.

The brain's computations unfold over time, with representations evolving, transforming, and interacting in complex ways. To truly understand these processes, we must move beyond static analyses and develop methods that can capture the temporal dynamics of neural representations. This chapter introduces novel techniques that merge the insights of topological data analysis with the challenges of time series data.

Our journey in this chapter is twofold. First, we'll explore methods for visualizing and analyzing how representational geometries change over time, providing a dynamic view of neural computations. Second, we'll introduce advanced techniques for performing topological data analysis on time series data, with a particular focus on applications to single-cell genomics.

These approaches not only enhance our ability to analyze neural data but also open new avenues for understanding developmental trajectories and longitudinal changes in biological systems. By the end of this chapter, readers will have a comprehensive understanding of how to incorporate temporal dynamics into topological analyses of complex biological data.

This chapter leads to the following publications: 

\cite{lin2019visualizing} \fullcite{lin2019visualizing}.

\cite{ttda} \fullcite{ttda}.

\cite{bibm} \fullcite{bibm}.

\cite{www2022} \fullcite{www2022}.

\section{Understanding and Visualizing Representational Dynamics}




In recent years, technological innovations in computer vision have produced biologically-plausible models for human visual information processing. Among these models are goal-driven deep feedforward hierarchical neural networks, which have been proposed to model the ventral stream of visual cortex, the ``what pathway'' in the brain thought to underlie object recognition \cite{yamins2016using,khaligh2014deep,kriegeskorte2015deep}. However, there is a discrepancy in the hierarchical depth between the primate ventral visual stream ($<$ 10 stages of representation) and state-of-the-art computer-vision models ($>$ 100 layers). The primate visual system might make up for its limited hierarchical depth by recycling its resources through time, via recurrent connections and attention mechanisms, all of which require the analysis of the entire time-series of the dynamics in the brain. Few studies have investigated the dynamics of visual perception (for instance, object identity and categorization) and their representational changes \cite{hung2005fast,freiwald2010functional}.

The scarcity of methods to characterize the representational dynamics creates a major barrier to answer interesting questions such as: how are objects represented in the brain over the time course from early perception to categorical decision making, does the object identification or visual categorization follows a hierarchical classification paradigm; do different classes of objects merge and branch at different time points based on different tasks or recurrence paradigm; are these representational dynamics oscillatory or recurrent? 

A central challenge is exactly to test computational theories implemented in deep neural network models with exactly this type of time-stamped brain-activity data. Analyses of representational geometry can help us to compare representations between biological brains and computational models, and to understand brain computation as the transformation of representational similarity structure across stages of processing \cite{kriegeskorte2013representational}. 

Traditional RSA usually considers the entire time series of the neural measurement as the response pattern. In that sense, the dynamics is collapsed into one data point to characterize the brain-activity corresponding to a specific stimulus. However, it is unclear how to properly capture the representational dynamics, i.e., how the representations evolve over time. 

Neural computations unfold over time. Visualizations of the representational dynamics can help us understand how representations are transformed and gain insight into the computational process. Static visualizations of individual time frames can be animated by applying them in a time-windowed style to render neural dynamics as a movie. However, it is also important to visualize neural dynamics in static images. Classical studies have used dimensionality reduction methods (notably PCA and supervised variations including jPCA \cite{churchland2012neural}, dPCA \cite{kobak2016demixed}, and HDR \cite{lara2018a}) to embed high-dimensional trajectories of neural population activity in 2d and 3d visualizations. This is a powerful approach. However, for rich sets of representational states (e.g. many visual stimuli or distinct movements), linear dimensionality reduction may collapse neural states that are untangled in the high-dimensional representation \cite{russo2018}. 
This motivates alternative visualization methods that aim to represent the pairwise discriminabilities of the representational states (e.g. of sensory stimuli). MDS offers one such method, which optimizes the accuracy with which pairwise distances in the high-dimensional space are reflected in the visualization.

\begin{figure}[H]
\centering
    \includegraphics[width=0.75\linewidth]{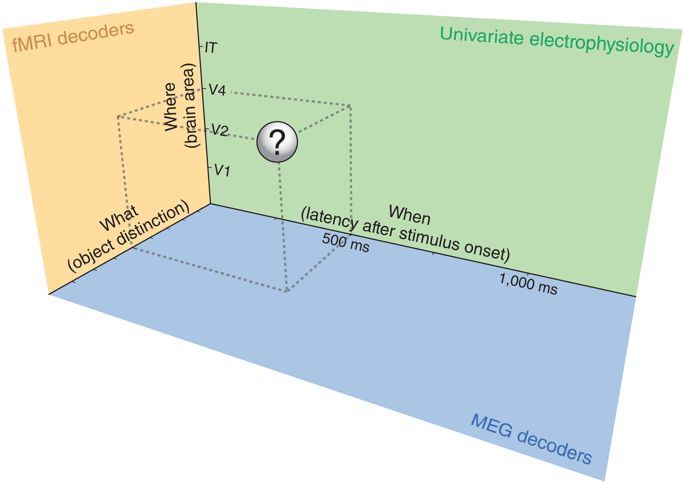}
    \vspace{.25em}
    \includegraphics[width=\linewidth]{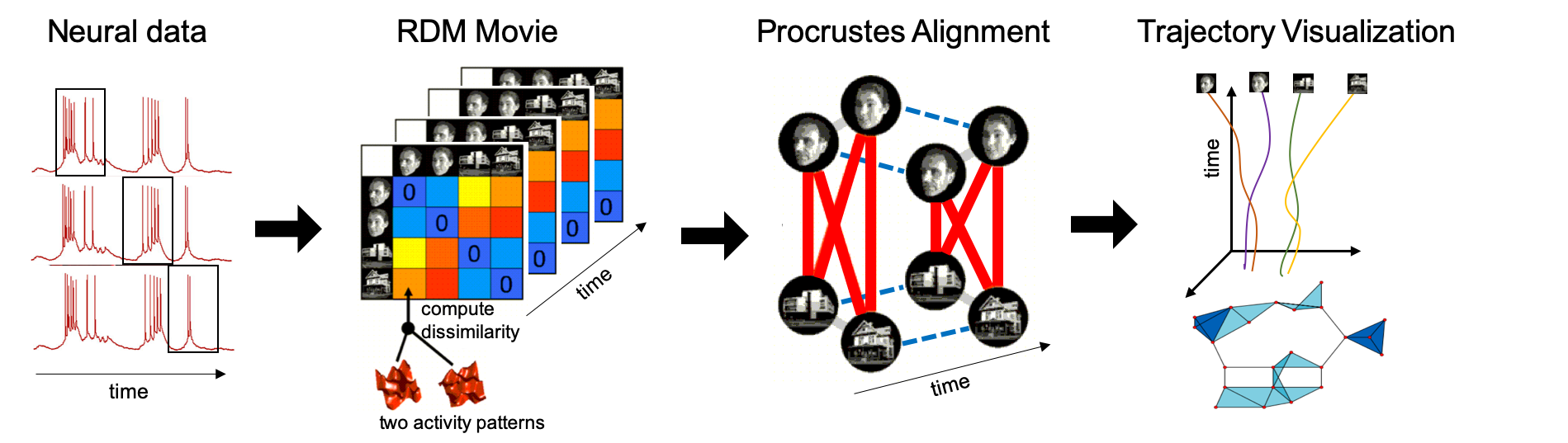}
\caption{\textbf{Motivation and pipeline of representational dynamics analysis.} From earlier chapters, we know that the representational dissimilarity of activity patterns within a brain region encapsulate the essence of what is represented in response to each pair of images or categories. The goal is to systematically quantify all such pairwise dissimilarities—capturing the ``what'' aspect of representation—for each specific brain area, thereby addressing the ``where'', as well as at each moment post-stimulus onset, which tackles the ``when'' \cite{mur2014s}. Presented here is a pipeline of the 2d+time representational dynamic analysis. We process the neural data in sliding window fashion and compute RDM ``movies''. These movies are then Procrustes-aligned before downstream analysis and visualizations \cite{www2022}.}\label{fig:repDyn}
\end{figure}

We develop extensions of MDS for visualizing dynamic representational geometries. The analyses of representational dynamics will use the summary statistics of geometry and topology developed in chapter 2 and 4. We can relax the requirement of MDS for models to accurately reflect the geometry (e.g. pairwise distances) and require only that models accurately reflect the topology (e.g. neighborhoods) of representational space. 

In this study, we propose a framework to first extract the snapshots of representational spaces with sliding-window RSA, and then align each frame of this RDM ``movie'' (as snapshots of the representational space) with Generalized Procrustes Analysis (GPA). We presented the visualizations on the data of monkey's stimulus-driven single-electrode recording, and demonstrated several neuroscience insights on visual object categorization revealed from the proposed method.

\subsection{Ride along with a dynamic representational geometry}

Einstein famously imagined riding alongside a beam of light. This thought experiment, a mental visualization, rendered stationary the most rapid dynamic process, and was a major inspiration that contributed to a scientific breakthrough. To visualize neuronal population dynamics, we may similarly choose either a static or a moving reference frame. 
\cite{mante2013context,churchland2012neural,hung2005fast,freiwald2010functional} 
are often visualized using dimensionality reduction methods such as PCA and its extensions \cite{mante2013context,churchland2012neural,lara2018conservation,kobak2016demixed} to find a single 2d or 3d linear subspace in which to view the trajectories. This approach of linear-subspace trajectories is well motivated by its simplicity and interpretability. Certain distinctions may be collapsed, but an advantage is that the approach equally represents \textit{within}-time-point and \textit{between}-time-point relationships among representational patterns. This enables us to visualize neural population response for even a single experimental condition as a trajectory. 

Here we introduce an alternative approach, in which we travel along with the ensemble of high-dimensional points that correspond to different experimental conditions (e.g., different stimuli or different movements performed by the animal). Rather than embedding all neural activity patterns across all time points in a single space, we first seek to represent the relationships between the patterns separately at each time point. We then align the embeddings across time points so as to \textit{minimize} motion of the ensemble as a whole and render more apparent the changes of the representational geometry or topology, i.e. which stimuli become distinct and which collapse and how their global distances develop. 
Any visualization method including linear subspace methods and the methods from Aim 1 could be used to obtain a separate configuration for each time frame. To illustrate the concept of a dynamic visualization with a moving frame of reference, let us consider MDS. MDS is a natural choice for visualizing geometries because it minimizes the distortion of the distances in the visualization (using the metric stress cost function here). The MDS cost function explicitly penalizes the collapsing of points that are distant in the the original space, e.g., neural patterns that represent distinct states \cite{russo2018}.

Fig. \ref{fig:repDyn} shows the sequence of steps of the proposed procedure. First each time frame of neural activity is analyzed separately. This could involve, for example, spike counts performed in a temporal sliding window to obtain a neural population activity pattern for each experimental condition. Second, for each time frame, the representational dissimilarity is estimated \cite{kriegeskorte2019peeling} for each pair of experimental conditions and assembled in the RDM. Stacking these RDMs along the time dimension results in an RDM movie. Third, each RDM is separately subjected to MDS to obtain, for each time frame, a 2d or 3d configuration of the points for visualization. Fourth, the configurations are aligned to each other using Generalized Procrustes Analysis (GPA), and finally, the aligned configurations are presented together. The alignment implements the moving frame of reference, which minimizes the motion of the ensemble as a whole in the visualization from one time point to the next. We can present the low-dimensional embedding of the dynamics as a movie (using visualization time to represent neural activity time) or as a static figure. For a static figure, we must choose how to indicate the time frame. In Fig. \ref{fig:repDyn}, we designate one dimension of the visualization space as representing time and used the remaining two orthogonal dimensions for the relationships among stimuli within a time frame.

\subsection{Procrutes-aligned multidimensional scaling (pMDS)}



\textbf{Multidimensional Scaling (MDS).}
As a popular non-linear dimensionality reduction method, Multidimensional Scaling (MDS) rearrange the location of a set of data points from a set, given a distance matrix characterizing the distances between each pair of objects in a set (for instance, the RDM computed from the response patterns of different stimuli), into an N-dimensional space such that the between-object distances are preserved \cite{buja2008data}. 

\textbf{Generalized Procrustes Analysis (GPA).}
In computer vision and signal processing, Procrustes analysis is usually used to analyse the statistical distribution of a set of shapes. The Generalized Procrustes analysis (GPA) compares three or more shapes to an optimally determined "mean shape" and can align all the shapes according to this mean shape as a reference frame \cite{gower1975generalized}. GPA solves the mean shape iteratively by optimizing against the Procrustes distance, a metric to minimize in order to superimpose any pair of shape or time frame instances annotated by landmark points. The analysis starts from choosing an arbitrary reference frame, and then superimpose all instances to current reference shape. If Procrustes distance between the reference shape and the computed mean shape of the current set of superimposed shapes is above a certain threshold, the reference shape is set to the mean shape to continue above steps iteratively, until the Procrustes distance between the two is small enough within the trivial threshold.

\textbf{Procrustes-aligned MDS (pMDS).}
In our specific problem, the RDM movie consists of the representational shapes of each time point without any intertemporal information. We apply GPA to the MDS embeddings computed from RDMs at each time point, such that each frame are optimally aligned to all other time frames. The Procrustes analysis has the option to constrain rotation, scaling and reflection, but in our case, we only allow the rotation and reflection, because the scaling contains information about the how the representations diverge and converge over time. Because Procrustes alignment doesn't distort the geometrical information between each stimuli (constrained by the individual RDM at each time point), the Procrustes-aligned MDS (pMDS) can offer us a genuine and illustrative visualization of the representational dynamics over time.

\subsection{MDS alignment reveals smooth transition over time}

The data used to demonstrate the proposed method are the monkey single-electrode recordings from the inferior bank of the ST segment \cite{bell2011relationship}. Two adult male rhesus monkeys were shown 100 grayscale object images from five different categories each with 20 instances (faces, fruits, places, body parts and objects) in a serial visual presentation. RSA was further applied to select visually-responsive neurons and extract single-trial response patterns from spike-density function. The recordings were truncated into sections of 821 ms (starting from 100ms before stimulus onset). RDM movies were generated using a sliding window of 21 ms with cross-validated spike rate distance (SRD) as the reponse-pattern dissimilarity measure.

\begin{figure}[tb]
\centering
    \includegraphics[width=1\linewidth]{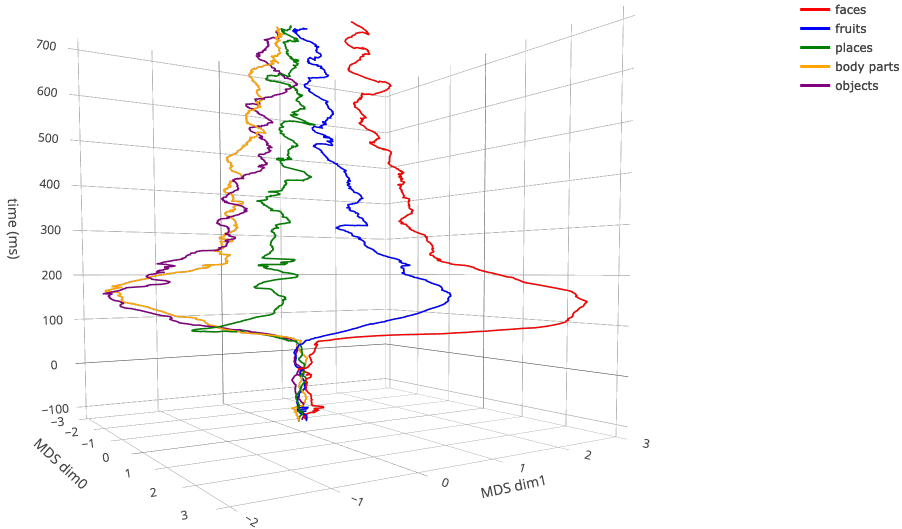}\par\caption{\textbf{Empirical results of pMDS.} 2d+time plot of Procrustes-aligned MDS of the neural representational patterns, averaged and colored by stimulus categorical information.}\label{fig:3d}
\end{figure}

\begin{figure}[tb]
\centering
    \begin{minipage}{.32\textwidth}\centering
    \includegraphics[width=.95\linewidth]{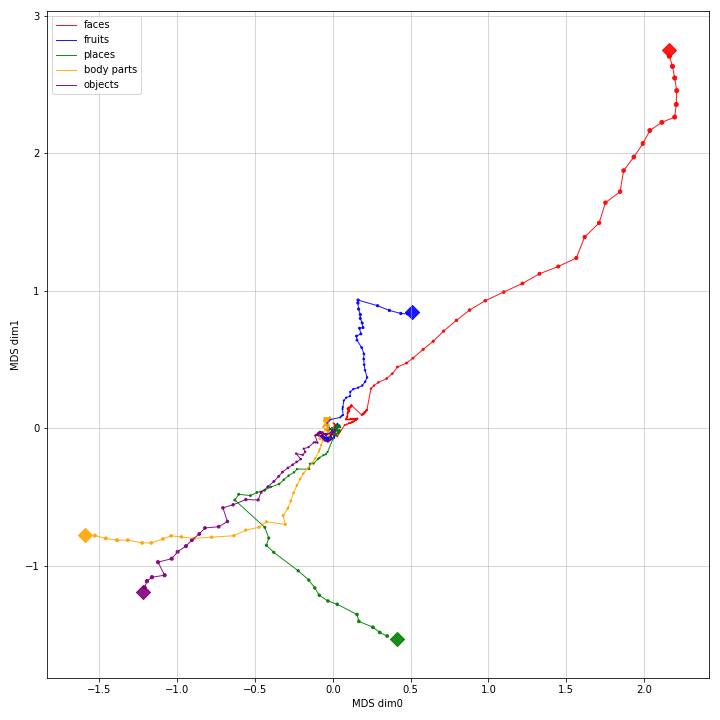}\par\caption{\textbf{Visualization of pMDS.} 2d temporal footprint of annotated representational space during onset (0-100ms).}\label{fig:2d_0_100}
    \end{minipage}\hfill
    \begin{minipage}{.33\textwidth}\centering
    \includegraphics[width=.93\linewidth]{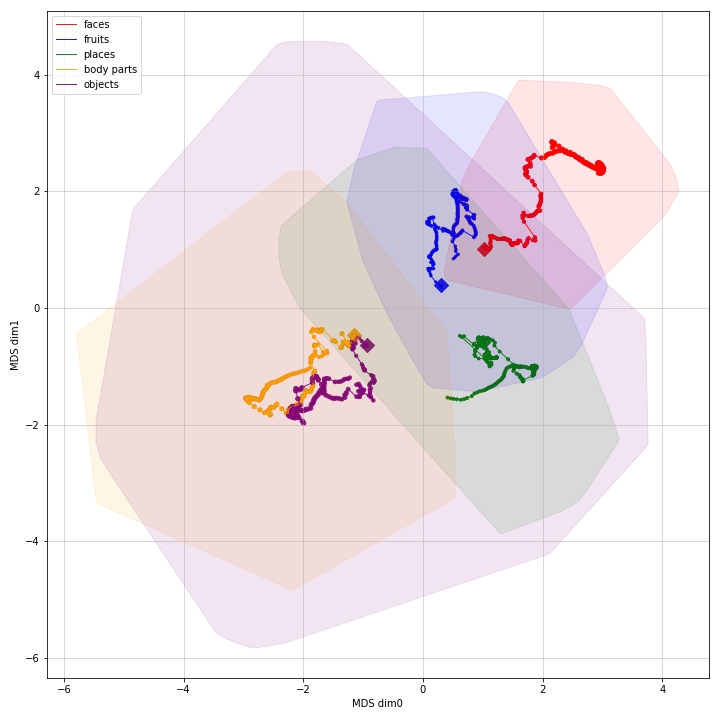}
    \vspace{-.1em}
    \par\caption{\textbf{Visualization of pMDS.} 2d temporal footprint of annotated representational space during onset (100-300ms).}\label{fig:2d_100_300}
    \end{minipage}\hfill
    \begin{minipage}{.32\textwidth}\centering
    \includegraphics[width=.95\linewidth]{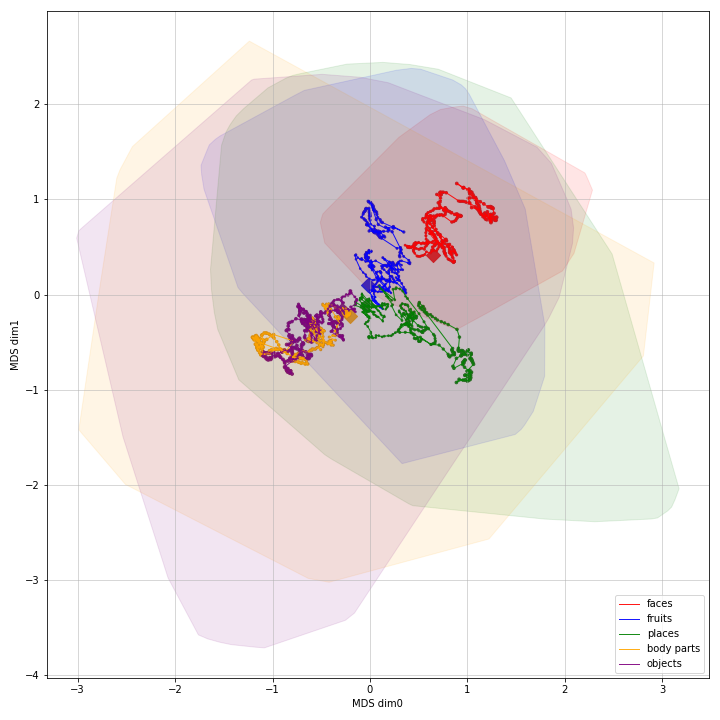}\par\caption{\textbf{Visualization of pMDS.} 2d temporal footprint of annotated representational space after offset (300-800ms).}\label{fig:2d_300_800}
    \end{minipage}
\end{figure}

We performed GPA on the MDS embeddings computed from each time frame of RDM movies MDS based on the stimulus label (not the category label). As shown in Fig. \ref{fig:3d}, the average trajectory of all the data points of the same category are plotted in the MDS space over time. The dynamics over time can be reasonably visualized while the separation of each categories (as between-category distances) is well preserved. From the 3d plot of the representational space, the separation of each categories happens around 80ms and reached a maximum distinction at around 150ms after stimulus onset, then the trajectories gradually converges over time in an oscillatory fashion after the stimulus offset (at 300ms).  A movie of Procrustes-aligned MDS plots is also generated and can be accessed at \url{https://youtu.be/WQbgDCq7Dhg}, where each data point (a stimulus instance) can be distinctively tracked as moving seamlessly across each time frame.

\begin{figure}[H]
\centering
    \includegraphics[width=1\linewidth]{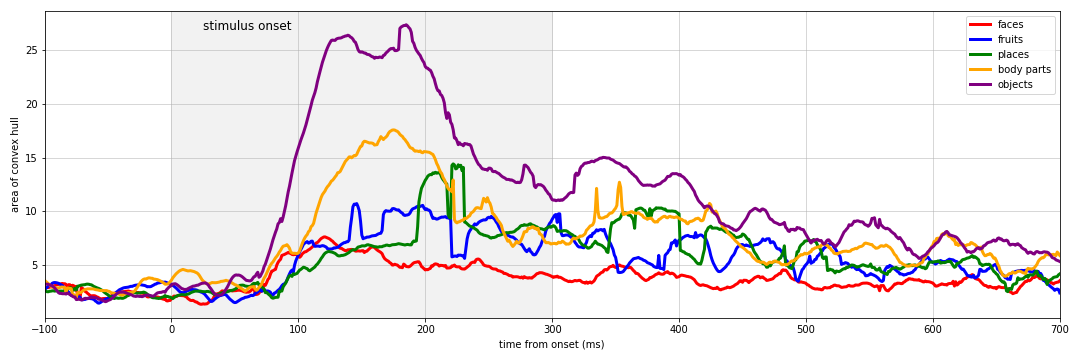}\par\caption{\textbf{Empirical results of pMDS.} Changing areas of convex hulls over time of the neural representational spaces of different stimulus categories.}\label{fig:2d_convex_hulls}
\end{figure}

\begin{figure}[H]
\centering
    \includegraphics[width=.48\linewidth]{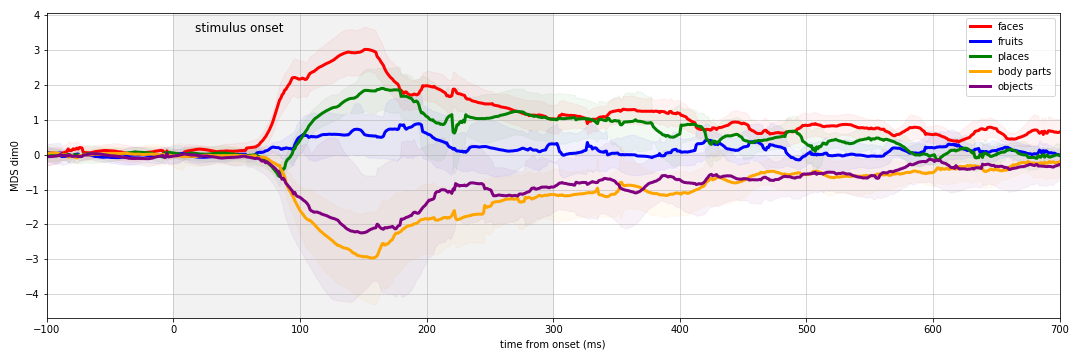}\hfill
    \includegraphics[width=.48\linewidth]{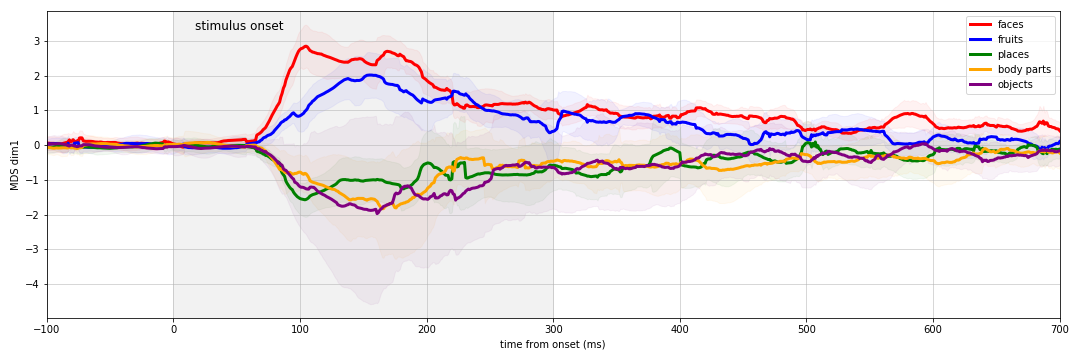}
    \includegraphics[width=.48\linewidth]{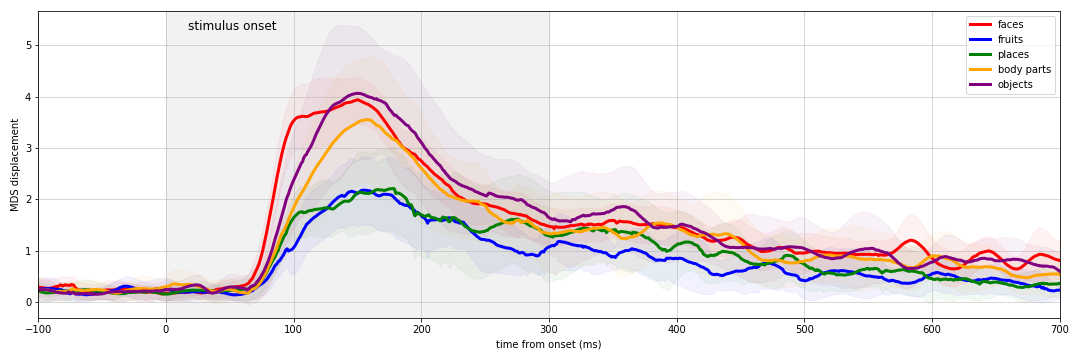}\hfill
    \includegraphics[width=.48\linewidth]{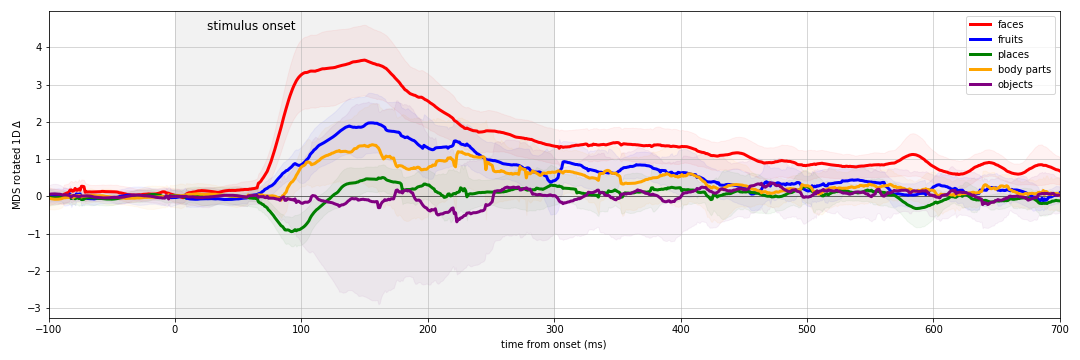}
    \par\caption{\textbf{Empirical results of pMDS.} MDS embeddings over time (dim0, dim1, the displacement from origin, maximum incremental changes  of the representational spaces of different stimuli.}\label{fig:mds_2d}
\end{figure}

\subsection{Hierarchical visual categorization with major stages}

Given the intuition from the 3d visualization, we further explored segments of the representational dynamics. Fig. \ref{fig:2d_0_100} demonstrated the average representational trajectories of each categories in the first 100ms after stimlus onset (with the end marked as square and dot size indicating the standard deviation across different stimulus instances). We see that the categories faces and fruits diverges become discriminable rapidly in the IT population code due to their distinct visual dissimilarity, while places, objects and body parts diverge much later in time. Among the three late classes, the separation of the objects and body parts happens even later in time, suggesting a hierarchical process of categorization. Later during the stimulus is on, the representations of each category seems to be dwelling around their own cluster in a slowly drifting fashion, as shown in Fig. \ref{fig:2d_100_300} (the convex hulls are plotted for all stimuli within the time range in the selected category). After the stimulus offset, the average trajectory of each category gradually converge into proximity, as shown in Fig. \ref{fig:2d_300_800}, where the convex hulls of each categories gradually merge into one. The representational space for each category (as indicated by the areas of the convex hull covering all stimulus instances within the category) also follows several major segments (Fig. \ref{fig:2d_convex_hulls}): a peak around 100 to 200 ms after onset and another bump after the stimlus offset around 300 to 400 ms. Further investigations can potentially illuminate the role of working memory and other factors that might contribute to the second rise of the representational areas.

\begin{figure}[tb]
\centering
\vspace{-2em}
    \includegraphics[width=0.32\linewidth]{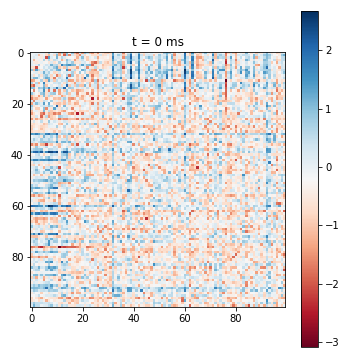}\hfill
    \includegraphics[width=0.32\linewidth]{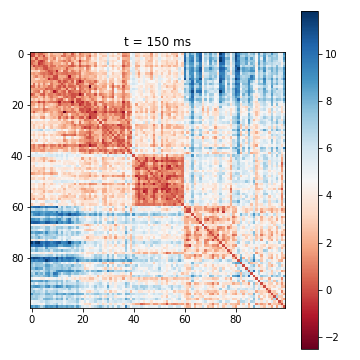}\hfill
    \includegraphics[width=0.32\linewidth]{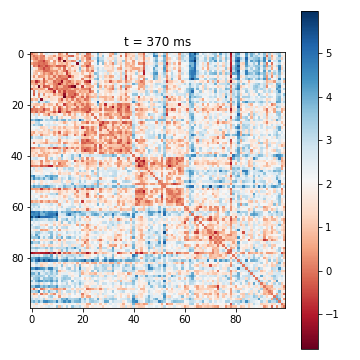}
    \includegraphics[width=0.32\linewidth]{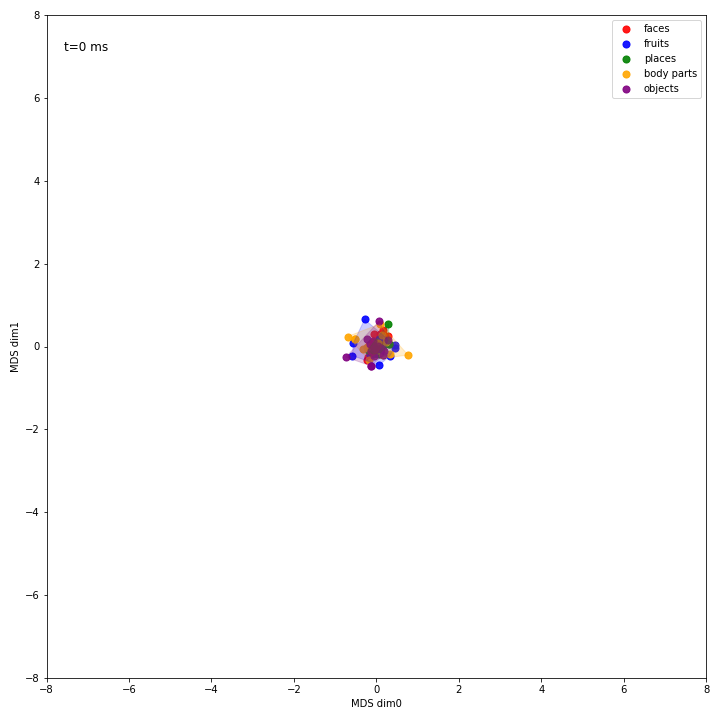}\hfill
    \includegraphics[width=0.32\linewidth]{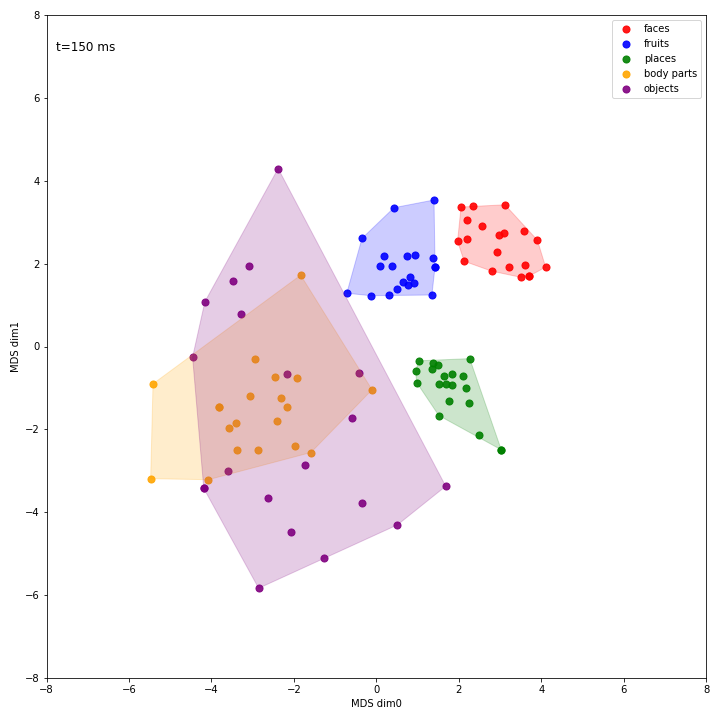}\hfill
    \includegraphics[width=0.32\linewidth]{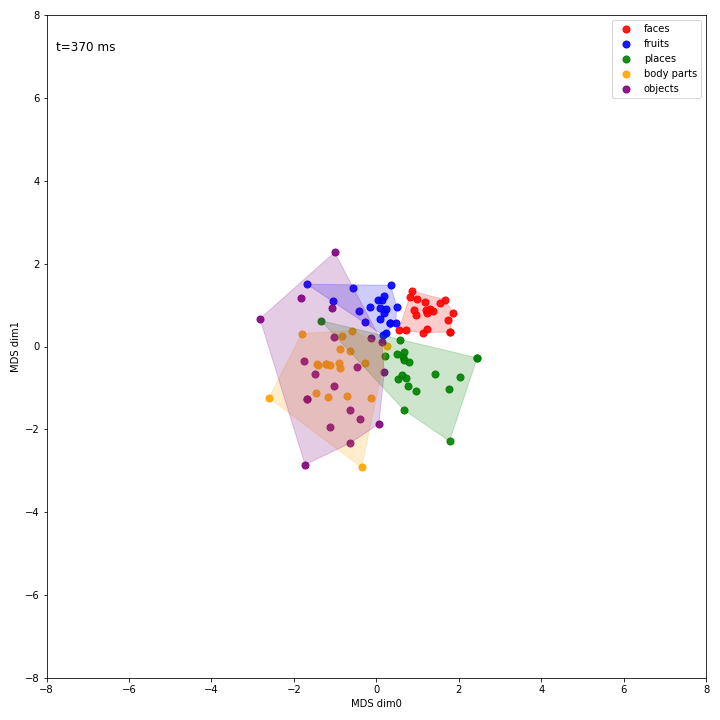}
    \caption{\textbf{Empirical results of pMDS.} Demonstrated are the corresponding RDM and MDS at example time points (onset at 0ms, during onset at 150ms, and after offset at 370ms).}\label{fig:2d_rdm_mds}
\end{figure}

\subsection{Temporal analysis of Procrustes-aligned representations}

With the aligned representations, other temporal analyses can be applied to compare between time points. Fig. \ref{fig:mds_2d} offers a subset of such inquires that can potentially offer neuroscience insights. For instance, the MDS displacement away from the origin (the third plot) indicates a similar dual bump feature during stimulus onset and offset, suggesting the drifting of the centroids of each categorical representation follows a unique dynamics that was not previously widely understood. In the fourth plot, we aligned the signal for each category onto the category-specific 1D plane where the signal amplitudes are maximized. Initial analysis reveals that there seems to be interesting oscillations worth further investigation. Due to GPA's preservation of geometrical information of the pairwise stimulus-specific response pattern, these temporal analyses offer none artificial insights on the variance or noise (due to the scaling restriction) and directionality (due to the minimized Procrustes distance).

\subsection{Future directions of representational dynamics}


We here proposed a representation alignment method to extend the RSA framework to analyze time-stamped brain-activity profiles as representational dynamics. From the neural data, RDM movies are computed with as sliding-window snapshots of representational geometry, and then aligned across all time points with generalized Procrustes analysis. We applied the proposed method to the single-electrode recording of monkey's IT cortex viewing 100 images of 5 categories. The results demonstrated that the alignment can reasonably capture the temporal dynamics of the representation space for each category, and reveal insights on the hierarchical separations of classes and possibly connection with other mechanisms such as working memory and oscillatory behaviors. 

Other than working with RDM movies, there are several alternative methods to study representational dynamics. One such approach under ongoing investigation is to work with raw data of the time-series measurements, by directly extracting the pairwise intertemporal distances into a full RDM of dimension $(N\times T)\times (N\times T)$ with N as number of stimuli and T as number of time points. However, there are clear advantages of our currently proposed multidimensional scaling alignment of RDM movie over this alternative approach of RSA of full RDM: (1) the algorithmic complexity of working with the full RDM is so expensive that it's computationally prohibitory, while Procrustes-aligned MDS is very scalable and light-weighted; (2) the pairwise distances of time series segment is an ongoing challenge and topic of interests in text mining and bioinformatics, that requires deeper understanding to apply in a logical way. Another alternative is to simply use the snapshots RDMs themselves. Fig. \ref{fig:2d_rdm_mds} compares these two methods, where RDM's grid-like and MDS's patch-like visualizations each offer a unique insight of the pattern. Future and ongoing work includes the application of this method to the neuroimaging data of different brain regions and time scales to explore whether the representations are also recurrent as the neural recordings \cite{kietzmann2019recurrence}, as well as visualizing the representational dynamics of deep neural networks to understand their behaviors. 



\section{Temporal Filtration and Simplicial Analysis for Time Series Data}

While the methods introduced in section 5.1 focus on visualizing and analyzing representational dynamics in neural data, the techniques we'll explore in section 5.2 take a more abstract topological approach to time series data. These complementary methods provide a comprehensive toolkit for understanding temporal dynamics in complex biological systems, from brain activity to cellular development.

The absence of a conventional association between the cell-cell cohabitation and its emergent dynamics into cliques during development has hindered our understanding of how cell populations proliferate, differentiate, and compete, i.e. the cell ecology. With the recent advancement of the single-cell RNA-sequencing (RNA-seq), we can potentially describe such a link by constructing network graphs that characterize the similarity of the gene expression profiles of the cell-specific transcriptional programs, and analyze these graphs systematically using the summary statistics informed by the algebraic topology. We propose single-cell topological simplicial analysis (scTSA). Applying this approach to the single-cell gene expression profiles from local networks of cells in different developmental stages with different outcomes reveals a previously unseen topology of cellular ecology. These networks contain an abundance of cliques of single-cell profiles bound into cavities that guide the emergence of more complicated habitation forms. We visualize these ecological patterns with topological simplicial architectures of these networks, compared with the null models. Benchmarked on the single-cell RNA-seq data of zebrafish embryogenesis spanning 38,731 cells, 25 cell types and 12 time steps, our approach highlights the gastrulation as the most critical stage, consistent with consensus in developmental biology. As a nonlinear, model-independent, and unsupervised framework, our approach can also be applied to tracing multi-scale cell lineage, identifying critical stages, or creating pseudo-time series.

\subsection{Challenges facing topological data analysis and single-cell/neuron data}

In recent years, technological developments in data visualizations, especially the subfield of topological data analysis (TDA), has illuminated the structure of biological data with features like clusters, holes, and skeletons across a range of scales \cite{carlsson2009topology}. The TDA approach has proven to be especially useful with recent advancements in experimental techniques at the single cell resolution, both in genomics and neuroscience, such as radiomics \cite{crawford2020predicting} and brain imaging \cite{saggar2018towards,phinyomark2017resting}. The utility of topology comes from the idea of persistence, which extracts the underlying structures within data while discarding noisy elements in the single cell data collection. Unlike graph-based data like human connectomes, most of the time, the high-dimensional data collected from single cell techniques are similiarity-based. Under the assumption that these data was sampled from underlying space $\mathcal{X}$, the goal is to first approximate $\mathcal{X}$ with a combinatorial representation, and then compute some sort of invariant features to recover the topology of $\mathcal{X}$. For interested readers, \cite{amezquita2020shape,topaz2015topological,offroy2016topological} are a few recent reviews of the applications of TDA in various field of biology; \cite{chazal2017introduction} is a practical introduction and guide on how to apply TDA to data science and understand its results; \cite{otter2017roadmap} is a gentle introduction and tutorial to the computation
of persistent homology. 

The single-cell topological data analysis (scTDA) is one of the first attempts to apply topology-based computational analyses to study temporal, unbiased transcriptional regulation given the single-cell RNA sequencing data \cite{rizvi2017single}. In order to visualize the most invariant features of the entire gene expression data, scTDA clusters low-dispersion genes with significant gene connectivity according to their centroid in the topological representation, and visualize the data points in low-dimension space with the Mapper algorithm \cite{carlsson2014topological}. Computing the library complexity as the number of genes whose expression is detected in a cell, scTDA observes a mild dependence of library complexity over the timescale of the single cell data of 1,529 cells collected at 5 time points. This is expected because the number of genes expressed by cells in early stages of a developmental process is larger than in the adult case, as pointed out in \cite{gulati2020single}. As a result, in scTDA the library complexity is not used for any purpose at the topological data analysis and not related to any topological properties. 

Intuitively thinking, if we were to introduce a definition for ``cell complexity'', that characterizes the behaviors of cell-cell coexpression or interactions, the quantities of cell complexity should be agnostic to the number of genes expressed by the cells, and should be different across differentiated cells and across the developmental process. Can we introduce a better summary statistic for the cell complexity that can capture the developmental trajectory with more distinctions between time points? To clarify, unlike the previous definition of ``library complexity'', which simply quantifies the number of genes expressed in a cell, we wish to define a cell complexity measure to better model higher-order networks and dynamic interactions in single-cell data. Understanding the cell-cell interactions can help identify intercellular signaling pathways and previous analytical studies have focused on computing a communication score between the ligand–receptor pair of interacting proteins \cite{armingol2021deciphering}. For instance, \cite{arneson2018single} and \cite{oh2015extensive} infer the intercellular signaling pathways of cell-cell communications by computing the coexpression of all genes or other cell markers. The alternative would be to compute the similarity between gene expression profiles as in \cite{han2018mapping}. In this work, we aim to focus directly on the cell level, and use the similarity between each cell's gene expression profiles as a graph to compute a topological descriptor of the complexity. The more connected a group of cells are in this similarity graph, the higher the complexity of this group of cell is. There are two major quests in this line of research:

    \textbf{Quest from topological data analysis.}
Existing TDA applications usually focus on the low-dimensional graph visualization and the persistent homology of the data (i.e. computing the Betti numbers or barcodes up to dimension 2), because interpreting the biophysical meaning of the geometry and higher dimensional persistent modules is a conceptual challenge. Others have proposed hybrid approaches to combine the merits of data geometry and topology by adaptively selecting the proper thresholds in the pairwise distance matrix of the data points \cite{lin2018adgtic,lin2022geometric}. Another alternative to these low-dimensional TDA methods is the simplicial analysis. Simplicial architecture was studied in biological data through the application on human brain connectomes \cite{reimann2017cliques}, where each connected pairs of neurons are considered an edge to create a graph and the numbers of Rips-Vietoris simplices in dimensions up to 7 are computed at the static graphs comparing with the random graphs. Likewise in our inquiry, we are interested in the intercellular interaction within the same type of cells, the cell complexity, rather than the relationships between different groups of cell, as in scTDA \cite{simpArch}. However, the filtration challenge of deriving a graph from the distance-based data by choosing the best threshold, hinders the practical application of such simplicial analysis in these point cloud data. 

\textbf{Quest from single-cell-resolution data.}
With the increasingly popular usage of single-cell genomic techniques, it might be possible to infer such cell-cell interaction (or cellular ecology) in a fine resolution. However, as far as we are aware, there are only a few literature exploring the cellular ecology from single-cell RNA sequencing data. For instance, \cite{gallaher2020cells} and \cite{amend2016ecological} apply the ecology and multi-agent models to model single-cell systems. We wish to complement this line of work by connecting it to the topological data analysis, where the focus is to model the shape or manifold of the data from the similarity of data points. For instance, simplicial complexes are high dimensional objects or generalizations of neighboring graphs to represent the cliques of data points, and in other words, a notion of \textit{ecology}. The ecology doesn't have to be the organisms within a physical system. In the field of data science where we represent biological cells by their measurements (e.g. gene expression profiles) as data points residing in high-dimensional feature spaces, the ecology can be how these data points are connected to one another in the feature space. If we adopt an ecology research point of view, in order to characterize the dynamic systems of a community, one need to have knowledge or priors regarding the causal relationships between the agents (e.g. how do preys and predates interact, and in what ways). In order to parse out causal relationships, the temporal sequence of these events matters. Thus, the property of the synchrony and asynchrony of the events is a key to translate the feature space (represented by a similarity graph) to an ecology, which has directed (e.g. causal) relationships among the agents. This is why a temporal take into the topological data analysis can potentially unlock the first step from finding a static representations of the overall shape of the data points to discovering the event-directed representations (i.e. a temporal skeleton) of the data points.

One challenge of this hybrid direction, is to conceptually understand the biological meaning behind the dissimilarity of the omic data. For instance, what does it mean if two cells have similar gene expression profiles from each other? Does that indicate a homogeneity if the two cells are from the same tissues, or is it an artifact that the manual labeling or classifications are not perfect? Can we measure the ``complexity'' of the cell populations based on the heterogeneity or diversity within populations? If we can, how to we evaluate and interpret lower-order versus higher-order ``complexity''? These are some open questions we wish to engage the field to discuss and investigate together, instead of answering them directly in this first work. 

The other challenge is the scalability and compariablity of the single-cell data. With the advances of multi-channel high-throughput data collection techniques in biological fields, how to compute the pairwise distances of the point clouds efficiently? In different trials of single-cell experiments, how to make sure that the persistent modules are comparable to one another? 

\begin{figure}[tb]
\includegraphics[width=\linewidth]{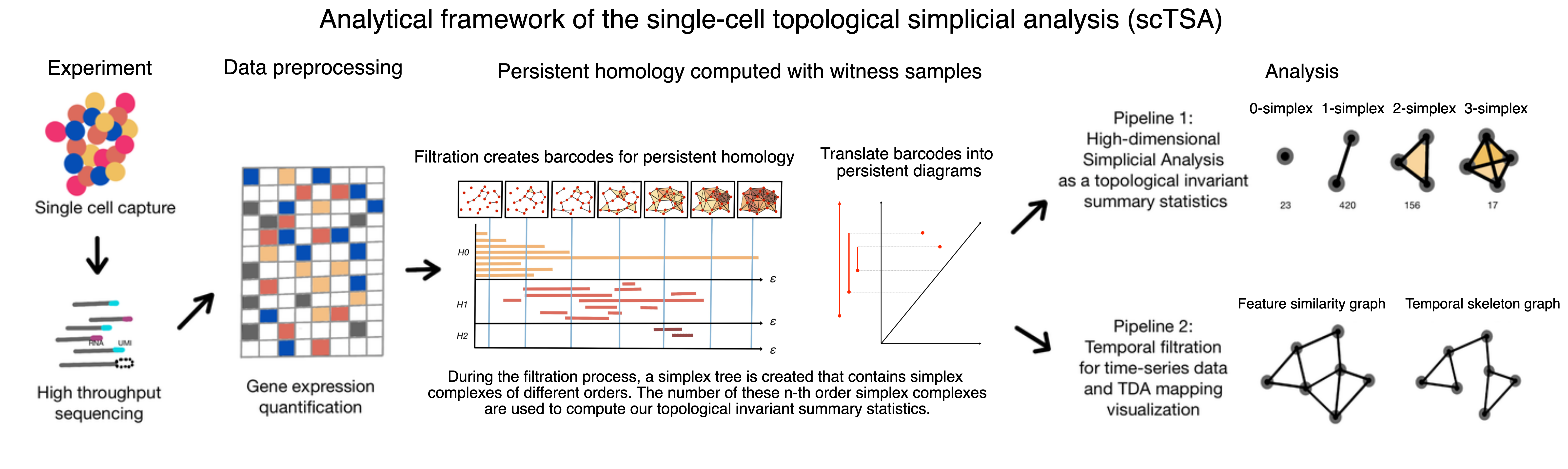} \hfill
\caption{\textbf{The analytical framework of the single-cell topological simplicial analysis (scTSA)}. The pipeline starts with the single-cell sequencing data, which is then preprocessed into gene expression profiles in a 2d matrix (rows are cells and columns are genes). This step can also go through another layer of dimension reduction. From the matrix, we compute filtrations over their feature space and temporal constraints. The persistent homology can be computed from the filtration process with either a persistent barcode or a persistence diagram. The filtrations obtained through the processes can also used for simplicial analysis, which groups the cells by time steps before the analysis. Finally, one can visualize the data using the Mapper algorithm, with or without the temporal constraints.}
\label{fig:pipeline}
\end{figure}

\subsection{Temporal topological data analysis (tTDA) and simplicial analysis (TSA)}

In this study, we propose a temporal topological data analysis (tTDA) and a topological simplicial analysis (TSA) pipeline (Fig. \ref{fig:pipeline}) as an exploratory inquiry to solve these three challenges: (1) with the algebraic geometry's definitions of forming higher-order simplices, we can potentially interpret that cliques of higher orders indicates operational units of higher order; (2) with the bootstrapping techniques to sample from the data points collected at each sub-level, we can scale the analysis to large single cell datasets and compare groups of cells quantitatively; (3) with a time delay constraint on the filtration process, we can sort the projected data points of cells into distinct groups of cells collected from the same time stamps. The framework first takes the measurements of the single-cell RNA sequencing data which generates a similarity matrix among the cells based on their gene expression profiles. Other than performing the persistent homology to obtain lower-order topological descriptors of the data, we compute additional higher-order topological descriptors by counting the number of the simplices emerged from the filtration process. In addition, we introduce a technique to extract the temporal skeleton of the developmental processes, called temporally filtrated TDA, and show that the developmental trajectories of cells can be better revealed in this approach comparing to existing TDA mapping techniques. 

We begin our presentation with a short overview of mathematical definitions of the single cell data visualization problem and introduction of necessary concepts and definitions in the language of computational topology. Later section formulates the topological simplicial analysis pipeline we are proposing as well as numerical tricks applied in the implementation to ensure the scalablity. We apply this single cell Topological Simplicial Analysis (scTSA) to the zebrafish single-cell RNA sequencing data with 38,731 cells, 25 cell types, over 12 time steps \cite{farrell2018single}. We select the top 103 genes based on the scTDA pipeline from the high-dimensional high-throughput transcriptomic data. In section \ref{sec:results}, we introduce the dataset used to benchmark the method and present the analysis results with their mathematical interpretations to the biological insights. In the last section, we discuss the validity of using our framework to understand the higher-order cellular complexity, and conclude our methods by pointing out several future work directions as the next step.


\subsubsection{Single-cell data in the point cloud space}

Genomic measurement and analysis at single-cell resolution has enabled new understandings of complex 
biological phenomena, such as revealing cellular composition of complex tissues and organisms \cite{kalisky2011single}. Single-cell RNA sequencing (scRNA-seq) techniques measure the gene expression profiles of individual cells through mechanisms like microfludics. For instance, the benchmark dataset of zebrafish embryogenesis \cite{farrell2018single} that we use in this study, applied Drop-seq, a massively parallel scRNA-seq method to profile the transcriptomes of tens of thousands of embryonic cells \cite{macosko2015highly}. These single cell data are usually point clouds in a finite metric space, a finite point set $S \subseteq {\mathbb R}^d$. Let $d(\cdot, \cdot)$ denote the distance between two points in metric space $\mathcal{Z}$. The assumption is that data was sampled from underlying space $\mathcal{X}$. The goal is to recover topology of $\mathcal{X}$. To accomplish the goal, one needs to first approximate X with a combinatorial representation (e.g. with the simplicial complex), and then compute a topological invariant summary statistics (e.g. with the persistent homology).

\begin{figure}[tb]
\centering
\includegraphics[width=\linewidth]{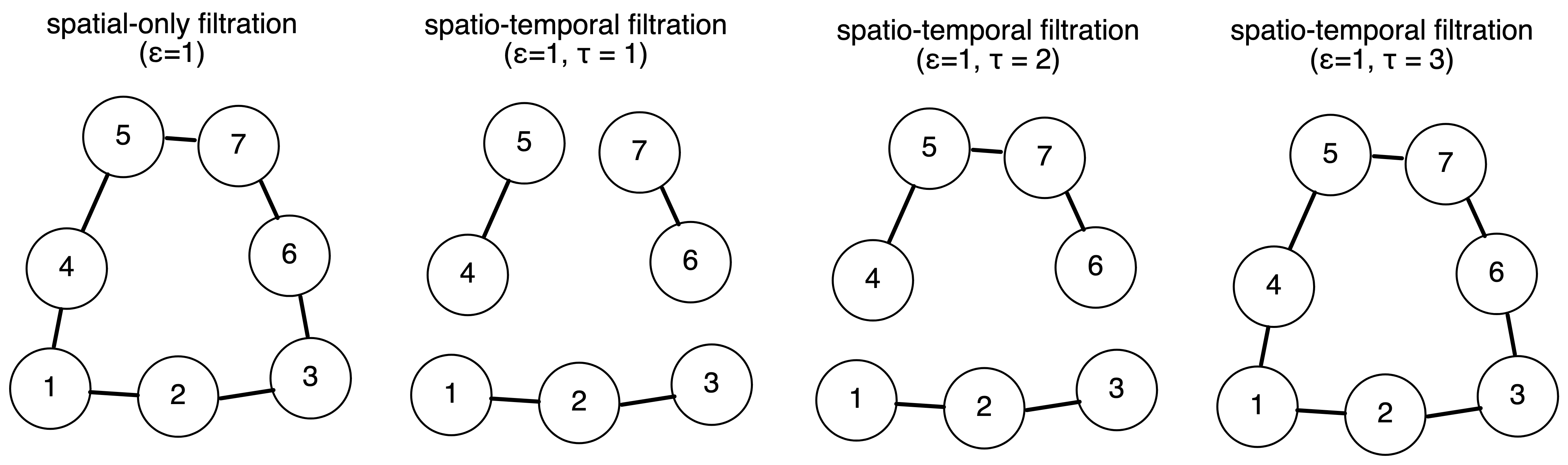} 
\caption{\textbf{Intuitive example of the temporal filtration}. Presented here are 7 data points, which are marked by their time stamps (when they are measured). In all four cases, we consider the case where the spatial threshold $\epsilon$ of the nerve ball around each data point is 1 (which in our case only contains every data point's nearest neighbor, but not their second nearest neighbor). If we only perform the spatial filtration, we would consider them all to be connected. However, that would not match the temporal skeleton. Instead, we can set a temporal constraint $\tau$ such that only if two data points that are spatially (in the feature space) proximal to each other are also measured temporally close to each other, their edge is included. If $\tau$ is small (say, 1 time step apart), we have a fine resolution temporal skeleton which separate the data points into three main phases. If $\tau$ is big (say, 3 time steps part), we have a crude temporal skeleton which groups them all in a connected components.
}
\label{fig:tfilter}
\end{figure}

\subsubsection{Definition of the simplicial and temporal filtration}
\label{sec:temporal_filtration}

Given the point cloud data, we then construct a continuous shape on top of the data to highlight the underlying topology and geometry. The process to build such a shape is through a mathematical filtration, which is often a simplicial complex or a nested family of simplicial complexes, that reflects the innate structure of the point cloud data at different scales \cite{chazal2017introduction}. 
If we consider all the points in the point cloud data each with a coordinate of their locations in certain embedding, they each occupy a spherical space with the same radius $\epsilon$ around them, which are called nerve balls. If the two nerve balls overlap or contact each other, we consider an edge to be formed between them in this graph. The filtration is a process to tune the parameter $\epsilon$ from $0$ to $\infty$ and record the families of simplicial complexes generated through the increasingly connected (or ``complex'') graph.

Usually, the challenge is to extract relevant information about the shape of the data through defining such simplicial complexes from the graph (generated through the filtration process). Rips-Vietoris complex is one of the common choices in practice to compute topological invariants of point clouds, defined as follows: given the vertex set $\mathcal{Z}$, for each pair of vertices $a$ and $b$ edge a-b is included in Rips-Vietoris complex $C(\mathcal{Z}, t)$ if $d(a, b) \leq t$, and a higher dimensional simplex is included in $C(\mathcal{Z}, t)$ if all of its edges are included. Since $C(\mathcal{Z}, t) \in C(\mathcal{Z}, t')$ whenever $t \leq t'$, the filtered Rips-Vietoris complex is a filtered simplicial complex, and also the maximal simplicial complex that can be built on top of its $1-$skeleton, thus a clique complex or a flag complex.
Unlike conventional low-dimensional topological data analysis, we computed simplices into high dimension (up to 7) during the entire filtration process. To record the number of cliques, we compute filtered simplicial complexes and record their cumulative counts across the full filtration process.

Since the topological data analysis usually only consider the graph constructed by the spatial proximity (i.e. the distance matrix) between the data points in the low-dimensional embedding, it is not clear how to incorporate timestamp information for meaningful inference and visualization when facing the time-series data streams. One approach would be to simply consider the time stamp as the meta data for post hoc labeling of the topological representations. Another alternative would be to consider time as an additional dimension in the filtration process. We present the Temporal Filtration as the following: alongside the conventional sweeping of the parameter $\epsilon$ from $0$ to $\infty$, we set another parameter $\tau$ to indicate a hard constraint in edge forming between two points. Alternatively and intuitively, the temporal filtration is equivalent to conventional filtration by using the composite norm: 

\begin{equation}\label{eq:composite}
d((x,t_x), (y,t_y)) = max( \frac{1}{\epsilon^*} |y-x|, \frac{1}{\tau^*} |t_y-t_x| )
\end{equation}

where the $\epsilon^*$ and $\tau^*$ are directly relates to the spatial threshold $\epsilon$ (in the feature space) and the temporal threshold $\tau$.
As a practical note from this notation, it can be used without additional specialist software.

in other words, only if the time stamp difference between the two data points is within the time delay limit $\tau$, can two nerve balls, if spatially proximal enough (less than $\epsilon$), form an edge in between. On the other hand, if the time stamp difference between the two data points is larger than $\tau$, even if they are spatially proximal enough (less than $\epsilon$), they cannot form an edge. Given the problem settings, one can either set a reasonable time delay limit $\tau$ given the domain knowledge, or tune $\tau$ from $0$ to $\infty$, similar to the filtration process on the spatial filtration parameter $\epsilon$. The later approach can potentially extract temporally invariant topological summary statistics.

Fig. \ref{fig:tfilter} is an intuitive example of the criterion of edge forming in the temporal filtration. In the example, we have 7 data points, which are marked by their time stamps (when they are measured). The node marked 1 would indicates it is collected the time step 1. There is a 1 time step difference between each data point of consecutive numbers. To illustrate the differences between the conventional and temporal filtration, the schematic is a snapshot of the full filtration process, frozen at a set of filtration thresholds.
In all four cases, we consider the case where the spatial threshold $\epsilon$ of the nerve ball around each data point is 1 (which in our case only contains every data point's nearest neighbor, but not their second nearest neighbor). If we only perform the spatial filtration, we would consider them all to be connected. However, that would not match the temporal skeleton. Instead, we can set a temporal constraint $\tau$ such that only if two data points that are spatially (in the feature space) proximal to each other are also measured temporally close to each other, their edge is included. If $\tau$ is small (say, 1 time step apart), we have a fine resolution temporal skeleton which separate the data points into three main phases. If $\tau$ is medium (say, 2 time step apart), we have a relatively crude resolution temporal skeleton which separate the data points into two main phases. If $\tau$ is big (say, 3 time steps part), we have the crudes temporal skeleton which groups them all in a connected components. This also demonstrates the possibility of using $\tau$ as a hierarchical mechanism to parse persistent features of different temporal resolutions.

\begin{figure}[H]
\centering
\includegraphics[width=\linewidth]{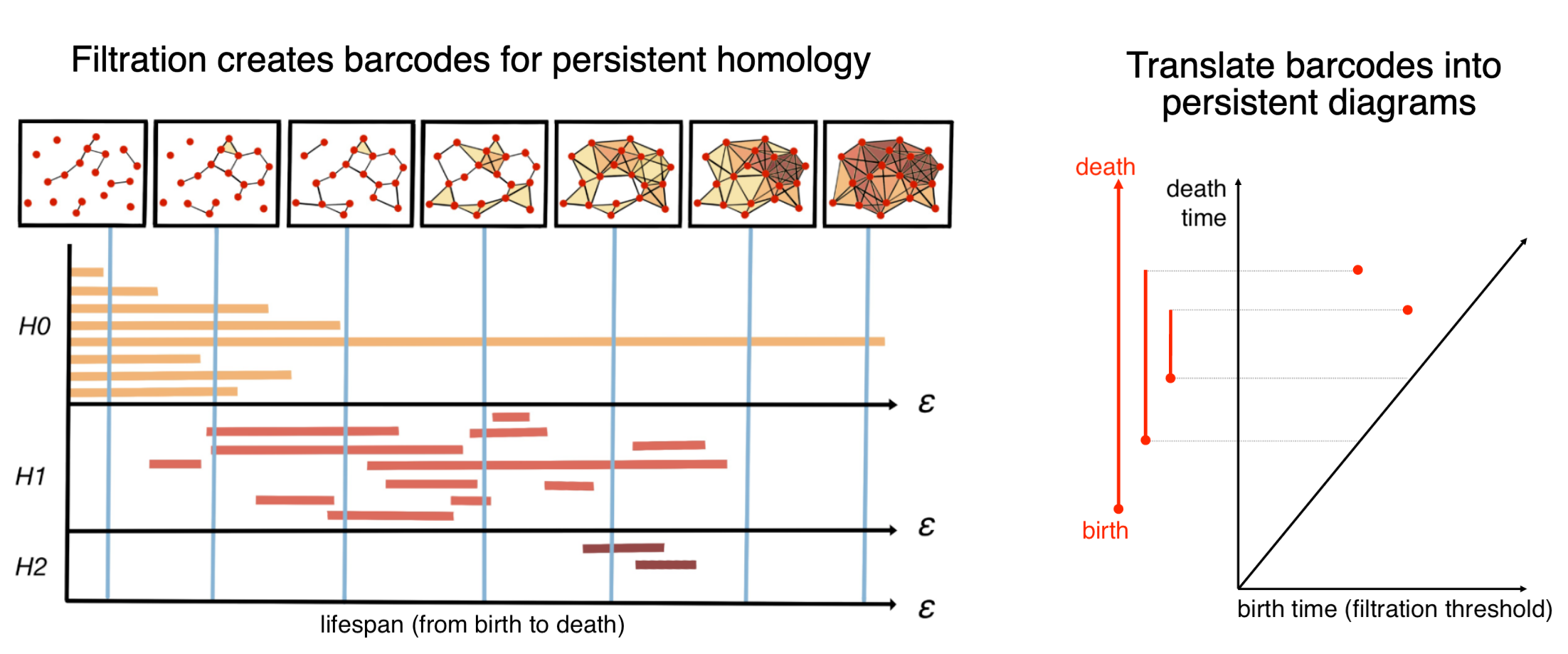} 
\caption{\textbf{Persistent homology via mathematical filtration.} In this schematic diagram, a point cloud of 19 data points are presented in a low-dimensional embedding space. In the filtration process, a parameter $\epsilon$ is swept from 0 to the maximum pairwise distance within the point cloud, indicating a distance threshold under which the two points can form an edge to become one connected component in the graph. For each value $\epsilon$, we obtain a space $S_\epsilon$ consisting of vertices, edges formed among the vertices, and higher-dimensional polytopes connected by these edges. For instance, a nerve ball of radius $\epsilon$ grows around each point cloud, and an edge will form if two nerve balls touch. Homology counts the number of essentially different cycles -- linear combinations of simplices that form a cycle (for example a loop formed by a sequence of edges) -- that are not the boundary of something that can fill in the hole (for example a combination of 2d simplices or triangles spanning the inside of the loop). We denote $H_n$ as the $n$-th homology group, i.e. the formation of the simplex complexes of order $n$, with 0-simplex to be the nodes (or clusters), 1-simplex to be the edges between two nodes, 2-simplex to be the triangles, 3-simplex to be the tetrahedrons and so on. We log the existence of a n-simplex if and only if all of its components (e.g. (n-1)-simplex, (n-2)-simplex, $\cdots$, 1-simplex, and 0-simplex) are all in $S_\epsilon$, and mark their demises when some of these topological cavities are filled with the additions of new edges (and potentially, nodes). 
Each colored line indicates the ``lifespan'' of a simplex, with its starting point to be its ``birth'' (or first appearance) and ending point to be its ``death'' (or disappearance due to the two nerve balls fully overlapping). 
In this example, the persistent homology of the data cloud can be presented in the form of a ``barcode'' representation, which is a finite collection of intervals. The birth and death of the simplicial complexes up to the order 2 are recorded when the filtration process gradually sweeps the distance threshold. The barcode representation is often replaced with the visualization of a 2d persistence diagram \cite{cohen2005stability}, in which the x-axis indicates the birth time (the distance threshold a filtration appears) and the y-axis indicates its death time (the distance threshold the filtration disappears). 
}
\label{fig:PH}
\end{figure}

\subsubsection{Topological data analysis with persistent homology}

Following the definition above, an abstract simplicial complex is given by a set $\mathcal{Z}$ of vertices or $0-$simplices, for each $k \leq 1$ a set of $k-$simplices $\sigma = [z_0,z_1,\dots,z_k]$ where $z_i \in \mathcal{Z}$, and for each $k-$simplex a set of $k + 1$ faces obtained by deleting one of the vertices. A filtered simplicial complex is given by the filtration on a simplicial complex $\mathcal{Y}$, a collection of subcomplexes $\{\mathcal{Y}(t)|t\in {\mathbb R}\}$ of $\mathcal{Y}$ such that $\mathcal{Y}(t) \subset \mathcal{Y}(t')$ whenever $t\leq t'$. The filtration value of a simplex $\sigma\in\mathcal{Y}$ is the smallest $t$ such that $\sigma\in\mathcal{Y}(t)$. 
Topological data analysis methods usually involve computing the persistent homology \cite{de2004topological}. The Betti numbers help describe the homology of a simplicial complex $\mathcal{Y}$. The Betti number value $BN_k$, where $k \in {\mathbb N}$, is equal to the rank of the $k-$th homology group of $\mathcal{Y}$. The Betti intervals over the filtration process help describe how the homology of $\mathcal{Y}(t)$ changes with $t$. A $k-$dimensional Betti interval, with endpoints $[t_{\text{start}}, t_{\text{end}})$, corresponds to a $k-$dimensional hole that appears at filtration value $t_{\text{start}}$, remains open for $t_{\text{start}} \leq t < t_{\text{end}}$, and closes at value $t_{\text{end}}$. 

Fig. \ref{fig:PH} is a schematic diagram outlining how to perform a filtration process (by sweeping the $\epsilon$), document the ``birth'' and ``death'' of each complexes (the colored lines of various length in the chart), and generate this as a barcode representation \cite{ghrist2008barcodes} or a persistence diagram \cite{cohen2005stability} for the downstream analyses. In this schematic diagram, a point cloud of 19 data points are presented in a low-dimensional embedding space. 
In the filtration process, a parameter $\epsilon$ is swept from 0 to the maximum pairwise distance within the point cloud, indicating a distance threshold under which the two points can form an edge to become one connected component in the graph. For each value $\epsilon$, we obtain a space $S_\epsilon$ consisting of vertices, edges formed among the vertices, and higher-dimensional polytopes connected by these edges. For instance, a nerve ball of radius $\epsilon$ grows around each point cloud, and an edge will form if two nerve balls touch. Homology counts the number of essentially different cycles -- linear combinations of simplices that form a cycle (for example a loop formed by a sequence of edges) -- that are not the boundary of something that can fill in the hole (for example a combination of 2d simplices or triangles spanning the inside of the loop). 
We denote $H_n$ as the $n$-th homology group, i.e. the formation of the simplex complexes of order $n$, with 0-simplex to be the nodes (or clusters), 1-simplex to be the edges between two nodes, 2-simplex to be the loops (or triangles in this case), 3-simplex to be the tetrahedrons and so on. We log the existence of a n-simplex if and only if all of its components (e.g. (n-1)-simplex, (n-2)-simplex, $\cdots$, 1-simplex, and 0-simplex) are all in $S_\epsilon$, and mark their demises when some of these topological cavities are filled with the additions of new edges (and potentially, nodes). 
Each colored line indicates the ``lifespan'' of a simplex, with its starting point to be its ``birth'' (or first appearance) and ending point to be its ``death'' (or disappearance due to the two nerve balls fully overlapping). 
In this example, the persistent homology of the data cloud can be presented in the form of a ``barcode'' representation, which is a finite collection of intervals. The birth and death of the simplicial complexes up to the order 2 are recorded when the filtration process gradually sweeps the distance threshold. The barcode representation is often replaced with the visualization of a 2d persistence diagram \cite{cohen2005stability}, in which the x-axis indicates the birth time (the distance threshold a filtration appears) and the y-axis indicates its death time (the distance threshold the filtration disappears). In most cases, only the first two orders of the filtrations are computed and included in persistent barcodes or diagrams.

By using the temporal filtration in place of the conventional filtration, we can extend the methods of persistent homology into one for temporal persistent homology. Our method is related to the research of multi-parameter persistence \cite{botnan2022introduction}, which aims to construct a topological space with more than one filtered spaces.
In other words, the computation of a persistent barcode or diagram can also be customized to use temporal filtration as its filtration criterion, by either using a composite norm function as in Eq. \ref{eq:composite}, or using a multi-parameter filtration with a temporal constraint.

\begin{figure}[tb]
\centering
\includegraphics[width=\linewidth]{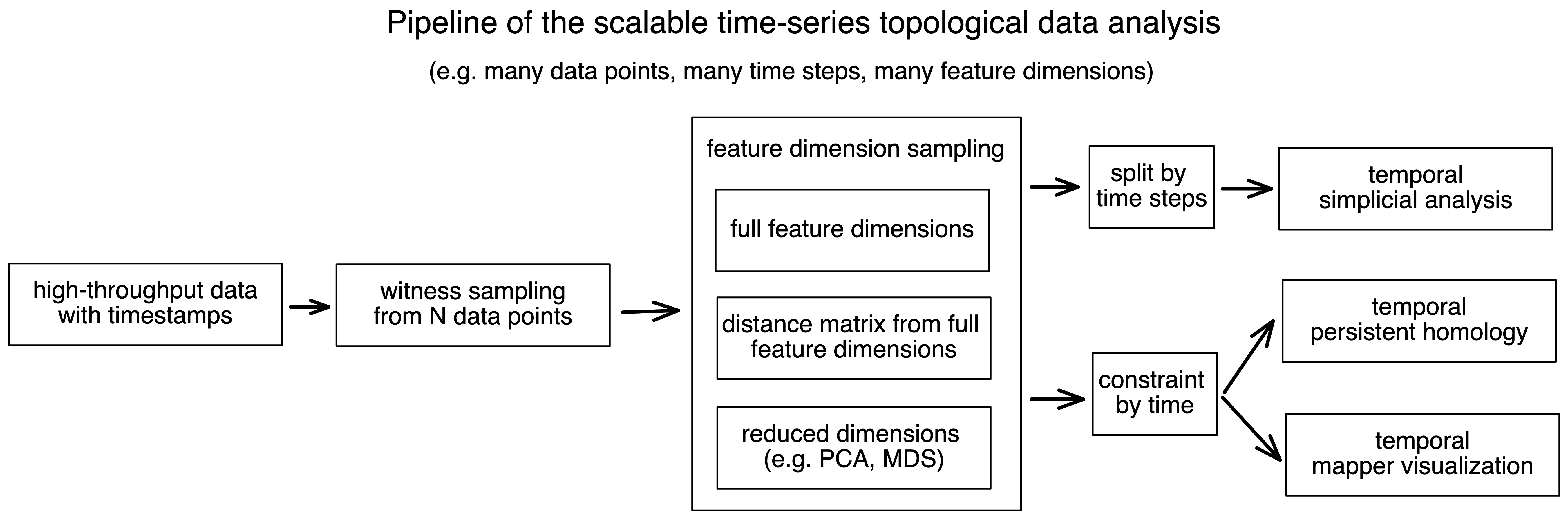} 
\caption{\textbf{Pipeline of the time-series topological data analysis for high-throughput data}. We start with the high-throughput data points marked with their timestamps. To decrease the number of data points for efficient computation (and also comparability across time points), a witness sampling is performed among these data points. Then one can choose to reduce the dimension or not given the noise and distribution properties of their data. To perform the temporal simplicial analysis, the data points are first separately grouped into different time points, and then computed their filtrations to obtain their number of simplicial complexes at different orders. To perform the temporal persistent homology and mapper visualization, one can apply the temporal constraint onto the sampled data points so far to obtain a temporal skeleton.}
\label{fig:scalable}
\end{figure}

\subsubsection{Empirical simplicial computation with witness sampling and dimension reduction}

Overall, the witness sampling is critical for two reasons: (1) The single cell data has different noise granularity across cell types and data collection procedures \cite{faure2017systematic}, and thus, the number of cells collected in each time points and different cell types (as in the analyzed developmental study \cite{farrell2018single}) can vary in different magnitude, making direct simplicial computation incomparable. (2) In large-scale high-throughput data, the large number of data points and feature sizes can make computation especially expensive and infeasible. For instance, the computation of filtration requires a comparison between a sweeping proximity threshold and the distance between two data points, and computing the distance matrix between all points is not only time-consuming and memory-exhaustive (e.g. 1M points would be 1T to just storing the distance matrix).

For these larger datasets, if we include every data point as a vertex, the filtrated simplicial complexes can quickly contain too many simplices for efficient computation. To solve this numerical inconsistency issue, we instead extract the lazy witness complexes by sampling $m$ data points \cite{de2004topological} with a sequential maxmin procedure \cite{adams2009nonlinear}, setting a nearest neighbor inclusion of 2 (as in the term ``lazy'')\footnote{The selection of $m$ depends on the scale of the dataset. The bigger the sample size $m$ is, the better the estimate. However, since different partitions of the data points have varying sizes. For instance, if there are only 50 data points collected in time step 1, while there are more than 100 points in other times steps, then the maximum of $m$ that can be picked is 50.}. The computation of the witness complex in high dimensions can be implemented with GUDHI \cite{maria2014gudhi}, Ripser \cite{bauer2021ripser}, and JPlex software \cite{sexton_jplex_2008}. 
The codes to reproduce the empirical results can be accessed at \href{https://github.com/doerlbh/scTSA}{\underline{https://github.com/doerlbh/scTSA}}. 

Fig. \ref{fig:scalable} outlines our scalable time-series topological simplicial analysis pipeline. We start with the high-throughput data points marked with their timestamps. To decrease the number of data points for efficient computation (and also comparability across time points), a witness sampling is performed among these data points. Then one can choose to reduce the dimension or not given the noise and distribution properties of their data. The usage of dimension reduction is a useful step before the filtration. Due to the ``Curse of Dimensionality'', the data points in a very high-dimensional space can be very sparse and thus the distances between them usually collapses to a constant, i.e. residing at a hyperspherical space. As a result, the filtration computation around them can be ineffective and unstable. Mapping them onto a low-dimensional space can partly solve this issue.

Then to perform the temporal simplicial analysis, the data points are first separately grouped into different time points, and then computed their filtrations to obtain their number of simplicial complexes at different orders. To perform the temporal persistent homology and mapper visualization, one can apply the temporal constraint onto the sampled data points so far to obtain a temporal skeleton.

\subsubsection{Topological simplicial analysis}

Given the simplicial complexes of different orders from the witness sampling approach, we need to correct for the effect of sampling. The larger the sample size, the more likely the higher-order simplicial complexes emerge. One way to correct for this amplification effect is to normalize this quantity directly to the quantity collected from a null distribution of the data.
Usually for a graph, network or more generically, data with a binary connectivity format (e.g. brain connectome), the Erd{\H{o}}s-R{\'e}nyi random graph \cite{erdHos1960evolution} can be used as control models. However, in fully connected similarity-based data, the average connectivity probability is entirely dependent on the filtration factor. To avoid this caveat, we take a different approach by permuting the pairwise distances of the data points, which is equivalent to a weighted version of the Erd{\H{o}}s-R{\'e}nyi random graph. Another strategy would be permuting the feature at each dimension. In this way, the low-dimensional embeddings computed by the multidimensional scaling can form different connectivity profiles while maintaining the same distance distribution. Then we apply the same topological data analysis pipelines to the embeddings computed from the pairwise distance matrices from both the actual data and the control models. 

To this point, we propose a formal definition of cellular complexity, as the \textit{normalized n-simplicial complexity}, $NSC_n$, a family of summary statistics with an increasing order $n$:

\begin{equation}
    NSC_n = \frac{SC_n^{data}}{SC_n^{null}}
\end{equation} 

where $NSC_n$ is computed by taking the ratio between the number of the simplicial complexes for a certain order $n$ computed from the actual data (which we denote $SC_n$), to the sum of the number of those computed from the control models and from the actual data. An alterantive would be $NSC_n = \frac{SC_n^{data}}{(SC_n^{data}+SC_n^{null})}$. A value of 0.5 would indicates that the simplicial complexity at order n is the same in the data and the null models. Empirically, we compute the $NSC_n$ with the order \textit{n} from 1 to 7, as the summary statistics characterizing the ecology among the data points with cliques and cavities of increasing modularities.

\subsubsection{Topological data visualization with low-dimensional mapping}

To build and visualize the topological representation of the point cloud data, we use the Mapper algorithm \cite{singh2007topological} through the implementations provided by Kepler-Mapper \footnote{https://github.com/scikit-tda/kepler-mapper} with modifications for temporal filtration
at \href{https://github.com/doerlbh/tkMapper}{\underline{https://github.com/doerlbh/tkMapper}}. 
In brief, a dissimiliarity matrix is computed from the preprocessed RNA-seq data by taking the pairwise correlation distance. This metric space was then reduced to a low-dimensional embeddings with the multi-dimensional scaling \cite{mead1992review}. Given this embedding, the point cloud data are chopped into coverings of hypercubes with a 50\% percentage of overlapping between the cubes\footnote{The choice of 50\% is empirically determined by our dataset. We vary the overlap parameter among 25\%, 50\% and 75\%, and 50\% gives the best clustering effect.}. Then for each hypercube, the data points within the cube are then clustered with single-linkage rule. This step further aggregates all the points into a network in which each vertex corresponds to a cluster and each edge corresponds to a non-vanishing intersection between the clusters. As defined in section \ref{sec:temporal_filtration}, if temporal filtration is applied, then edge forming is also controlled by the additional time delay constraint that the clusters are formed with both spatial and temporal proximity, and the edges would only exist between two clusters if all points in the two clusters are within the time delay limit $\tau$. In other words, the filter function is the same that we apply to persistent homology, which can either be a single filtration with the temporal constraint, a single filtration with the temporal composite norm, or a multi-parameter filtration.
Once we reach a network representation, the network can eventually be visualized with force-directed algorithms for insights.


\subsection{Empirical evaluations of tTDA and scTSA}
\label{sec:results}

\begin{figure}[tb]

\includegraphics[width=\linewidth]{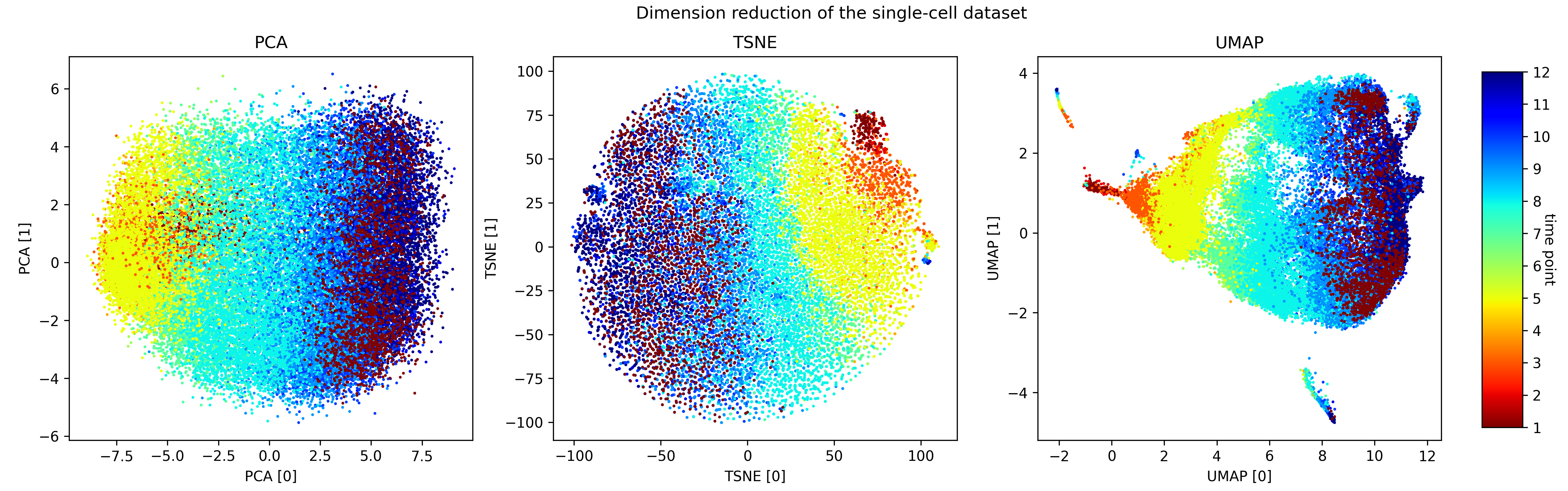} \hfill
\caption{\textbf{Dimension reduction of the dataset}. PCA, TSNE and UMAP applied to our preprocessed gene expression profiles.}
\label{fig:dr}

\end{figure}

\begin{figure}[tb]

\includegraphics[width=\linewidth]{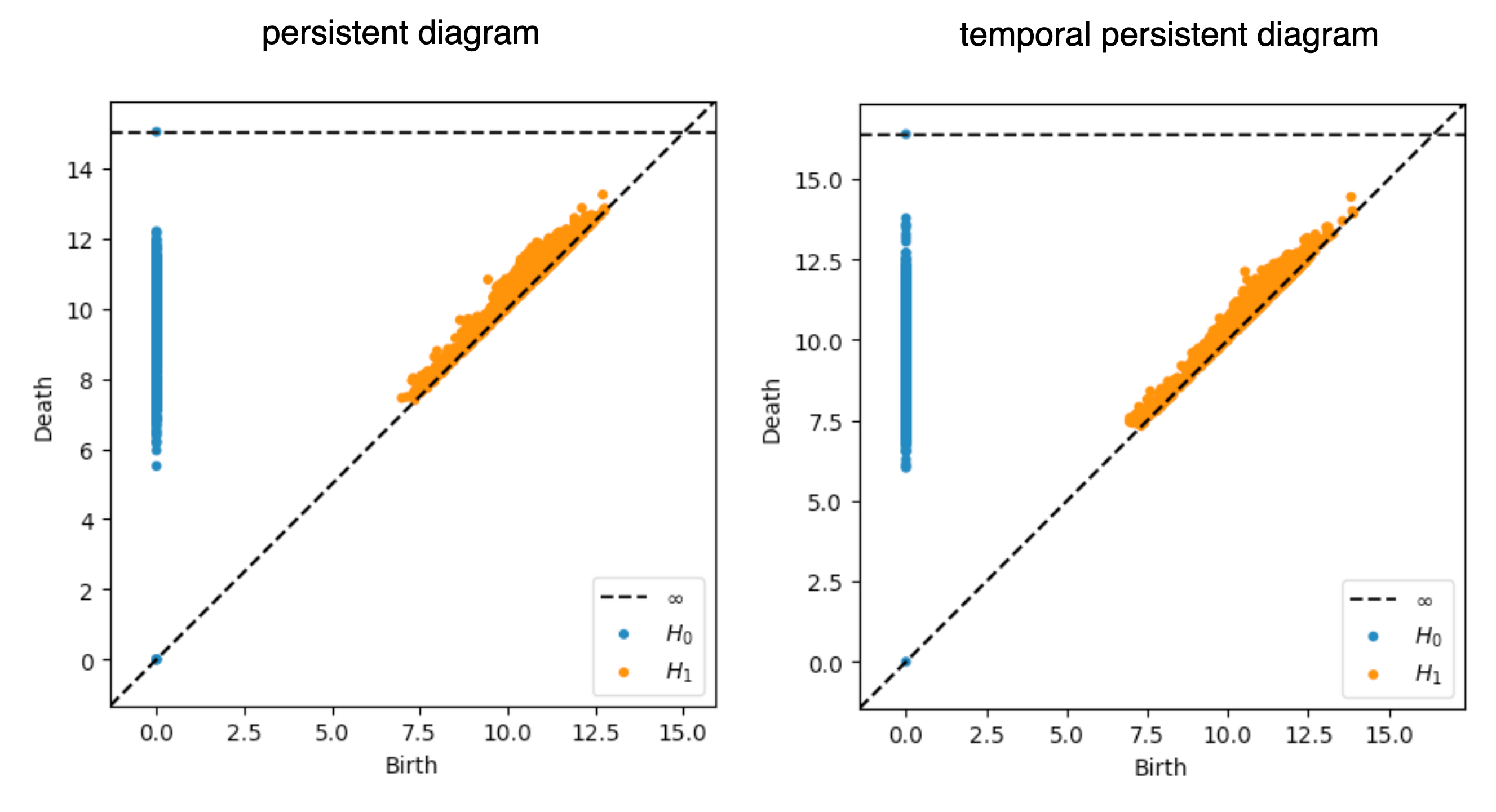} \hfill
\caption{\textbf{persistence diagrams}. The persistence diagrams computed from the persistent homology and temporal persistent homology is shown here. The x-axis corresponds to the birth of all the persistent modules arises in the filtration process, and the y-axis corresponds to their death.}
\label{fig:tph}

\end{figure}

\begin{figure}[tb]

\centering
\includegraphics[width=\linewidth]{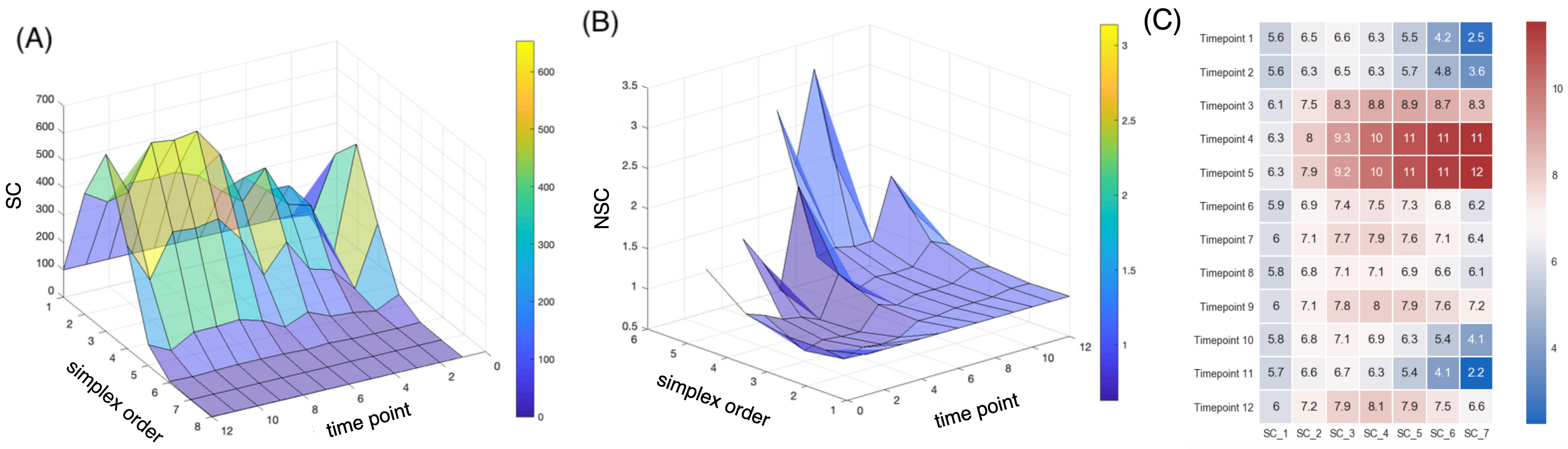}
\vspace{0.1em}
\caption{\textbf{Simplicial dynamics across developmental stages.} (A) The number of $n$-simplices is computed from the sampled data points in each time points. (B) The normalized $n$-simplicial complexity, i.e. the normalized number of $n$-simplices, is computed as the ratio of the number of the $n$-th order simplicial complexes from the data over the number of those from the null models. The normalized simplicial complexity of higher order appears to be well above 1 in certain developmental stages with a distinctive separation between the 5th and 6th time points. (C) The heatmap of the normalized $n$-simplicial complexity across the time points supports the observation. 
}
\label{fig2}

\end{figure}

\begin{figure}[tb]

\centering
\includegraphics[width=0.48\linewidth]{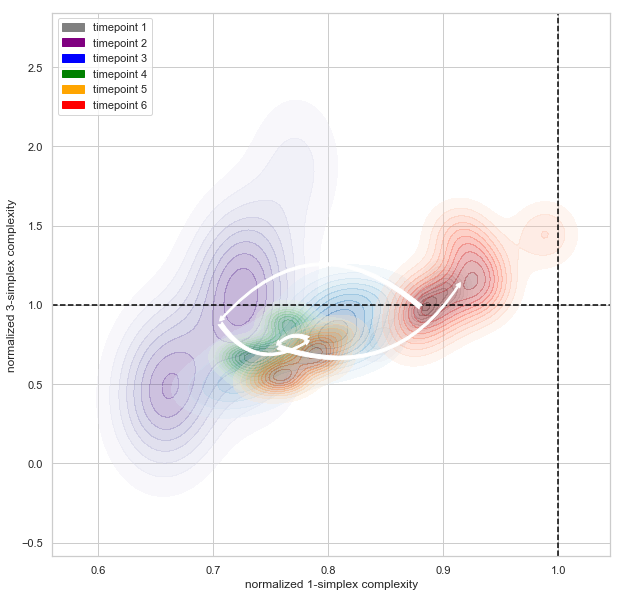}
\includegraphics[width=0.48\linewidth]{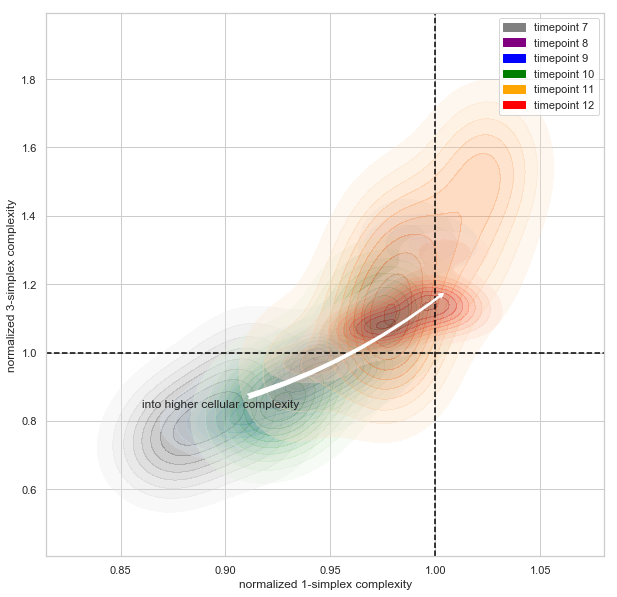}
\vspace{1em}
\caption{\textbf{Simplicial dynamics across developmental stages.} To investigate the tradeoff between the higher-order and the lower-order simplicial complexity in the developmental stages, the normalized 3-simplicial complexity is mapped against the normalized 1-simplicial complexity. The color indicates different time points. The arrow indicates the transition between the centroids in each groups of time points. A transition of lower-order and higher-order normalized cell complexity is marked with the white trajectories across sequential time points.}
\label{fig2C}

\end{figure}

%
%

\begin{figure}[tb]

\includegraphics[width=1\linewidth]{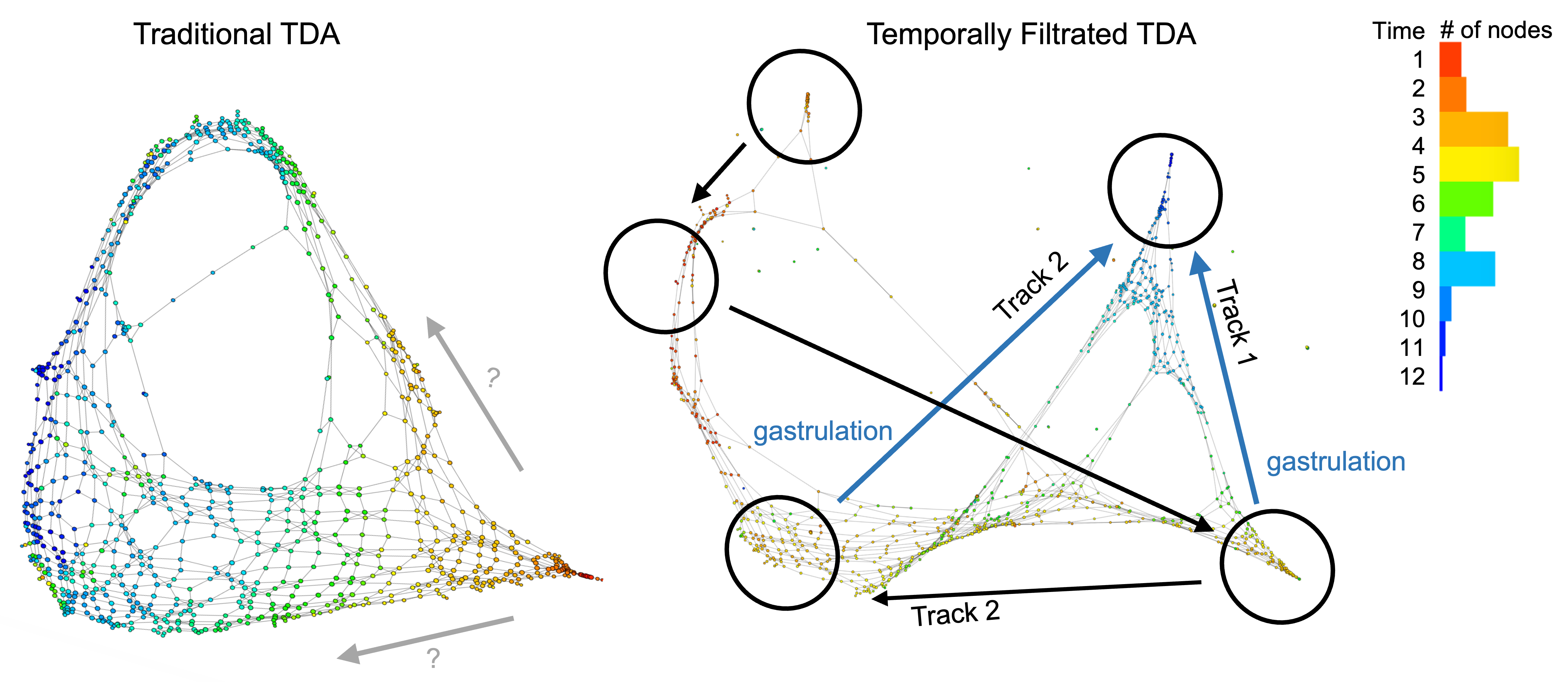}
\vspace{0.1em}
\caption{\textbf{Temporal filtration identifies the critical stage of cellular complexity change.} The color indicates the time points and each node corresponds to a small cluster of cells collected at the same time points. The conventional TDA mapping (the left panel) identifies a bifuraction structure, but there are spatial locations that has a mixture of clusters that belong to non-consecutive time points. This makes the identifications of a developmental pathway challenging. When applying the temporal filtration (the right panel), the mapping identifies a clean separation of two tracks, or two subpopulations of cells that evolves in the gastrulation stage, matching the observation in our summary statistics from the algebraic topology.}
\label{fig3C}

\end{figure}

\begin{figure}[tbh]

\includegraphics[width=1\linewidth]{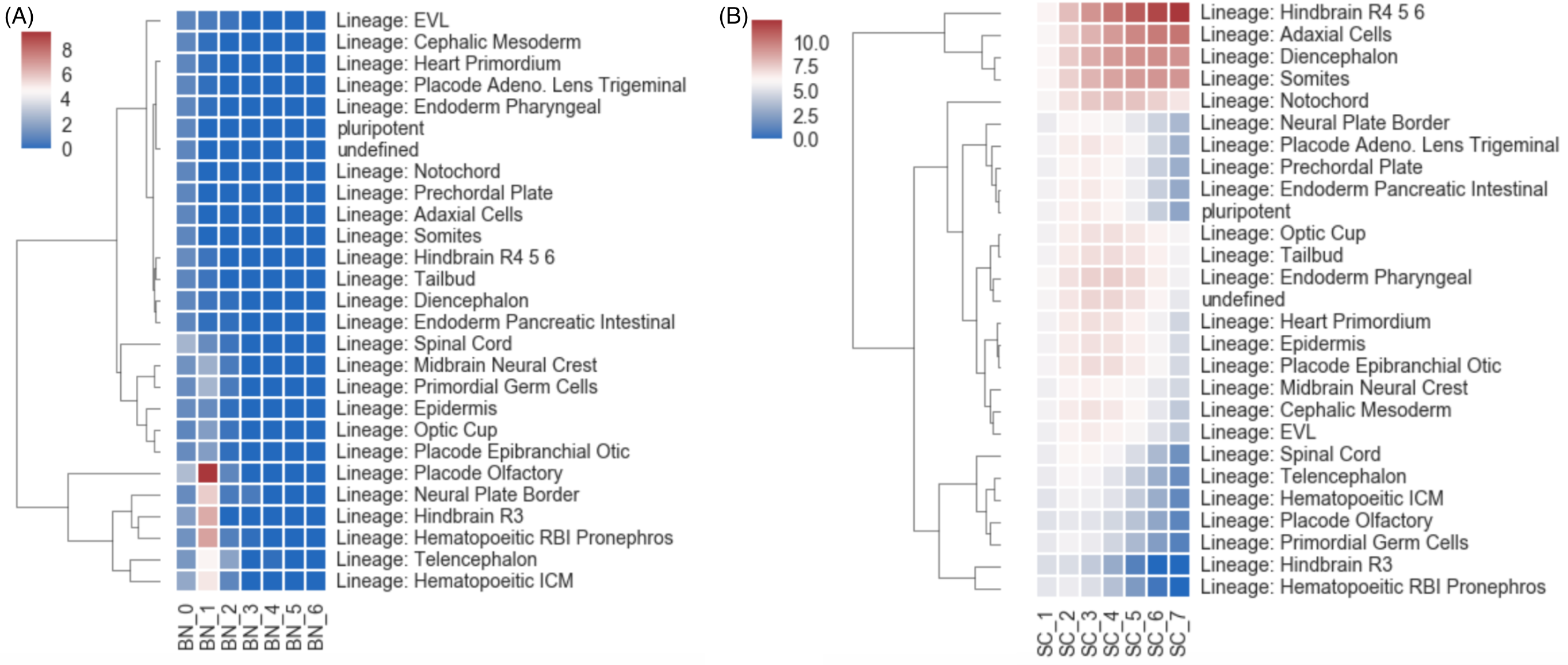}
\vspace{0.1em}
\caption{\textbf{Cell lineage tracing with the simplicial statistics.} In this analysis, the hierarchical clustering is performed on the summary statistics of transcriptomic data of different cell types. (A) The heatmap and clustering result using the Betti numbers as the clustering features. (B) The heatmap and clustering result using the normalized simplicial complexity as the features for the hierarchical clustering.}
\label{fig:lineage}

\end{figure}

\begin{figure}[tb]
\centering
\includegraphics[width=\linewidth]{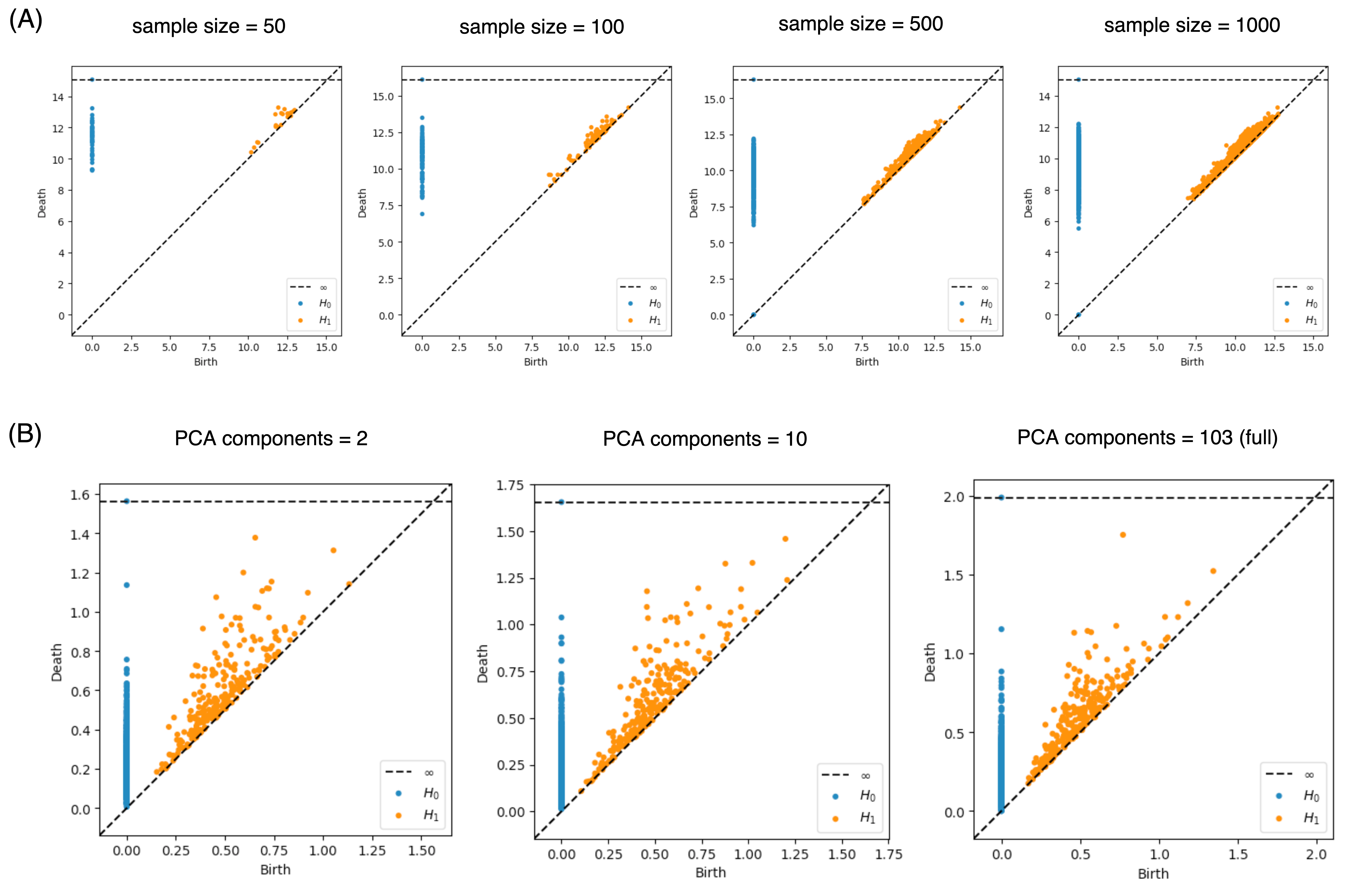} 
\caption{\textbf{Sensitivity analysis of tTDA.} Demonstrated is the effect of witness sampling and PCA dimension reduction to the persistent homology and simplicial analysis. (A) persistence diagrams of the dataset when sampling 50, 100, 500 and 1,000 data points. (B) persistence diagrams when choosing the first 2, 10 and 103 (all dimensions) principal components. 
}
\label{fig:sensitivity1}
\end{figure}

\begin{figure}[tb]
\centering
\includegraphics[width=\linewidth]{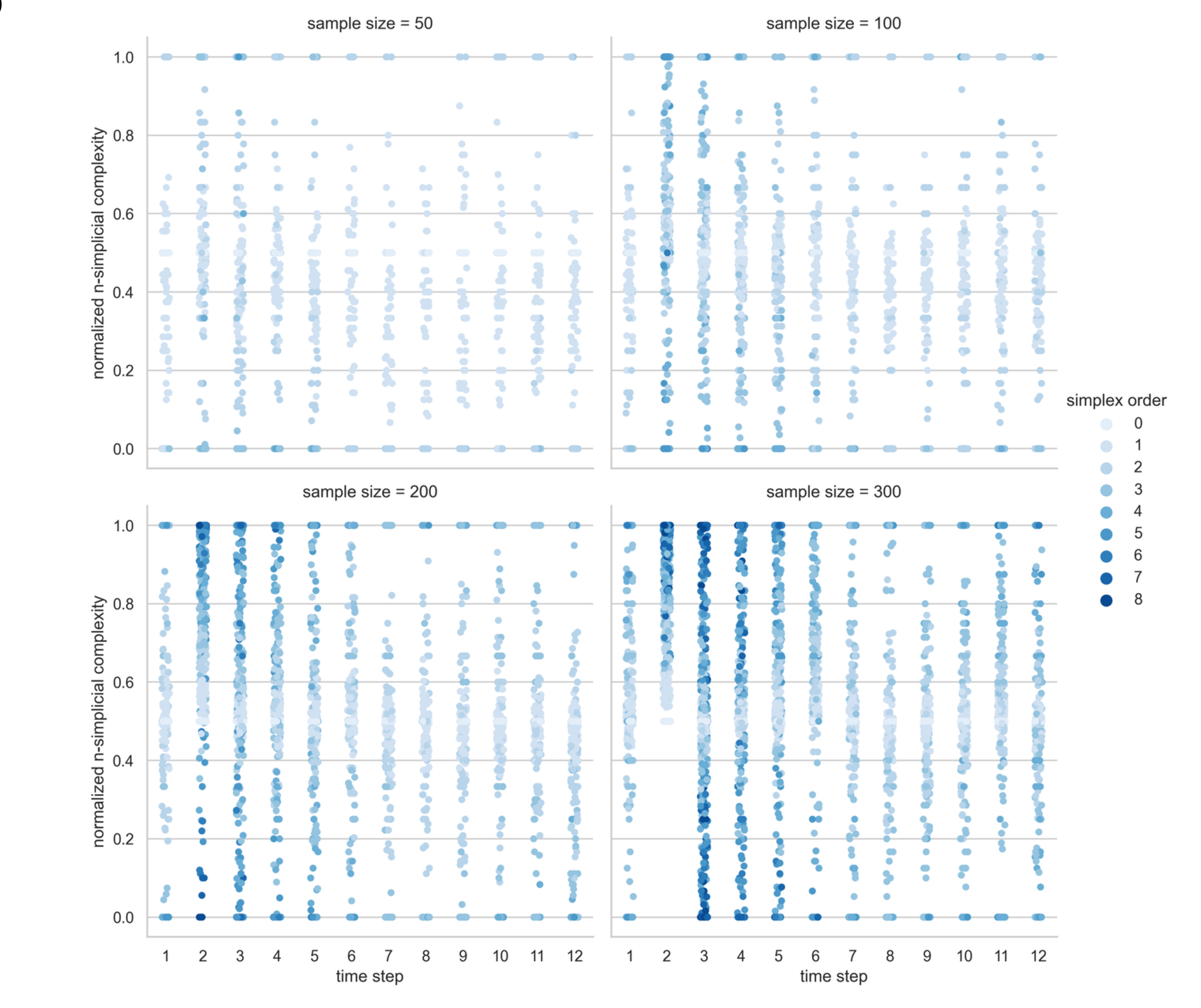} 
\caption{\textbf{Sensitivity analysis of tTDA.} Demonstrated is the effect of witness sampling and PCA dimension reduction to the persistent homology and simplicial analysis. The overall distribution of the normalized simplicial complexity doesn't change much when the sampling size at each time step arises from 10, 100, 200 to 300.
}
\label{fig:sensitivity2}
\end{figure}

We benchmark the scTSA method on the zebrafish single-cell RNA sequencing data with 38,731 cells, 25 cell types, over 12 time steps \cite{farrell2018single}. The dataset studies the embryogenesis, which is the process where the cells gradually differentiate into distinct fates through stages of transcriptional changes. The goal of this study is to facilitate a comprehensive identification of cell types with their time stamps in order to reconstruct their developmental trajectories (e.g. transcriptional states, branch points and asyncrhony). As the gene expression profiles they obtained from the vertebrate embryo are time stamped spanning 3.3–12 hours post-fertilization (hpf), it provides a perfect testbeds for time-series analysis to reconstruct transcriptional trajectories and characterize time-dependent development properties. 

We process the scRNA sequencing data into entries of 103 dimensions corresponding to the expression levels of 103 significant genes (which we select using scTDA). We then standardize the features by removing the mean and scaling to unit variance. Before we perform the persistent homology, we first embed the dataset into low-dimensional space using dimension reduction. In Fig. \ref{fig:dr}, we embed the data using Principal Component Analysis (PCA), t-distributed Stochastic Neighbor Embedding (TSNE) \cite{van2008visualizing} and Uniform Manifold Approximation and Projection (UMAP) \cite{mcinnes2018umap} and color them by time steps. We observe that they all demonstrate a temporal gradient. 

We perform the persistent homology and temporal persistent homology on the scaled dataset. The $\tau$ in this case is set to be 1 (meaning that we only care about the linkage formed between consecutive time points). Fig. \ref{fig:tph} compares the persistence diagram for the two approaches. From the persistence diagrams, the persistent features detected by the persistent homology are not noticeably different from the temporal persistent homology. While not a focus of this work, further study using downstream machine learning tasks can potentially pinpoint the benefits of these temporal persistent features.

For the simplicial analysis, we first group the data by their time steps. The data collected at the 12 time steps are highly imbalanced: 1 (2225 data points), 2 (200),  3 (1158),  4 (1467),  5 (5716), 6 (1026), 7 (4101), 8 (6178), 9 (5442), 10 (5200), 11 (1614) and 12 (4404). For each time points, we perform a witness sampling of 200 data points, since it is the lowest number of samples among all time points. We identifies the simplicial complexity to vary over the time, suggesting a potential better summary statistic with better distinction among time steps (Fig. \ref{fig2}). The normalized simplicial complexity (computed as the ratio of the number of simplicial complexes discovered within the data over the number of those discovered within the null model) suggests an abundance of high-dimensional simplices over the null models. The existence of a significant number of high-dimensional simplices is observed for the first time in the single cell level. In all time points, the number of simplices of dimensions larger than 1 in the null model was far smaller than those found in the actual data. In addition, we observe this relative differences between what we discover in null models and the actual data increase drastically when the dimensions are higher. Furthermore, the number of low-dimensional simplices (up to dimension 3) of the data appears to be equal or smaller than the null models (with normalized complexity less than 1), suggesting a possible transfer from lower order clique structure to a higher-order structure. 

In order to investigate the tradeoff between the higher-order and the lower-order simplicial complexity in the developmental stages, we map the normalized 3-simplicial complexity against the normalized 1-simplicial complexity. Fig. \ref{fig2C} suggests an a gradually increasing higher-order complexity starting from the 5th time point, and an overall below-null lower-order complexity in a monotonically increasing direction since the 2nd time point. Comparing to the null model, the presence of a much larger numbers of cliques across a range of dimensions in the single cell data suggests that the connectivity between these cells might be highly organized into numerous fundamental building blocks (e.g. proto-cell types) with increasing complexity.
These two figures both suggest that the gastrulation stage (time point 5 to 6) is a very critical stage in vertebrate development, matching the established understanding in the developmental biology that it is a process where the embryo begins the differentiation process to develop into different cell lineages \cite{gilbert2016developmental}. Before gastrulation, the embryo is a continuous epithelial sheet of cells. After the gastrulation stage, organogensis starts where individual organs develop within the newly formed germ layers.

This observation is further supported by the visualization of topological data analysis mapping. Fig. \ref{fig3C} compares the network visualizations with and without the temporal filtration. We observe that, when color-labelled with the time points, the conventional topological data analysis outlines a progression of cellular development, but there are many subsequent time points in the middle of earlier timesteps. For instance, we see there are many dark blue nodes from the 11th or 12th time points in the middle of web where the majority of the nodes are earlier stages from the 5th to 7th. When using the temporal filtration (with $\tau$ set to be just 1 time step), we observe that the network has much more skeleton and branches, where each branching nodes consist only of points of the same time stamp. The gastrulation stage, which happens between the 5th and 6th time points, appears to belong to two separate tracks, supporting the hypothesis that after the notochord and prechordal plate territories become transcriptionally distinct, the gastrulation process refines the boundary between the two cellular populations \cite{farrell2018single}.

These filtrated simplicial architectures may also offer insights in cell lineage tracing. We perform the hierarchical clustering of the summary statistics computed from the transcriptome data of different cell types. We compare the result using the proposed normalized simplicial complexity versus the one using the Betti numbers (which is more conventionally used in many downstream topological data analyses). As shown in Fig. \ref{fig:lineage}, the normalized simplicial complexity offers a more reasonable clustering performance as a more distinctive summary statistics than the Betti numbers by themselves.

\subsubsection{Future directions of tTDA}
\label{sec:conclusion}

What is cellular complexity and what does the higher-order complexity mean? As an inquiry to this question, we explore the possibility of introducing the mathematical notion of higher-order simplicial complexes into analyzing distance-based single cell data. Benchmarked on a single cell gene expression data with multiple developmental stages, we propose the single-cell Topological Simplicial Analysis, and demonstrate that the simplicial complexity can be a well-defined summary statistic for celluar complexity. 

This investigation provides a scalable, parameter-free\footnote{By ``parameter-free'', we mean that it doesn't have arbitrary hyper-parameters that the users have to set in order to perform the analysis. The parameter $\tau$, instead, is a user-specified parameter that is relevant to the specific application and problem of interest. An analogy to a prediction model would be, the learning rate is an arbitrary hyper-parameter, and the prediction window would be a user-specified parameter relevant to the application.}, expressive and unambiguous mathematical framework to represent the cellular complexity with its underlying structure. Locally, these structures are characterized in terms of the simplicial complexes. Globally, these structures are characterized in terms of the cavities formed by these simplices. Topological cavities are usually formed and then later filled with the additions of new edges (and potentially, nodes). When computing the persistent homology, we perform a filtration process which innately tracks the formation and later filling of topological cavities of different dimensions. The temporal persistent homology characterizes the information of cavities with the lifespan of these topological objects. 
This framework reveals an intricate topology of cellular similarity which includes a vast number of cliques of cells and of the cavities that bind these cliques together. These topological summary statistics that captures the relationships among the high-dimensional cliques uncover the transcriptional differences in the connectivity of cells of different types during the graph reconstructions process.

From the scTSA visualization, we discover, for the first time in any single cell data, an abundant number and variety of higher-order cliques and cavities. Comparing to the control models, the framework measures a much higher number of high-dimensional cliques and cavities in the graph construction filtration process. The critical stage identified by the framework matches the current understanding in the developmental biology. Comparing with the statistics of Betti numbers, the NSC demonstrates better distinctions between time points and cell types. 

Topological data analysis, like many other machine learning methods, have many empirical considerations related to sample sizes and dimensionality selections. 
To demonstrate the sensitivity of persistent homology to sampling size and reduced dimensions, we perform the following experiment. We use the full dimensions of the standard scaled dataset, vary the sampling size from 50, 100, 500 to 1,000 data points and compute their persistence diagrams. We then set the sample size to 1,000, vary the PCA dimensions to be the first 2, 10 and 103 (full) dimensions, and compute their persistence diagrams. We observe no clear difference. Then, we perform the simplicial analysis with witness sampling using sample size from 10, 100, 200 to 300. In this case, we observe a slightly higher numbers of higher order simplical complexes, but the overall shape and distinction between the time steps are maintained (Figs. \ref{fig:sensitivity1} and \ref{fig:sensitivity2}). Future study can investigate strategies of increasing the stability of the simplicial analysis to sample size.

In the introduction, we pose some open questions we wish to engage the field to discuss and investigate together, instead of answering them directly in this first work. Here we will briefly share our preliminary take on some of the specific ones:

Why does expression similarity deserve the name of complexity? To clarify, the expression similarity may not be a measure of complexity. However, the temporally connected higher order co-expression structure characterized by similarity can be a useful measure of complexity. If the task requires several agents to co-work together at the same time, or follow a specific sequence of actions by different agents, then it is more complex than a task which only requires a few agents or doesn't need to follow a specific sequence. The notion of similarity is usually related to clustering and thus, the separation of homogeneous groups. To extend on this understanding, the similarity relationships that are further constrained by temporal sequences would relate to functionally separating groups of homogeneous agents, and thus, potentially informative to their interactions. 

Is there a reason to believe gene expression similarity has something to do with interactions rather than reflecting the number of similar cells that happen to be present in the sample? These temporally constrained gene expression similarity can both reflect the number of similar cells that coexist at the same time, but also potentially related to the some level of functional interactions, as discussed above. We wish to leave further investigations on what type of interactions for future work, and welcome discussions and critique in these interpretations.

Finally, there are other potentially applicable questions we can explore: Can we determine developmental stages without physiological features? Can we generate pseudo-time series based on single cell sequencing data? And most importantly, does the vast presence of high-dimensional cliques suggest that the interaction between these cells is organized into fundamental building blocks of increasing complexity? Through this inquiry with topological simplicial analysis, we can form such hypothesis that the cells organize themselves into high-dimensional cliques for certain functional or developmental reasons. Further research includes developing mechanistic theories behind the emergence of such high-dimensional cellular cliques and experimentally testing these hypotheses to reveal the missing link between functions and cellular complexity.


Here our work describes a novel scalable and unsupervised machine learning\footnote{By ``machine learning'', we refer to the general goal of building a model that learns from the data. The topological data analysis is a class of unsupervised learning method. The topological features identified from the process can be further applied to downstream machine learning tasks, such as the hierarchical clustering of cellular lineage.} method that facilitate the understanding and solutions three main technical challenges (TC) in bioinformatics:

\textit{(\textbf{TC1}) A lack of time-series analytical methods in quantifying the underlying temporal skeleton within the manifold of the similarities among data points.}
In persistent homology and mapper visualization, our temporal filtration uses a user-specified time separation parameter $\tau$, which can be either discrete (consecutive time steps) or continuous (by a time delay quantity). This enables the computation of persistent components that is computed only on data points that are temporally proximal, and thus, provides a temporal skeleton representation. In the simplicial analysis, we can group the data points by time steps, and compute the normalized simplicial complexity as a quantity to inform the ecology of cells in the transcriptomic feature space.

\textit{(\textbf{TC2}) A lack of scalable computational methods to characterize single-cell sequence signals in the scale of 10k+ data points, while the single-cell sequencing data are dominating the bioinformatics in recent few years.}
The usage of witness sampling and dimension reduction enable the computation of persistent homology to large numbers of high-dimensional data points. Sampling is also a required step to compare the topological features in groups of data points with different count numbers. The normalization against null distribution of the data sample partly corrects for the amplification effect of higher-order topological quantities. The usage of dimension reduction techniques such as PCA help with data management and computation without a significant loss of performance.

\textit{(\textbf{TC3}) A lack of insight and interpretation that connects the mathematical language of algebraic topology to the physical references to the biological phenomena.}
In the introduction and discussion, we initiate the discussion of the interpretations of the topological properties. More specifically, we point out how the temporally directed relationships among data points can be related to functionally separating groups of homogeneous agents in the feature space, and thus,  potentially informative to their interactions. With our temporal-directed treatment of filtration or grouping techniques, our study is a small but first step to use topological data analysis as not only a descriptor tool for static manifold, but also, in the future, a discovery tool of dynamic or mechanistic components. Our goal in this work is not to fully answer the question of interpreting the biological insights topological properties, but to further motivate and facilitate our understanding to the question. As more techniques of topological data analysis are applying to biological problems, we wish to encourage the discussion and critique from the biology and machine learning research community. 

\subsection{Summary of tTDA and scTSA}

In summary, we propose a new family of filtrations for longitudinal time-series multidimensional data along with auxiliary data analysis tools. We demonstrate our application to the temporal inference problems using a set of time-resolved gene expression data. The key technique, called \textit{temporal filtration}, substitutes a conjunctive distance and time threshold for the conventional distance threshold for point cloud data augmented with time stamps. In addition to persistent homology, mapper constructions, and the use of witness sampling with this technique, an original set of standardized summary statistics, the \textit{normalized simplicial complexities}, are proposed. These techniques are used to conduct an exploratory analysis of zebrafish embryonic development through the lens of longitudinal single-cell RNA sequencing data. The applications showcase clear improvements in the interpretability of visualizations compared with a cross-sectional approach and suggest that the key events in the evolution of a biological system can be more effectively detected using normalized simplicial complexity than using Betti numbers. Other than the biological application in single-cell genomics, the time-series problem is especially a topic that is applicable beyond the application proposed in our work, and thus a major interest in the unsupervised machine learning communities dealing with high-dimensional time series signals.

The techniques introduced in this chapter have implications far beyond neuroscience and genomics. By providing tools to analyze the topological structure of time-evolving systems, these methods open new possibilities in fields ranging from climate science to financial modeling. Future work should explore how these approaches can be adapted to diverse types of time series data across scientific disciplines.



\clearpage
\phantomsection
\addcontentsline{toc}{chapter}{Conclusion}

\begin{center}
\pagebreak
\vspace*{5\baselineskip}
\textbf{\large Conclusion}
\end{center}


As we conclude our journey through the intricate landscape of neural representations, we find ourselves standing at the edge of a new frontier in computational neuroscience. This thesis has not merely added tools to our analytical arsenal, but also has potentially reshaped how we perceive and probe the inner workings of the brain and artificial neural networks.

Our exploration began with a simple yet profound question: How can we better understand the language of neurons? Classical Representational Similarity Analysis (RSA) provided a window into this neural dialogue, but it was as if we were listening to a symphony while hearing only a single instrument. The introduction of Topological RSA (tRSA) has allowed us to hear the full orchestra, revealing harmonies and counterpoints in neural activity that were previously hidden from view.

This thesis introduced tRSA as a novel framework that combines geometric and topological properties to characterize neural representations. tRSA applies nonlinear monotonic transforms to representational dissimilarities, emphasizing the topology while retaining critical geometric information. The resulting geo-topological matrices enable more robust model comparisons, as demonstrated through analyses of neural recordings and simulations.

The power of tRSA lies not just in its ability to capture the shape of neural representations, but in its robustness to the noise and individual variations that often confound neuroscientific research. As we applied tRSA to both biological brains and artificial networks, we uncovered a striking parallel: the topology of representations often carries the essence of computation, transcending the specific geometry of neural activity patterns. This insight opens new avenues for comparing and contrasting biological and artificial intelligence, potentially bridging the gap between neuroscience and machine learning in ways we have only begun to explore.

A key finding is that tRSA can identify the unique computational signatures of different brain regions and neural network layers as accurately as traditional RSA, while compressing unnecessary variations due to noise and individual idiosyncrasies. The topology-sensitive statistics discard excessive global geometry while preserving critical local geometry that determines discriminability and neighborhood structure. This confirms the hypothesis that variations in large representational distances may reflect arbitrary differences, rather than computation.

But the brain is not a static entity, and our journey led us to confront the dynamic nature of neural computations. By developing methods to track the evolution of representational geometries over time, we have begun to unveil the choreography of thought itself. Our analysis of visual object recognition revealed distinct processing stages, each with its own representational signature. This dynamic view of neural activity promises to revolutionize our understanding of cognitive processes, from perception to decision-making and beyond.

This is indeed what we discovered in the suite of temporal analytical methods introduced in this thesis. Aligning time-resolved representational geometries in multi-dimensional scaling space (pMDS) provided illuminating visualizations of temporal processing stages in brains. Temporal filtration of developing cell populations revealed their maturation trajectories. Topological simplicial analysis (TSA) characterized cell population complexity across development. 

As we delved deeper into the temporal dimension, we found ourselves at the intersection of neuroscience and developmental biology. The application of topological time series analysis to related fields such as single-cell genomics data revealed hidden patterns in cellular differentiation, offering a new lens through which to view the unfolding drama of development. This unexpected convergence of neuroscience and genomics underscores a broader truth: the methods we develop to understand the brain often have profound implications far beyond neuroscience.

Indeed, our quest into developing an Adaptive Geo-Topological Dependence Measure (AGTDM) hints at the broader applicability of these ideas. Along with the same motivation behind the tRSA framework, AGTDM enables a statistically principled test to detect complex dependencies between multivariate variables. This dependence measure adapts to the relationship's structure, maximizing its sensitivity. Exploring this family of dependence measure revealed its robustness and potential interpretability across varying univariate and multivariate relationships. While still in its infancy, this approach suggests new ways of detecting and characterizing complex relationships in high-dimensional data across various scientific disciplines. 

Together, these advances demonstrate the power of topological techniques to advance representational similarity analysis and provide deeper insights into neural computation in both biological and artificial systems. The findings contribute to the theoretical understanding of how brains solve challenging computational problems. Combining the three aspect, we introduced a broader framework for inferential comparison of complex networks using topological and geometric summary statistics. The adaptive dependence measure offers a robust statistical test for multivariate relationships, demonstrating the generalizability of tRSA.

While this thesis focuses on the topological and geometric aspects of neural representations, an exciting avenue for future research lies in connecting these insights with information-theoretical approaches. Exploring how the topological structures we've uncovered relate to information flow and efficiency in neural networks could provide a more comprehensive understanding of neural computation \cite{lin2019neural,lin2021regularity}. This intersection of topology and information theory might reveal new principles of how the brain optimizes its representational structures for efficient information processing and could inspire novel architectures in artificial neural networks.

As we look to the future, the horizons before us are both exciting and daunting. The methods developed in this thesis provide a foundation for ambitious new directions in neuroscience and beyond:

\begin{itemize}
    \item \textit{Decoding the Neural Code}: Armed with tRSA and our temporal analysis techniques, we are now poised to tackle one of the grandest challenges in neuroscience – cracking the neural code. By applying these methods to large-scale recordings across multiple brain areas, we may finally begin to decipher how information is encoded, transformed, and transmitted throughout the brain.
    \item \textit{Bridging Scales in Neuroscience}: Our work on temporal topological analysis offers a unique opportunity to bridge the gap between cellular and systems neuroscience. By applying these techniques to data ranging from single-neuron recordings to whole-brain imaging, we could uncover principles of neural computation that span multiple scales of brain organization.
    \item \textit{Advancing Artificial Intelligence}: The insights gained from tRSA could inspire new architectures for artificial neural networks that more closely mimic the topological properties of biological brains. This could lead to AI systems that are more robust, efficient, and capable of the kind of flexible cognition that characterizes human intelligence.
    \item \textit{Personalized Neuromedicine}: The ability to track representational dynamics over time opens new possibilities for personalized medicine in neurology and psychiatry. By characterizing individual trajectories of neural representations, we might be able to develop early diagnosis methods and tailored treatments for neurological and psychiatric disorders.
    \item \textit{Unifying Theories of Brain Functions}: Our topological approach to neural representations provides a new language for describing brain function. This could serve as a foundation for developing unifying theories that bridge currently disparate areas of neuroscience, from sensory processing to high-level cognition.
\end{itemize}

The work presented in this thesis has laid the foundation for several exciting research directions. As I embark on my career as a tenure-track professor, I look forward to expanding on these ideas, particularly through a novel concept I am introducing: \textit{Representational Pathology}. This new field aims to study the representational space in individuals with neurological or psychiatric disorders, potentially revolutionizing our understanding of brain dysfunction.

Representational Pathology, a term I am coining here, goes beyond merely applying RSA in clinical settings. It leverages the topological insights developed in this thesis to examine not just the geometry, but the topology of neural representations in diseased brains, as opposed to healthy ones. This approach emphasizes the study of how representational spaces evolve over time in pathological conditions, providing a dynamic view of brain dysfunction that integrates insights across multiple scales.

In schizophrenia, for instance, Representational Pathology might reveal a fragmentation of the topological structure in conceptual spaces, reflecting the disorganized thinking characteristic of the disorder. We might observe disconnected clusters in the representational space where a healthy brain would show a continuous manifold, potentially explaining the jumbled associations and loose conceptual boundaries often seen in schizophrenic thought.

For neurodegenerative diseases like Alzheimer's, this approach could uncover a progressive collapse of the dimensionality in memory representational spaces. As the disease advances, the rich, high-dimensional topology of memory representations might simplify and contract, correlating with the gradual loss of nuanced and detailed memories.

In autism spectrum disorders, Representational Pathology might reveal an altered topology in social cognition spaces. We might find isolated, high-dimensional representations of specific interests, contrasting with a simplified, low-dimensional structure for social concepts, reflecting the characteristic pattern of restricted interests and social difficulties.

By providing this unique perspective, Representational Pathology could uncover new diagnostic markers and track disease progression in ways that conventional analyses might miss. It aims to develop predictive models of disease progression and treatment response based on changes in representational topology, emphasizing individual trajectories in representational space.

This innovative approach bridges the gap between computational neuroscience and clinical practice by providing a quantitative framework for understanding cognitive deficits based on topological features of neural representations. It could inform the development of targeted interventions and personalized treatment strategies that are more precisely tailored to individual patients' neural dynamics.

My research program in neuro-AI and computational psychiatry aims to establish this new field, leveraging the power of topological analyses to decode the neural bases of mental illness and pave the way for more effective treatments. By introducing Representational Pathology, we open a new frontier in neuroscience that combines the theoretical insights of tRSA with the practical needs of clinical neurology and psychiatry, exemplifying how the methodologies developed in this thesis can have far-reaching implications in critical areas of medical research and practice.

In conclusion, this thesis integrates mathematical tools from algebraic topology to analyze the temporal dynamics and latent structure of neural representations, revealing a landscape richer and more intricate than we ever imagined. By advancing RSA with topological methods, we've developed new maps and tools for navigating the complexities of brain function. This work underscores the importance of interdisciplinary thinking in neuroscience, demonstrating the profound impact that novel mathematical approaches can have on our understanding of cognition.

As we move forward, we carry with us the excitement of discovery and the humbling realization of how much remains to be explored. The brain, in all its complexity and beauty, continues to inspire and challenge us. The journey through the topology of neural representations is just beginning, promising further adventures and potential breakthroughs in our quest to understand the most complex object in the known universe: the human brain.



\clearpage
\phantomsection 
\titleformat{\chapter}[display]
{\normalfont\bfseries\filcenter}{}{0pt}{\large\bfseries\filcenter{#1}}  
\titlespacing*{\chapter}
  {0pt}{0pt}{30pt}

\begin{singlespace}  
	\setlength\bibitemsep{\baselineskip}  
	\addcontentsline{toc}{chapter}{References}  
	\printbibliography[title={References}]
\end{singlespace}


\titleformat{\chapter}[display]
{\normalfont\bfseries\filcenter}{}{0pt}{\large\chaptertitlename\ \large\thechapter : \large\bfseries\filcenter{#1}}  
\titlespacing*{\chapter}
  {0pt}{0pt}{30pt}	
  
\titleformat{\section}{\normalfont\bfseries}{\thesection}{1em}{#1}

\titleformat{\subsection}{\normalfont}{\thesubsection}{0em}{\hspace{1em}#1}

\begin{appendices}

\addtocontents{toc}{\protect\renewcommand{\protect\cftchappresnum}{\appendixname\space}}
\addtocontents{toc}{\protect\renewcommand{\protect\cftchapnumwidth}{6em}}


\chapter{Reproducibility and Open-Source Repositories}

In the spirit of open science and to facilitate reproducibility, the code and datasets supporting the experimental results presented in this thesis are openly available. This ensures that other researchers can validate the findings, build upon the work, and use the resources for related research endeavors. Below are the links to the GitHub repositories where the code and relevant data for this thesis can be accessed:

\begin{itemize}
    \item \href{https://github.com/doerlbh/TopologicalRSA}{\texttt{https://github.com/doerlbh/TopologicalRSA}}. \fullcite{lin2023topology}


    \item \href{https://github.com/doerlbh/scTSA}{\texttt{https://github.com/doerlbh/scTSA}}. \fullcite{ttda}; \fullcite{bibm}.

    \item \href{https://github.com/rsagroup/rsatoolbox/}{\texttt{https://github.com/rsagroup/rsatoolbox}}. \fullcite{van2023rsa}; \fullcite{lin2019visualizing}.


\end{itemize}

Researchers are encouraged to use these resources in compliance with the licenses specified in the repositories and cite the respective publications when utilizing these tools in their work.



\end{appendices}

\end{document}